\documentclass[prd,12pt,reprint, twocolumn,noshowpacs,nofootinbib,amsmath,amssymb,superscriptaddress,preprintnumbers]{revtex4-1}
\usepackage{graphicx,color}
\usepackage{tikz-feynman} 
\usepackage{caption}
\usepackage{amssymb,amsmath,mathrsfs}
\usepackage{cancel}
\usepackage{subfig}
\usepackage{hyperref}
\usepackage{verbatim,slashed}
\usepackage{xspace}
\usepackage{afterpage}
\newcommand{\bp}{{\bf p}}
\newcommand{\bk}{{\bf k}}
\newcommand{\bq}{{\bf q}}
\newcommand{\beq}{\begin{equation}}
\newcommand{\eeq}{\end{equation}}
\newcommand{\be}{\begin{eqnarray}}
\newcommand{\ee}{\end{eqnarray}}

\def\lsim{\mathrel{\rlap{\lower4pt\hbox{\hskip1pt$\sim$}}
    \raise1pt\hbox{$<$}}}
\def\gsim{\mathrel{\rlap{\lower4pt\hbox{\hskip1pt$\sim$}}
    \raise1pt\hbox{$>$}}} 

\begin{document}
\title{Freeze-in Leptogenesis via Dark-Matter Oscillations}
\author{Justin Berman}
\email{jdhb@umich.edu}
\affiliation{Department of Physics, Williams College, Williamstown, MA 01267, USA}
\author{Brian Shuve}
\email{bshuve@g.hmc.edu}
\affiliation{Harvey Mudd College, 301 Platt Blvd., Claremont, CA 91711, USA}
\author{David Tucker-Smith}
\email{dtuckers@williams.edu}
\affiliation{Department of Physics, Williams College, Williamstown, MA 01267, USA}

\date{\today}

\begin{abstract}
We study the cosmology and phenomenology of freeze-in baryogenesis via dark-matter oscillations, taking the dark matter to couple to Standard Model leptons. 
We investigate viable models both with and without a $Z_2$ symmetry under which all new fields are charged. Lepton flavor effects are important for leptogenesis in these models, and we identify scenarios in which the baryon asymmetry is parametrically distinct from and enhanced relative to leptogenesis from sterile neutrino oscillations. The models we study predict the existence of new, electroweak-charged fields, and can be tested by a combination of collider searches, structure-formation studies, X-ray observations, and terrestrial low-energy tests.

\end{abstract}
\maketitle

\section{Introduction}\label{sec:intro}

The nature of dark matter (DM) and the origin of the baryon asymmetry are two of the most important open questions in particle physics.  In this paper we study an extension of the Standard Model (SM) that simultaneously addresses both questions.   We consider a  model of freeze-in DM \cite{McDonald:2001vt,Choi:2005vq,Kusenko:2006rh,Petraki:2007gq,Hall:2009bx,Bernal:2017kxu} involving two DM mass eigenstates, the mass splitting between which is sufficiently small that DM production, propagation,  and annihilation  are coherent processes in the early universe.  In this situation, the different propagation phases associated with the two DM mass eigenstates can lead to SM particle/antiparticle asymmetries \cite{Shuve:2020evk}, along lines similar to asymmetry generation via oscillations of right-handed neutrinos in ARS leptogenesis \cite{Akhmedov:1998qx,Asaka:2005pn}.

In the minimal version of the model we consider, the particles beyond the Standard Model (BSM) are a pair of gauge-singlet Majorana fermions $\chi_i$ ($i = 1,2$), which constitute the DM, and a complex scalar $\Phi$ with charges $({\bf 1},{\bf 1},-1)$ under the $SU(3)_c\times SU(2)_w \times U(1)_y$ SM gauge group. 
While $\Phi$ is in equilibrium with the SM due to its gauge interactions, we assume that one or both of the $\chi_i$ are feebly interacting and never come fully into equilibrium.   In two-component notation, the interaction term responsible for DM production is
\be\label{eq:L}
\mathcal{L} \supset -F_{\alpha i} e^c_\alpha \chi_i \Phi + \text{h.c.},
\ee
where $e^c_\alpha$ are left-handed spinors with hypercharge $+1$, representing the $SU(2)_w$-singlet charged leptons of the SM, with flavor index $\alpha$.  We work in the mass basis for both the DM and the SM leptons. Since $\Phi$ carries only hypercharge, it can be as light as $\mathcal{O}(100\,\,\mathrm{GeV})$ depending on its couplings to DM and SM fermions. This simple model is sufficient to obtain both the observed baryon and DM abundances.

As in the ARS mechanism, the production and oscillation of DM can generate asymmetries in individual flavors of SM leptons. Although the leading-order lepton flavor asymmetries sum to zero, flavor-dependent washout of these asymmetries can lead to a non-zero total lepton number asymmetry.
However, our model also features potential sources for a flavor-summed asymmetry that are distinct from the ARS mechanism, arising from the asymmetry that can be stored in $\Phi$.  
In particular, the particle content allows  $\Phi$ to couple to the SM lepton doublets,
\be
{\mathcal L} \supset - \frac{\lambda_{\alpha \beta}}{2} l_\alpha l_\beta \Phi^* + \text{h.c.},
\ee
which can significantly  impact both the asymmetry calculation and  the collider phenomenology.  This interaction violates the $Z_2$ symmetry under which the BSM particles $\chi$ and $\Phi$ are odd, leading to astrophysical signatures of DM decay such as X-ray lines.

Ref.~\cite{Shuve:2020evk} first established the mechanism of freeze-in baryogenesis via DM oscillations by studying a related model,  in which the DM couples to a QCD-charged scalar and SM quarks.  We briefly summarize the main findings of that paper. In the quark-coupled case, flavor mixing  prevents the quarks from having flavor-dependent chemical potentials.  This spoils the ARS mechanism, which depends critically on the presence of flavor-dependent asymmetries to generate a flavor-summed one.  To find a non-zero asymmetry in the minimal realization, with a single QCD-charged scalar and two DM states,  one needs to take into account the flavor-dependence of the quark thermal masses, particularly of the top quark.  The DM must have a substantial coupling to the top quark for the asymmetry to be large enough, and the viable parameter space is tightly constrained, with the scalar having a mass of at most a few TeV.  The parameter space broadens in the presence of an additional source of DM production, for example a second, heavier scalar whose decays leave behind a ``primordial'' coherent DM background.  Even in this second scenario, the lightest scalar is typically not far above the TeV scale for parameters that work for both DM and baryogenesis.  In either scenario, the DM mass is in the $\sim10-1000$ keV range, and the lifetime of the $\sim$TeV-mass scalar typically satisfies $c\tau \gsim$ cm, potentially leading to events with displaced jets plus missing transverse momentum at colliders.  Finally, $Z_2$-violating terms  for  the quark-coupled case are  tightly constrained by proton decay, making it more challenging for those interactions to be relevant for baryogenesis
\footnote{More precisely,  the $Z_2$-violating terms can be relevant in quark-coupled models only for certain matter content and coupling choices.  Ref.~\cite{Shuve:2020evk} focused on the case in which the DM couples to the 
$u^c$ quarks of the SM, ${\mathcal L} \supset -F_{\alpha i} \Phi_i u^c_\alpha \chi_i$.  With this choice, the relevant $Z_2$-violating term, $\Phi^* d^c d^c$ is  $B$-violating (with the DM assigned $B=0$).  If the DM instead couples to $d^c$ (and always defining $\Phi$ to be an $SU(3)_c$  triplet), the $Z_2$-violating couplings $\Phi^* q l$ and $\Phi u^c e^c$   are $B$-conserving but $\Phi^* u^c d^c$ and $\Phi qq$ are not.  Finally, if the DM couples to $q$, the only relevant $Z_2$-violating coupling is $\Phi d^c l$, which is $B$-conserving.}.  

As already mentioned, 
the mechanism of freeze-in leptogenesis has most commonly been discussed in the context of the production and oscillation of right-handed neutrinos in SM neutrino mass models \cite{Akhmedov:1998qx,Asaka:2005pn} (see Ref.~\cite{Drewes:2017zyw} for a review). There has also been a recent proposal in which the freeze in of DM and baryogenesis are simultaneously achieved through the interference of tree and loop processes in the decay of a heavy mediator particle \cite{Goudelis:2021lra}, and there exist earlier proposals of asymmetric DM models in which SM and DM asymmetries are simultaneously generated through out-of-equilibrium scattering \cite{Hall:2010jx,Hook:2011tk,Unwin:2014poa}.
%
\subsection{Generation of flavor-dependent asymmetries}\label{sec:introasym}
%
\begin{figure}
        \includegraphics[width=2.7in]{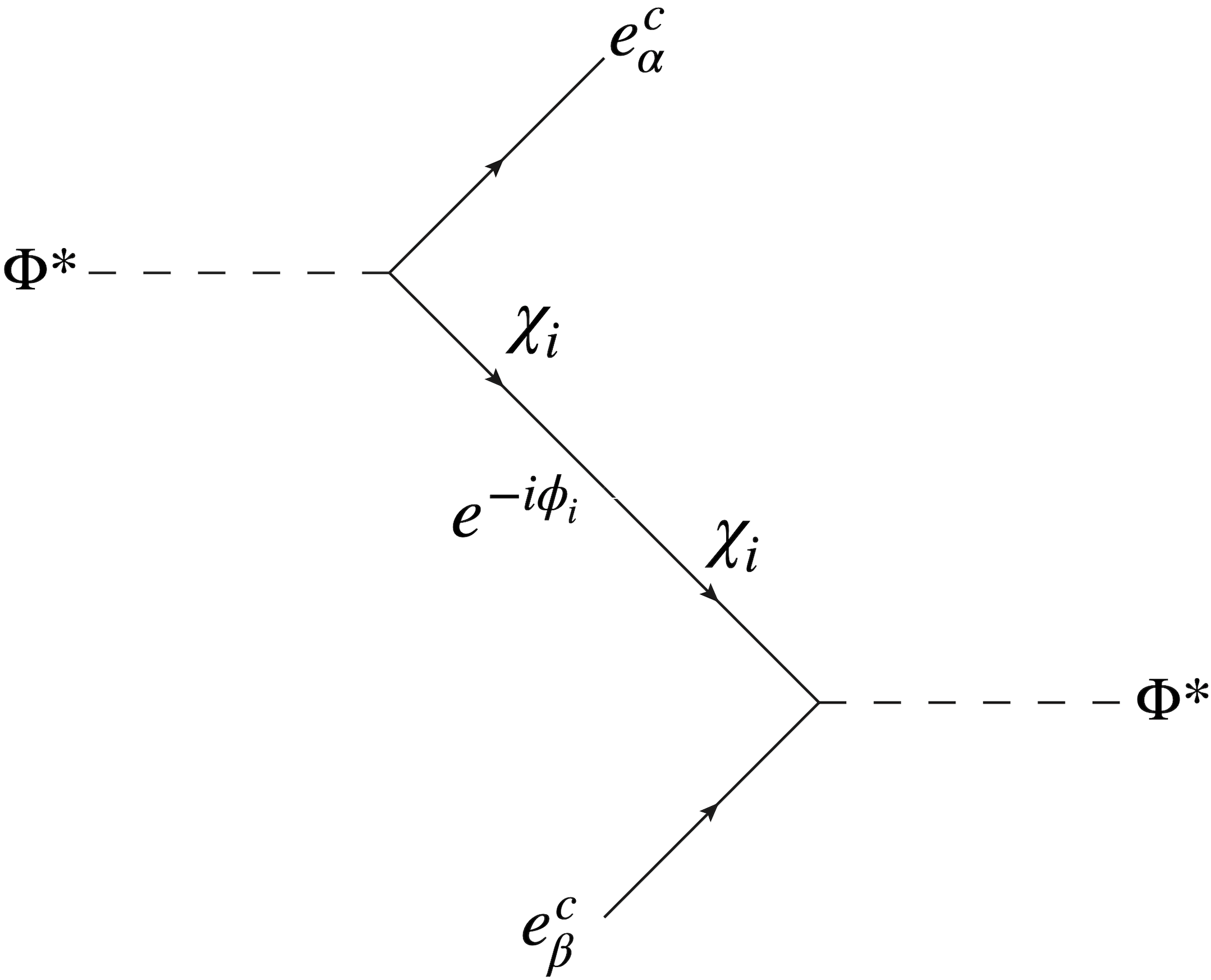}
   \caption{
Feynman diagram illustrating the production of DM ($\chi_i$), its propagation, and subsequent annihilation. First, the scalar $\Phi^*$ decays into $\chi_i$ and the SM particle $e^c_\alpha$; following propagation, the $\chi_i$ field annihilates with another SM field $e^c_\beta$ to reconstitute $\Phi^*$. The net reaction is $e^c_\beta\Phi^* \rightarrow e^c_\alpha\Phi^*$. The process is a coherent sum over $\chi_i$ mass eigenstates since  $\chi_i$ is out of equilibrium. 
    }
    \label{fig:feynman_diagram}
\end{figure}

In this paper, we study three model benchmarks, 
each with a different mechanism that ultimately generates a baryon asymmetry.  However, at leading-order in the DM couplings, the process that generates asymmetries in {\em individual lepton flavors}, depicted in Fig.~\ref{fig:feynman_diagram},  is always essentially  the same.

Consider the DM interaction of Eq.~(\ref{eq:L}), involving a single BSM scalar $\Phi$.   For a generic DM coupling matrix $F_{\alpha i}$, each $\Phi^{(*)}$ decay produces a coherent superposition of DM mass eigenstates that  depends on the flavor of the lepton produced in association.  The DM abundance thus arises at ${\mathcal O}(F^2)$.   Subsequent DM oscillations and inverse $\Phi^{(*)}$ decays generate flavor-dependent $e^c_\alpha/\overline{e^c_\alpha}$ asymmetries at ${\mathcal O}(F^4)$. SM-Yukawa interactions and sphalerons then produce asymmetries in other SM species as well.  Neglecting neutrino masses, the three charges 
\beq\label{eq:Xdef}
X_\alpha \equiv B/3-L_\alpha
\eeq
are conserved in the SM, where $B$ is baryon number and $L_\alpha$ is lepton flavor number.     The final baryon asymmetry is proportional to $X \equiv \sum_\alpha X_\alpha$, the  $B-L$ charge stored in the SM sector, at the sphaleron decoupling temperature $T_\text{ew} \simeq 131.7$ GeV \cite{DOnofrio:2014rug}.  In our bookkeeping, we always define the $X_\alpha$ charges of the BSM particles to be zero, $X_\alpha(\Phi) = X_\alpha(\chi_i) = 0$, even when we find it useful to regard one or both of these particles as carrying lepton number.

For sufficiently small $F_{\alpha i}$, and assuming that the universe starts with $X_\alpha = 0$ after reheating,
a perturbative calculation of the $X_\alpha$ asymmetries is appropriate.   
The leading-order
asymmetries turn out to be proportional to the coupling combination \cite{Akhmedov:1998qx,Asaka:2005pn}
\beq\label{eq:F4coup}
\mathrm{Im}\left[
F_{\alpha 1}  
F_{\alpha 2}^*
\left(F^\dagger F \right)_{12}
\right], 
\eeq
which means that the flavor-summed asymmetry $X$ vanishes at this order,
\beq\label{eq:zeroYBatF4}
\sum_\alpha
\mathrm{Im}\left[
F_{\alpha 1}
F_{\alpha 2}^*
\left(F^\dagger F \right)_{12}
\right] = 
\mathrm{Im}\left[
\left|\left(F^\dagger F \right)_{12} \right|^2
\right] = 0.
\eeq
In the absence of additional model ingredients, then, no baryon asymmetry is generated at $\mathcal{O}(F^4)$.  We now define our three benchmark models, specifying how a final baryon asymmetry arises in each.
%
\subsection{Model benchmarks}\label{sec:introbenchmarks}
%
(1) {\it The Minimal Model:}
Without the need for additional model ingredients, effects of $\mathcal{O}(F^6)$ and higher spoil the cancelation that leads to zero net $X$ charge at $\mathcal{O}(F^4)$ \cite{Akhmedov:1998qx,Asaka:2005pn}.   As in ARS leptogenesis, flavor-dependent washout can deplete the asymmetry in certain lepton flavors more than others, leading to 
an $X$ charge density that is equal to the asymmetry in DM.
We take these effects into account  at the perturbative level in Sec.~\ref{sec:pertMM}, and in Sec.~\ref{sec:MM} we use the network of quantum kinetic equations (QKEs) presented in Appendix~\ref{sec:MIQKEs} to solve for the cosmological evolution of the various flavor asymmetries and identify the viable parameter space for DM and leptogenesis.  

(2) {\it The UVDM Model:} 
The cancelation of the baryon asymmetry at fourth-order in DM couplings assumes that the interaction of Eq.~(\ref{eq:L}) is entirely responsible for DM production and annihilation.  As for the case of DM couplings to QCD-charged states~\cite{Shuve:2020evk}, an additional source of coherent $\chi$ production can strongly enhance the asymmetry \cite{Asaka:2017rdj,Shuve:2020evk}.  There are many possibilities for this additional DM interaction; for concreteness, in Sec.~\ref{sec:UVDM} we follow Ref.~\cite{Shuve:2020evk} by adopting a model with a second, heavier scalar $\Phi_2$, leading two coupling matrices, $F^1_{\alpha i}$ for $\Phi_1$ and $F^2_{\alpha i}$ for $\Phi_2$.  In this scenario the $\mathcal{O}(F^4)$ baryon asymmetry is proportional to 
\beq\label{eq:F4coupUVDM}
\mathrm{Im}\left[
\left({F^1}^\dagger F^1 \right)_{21}
\left({F^2}^\dagger F^2 \right)_{12}
\right],
\eeq
which does not vanish in general.  

(3) {\it The Z2V Model:} 
Even if the interaction of Eq.~(\ref{eq:L}) is the only coupling of the DM to the SM sector, additional $\Phi$ interactions with SM fields can also qualitatively impact the asymmetry calculation, as we explore in Sec.~\ref{sec:Z2V}.  
The $Z_2$-violating case admits two additional renormalizable interaction terms,
\beq\label{eq:LZ2V}
\mathcal{L} \supset
-h_{\alpha i} l_\alpha \chi_i H - \frac{\lambda_{\alpha \beta}}{2} l_\alpha l_\beta \Phi^* + \text{h.c.},
\eeq
where $H$ is the SM Higgs doublet, $l_\alpha$ are the SM lepton doublets, and we assume only a single scalar $\Phi$.  X-ray line constraints on DM decay prevent the neutrino-portal couplings $h_{\alpha i}$ from playing a role in leptogenesis if $\chi$ is taken to be the DM \cite{Asaka:2005an,Asaka:2005pn}. By contrast, the Z2V couplings $\lambda_{\alpha \beta}$ can be large enough to significantly modify the asymmetry calculation while being consistent with all experimental and observational constraints.    These interactions  violate $X_\alpha$, and they shift the $\mathcal{O}(F^4)$ $X_\alpha$ charge densities produced by the DM interactions to produce a baryon asymmetry at $\mathcal{O}(F^4 \lambda^2)$.  Moreover, we will see that the Z2V couplings can be large enough that the baryon asymmetry is dramatically enhanced relative to the Minimal Model.    To be more precise, if all three independent Z2V couplings come into equilibrium, the lepton chemical potentials are driven to be flavor universal, which in turn drives all asymmetries to zero.   If, however, only one or two of the Z2V couplings come into equilibrium, the baryon asymmetry is not washed out and effectively arises at $\mathcal{O}(F^4)$.

%
\subsection{Structure formation constraints}\label{sec:introstructure}
%

With the DM abundance generated at ${\mathcal O}(F^2)$ and the baryon asymmetry arising at ${\mathcal O}(F^4)$ or higher, a general challenge in these models is to produce a large enough baryon asymmetry without overproducing DM.  Because we require the DM energy density to match the observed value, lighter DM means a larger DM number density and larger DM couplings, leading to a larger asymmetry.  The DM/leptogenesis tension is therefore minimized by taking the DM to be as light as allowed by observational probes of structure formation.   Constraints from Lyman-$\alpha$ forest data are often expressed as a lower bound on $M_\text{wdm}$, the mass of a warm thermal relic.  Recent studies have obtained lower bounds on $M_\text{wdm}$ ranging from 1.9 keV \cite{garzilli2019warm}  to 5.3 keV \cite{Palanque_Delabrouille_2020} at 95\% confidence level; see also Refs.~\cite{Baur_2016, Baur_2017, Ir_i__2017}.  A more stringent constraint on the dark matter mass applies in the type of freeze-in model we consider, with a heavy particle in thermal equilibrium decaying to out-of-equilibrium DM plus an additional light state.  By matching matter power spectra, Refs.~\cite{Kamada_2019, Ballesteros_2021} find that the constraint $M_\text{wdm} > 5.3$ keV translates to $M_\text{dm}\gtrsim 16$ keV for  freeze-in via two-body decay.

In the models we consider, with two DM mass eigenstates, it is possible possible for $\chi_1$ to be much lighter than this $\sim 16$ keV lower bound, provided that the DM energy density is dominated by $\chi_2$.  Moreover, this type of scenario is particularly advantageous for getting a large asymmetry, because $\chi_1$ can have larger couplings than would otherwise be allowed by the observed DM energy density.
The authors of Refs.~\cite{Boyarsky_2009, Kamada_2016, Baur_2017} present constraints on mixed cold/warm dark matter in the $M_\text{wdm}-r$ plane, where $r$ is the fraction of DM  energy density in the warm state.  The most stringent constraints are obtained in Ref.~\cite{Baur_2017}, which uses Lyman-$\alpha$ forest data  to find,  at 2$\sigma$ CL,  $r \lsim 0.3$  for $M_{\text{wdm}}\simeq 2 \text{ keV}$, going down to $r \lsim 0.08$  for $M_{\text{wdm}}\simeq 0.7 \text{ keV}$, at which point the bound on $r$ appears to have leveled off. 

We take these findings into account in an approximate way.  We require the mass of $\chi_2$, the heavier DM particle, to satisfy $M_{2} > 15$ keV.   The lighter DM particle, $\chi_1$, can be arbitrarily light provided that the fractional $\chi_1$ contribution to the DM energy density is sufficiently small.   For Majorana-fermion DM that  decouples at temperatures around or above the electroweak scale (as is the case in the models we consider), and which comes fully into equilibrium before decoupling, the observed DM energy density is realized for a DM particle mass of $\simeq 0.1$ keV.   For $M_1 \ll 0.1$ keV, then, $r$ is acceptably small regardless of the sizes of the $\chi_1$ couplings.  We call $M_1 \ll 0.1$ keV the massless $\chi_1$ limit\footnote{Even if it into equilibrium, a Majorana fermion of negligible mass that decouples at $T\gsim T_\text{ew}$ gives a contribution to $N_\text{eff}$ well within the BBN and CMB constraints~\cite{Brust:2013ova}.}.  
 
We will study the massless $\chi_1$ limit to determine, for example, the full range of $\Phi$ masses and lifetimes that work for DM and leptogenesis.    We will will also identify viable parameter space with larger $M_1$, under the assumption that points with $r \lesssim 0.1$ give acceptable matter power spectra for arbitrary $M_1$.   A detailed and robust determination of the $(M_1, M_2, r)$ parameter space allowed by structure-formation constraints is work in progress, and  beyond the scope of this paper.  It is clear,  however, that there exists abundant parameter space for DM and leptogenesis that does satisfy these constraints.
%
\subsection{Outline of our analysis}\label{sec:introoutline}
%
The outline of the rest of the paper is as follows.  In Sec.~\ref{sec:pertMM}, we present a perturbative analysis of DM and leptogenesis in the Minimal Model.  The analysis provides useful context for understanding the results of subsequent sections while also motivating the need to work with the full system of QKEs to map out the viable parameter space more completely.  

We present DM and leptogenesis results for the Minimal, UVDM, and Z2V models in Secs.~\ref{sec:MM}, ~\ref{sec:UVDM}, and~\ref{sec:Z2V}, respectively.  
For each model, we show results for the massless-$\chi_1$ limit to determine the range of allowed masses and lifetimes for the collider target $\Phi$ (or lightest scalar, $\Phi_1$, in the UVDM model), along with the allowed range of DM masses.  We  also investigate what parts of the parameter space survive departure from the massless-$\chi_1$ limit:  how heavy is $\chi_1$ allowed to be, and to what extent (if at all) must the couplings of the lighter DM state $\chi_1$ dominate over the couplings of $\chi_2$?

We find viable parameter space for all three models.  
In the Minimal Model, the upper bound on $M_\Phi$ is $\sim 1.5$ TeV, and $\Phi$ decays promptly in much of the viable parameter space.  Moreover, the Minimal Model is constrained to be near its massless-$\chi_1$ limit:  we need $M_1 \lesssim 0.05$ keV, and the $\chi_1$ couplings must be much larger than those of   $\chi_2$. 
 In contrast, the UVDM Model has ample parameter space with $M_1 \gsim 15$ keV and a long-lived $\Phi_1$ particle.  To a lesser degree, the Z2V Model also has viable parameter space with $M_1 \gsim 15$ keV, with the scalar typically decaying promptly for scenarios in which the Z2V couplings significantly impact  leptogenesis.  Although larger $\Phi$ masses are viable in the UVDM and Z2V Models, the largest asymmetries are realized for $M_\Phi \lesssim 1$ TeV.  We discuss implications for collider searches and other experimental probes, including $g_\mu-2$, in Sec.~\ref{sec:pheno}.

We relegate certain technical details to a series of appendices.  These include benchmark DM coupling matrices (Appendix~\ref{sec:appendixFbench}), equilibrium chemical potential relations (Appendix~\ref{sec:appendixCPs}), reaction density calculations (Appendix~\ref{sec:appendixRDs}), background for our adopted system of QKEs (Appendix~\ref{sec:appendixQKE}), and a discussion and collection of perturbative results (Appendix~\ref{sec:appendixpert}).

\section{Perturbative analysis of the Minimal Model}\label{sec:pertMM}

\subsection{DM versus ${\mathcal O}(F^4)$ flavor-dependent asymmetries}\label{sec:F4pert}

In this section, we quantitatively study the $\mathcal{O}(F^4)$ $X_\alpha$ asymmetries  alongside the $\mathcal{O}(F^2)$ DM abundance, all within the Minimal Model.    In addition to  highlighting certain qualitative aspects of asymmetry generation, this perturbative analysis also illustrates how the combined DM and leptogenesis requirements predict upper bounds on the masses of the new particles, making $\Phi$  in particular a promising target for colliders.  Our three model benchmarks share the same basic mechanism for the leading-order $X_\alpha$ asymmetries, so this discussion  is also a useful starting point for understanding  our final DM and leptogenesis results.  The reader more interested in those final results should skip ahead to Secs.~\ref{sec:Z2P} and~\ref{sec:Z2V}.

This section draws from the perturbative results derived and collected in Appendix~\ref{sec:QKEpert} and Appendix~\ref{sec:appendixpert}.  For those results to apply, two conditions must be satisfied.  First, the abundances of both $\chi$ mass eigenstates must remain well below their equilibrium values.  Second, we need $\Gamma_\Phi/H_\text{ew} \lesssim 2$, where $H_\text{ew}$ is the Hubble parameter at sphaleron decoupling and where 
\beq\label{eq:width}
\Gamma_\Phi = \frac{{\rm Tr} \left[F^\dagger F \right]}{16 \pi} M_\Phi
\eeq
is the $\Phi$ decay width in the Minimal Model, at leading order ({\it e.g.} neglecting thermal mass effects).  This second condition ensures that washout processes, including those that do not depend on the DM abundance, have at most an order-one effect on $Y_\alpha$ \footnote{For $M_\Phi \gg T_\text{ew}$, a $\Phi/\Phi^{*}$ asymmetry generated at $T\sim M_\Phi$ has decayed by a factor $\sim e^{-\Gamma_\Phi/(2H_\text{ew})}$ by the time of sphaleron decoupling.   We therefore adopt $\Gamma_\Phi/(2H_\text{ew}) \gsim 1$ as our criterion for washout effects to be important, although this is of course based on a rough estimate.}.    The leading-order calculation of the DM abundance in Appendix~\ref{sec:MMpert} leads to
\beq\label{eq:YGammaMainBody}
\frac{Y_\chi^{(2)}}{Y_\chi^\text{eq}} \simeq 
0.16 \times
\left(
\frac{500 \text{ GeV}}{M_\Phi} 
\right) ^2
\left(
\frac{\Gamma_\Phi}{H_\text{ew}} 
\right),
\eeq
where $Y_\chi^{(2)}$ is the  $\mathcal{O} (F^2)$ DM number-density  divided by entropy density  (defined to include  both DM mass eigenstates but only one helicity state:~$\chi$ or ${\overline \chi}$, not both), and $Y^\text{eq}_\chi \simeq 1.95\times 10^{-3}$ is the equilibrium abundance for an individual helicity and mass eigenstate of $\chi$ particle.  Eq.~(\ref{eq:YGammaMainBody}) shows that the $\Gamma_\Phi/H_\text{ew} \lesssim 2$ perturbativity condition is the limiting one for larger $\Phi$ masses.  

The $\mathcal{O}(F^2)$ DM energy density, $\rho_\text{dm}^{(2)}$, is determined by $Y_\chi^{(2)}$, the DM masses $M_1$ and $M_2$, and a mixing angle 
\beq\label{eq:theta}
\theta = \arctan \left[ (F^\dagger F)_{22}/(F^\dagger F)_{11}   \right],
\eeq
which controls the DM composition produced by $\Phi$ decays:~at $\mathcal{O}(F^2)$,  the $\chi_1$ and $\chi_2$ number densities are proportional to $\cos^2\theta$ and $\sin^2\theta$, respectively.
In Appendix~\ref{sec:MMpert} we find 
\beq\label{eq:rhoDMnum}
\frac{\rho_\text{dm}^{(2)}}{\rho_\text{dm}^\text{obs}}
\simeq
22
\left(
\frac{\Gamma_\Phi}{H_\text{ew}}
\right)
\left(
\frac{{\overline M}}{15 \text{ keV}}
\right)
\left(
\frac{500 \text{ GeV}}{M_\Phi}
\right)^2,
\eeq
where 
\beq\label{eq:aveDMmass}
{\overline M} \equiv M_1 \cos^2\theta+M_2 \sin^2 \theta.
\eeq
is the average mass of the $\chi$ particles, weighted by abundance.

In Appendix~\ref{sec:MMpert}, we also show that the  $\mathcal{O}(F^4)$ $X_\alpha$ asymmetry (charge density divided by entropy density), evaluated at the sphaleron decoupling temperature $T_\text{ew}$, satisfies
\begin{multline}\label{eq:Y4num}
\frac{Y_\alpha^{(4)}}{Y_B^\text{obs}}
\lesssim
(1.5\times 10^4)
\;
\sin^2 2\theta\\
\times
\left(
\frac{\Gamma_\Phi}{H_\text{ew}}
\right)^2
\left(
\frac{500 \text{ GeV}}{M_\Phi}
\right)^4
\mathcal{I}^{(4)}(x_\text{ew},\beta_\text{osc}),
\end{multline}
where $x_\text{ew} \equiv M_\Phi/T_\text{ew}$, $Y_B^\text{obs}=8.7\times 10^{-11}$ is the observed baryon asymmetry \cite{Zyla:2020zbs}, and the function $\mathcal{I}^{(4)}$ is defined in Eq.~(\ref{eq:final_I4_fn_1}) and plotted in Fig.~\ref{fig:I4}(a).
It is at most of order one, and depends on the oscillation parameter
\be\label{eq:betanum}
\beta_{\rm osc}\equiv \frac{M_0\Delta M^2}{6M_\Phi^3} \simeq
0.2
\times
\left( \frac{500 \text{ GeV}}{M_\Phi}\right)^3
\frac{\Delta M^2}{\left( 15 \text{ keV}\right)^2},
\quad\quad
\ee
where $\Delta M^2 \equiv M_2^2 -M_1^2$ is the DM mass-squared splitting and $M_0 \simeq 7.1 \times 10^{17}$ GeV is defined so that the relation between Hubble parameter and the temperature is $H = T^2/M_0$ at early times. The inequality of Eq.~(\ref{eq:Y4num}) is saturated when 
the phases and additional mixing angles (besides $\theta$)  that parametrize the DM coupling matrix $F_{\alpha i}$  take on appropriate values; see Eqs.~(\ref{eq:Ya4}) and (\ref{eq:F4coupbound}) in Appendix~\ref{sec:appendixpert}.  

\begin{figure}
          \includegraphics[width=3.3in]{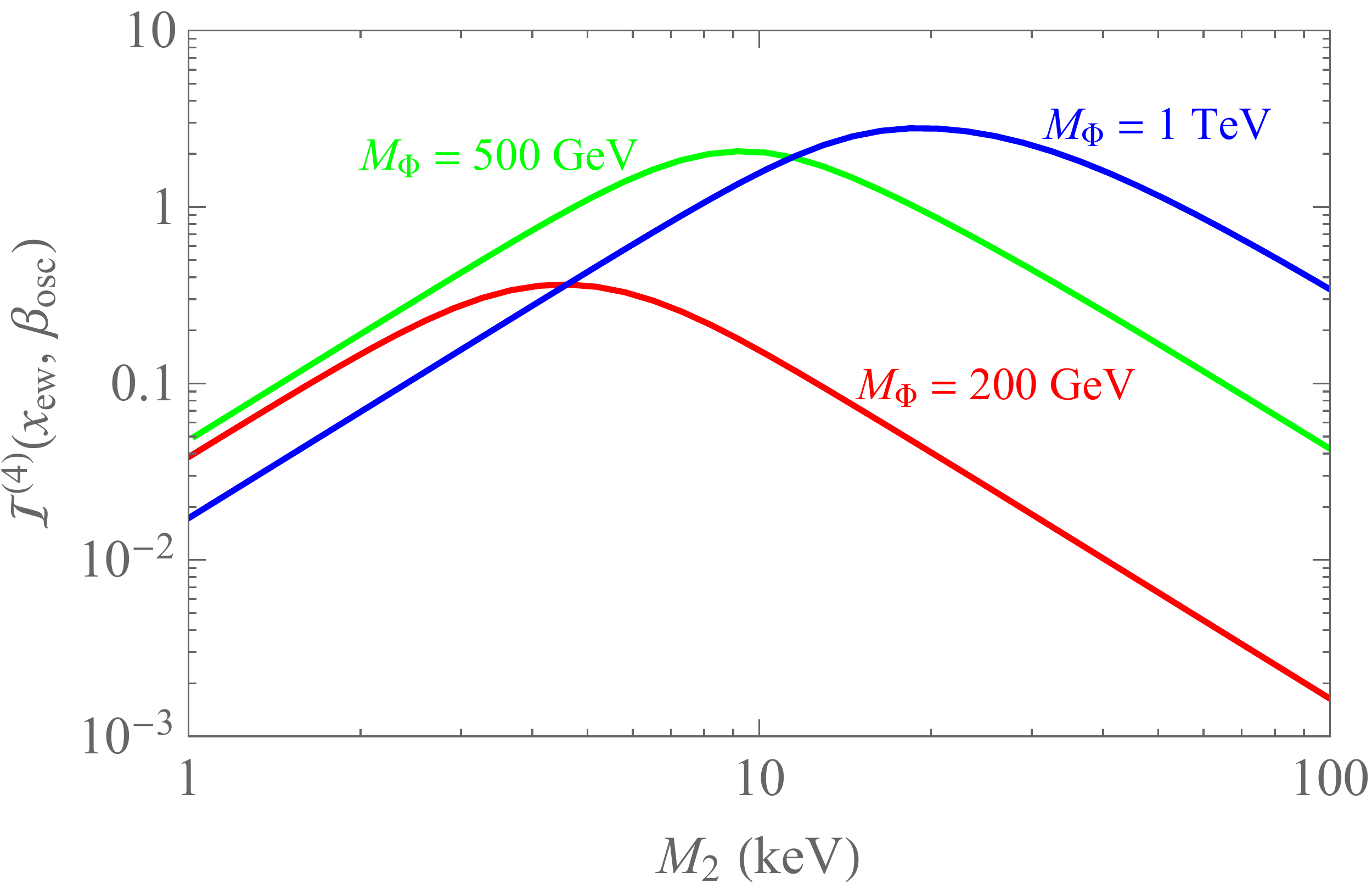}
\caption{
In the massless-$\chi_1$ limit, ${\mathcal I}^{(4)}(x_\text{ew},\beta_\text{osc})$ versus $M_{2}$ for various $M_\Phi$.  
}
\label{fig:I4vsMchi2}
\end{figure}

A typical $\chi$ state produced at high temperatures $T \gg M_\Phi$ undergoes $\sim \beta_\text{osc}/10$ oscillations by the time the temperature drops to $T = M_\Phi$, at which point the $\Phi$ abundance begins to become Boltzmann-suppressed. In Fig.~\ref{fig:I4}(a), we see that ${\mathcal I}^{(4)}(x_\text{ew}, \beta_\text{osc})$, and therefore the asymmetry, is suppressed at large and small values of $\beta_\text{osc}$.  
 For small $\beta_\text{osc}$,  the oscillations do not have enough time to develop before the temperature becomes too small to have an appreciable rate for inverse $\Phi$ decay.  For large $\beta_\text{osc}$, the  oscillations become rapid at early times, which cuts off the asymmetry growth prematurely.   In  Fig.~\ref{fig:I4}(a), ${\mathcal I}^{(4)}(x_\text{ew}, \beta_\text{osc})$ is peaked at $\beta_\text{osc}\sim 1$ for $M_\Phi \sim T_\text{ew}$ and $\beta_\text{osc} \sim 4 \times 10^{-2}$ for $M_\Phi \gg T_\text{ew}$.  The smallness of these optimal  $\beta_\text{osc}$ values reflects the importance of lower-energy $\chi$ particles (which oscillate more rapidly)  and oscillations that occur at temperatures well below $M_\Phi$ (whose effects are enhanced by the larger integrated time at lower temperatures).  

We now take the massless-$\chi_1$ limit to see how large $Y_\alpha^{(4)}$ can be, consistent with the DM constraint.  For fixed masses and fixed $\theta$, the DM constraint allows us to determine $\text{Tr}[F^\dagger F]$ and therefore $\Gamma_\Phi$.   Taking $M_1\rightarrow 0$ and $\rho_\text{dm}^{(2)} = \rho_\text{dm}^\text{obs}$ in Eq.~(\ref{eq:rhoDMnum}), we find
\beq\label{eq:gamoverHfromDM}
\frac{\Gamma_\Phi}{H_\text{ew}}
\simeq
\left(\frac{0.21}{\sin\theta} \right)^2
\left(
\frac{15 \text{ keV}}{M_2}
\right)
\left(
\frac{M_\Phi}{500 \text{ GeV}}
\right)^2,
\eeq
which we can use to rewrite Eq.~(\ref{eq:Y4num}) as
\beq\label{eq:Y4constraint1}
\frac{Y_\alpha^{(4)}}{Y_B^\text{obs}}
\lesssim
125
\times
\cot^2\theta
\left(
\frac{15 \text{ keV}}{M_2}
\right)^2
\mathcal{I}^{(4)}(x_\text{ew},\beta_\text{osc}).
\eeq
Fig.~\ref{fig:I4vsMchi2} shows $\mathcal{I}^{(4)}(x_\text{ew},\beta_\text{osc})$ versus $M_2$ for various $M_\Phi$ in the massless-$\chi_1$ limit. In this figure, the location of the peak shifts to higher $M_2$  as $M_\Phi$ is increased.  As the  available time for oscillations is reduced (by increasing $M_\Phi$), a shorter oscillation timescale (realized for larger $M_2$)  is preferred.  

If $\Phi$ decays produce equal abundances of $\chi_1$ and $\chi_2$ ($\theta = \pi/4$, {\em i.e.} maximal mixing), one can use Eq.~(\ref{eq:Y4constraint1}) and the properties of the $\mathcal{I}^{(4)}$ function to show that requiring $Y_\alpha^{(4)} > Y_B^\text{obs}$ leads to the upper bounds $M_2 \lesssim 300$ keV and $M_\Phi \lesssim 8$ TeV.  

Eq.~(\ref{eq:Y4constraint1}) shows a $1/\theta^2$ enhancement of $Y_\alpha^{(4)}$ for small mixing, $\theta \ll 1$.  We can understand this enhancement as follows.  In the massless-$\chi_1$ limit, the $\chi_1$ contribution to the energy density is negligible, and imposing the DM constraint  amounts to choosing the couplings of $\chi_2$, which are proportional to $(\text{Tr}[F^\dagger F])^{1/2} \sin \theta$, in such a way that the $\chi_2$ energy density matches $\rho_\text{dm}^\text{obs}$.  That is, decreasing $\theta$ means increasing $(\text{Tr}[F^\dagger F])^{1/2}$ to keep the $\chi_2$ couplings held fixed.  This causes an increase in the $\chi_1$ couplings, which are proportional $\sqrt{\text{Tr}[F^\dagger F]} \cos\theta$, and therefore an increase in the  asymmetry.  Exploiting this effect allows $Y_\alpha^{(4)} \sim Y_B^\text{obs}$ to be realized  for somewhat larger $\Phi$ and $\chi_2$ masses than for the case of maximal mixing.  However, the perturbativity condition $\Gamma_\Phi/H_\text{ew} \lesssim 2$ provides a ceiling.   One can use Eqs.~(\ref{eq:gamoverHfromDM}) and (\ref{eq:Y4constraint1})  to show that the DM constraint, 
 $\Gamma_\Phi/H_\text{ew} <2$, and $Y_\alpha^{(4)} > Y_B^\text{obs}$ can only be simultaneously satisfied for $M_2 \lesssim 700$ keV and $M_\Phi \lesssim 9$ TeV.

Those are the largest masses that can give an asymmetry in an individual lepton flavor that is comparable to the observed baryon asymmetry, consistent with the DM constraint and roughly consistent with the perturbative assumption.  A rather extreme optimization is required to realize these values, for example a particular texture for the $F$ matrix that that leads to $\Phi$ decaying predominantly to a lighter DM mass eigenstate with $M_1 \ll 0.1$ keV.  Moreover, we remind the reader that the {\em flavor-summed} asymmetry, and therefore the baryon asymmetry, is in fact zero at $\mathcal{O}(F^4)$ in the Minimal Model.

\subsection{DM versus ${\mathcal O}(F^6)$ baryon asymmetry.}\label{sec:MMF6}

We now extend our perturbative study of the Minimal Model to $\mathcal{O}(F^6)$,  the order at which a baryon asymmetry arises.  Throughout this section, we adopt an $F$ matrix of the form given in Eq.~(\ref{eq:FMMbench}), with $\theta$ and the overall scale $\text{Tr} \left[F^\dagger F\right]$  free to vary.  The remaining parameters that define $F$ are fixed at values that are favorable for getting a large asymmetry (but not exactly optimal; see Appendix~\ref{sec:MMFbench} for details).  For this choice of $F$ matrix, we find in Appendix~\ref{sec:MMpert} that the final $\mathcal{O}(F^6)$ baryon asymmetry is
\begin{multline}\label{eq:Y6constraint1}
\frac{Y_B^{(6)}}{Y_B^\text{obs}}
\simeq
23
\;
\sin^2 2\theta \\
\times
\left(
\frac{\Gamma_\Phi}{H_\text{ew}}
\right)^3
\left(
\frac{500 \text{ GeV}}{M_\Phi}
\right)^6
\mathcal{I}^{(6)}(x_\text{ew},\beta_\text{osc}),
\end{multline}
where the function $\mathcal{I}^{(6)}$ is defined in Eq.~(\ref{eq:MMYB6}) and plotted in Fig.~\ref{fig:I4}(b). 
We get the maximum asymmetry, subject to the DM constraint, by taking the massless-$\chi_1$ limit.  In that case we can use Eq.~(\ref{eq:gamoverHfromDM}) to eliminate $\Gamma_\Phi/T_\text{ew}$ in Eq.~\eqref{eq:Y6constraint1}, leading to the upper bound
\be\label{eq:Y6constraint2}
\!\!\!
\frac{Y_B^{(6)}}{Y_B^\text{obs}}
\lesssim
(8.6 \times 10^{-3})\;
\frac{\cos^2\theta}{\sin^4\theta}
\left(
\frac{15 \text{ keV}}{M_2}
\right)^3
\mathcal{I}^{(6)}(x_\text{ew},\beta_\text{osc})
\quad\quad
\ee
for our benchmark $F$ matrix.

Keeping in mind that ${\mathcal I}^{(6)}$ is never larger than $\sim 1/2$, it is clear at this point that $\Phi$ must couple preferentially to the lighter DM state, $\chi_1$, for the asymmetry to be large enough.  If $\Phi$ decays instead produce equal  $\chi_1$ and $\chi_2$ abundances ($\theta = \pi/4$), the asymmetry of Eq.~(\ref{eq:Y6constraint2}) is 
maximized for $M_2 = 15$ keV (which saturates our adopted structure-formation bound) and  $M_\Phi \simeq 700$ GeV.  These additional optimizations only get the asymmetry  up to  $Y_B^{(6)}/Y_B^\text{obs}\simeq 8 \times 10^{-3}$.   

To get a sense of the  parameter space that opens up for smaller $\theta$, we set $\Gamma_\Phi/H_\text{ew} =2$, at the high end of what is reasonable for our perturbative analysis, and use  Eq.~(\ref{eq:gamoverHfromDM}) to eliminate $\theta$ in Eq.~(\ref{eq:Y6constraint2}); the bounds we obtain in this way are guaranteed to apply only in the perturbative regime.  We find that the DM constraint and $Y_B^{(6)}=Y_B^\text{obs}$ can only be simultaneously satisfied for $M_2 \lesssim 30$ keV and $M_\Phi \lesssim 800$ GeV, with  a maximum $Y_B^{(6)}$ of about about six times the observed value for the optimal masses $M_\Phi \simeq 400$ GeV and $M_2 = 15$ keV.   For these optimal masses, Eq.~(\ref{eq:gamoverHfromDM}) gives $\theta \simeq 0.12$.  So, our perturbative analysis suggests that the Minimal Model works for DM and leptogenesis within a rather constrained parameter space.  In particular,  $\Phi$ must decay mostly to $\chi_1$, which must be quite light -- otherwise there would be nothing  gained by $\chi_1$ having  larger couplings than  $\chi_2$.  

We now turn to a more detailed and comprehensive analysis of DM versus leptogenesis in the Minimal, UVDM, and Z2V Models.

\begin{figure*}
          \includegraphics[width=3in]{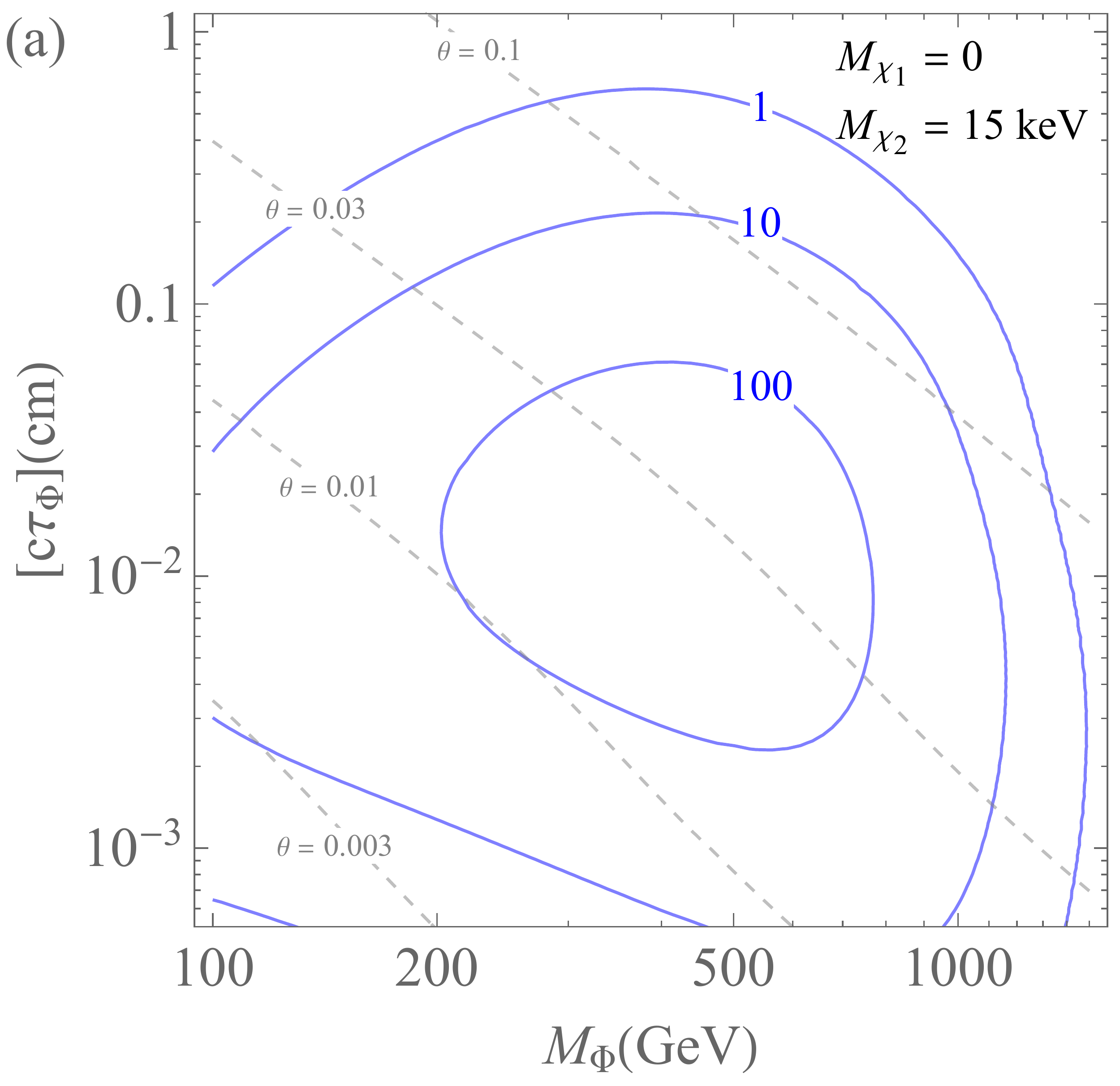}
           \quad \quad \quad
    \includegraphics[width=3in]{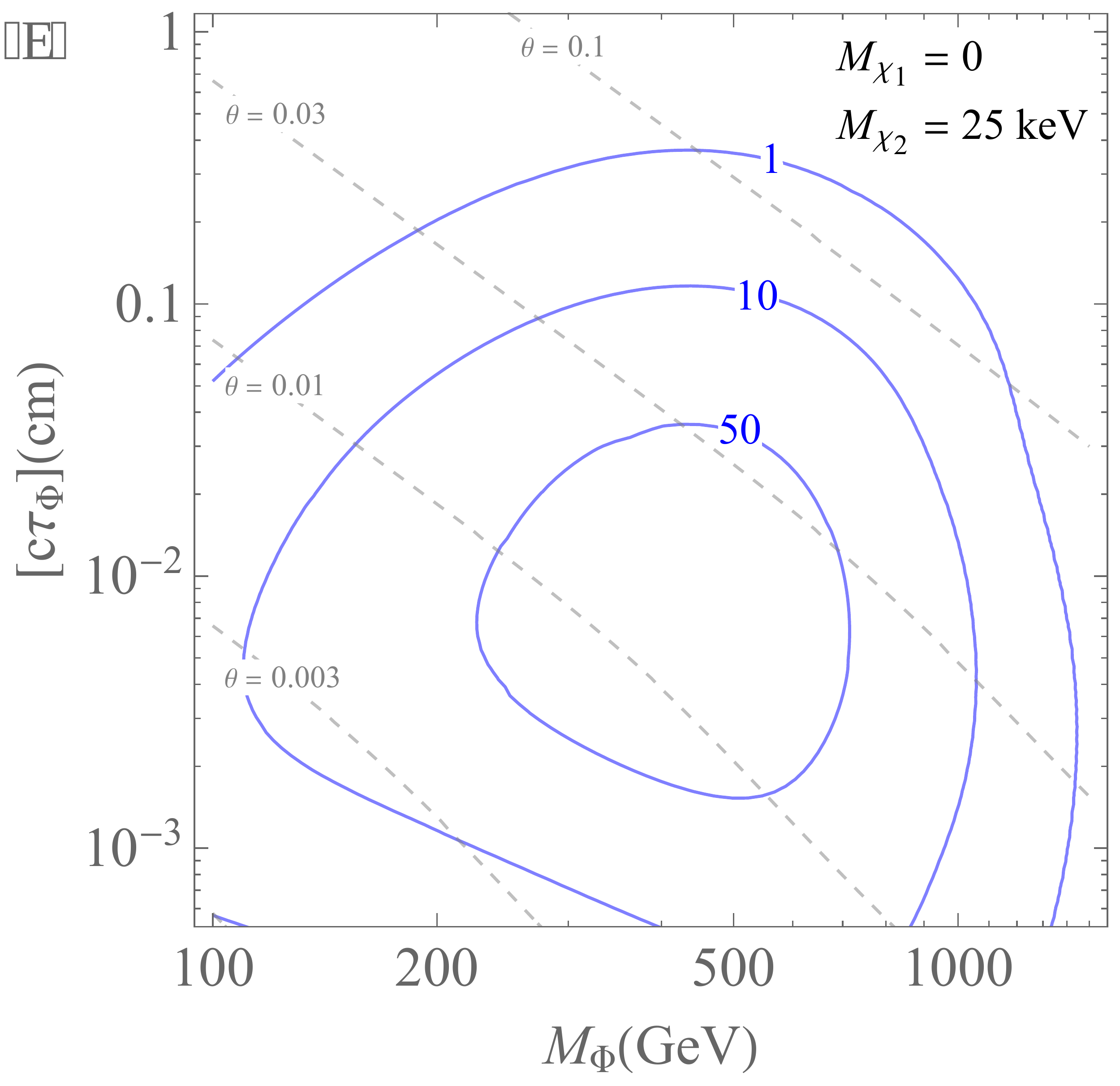}\\
    \includegraphics[width=3in]{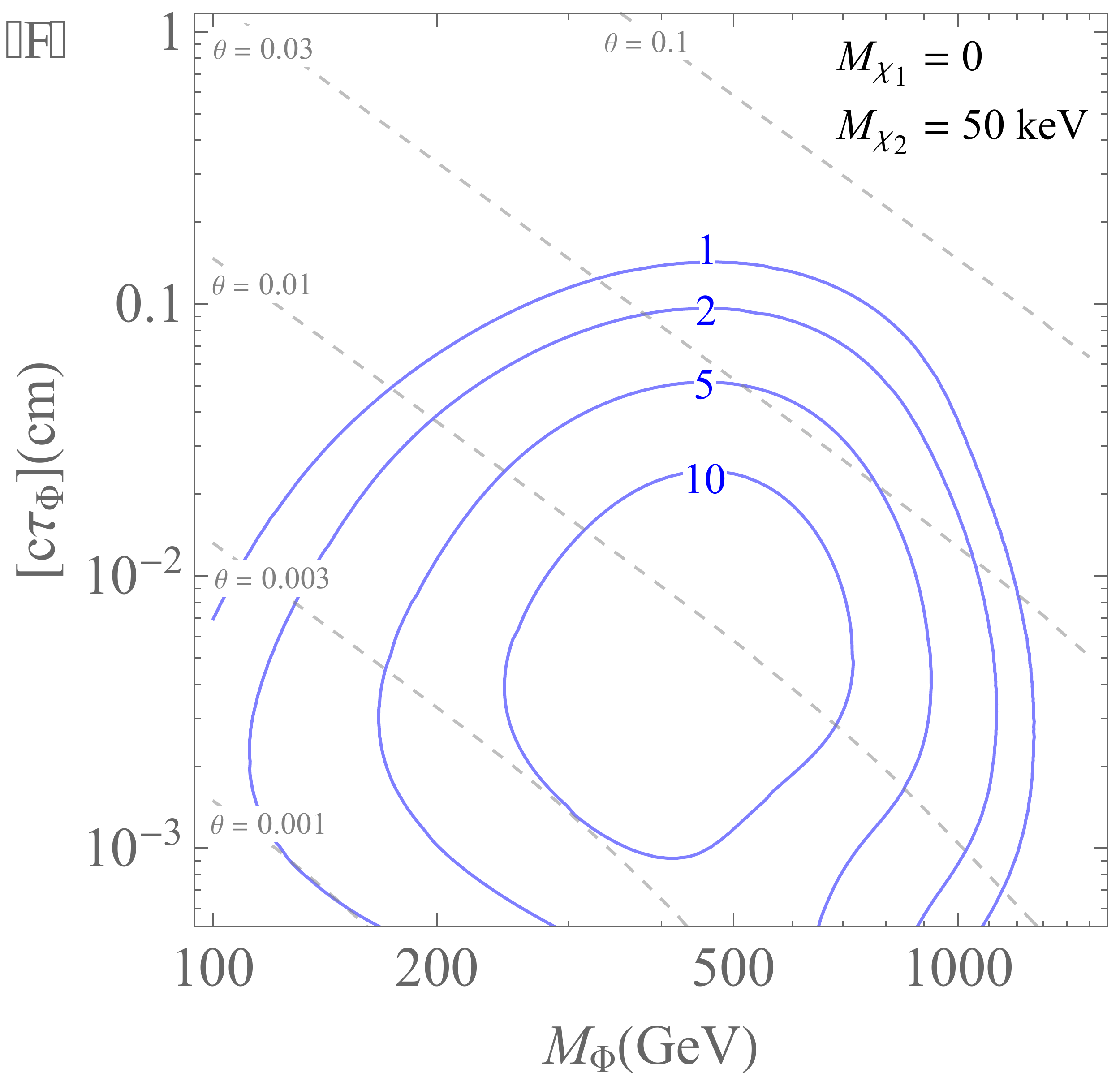}
           \quad \quad \quad
    \includegraphics[width=3in]{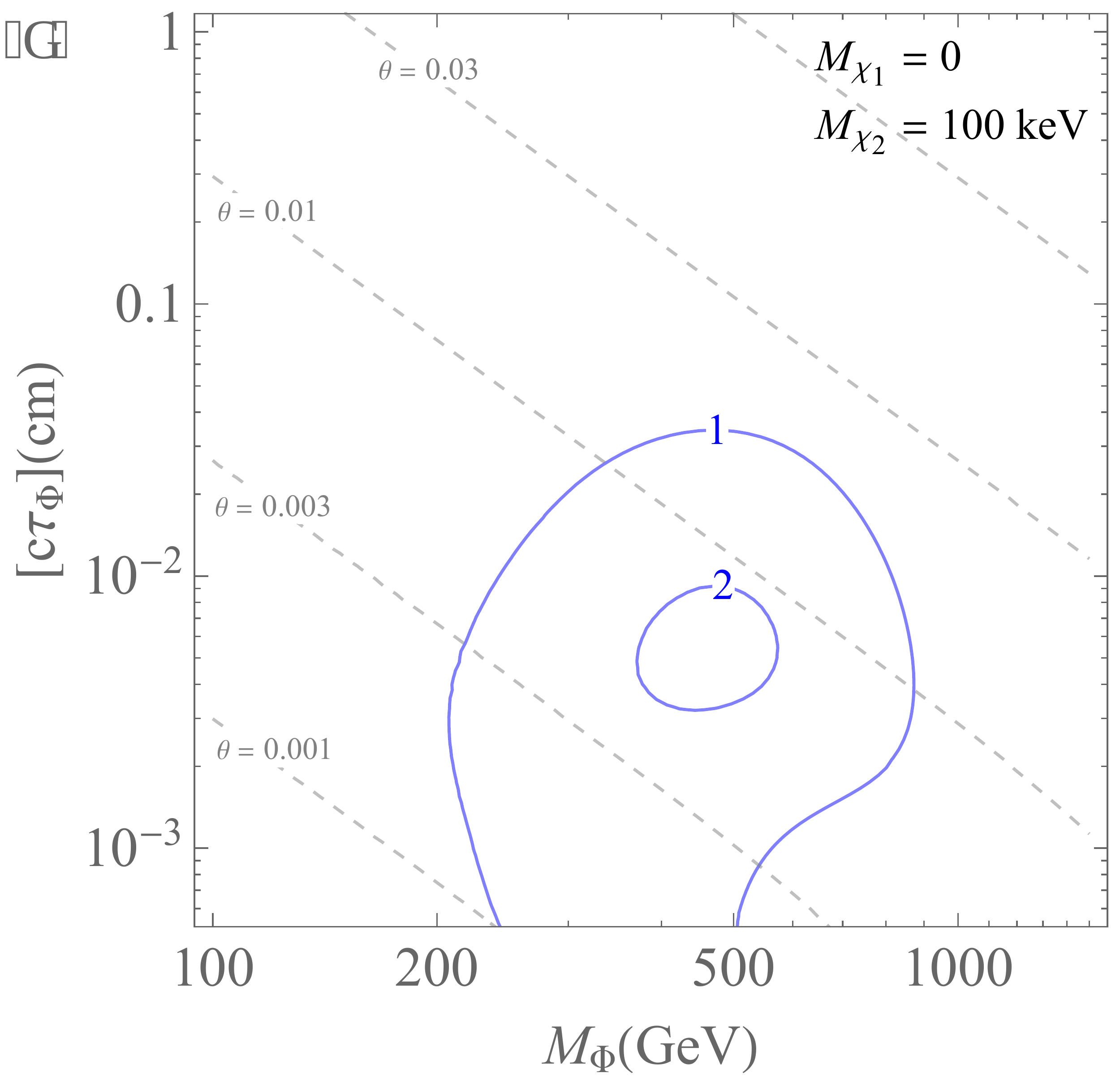}
\caption{
For the Minimal Model in the massless-$\chi_1$ limit, contours of $Y_B/Y_B^\text{obs}$ (blue, solid) and $\theta$ (gray, dashed) in the $(M_\Phi, c\tau_\Phi)$ plane, for various $\chi_2$ masses, with the $F$ matrix set to the Minimal-Model benchmark form of Eq.~(\ref{eq:FMMbench}).  At each point in the plane, $\theta$ is chosen to satisfy the DM constraint.  
}
\label{fig:MM1}
\end{figure*}
\begin{figure*}
          \includegraphics[width=3in]{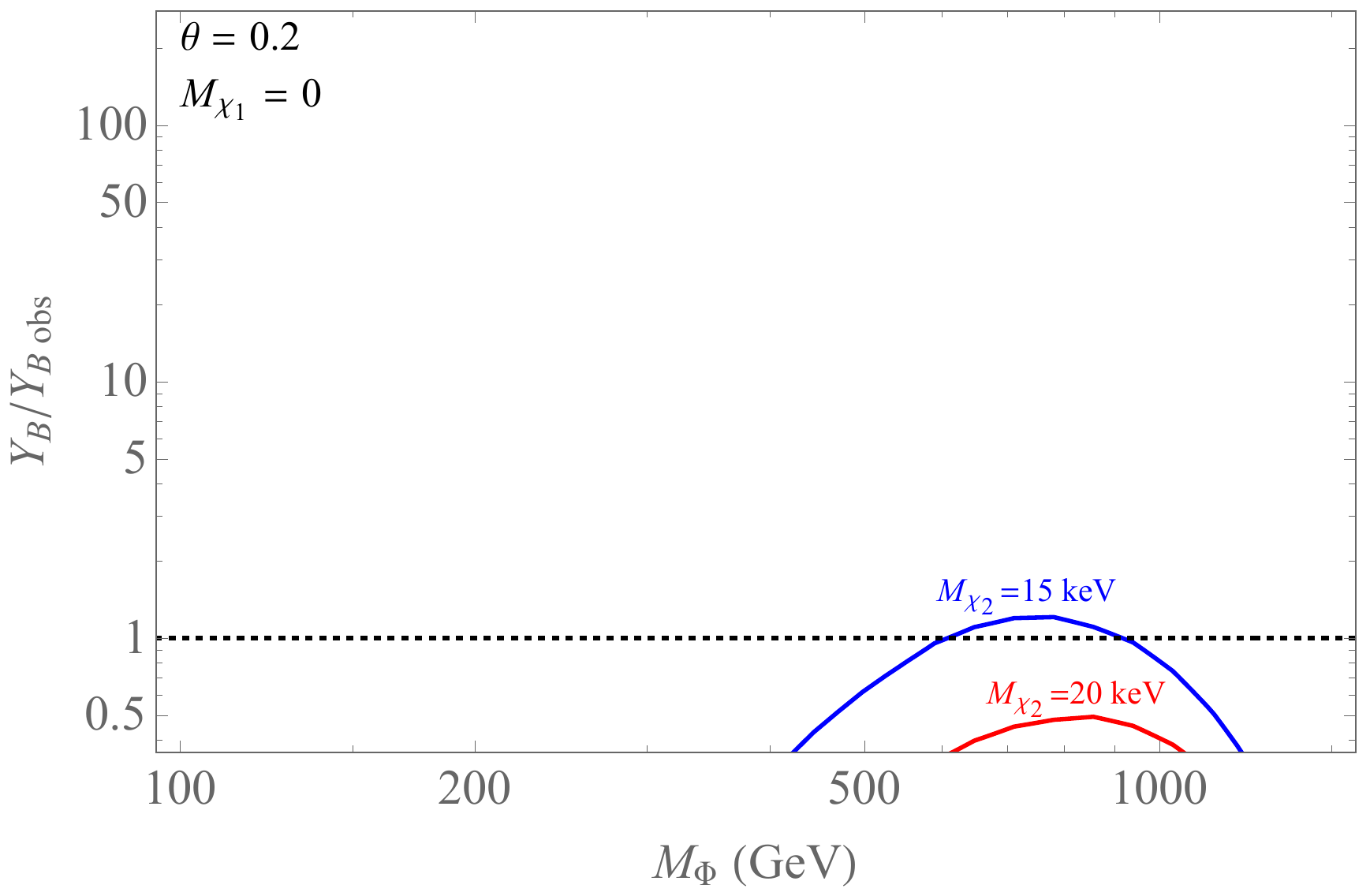}
           \quad \quad \quad
          \includegraphics[width=3in]{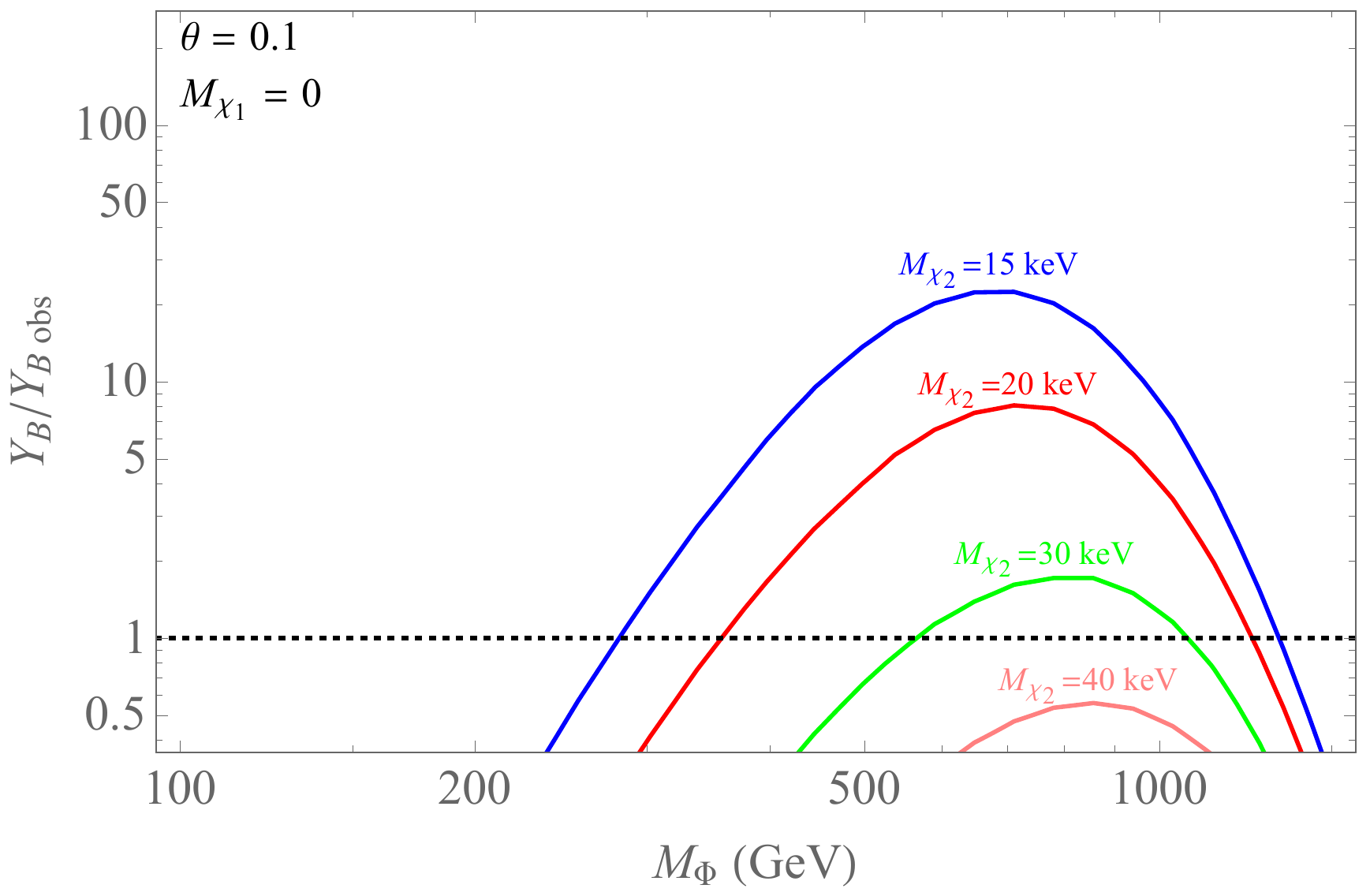}\\
          \includegraphics[width=3in]{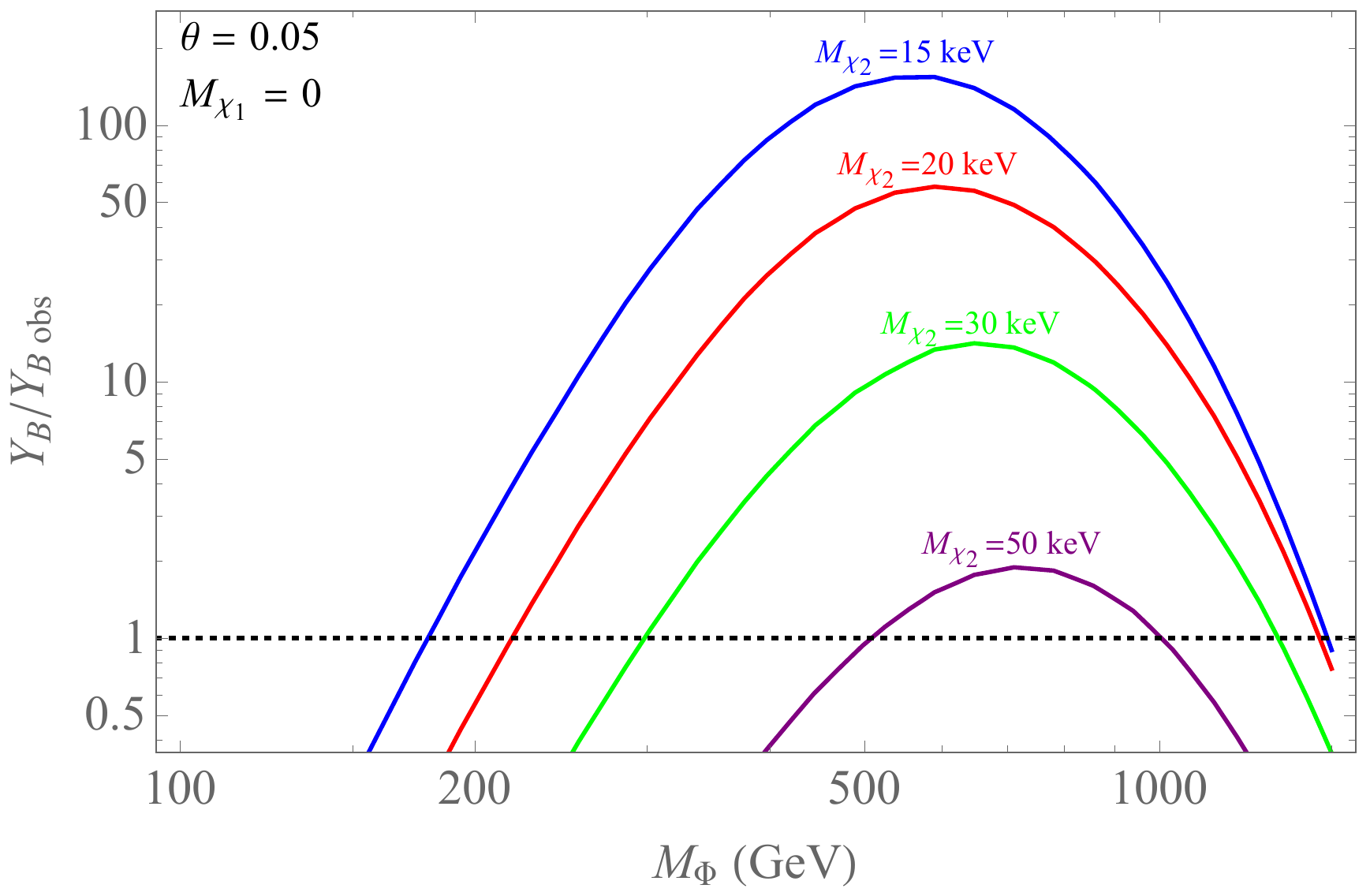}
           \quad \quad \quad
          \includegraphics[width=3in]{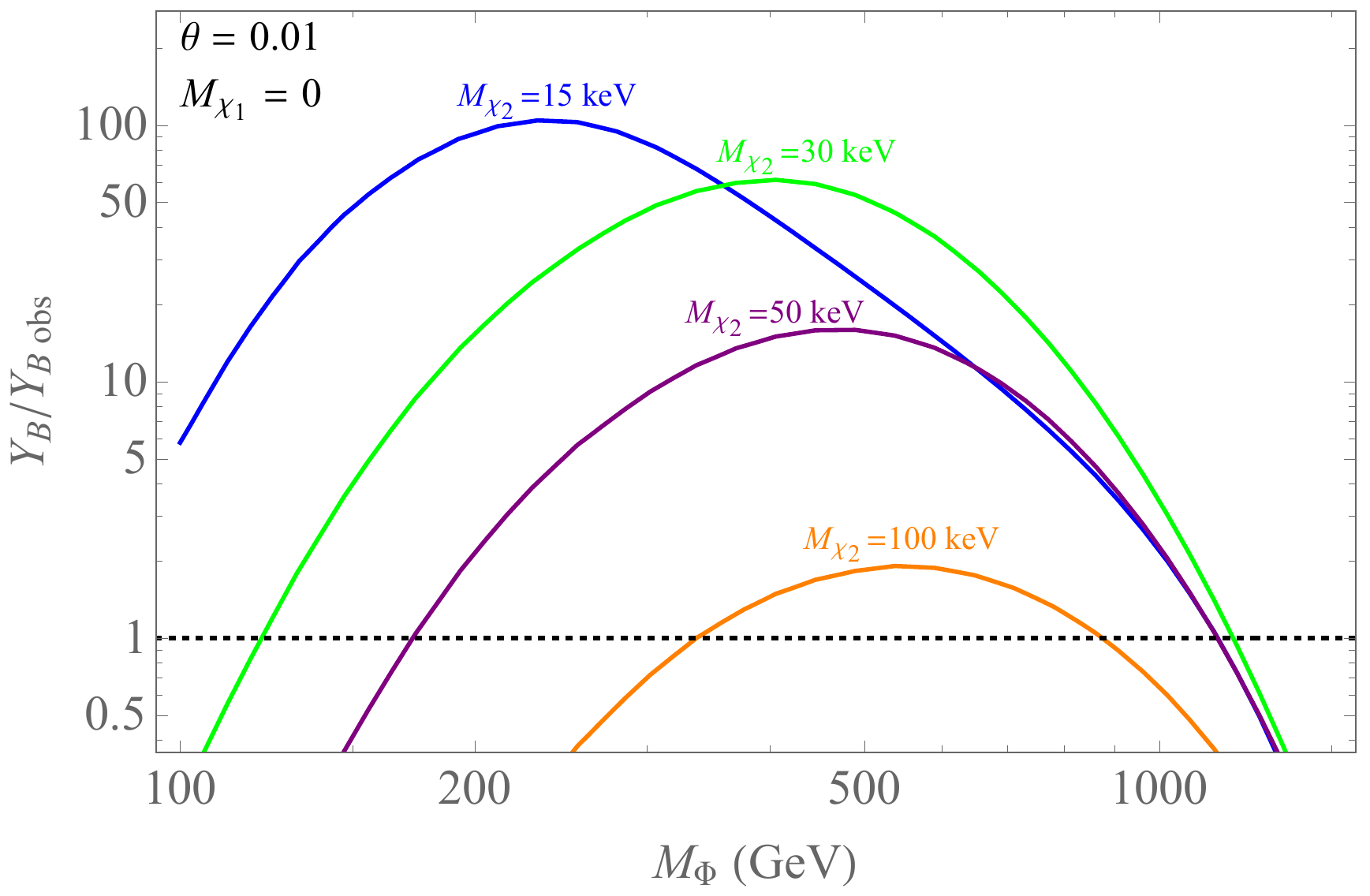}
\caption{
For the Minimal Model in the massless-$\chi_1$ limit, $Y_B$ versus $M_\Phi$ for various $\theta$ and $M_2$, with the $F$ matrix set to the Minimal-Model benchmark form of Eq.~(\ref{eq:FMMbench}).   For each combination of parameters, ${\rm Tr} \left[ F^\dagger F \right]$ is chosen to satisfy the DM constraint.  
}
\label{fig:MM2}
\end{figure*}
\begin{figure}
\includegraphics[width=3in]{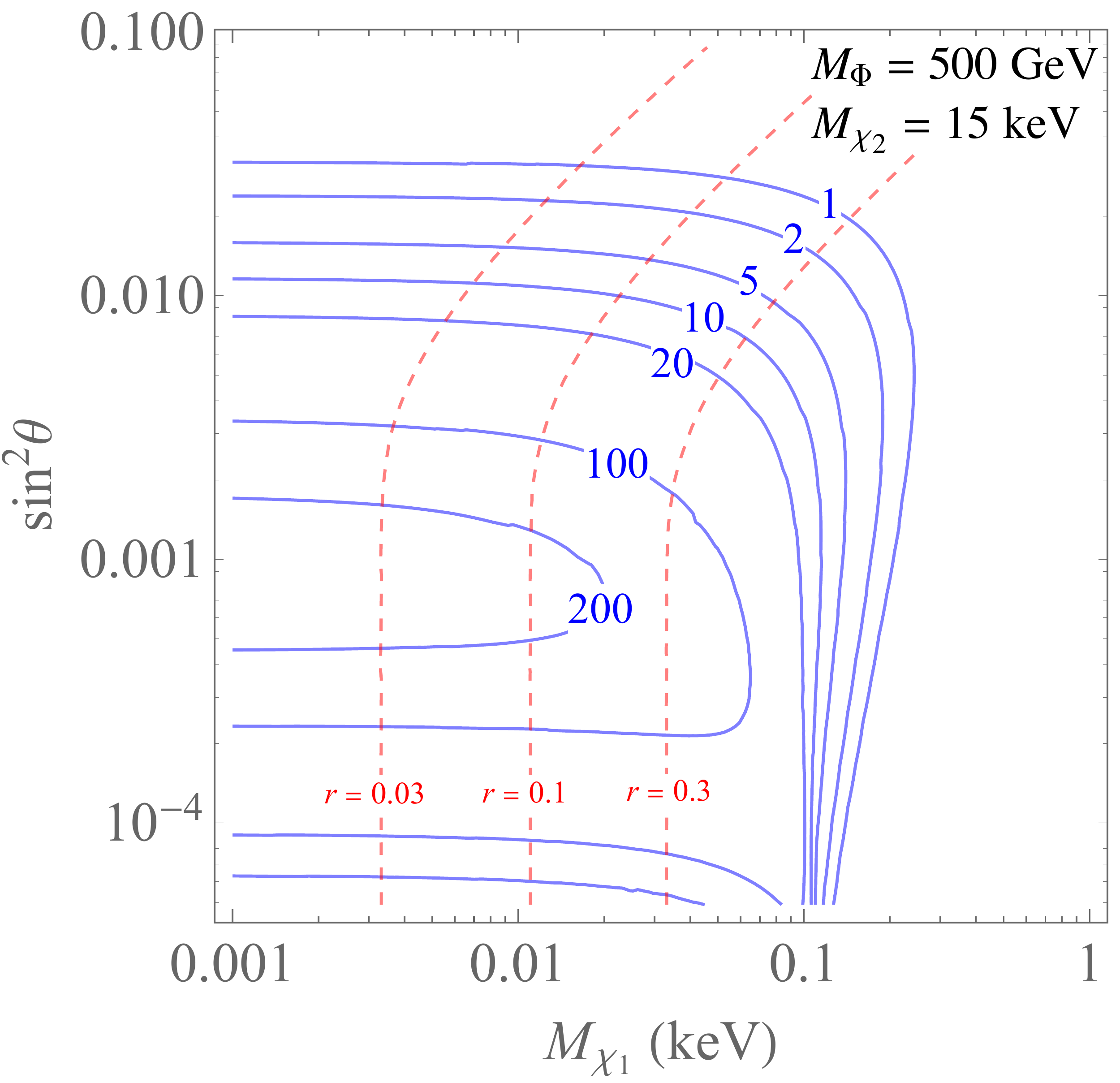}
\caption{
For the Minimal Model, contours of $Y_B/Y_B^\text{obs}$ (blue, solid) and $r$ (red, dashed) in the $(M_1, \sin^2\theta)$ plane, for $M_\Phi = 500$ GeV and $M_2 = 15$ keV, with the $F$ matrix set to the Minimal-Model benchmark form of Eq.~(\ref{eq:FMMbench}). 
At each point in the plane, ${\rm Tr} \left[ F^\dagger F \right]$ is chosen to satisfy the DM constraint.  
}
\label{fig:MM3}
\end{figure}
\begin{figure*}
          \includegraphics[width=2.3in]{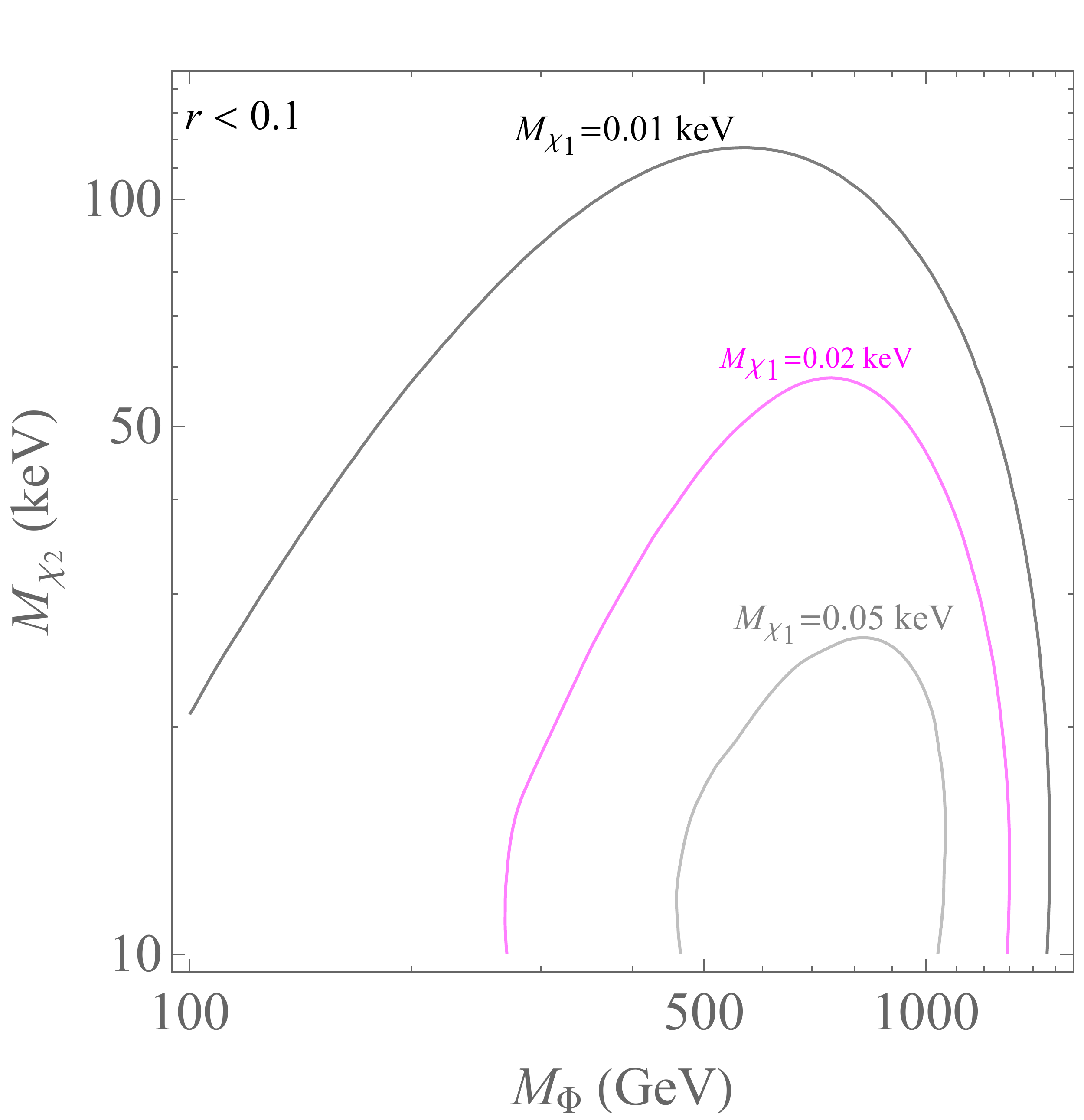}
          \includegraphics[width=2.3in]{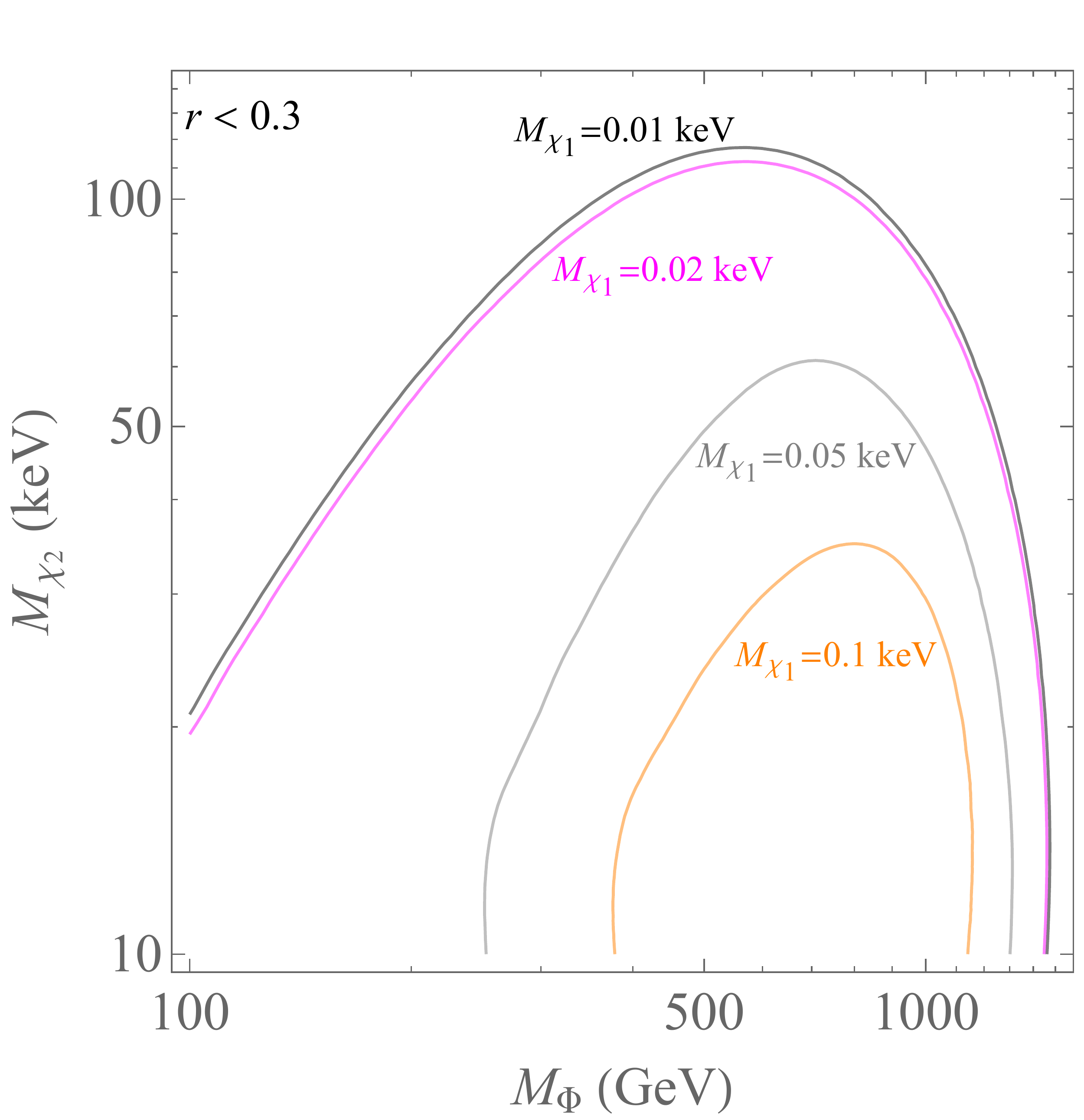}
          \includegraphics[width=2.3in]{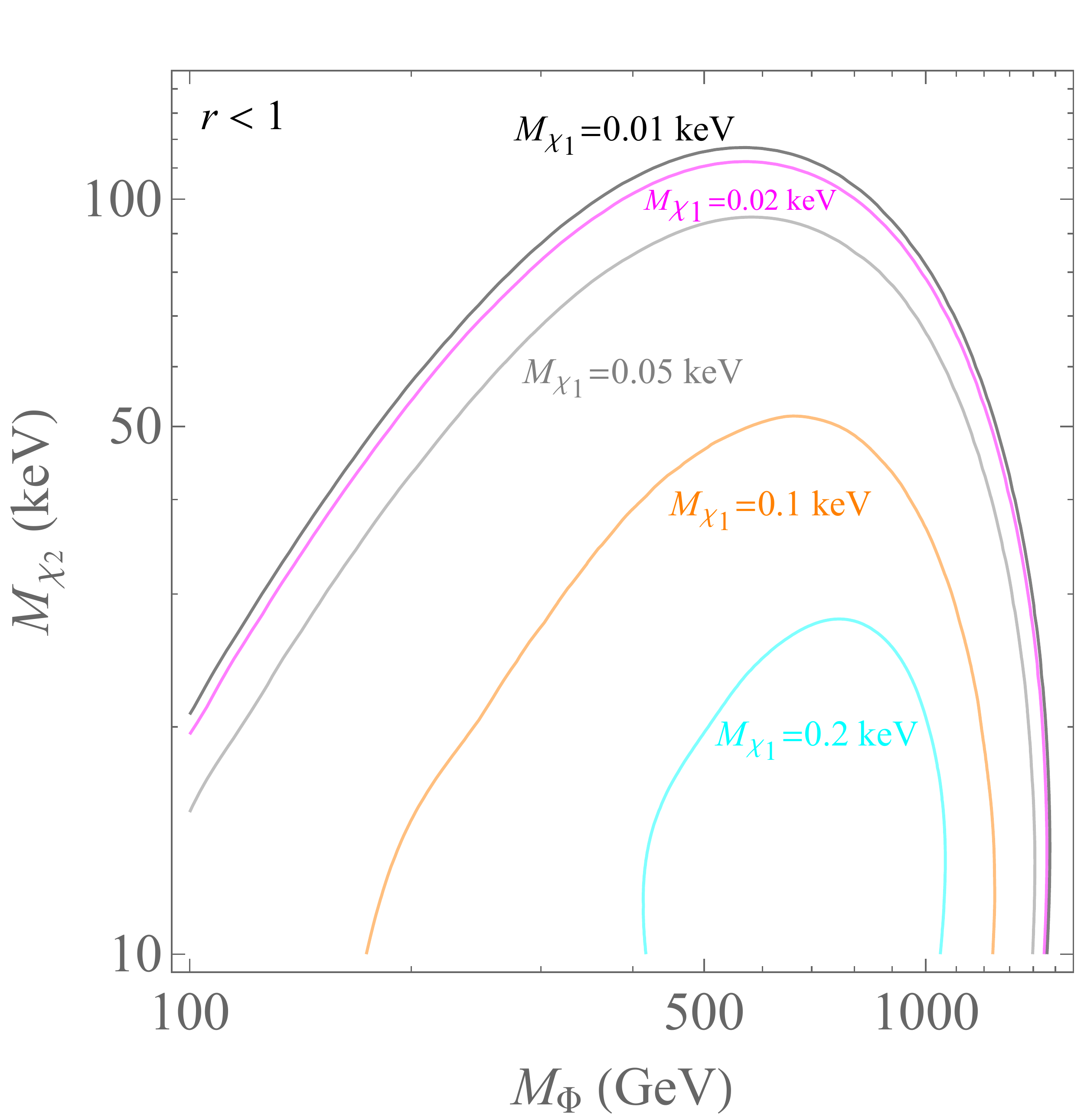}
\caption{
For various $M_1$, the contours enclose the $(M_\Phi,M_2)$ space that is viable for DM and leptogenesis in the Minimal Model, with the $F$ matrix set to the Minimal-Model benchmark form of Eq.~(\ref{eq:FMMbench}).   At each point in the plane, ${\rm Tr} \left[ F^\dagger F \right]$ and $\theta$ are chosen  to maximize $Y_B$ subject to both the DM abundance constraint and the upper bound on $r$ indicated; for the contours shown, that maximum $Y_B$ is equal to $Y_B^\text{obs}$. 
}   
\label{fig:MM4}
\end{figure*}
%
%

\section{Results for $Z_2$-preserving scenarios}\label{sec:Z2P}
\subsection{Results for the Minimal Model}\label{sec:MM}

To extend our study of the Minimal Model to larger values of $\Gamma_\Phi/H_\text{ew}$, we turn to the system of quantum kinetic equations (QKEs) presented in Appendix~\ref{sec:MIQKEs}.  In our implementation of the QKEs, which follows that of Refs.~\cite{Hambye:2017elz,Abada:2018oly}, the equations are integrated over momentum using a thermal ansatz for the DM momentum distribution;  see Eq.~(\ref{eq:TA}).  The QKEs track the $\chi$ and ${\overline \chi}$ density matrices and the $X_\alpha$ densities, all of which evolve slowly because only $F$-induced interactions change them.  The collision terms in the QKEs involve the $\Phi$ and $e^c_\alpha$ chemical potentials, which can be expressed in terms of the $X_\alpha$ densities and the $\chi$ and ${\overline \chi}$ density matrices using the asymmetry relations derived in Appendix~\ref{sec:appendixCPs}.  For further details we refer the reader to the appendices.

To present results for the Minimal Model, we will first take the massless-$\chi_1$ limit to see the full range of $\Phi$ masses and lifetimes and DM masses that work for DM and leptogenesis.  Then we will see what happens to the viable parameter space as we increase $M_1$.  We will take essentially the same approach when we present results for the UVDM and Z2V Models.

For the benchmark $F$ matrix given in Eq.~(\ref{eq:FMMbench}), $\rho_\text{dm}$ and $Y_B$  are determined once we specify $\text{Tr} \left[F^\dagger F \right]$, $\theta$, and the BSM particle masses $M_1$, $M_2$, and $M_\Phi$.  
For Figs.~\ref{fig:MM1} and~\ref{fig:MM2}, we take the massless-$\chi_1$ limit.
Fig.~\ref{fig:MM1} shows contours of $Y_B/Y_B^\text{obs}$ in the $\Phi$ mass-lifetime space for various $M_2$, with $\theta$ chosen to satisfy the DM constraint at each point.  The relation
\beq
\frac{\Gamma_\Phi}{H_\text{ew}} \simeq 
 0.8  \left(\frac{c\tau_\Phi}{\text{cm}}\right)^{-1}
\eeq
is useful when connecting to our earlier perturbative analysis,
in which we adopted $\Gamma_\Phi/H_\text{ew} \sim 2$  as a rough cutoff on the perturbative regime, corresponding to $c \tau_\Phi \sim 0.4$ cm.  For the DM masses chosen for Figs.~\ref{fig:MM1}(a-d), we find that the QKE asymmetry is about a factor $\sim 2-3$ smaller than the perturbative one for  $M_\Phi = 500$ GeV and $\Gamma_\Phi/H_\text{ew} = 2$.   We will present a more detailed comparison of QKE and perturbative results for the UVDM and Z2V Models, in which a more significant proportion of the viable parameter space lives in the perturbative regime.  

In Fig.~\ref{fig:MM1}(a), we take $M_2$ to be equal to 15 keV, our lower bound on the DM mass based on structure-formation considerations.  The $Y_B/Y_B^\text{obs} = 1$ contour of this plot, which lies within $M_\Phi \lesssim 1.5$ TeV and $c\tau_\Phi \lesssim 0.6$ cm, therefore represents our estimate of the full mass-lifetime parameter space for $\Phi$ in the Minimal Model. The $\Phi$ particle decays promptly (as far as collider searches are concerned) in much of the viable parameter space, and the 
relatively low ceiling on its mass makes $\Phi$ a promising target for colliders.
Baryon asymmetries over 100 times the observed value are in principle consistent with the DM constraint, but those large asymmetries require $\theta \ll 1$.  Much of the viable parameter space has $\theta < 0.1$, consistent with expectations based on the perturbative analysis.   In Figs.~\ref{fig:MM1}(b-d) we see the mass-lifetime  space contract for larger $M_2$, going up to 100 keV, which is roughly the largest viable DM mass in the Minimal Model.  

To obtain these these results, we neglect thermal corrections to the DM masses. A proper treatment of these effects is quite involved due to the challenges of modeling the $k \sim M_\Phi$ momentum modes that are important at $x \sim 1$. However, we have performed some numerical estimates of DM thermal mass effects that suggest that $Y_B$  may be suppressed below the observed value in the  small-$M_\Phi$, small-$c\tau_\Phi$ corners of the plots in Fig.~\ref{fig:MM1}.  The impact is most significant for the $M_2 =15$ keV plot, where our estimates indicate that lifetimes below $c\tau_\Phi \sim 0.1$ cm for $M_\Phi \sim 100$ GeV, and below $c\tau_\Phi \sim 10^{-3}$ cm for $M_\Phi \sim 300$ GeV, may not be viable. The impacted lifetimes shift to lower values for larger $M_2$, and we still find viable parameter space with $M_\Phi$ as low as 100 GeV, without significant suppression in the the peak asymmetries realized at higher $M_\Phi$.  Apart from the lower $(M_\Phi, c\tau_\Phi)$ region just described, the viable parameter space is largely unaffected. We leave a more careful study of these effects for future work.

Still working in the massless-$\chi_1$ limit, Fig.~\ref{fig:MM2} shows how the viable ranges of  $M_\Phi$ and $M_2$ expand as $\theta$ decreases. At fixed $M_\Phi$ and $\theta$, the asymmetry is suppressed as $M_2$ increases due to the smaller DM couplings required to match the observed DM energy density.  At fixed $M_2$ and $\theta$, the suppression of the asymmetry at large $M_{\Phi}$ is due to the oscillations not having time to develop, while the suppression at small $M_{\Phi}$ arises because significant dark matter production continues after asymmetry growth has slowed, requiring smaller DM couplings.  Finally, at fixed $\theta$, larger $M_2$ means a shorter oscillation timescale and a somewhat larger  $M_\Phi$ that maximizes the asymmetry.  The more dramatic shift of the $\theta = 0.01$, $M_2 = 15$ keV contour to lower $M_\Phi$ is a strong-washout effect.  Washout suppression of the asymmetry depends on $\Gamma_\Phi/H_\text{ew}$,   while the DM energy density is proportional to $\theta^2 M_2 \Gamma_\Phi/M_\Phi^2$ for small $\theta$.  Once the DM constraint is imposed, $\Gamma_\Phi/H_\text{ew}$ is proportional to $M_\Phi^2/(\theta^2 M_2)$, leading to stronger washout at larger $M_\Phi$ and lower $M_2$.

Figs.~\ref{fig:MM3}  and~\ref{fig:MM4} show the impact on the parameter space when we depart from the massless-$\chi_1$ limit, allowing a non-negligible fraction of the DM energy density to be stored in $\chi_1$.  In Fig.~\ref{fig:MM3} we take $M_\Phi = 500$ GeV and  $M_2 = 15$ keV, favorable for producing a large asymmetry, and show baryon asymmetry contours in the $M_1 -\sin^2\theta$ plane.  We see that DM and leptogenesis require $M_1 \lesssim 0.2$ keV, even before we impose a constraint on $r$, the fraction of dark matter energy density in $\chi_1$.  Once we impose $r < 0.1$, as favored by Lyman-$\alpha$ forest observations, the constraint tightens to $M_1 \lesssim 0.05$ keV.  Fig.~\ref{fig:MM4} shows how the $M_\Phi-M_2$ space shrinks as $M_1$ increases, with the effect becoming more pronounced for a more stringent upper bound on $r$.   

Our study of the Minimal Model shows that the viable parameter space is quite constrained, with the various upper bounds $M_\Phi < 1.5$ TeV, $c \tau_\Phi \lesssim 0.6$ cm, $M_2 \lesssim 100$ keV, $M_1 \lesssim 0.05$ keV and $\theta \lesssim$ 0.2.  For much of the parameter space, the $\Phi$ particle is a realistic discovery target for future runs of the LHC, as we discuss in Sec.~\ref{sec:collider}.

\subsection{Results for the UVDM Model}\label{sec:UVDM}
%
%
\begin{figure*}
          \includegraphics[width=3in]{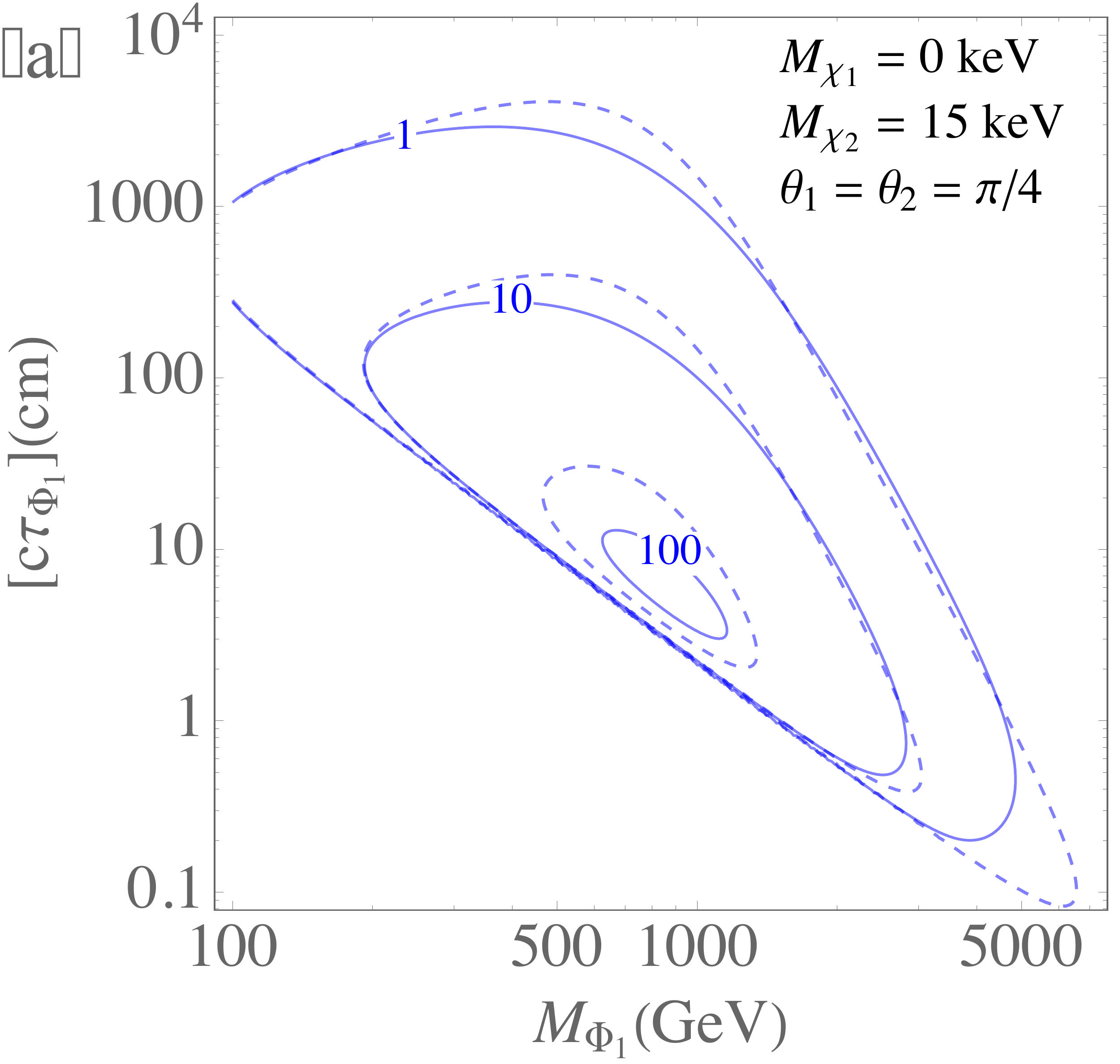}
           \quad \quad \quad
         \includegraphics[width=3in]{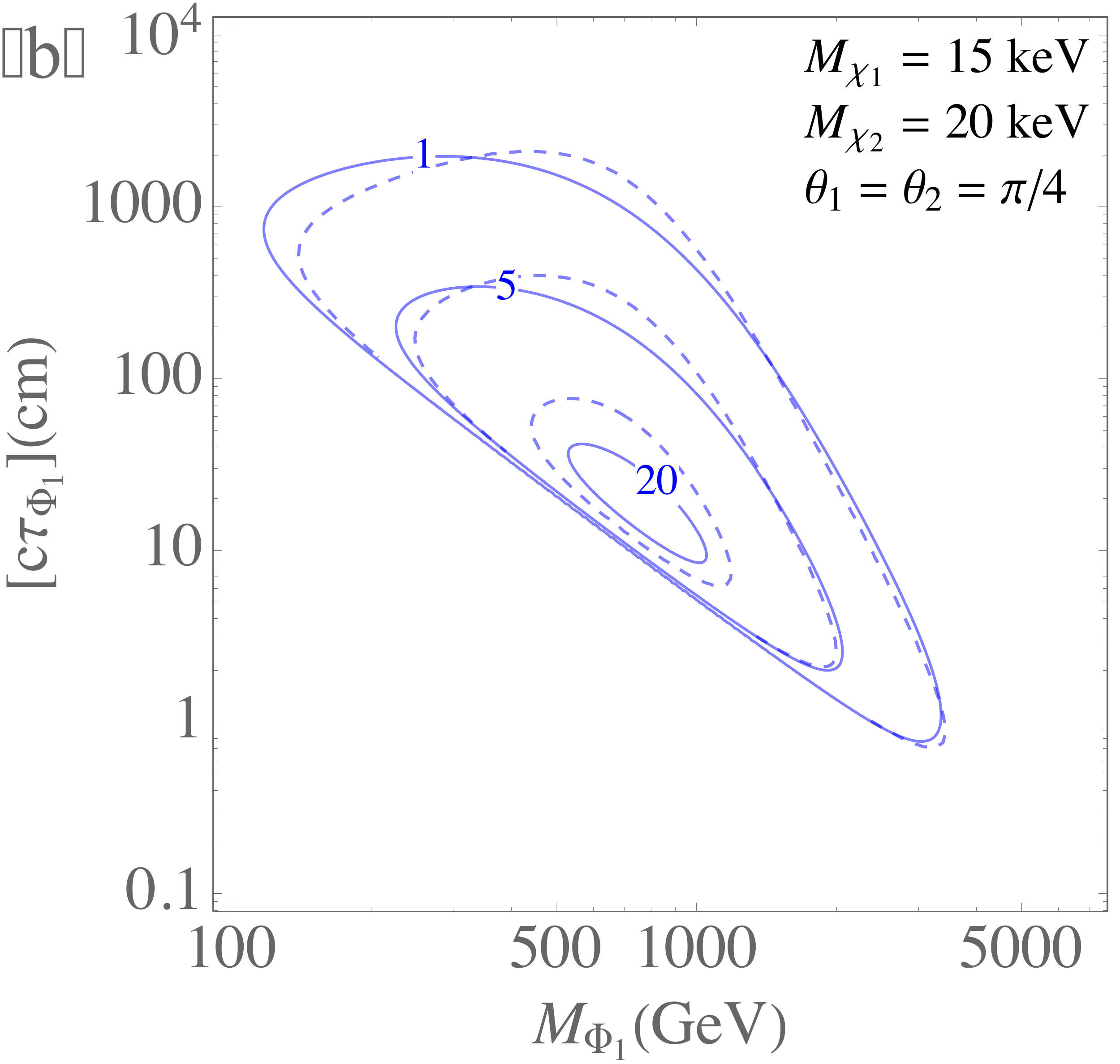}\\
          \includegraphics[width=3in]{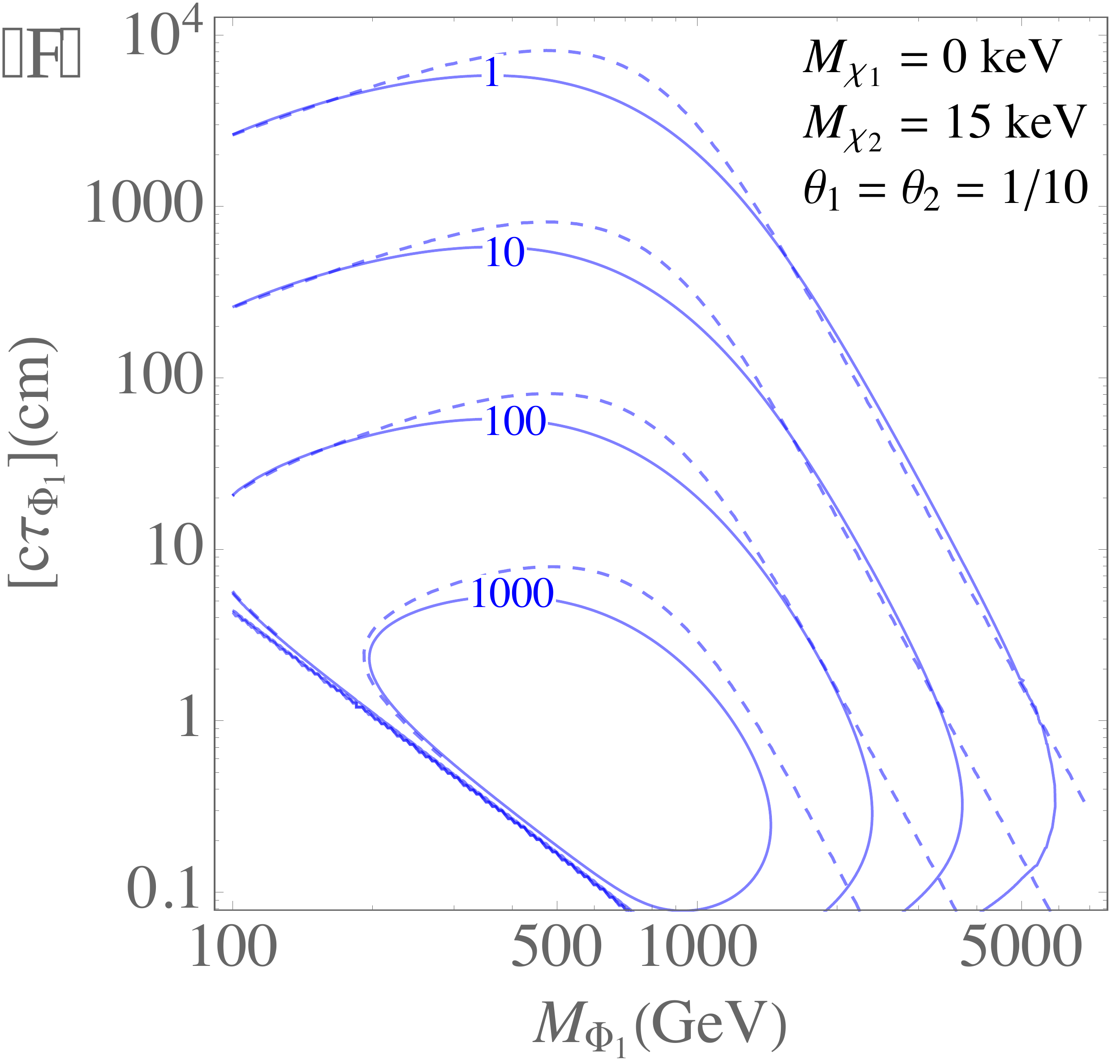}
            \quad \quad \quad
          \includegraphics[width=3in]{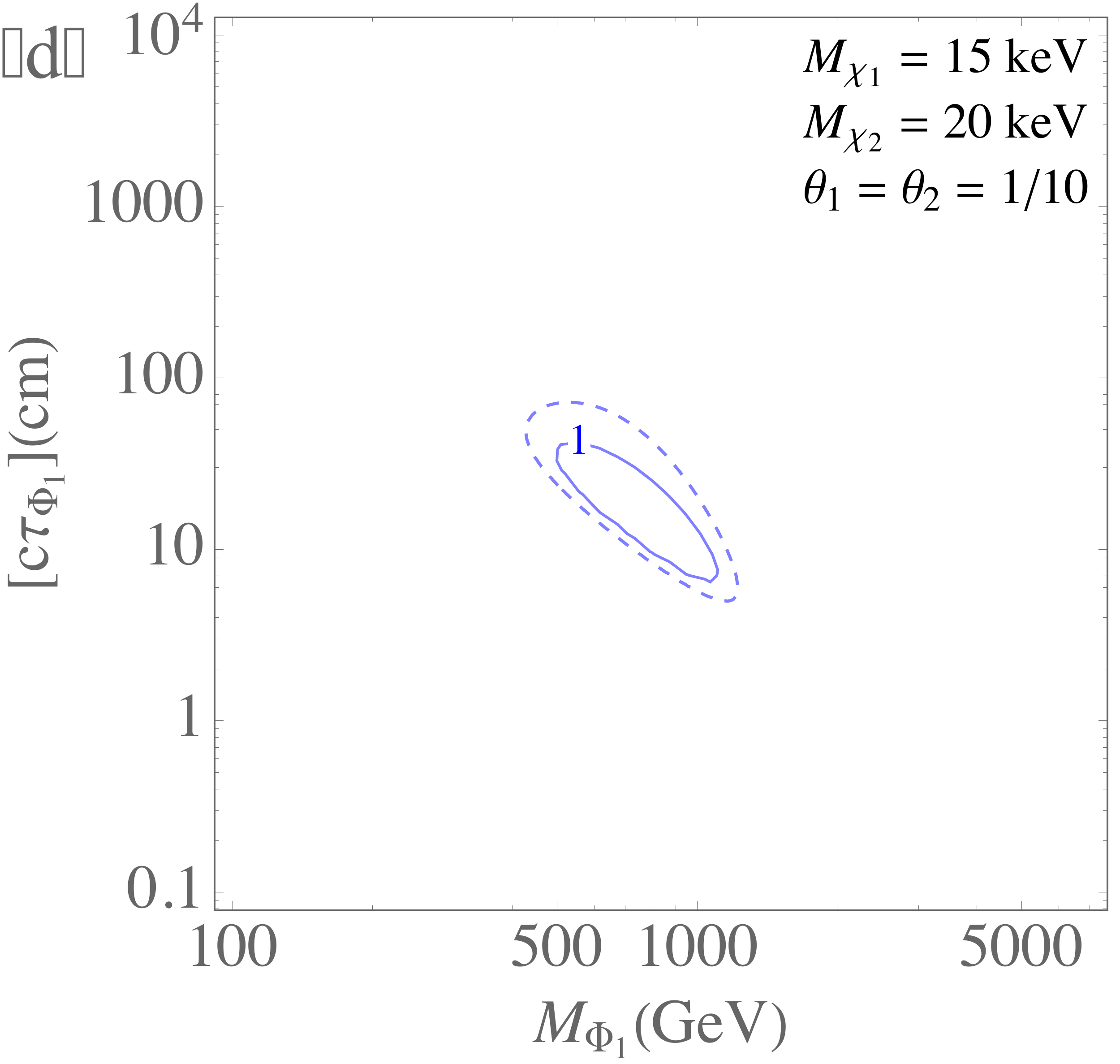}
\caption{
For the UVDM Model, contours of $Y_B/Y_B^\text{obs}$ in the $(M_{\Phi_1}, c\tau_{\Phi_1})$ plane, for various DM masses and mixings, with the $F$ matrices set to the UVDM benchmark of Eq.~(\ref{eq:UVDMbench}).  At each point in the plane,  the abundance of DM produced by $\Phi_2$ decays, $Y^\text{UV}_\chi$, is chosen to satisfy the DM constraint.  The dashed contours are based on the $\mathcal{O}(F^4)$ asymmetry of Eq.~(\ref{eq:YBnumUVDM_mainbody}), while the solid contours are based on numerical solution of the quantum kinetic equations presented in Appendix~\ref{sec:MIQKEs}.  The two calculations are in reasonable agreement in the perturbative regime, $c\tau_{\Phi_1} \gtrsim 1$ cm.  
}  
\label{fig:UVDM1}
\end{figure*}
\begin{figure*}
          \includegraphics[width=3in]{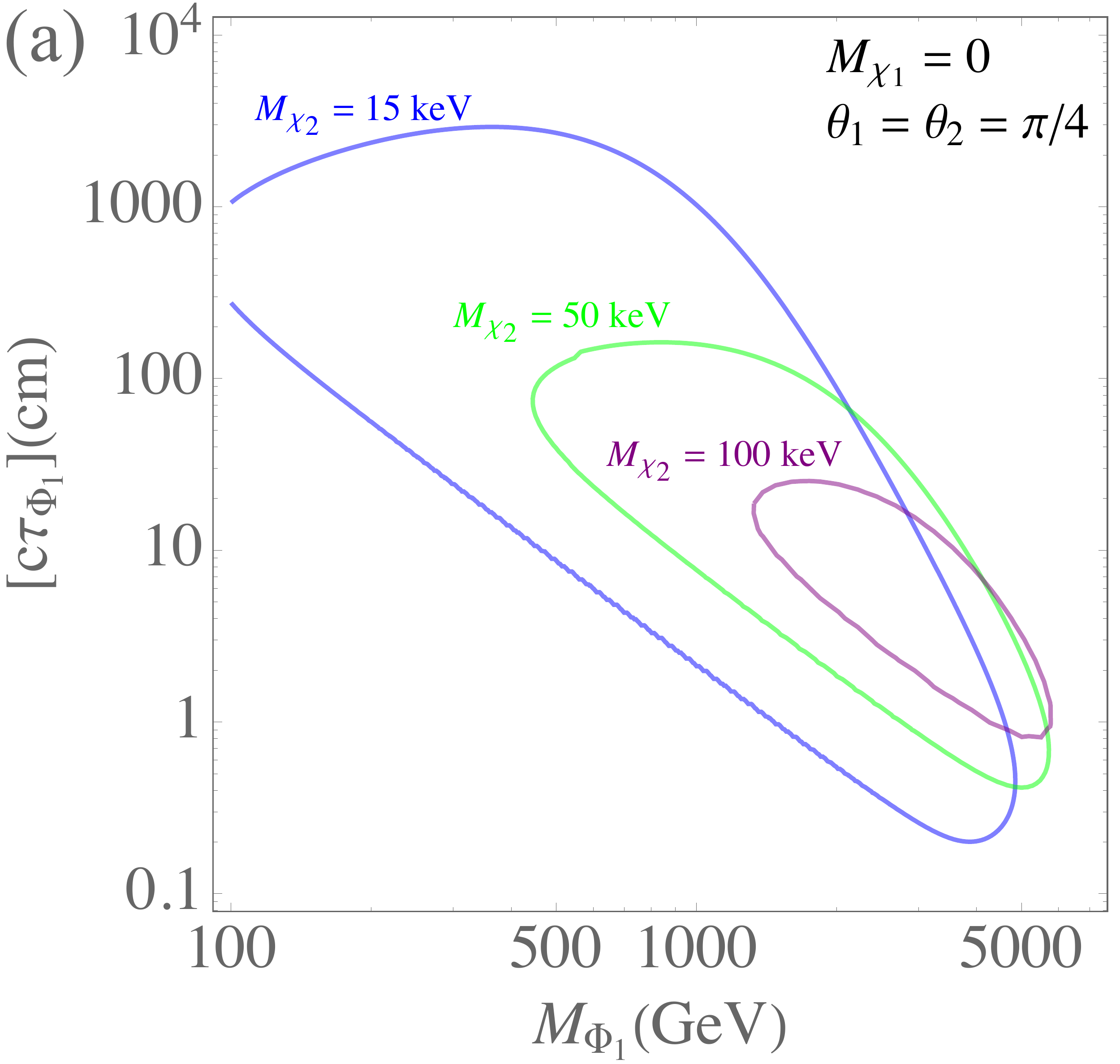}
           \quad \quad \quad
          \includegraphics[width=3in]{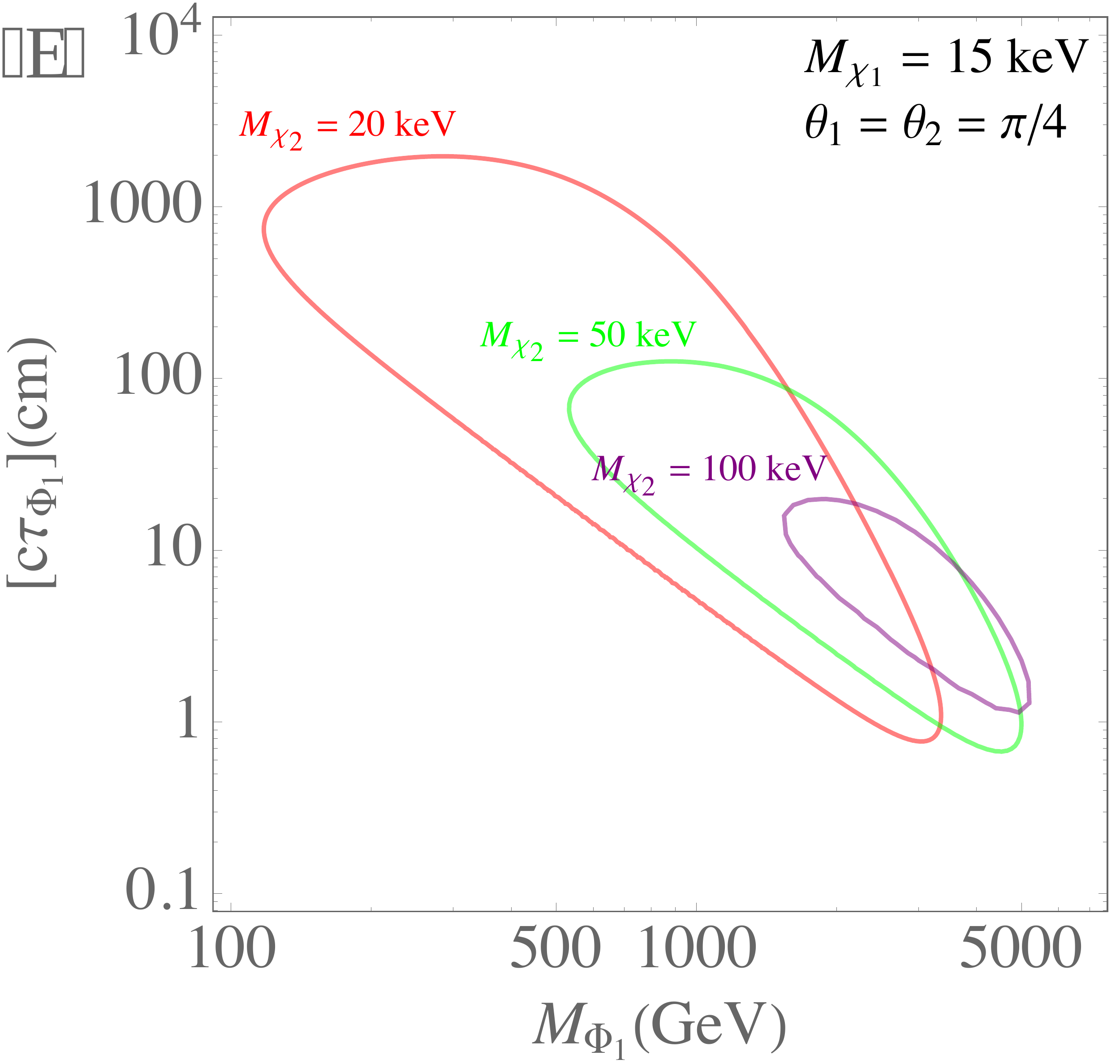}\\
          \includegraphics[width=3in]{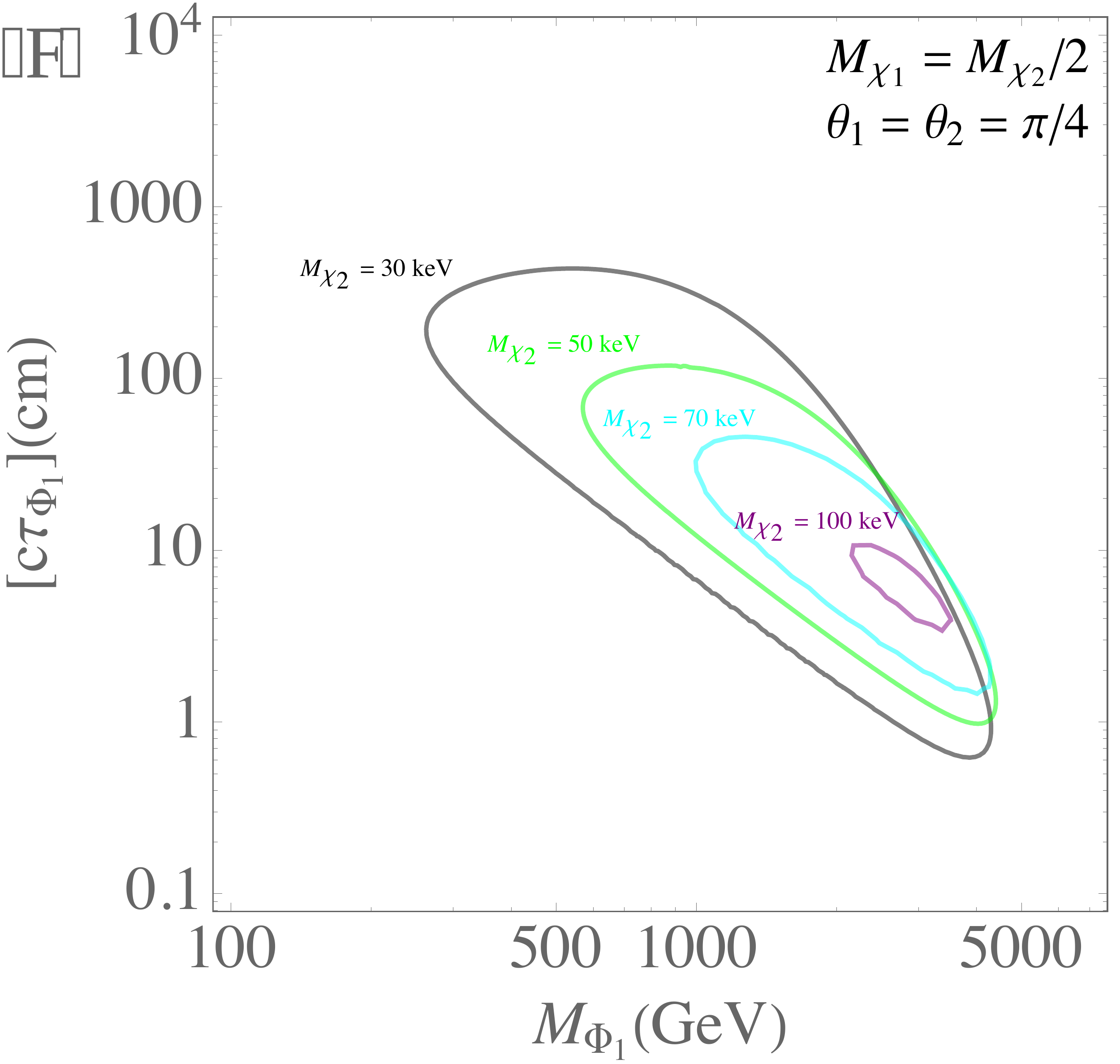}           
           \quad \quad \quad
\includegraphics[width=3in]{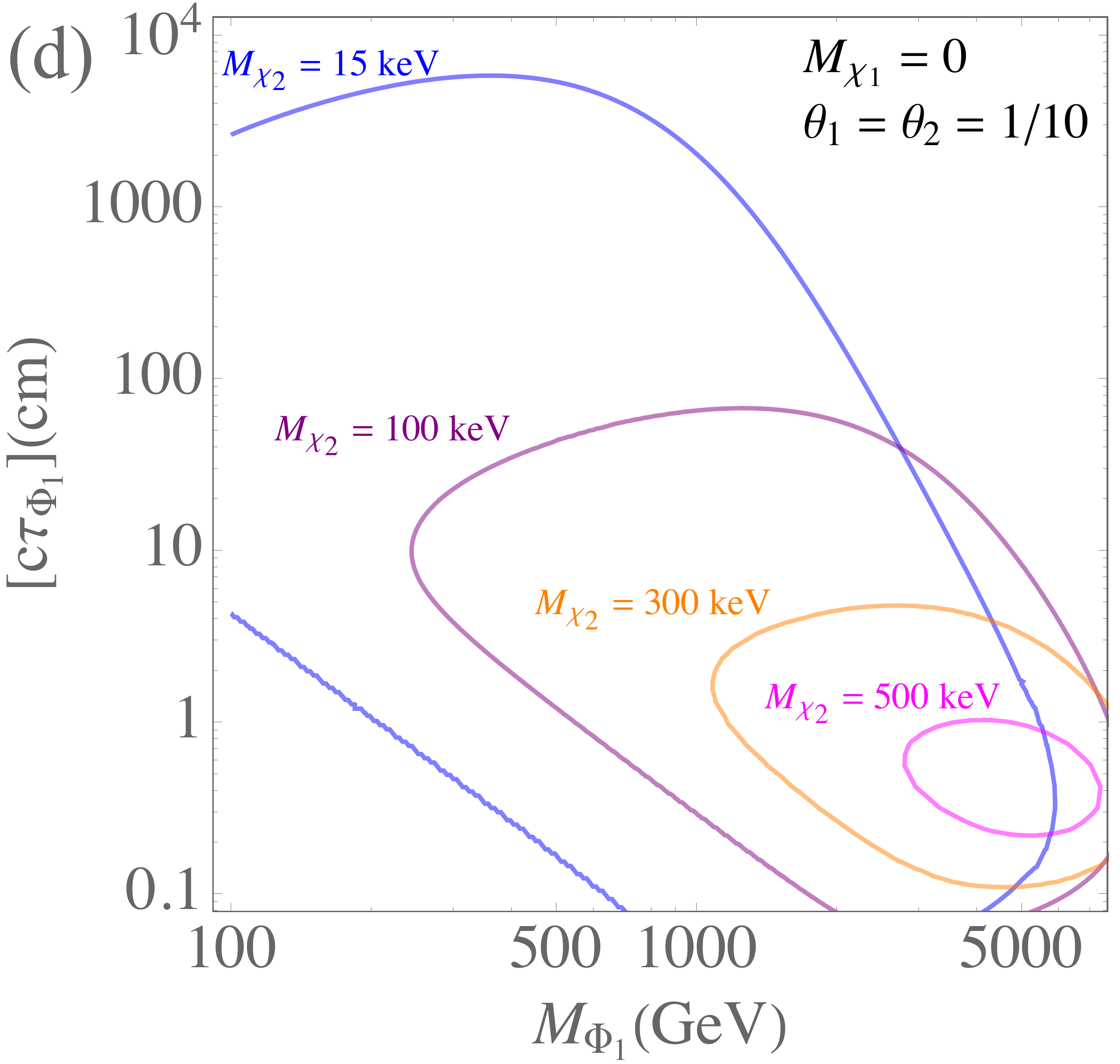}
  \caption{
The viable $(M_{\Phi_1}, c\tau_{\Phi_1})$ space for DM and leptogenesis in the UVDM Model, for various DM masses and mixings. The abundance of DM produced by $\Phi_2$ decays, $Y^\text{UV}_\chi$, is chosen to satisfy the DM constraint.  These contours are based on numerical solution of the quantum kinetic equations presented in Appendix~\ref{sec:MIQKEs}.
}     \label{fig:UVDM2}
\end{figure*}

The parameter space for DM and leptogenesis opens up significantly if we include an additional source of DM production  \cite{Asaka:2017rdj,Shuve:2020evk}.  For concreteness, we consider a model in which the DM couples to two scalars $\Phi_1$ and $\Phi_2$, both of which have the same SM quantum numbers as the $\Phi$ particle of the Minimal Model.  We focus on the case with $M_{\Phi_2} \gg M_{\Phi_1}$, so that $\Phi_2$ impacts the dark matter and leptogenesis calculations only through the coherent $\chi$ background its decays leave behind.   This ``decoupled-$\Phi_2$'' regime highlights the fact that the basic mechanism of asymmetry generation can work for any additional source of coherent DM production at high temperatures.  Our study of the UVDM Model in this section has close similarities to the analysis of the scenario with two QCD-charged scalars in Ref.~\cite{Shuve:2020evk}, although there are important differences as far as the associated collider phenomenology is concerned.  

In the UVDM, two interaction terms are relevant for DM production,
\beq\label{eq:LUVDM}
\mathcal{L} \supset -F^1_{\alpha i} e^c_\alpha \chi_i \Phi_1  -F^2_{\alpha i} e^c_\alpha \chi_i \Phi_2+ \text{h.c.}
\eeq
The cancelation of the flavor-summed asymmetry at $\mathcal{O}(F^4)$ is spoiled in the presence of the two coupling matrices $F^1$ and $F^2$.   Furthermore, for both the asymmetry and DM calculations, the dependence on $M_{\Phi_1}$ and $\Gamma_{\Phi_1}$ is different than for  $M_{\Phi}$ and $\Gamma_\Phi$ in the Minimal Model.  This is largely because the DM abundance produced by $\Phi_2$ decays is an additional free parameter, which we label as $Y_\chi^\text{UV}$ (defined to include both DM mass eigenstates but only one helicity state).  We also now have two separate mixing angles describing the relative production of the two DM mass eigenstates: $\theta_1$ controls the branching ratios of $\Phi_1$ to the lighter and heavier DM states, and $\theta_2$ does the same for $\Phi_2$.  As before, small $\theta$ means that decays to $\chi_1$ are favored over decays to $\chi_2$.

We present perturbative results for the UVDM Model  in Appendix~\ref{sec:UVDMpert}. Eqs.~(\ref{eq:DM2s}) and Eqs.~(\ref{eq:YB4_UVDM}) give the ${\mathcal O}(F^2)$ DM energy density and ${\mathcal O}(F^4)$ baryon asymmetry, respectively.    The $\mathcal{O}(F^4)$ asymmetry can be written as 
\begin{multline}\label{eq:YBnumUVDM_mainbody}
\frac{Y_B^{(4)}}{Y_B^\text{obs}} \simeq
(1.03 \times10^{5})
\;
\mathcal{J}
\;
\left(
\frac{Y^\text{UV}_{\chi}}{Y^\text{eq}_\chi}
\right)\\
\times
\left(
\frac{\Gamma_{\Phi_1}}{H_\text{ew}}
\right)
\left(
\frac{500 \text{ GeV}}{M_{\Phi_1}}
\right)^2
\;
\tilde{\mathcal{I}}^{(4)}(x_\text{ew},\beta_\text{osc}),
\end{multline}
where the $\tilde{\mathcal{I}}^{(4)}$ function is defined in Eq.~(\ref{eq:final_I4_fn_2}) and plotted in Fig.~\ref{fig:I4}(c), and we define $x_\text{ew} \equiv M_{\Phi_1}/T_\text{ew}$ in the context of the UVDM Model.  The factor ${\mathcal J}$ depends on the phases and mixing angles that determine the two DM coupling matrices $F^1$ and $F^2$; see Appendix~\ref{sec:UVDMFbench} for details.  We adopt an optimal benchmark for asymmetry generation with ${\mathcal J} =  \sin 2 \theta_1 \sin 2 \theta_2$, as realized for the coupling matrices given in Eq.~(\ref{eq:UVDMbench}).  

For selected DM masses and mixing angles, Figs.~\ref{fig:UVDM1} and~\ref{fig:UVDM2} show the viable mass-lifetime parameter space for $\Phi_1$.
At each point on the mass-lifetime plane, the DM energy density produced by $\Phi_1$ decays is determined, and the DM constraint is therefore satisfied by adjusting $Y_\chi^\text{UV}$.  In Fig.~\ref{fig:UVDM1}(a) for example, below and to the left of the diagonal line where the various $Y_B$ contours become very narrowly separated, the DM energy density produced by $\Phi_1$ exceeds $\rho_\text{dm}^\text{obs}$, and the DM constraint cannot be satisfied.

The dashed contours in Fig.~\ref{fig:UVDM1} are based on
 our perturbative results,  Eq.~(\ref{eq:YBnumUVDM_mainbody}) for $Y_B$  and Eq.~(\ref{eq:DM2s}) for $\rho_\text{dm}$, while the solid contours in both Figs.~\ref{fig:UVDM1} and~\ref{fig:UVDM2} are obtained by numerically solving the QKEs introduced in Appendix~\ref{sec:MIQKEs}, with  the initial condition taking into account the primordial abundance $Y_\chi^\text{UV}$.
 Fig.~\ref{fig:UVDM1} shows reasonable agreement between the perturbative and QKE calculations for $c\tau_{\Phi_1}\gtrsim 1$ cm;  the differences at long lifetimes are mainly due to the thermal ansatz adopted for the DM energy distribution in the QKE approach.

For $\theta_1 = \theta_2 = \pi/4$, the case of maximal mixing, $\chi_1$ and $\chi_2$ have equal overall coupling strengths and are produced with equal abundances.  In this case, $\Phi_1$ is long lived for most of the  parameter space that works for DM and leptogenesis, with $c \tau_{\Phi_1}$ as large as tens of meters, as Figs.~\ref{fig:UVDM1}(a,b) show.   Moreover, there is viable parameter space with comparable $\chi_1$ and $\chi_2$ masses, as we see in Fig.~\ref{fig:UVDM1}(b).  This is all in sharp contrast to the Minimal Model, where we needed $Y_{\chi_1} \gg Y_{\chi_2},\;M_1 \ll$ 1 keV, and $c \tau_\Phi \lesssim 1$ cm for DM and leptogenesis to work.  The UVDM parameter space extends out to larger $\Phi_1$ masses than colliders will probe in the near future, but the largest asymmetries, over 100 times the observed value, are realized for $M_{\Phi_1} \lesssim$ 1 TeV.  
 
Even larger $Y_B$ is  possible with small mixing ($\theta_1,\; \theta_2 \ll 1$), but only in the massless $\chi_1$ limit. This is illustrated in Figs.~\ref{fig:UVDM1}(c) and (d).   For $M_1 \ll 0.1$ keV,
the suppression of $Y_B$ from the small mixing angles is more than compensated by the larger overall size of DM couplings that match the observed DM energy density.  Comparing Figs.~\ref{fig:UVDM1}(a) and (c), we see that small mixing allows for shorter $\Phi_1$ lifetimes.  Indeed, for small enough 
$\theta_1$ and $\theta_2$, $c\tau_{\Phi_1}$ can be as small as in the Minimal Model.
When the $\chi_1$ and $\chi_2$ masses are comparable, on the other hand, nothing compensates for the suppression of $Y_B$ due to small $\theta_1$ and $\theta_2$, and the viable parameter space shrinks dramatically, as seen in Fig.~\ref{fig:UVDM1}(d). 

Fig.~\ref{fig:UVDM2} shows how the mass-lifetime parameter space shifts and contracts as $M_2$ is increased.  We take maximal mixing  for Figs.~\ref{fig:UVDM2}(a-c), with $M_1 = 0$, $M_1 = 15$ keV, and $M_1 = M_2/2$, respectively; the last of these three plots shows that in the UVDM Model, both DM masses can be well above those currently probed by Lyman-$\alpha$ forest observations.  Fig.~\ref{fig:UVDM2}(d) shows that even larger $\chi_2$ and $\Phi_1$ masses are possible in a small-mixing scenario with $\theta_1 = \theta_2 = 1/10$.  Here we restrict our attention to the massless-$\chi_1$ limit, because for these mixing angles and $M_1 \gsim 15$ keV, the small viable parameter space shown in Fig.~\ref{fig:UVDM1}(d) quickly disappears as $M_2$ is increased.

Figs.~\ref{fig:UVDM1}  and~\ref{fig:UVDM2} do not show the full extent of the UVDM mass-lifetime space one would find by maximizing the baryon asymmetry with respect to all other parameters.  It turns out that $\Phi_1$ masses of up to $\sim 20$ TeV  can work for DM and leptogenesis, but only in the massless-$\chi_1$ limit, and with $M_2$  and all mixing angles and phases tuned appropriately.  The required optimization is similar to what was presented in Appendix B of Ref.~\cite{Shuve:2020evk} for the case with QCD-charged BSM scalars.  Our results here for the $e^c$-coupled $\Phi$ are consistent with the results of Ref.~\cite{Shuve:2020evk}, once one takes into account the different spectator-effect factors and different $\Phi$ gauge multiplicities.

In summary,  broad ranges of parameters work for DM and leptogenesis in the UVDM Model, including scenarios with $M_1 > 15$ keV and comparably  sized  $\chi_1$ and $\chi_2$ couplings.  Unlike in the Minimal Model, $\Phi_1$ has a lifetime $c\tau_{\Phi_1} > $ cm for much of the viable parameter space.  We discuss the implications for LHC searches in Section~\ref{sec:collider}.

\section{The Z2V Model}\label{sec:Z2V}
\subsection{Qualitative discussion}\label{sec:Z2Vqual}

In this section we do not impose the $Z_2$ symmetry that guarantees absolute DM stability in the Minimal and UVDM Models.  There are now two interaction terms that can induce changes in the $X_\alpha$ charge densities:~the $Z_2$-preserving DM coupling and a new ``Z2V'' interaction term:
\beq\label{eq:fullLintZ2Vsec}
\mathcal{L} \supset -F_{\alpha i} e^c_\alpha \chi_i \Phi 
- \frac{\lambda_{\alpha \beta}}{2} l_\alpha l_\beta \Phi^* + \text{h.c.}
\eeq
The Z2V term has the same gauge and flavor structure as an $R$-parity-violating coupling often considered in supersymmetric theories;  because the $SU(2)_w$ indices of the SM lepton doublets are contracted antisymmetrically, the $\lambda_{\alpha \beta}$ is an antisymmetric matrix in lepton flavor, and there are therefore three independent Z2V couplings. 
We neglect a possible neutrino-portal coupling  $h_{\alpha i} l_\alpha \chi_i H$ because, as we discuss in Sec.~\ref{sec:xray}, X-ray line constraints prevent them from being large enough to be relevant for leptogenesis.  The $\lambda$ couplings, on the other hand, can significantly impact the asymmetry calculation.  The rough criterion for a Z2V coupling $\lambda$ to come into equilibrium in the early universe is that the $\lambda$-induced $\Phi$ decay width  should be at least comparable to  the Hubble parameter at $T = M_\Phi$, leading to
\be
\lambda \gtrsim 10^{-7} \times \left( \frac{M_\Phi}{500 \text{ GeV}}\right)^{1/2}.
\ee
This is a far smaller coupling than has been probed experimentally.  The X-ray line constraints on the $\lambda$ couplings depend on the lepton flavors involved, but we will show in Sec.~\ref{sec:xray} that they are rarely stronger than $\lambda < 10^{-4}$, and are often much weaker, for the parameter space that works for leptogenesis and DM.   Other constraints on $\lambda$ from low-energy experiments are never more stringent than $\lambda \lsim 10^{-2}$, as we discuss in Sec.~\ref{sec:otherZ2V}. 

To see how Z2V couplings might be relevant for leptogenesis, consider for simplicity a scenario with  $e^c$ and $\mu^c$  (but not $\tau^c$) coupled to DM, and  a single Z2V interaction involving $l_e$ and $l_\tau$:  $\mathcal{L} \supset -\lambda l_e l_\tau \Phi^{*} + \text{h.c.}$  At $\mathcal{O}(F^4)$, there are no asymmetries in $\Phi$, $\chi$, or the third-generation leptons,  but we have equal and opposite asymmetries in $e^c$ and $\mu^c$, and thanks to SM processes, equal and opposite asymmetries in $l_e$ and $l_\mu$.      Then the number of $l_e l_\tau \rightarrow \Phi$ inverse decays will differ from the number of 
${\overline l}_e {\overline l}_\tau \rightarrow \Phi^{*}$ inverse decays, and a $\Phi$ asymmetry is generated at $\mathcal{O}(F^4 \lambda^2)$.  That is enough to guarantee a net $B-L$ charge in the SM sector at the same order.   
To see this, note that the interactions of Eq.~(\ref{eq:fullLintZ2Vsec}) respect a generalized $B-L$  symmetry, with $L(\Phi) = 2$, $L(\chi) = -1$, and $B(\Phi) = B(\chi) = 0$ \footnote{The DM Majorana masses do not respect this symmetry, but the associated effects come with $\sim M_\chi^2/T^2$ suppressions and can be neglected.}.  No $\chi$ asymmetry arises at $\mathcal{O}(F^4)$ and, because the $\lambda$ coupling does not involve DM, the $\chi$ asymmetry remains zero at $\mathcal{O}(F^4 \lambda^2)$.   At this order, then, the SM sector has a  $B-L$ charge equal in magnitude to that stored in $\Phi/\Phi^*$. 

Two simple observations turn out to have important implications for the viable Z2V Model parameter space.  First,  larger Z2V couplings can produce a larger asymmetry without  increased DM production.  Second, even if a $\lambda$ coupling is large enough to invalidate the perturbative  $\mathcal{O}(F^4 \lambda^2)$ calculation of the asymmetry, that $\lambda$ coupling does not necessarily lead to washout of the asymmetry.  To illustrate the second point, we return to the simple example of the previous paragraph and consider the case in which $\lambda$ is large enough that it comes fully into equilibrium.  Imagine that  equal and opposite   number density asymmetries for $l_e$ and $l_\mu$ are first generated at $\mathcal{O}(F^4)$ (that is,  $\delta n^{(4)}_{l_e} = - \delta n^{(4)}_{l_\mu}$)  and subsequently processed by $\lambda$,  neglecting for simplicity spectator effects associated with SM processes.  By conservation of the generalized $B-L$, we have
\beq
\sum_\alpha\delta n_{l_\alpha} + 2 \delta n_\Phi  =   0,
\label{eq:BmLZ2V}
\eeq
where we set the $\chi$ asymmetry to zero because it arises only at higher order in $F$.  
With $\lambda$ in equilibrium we also have the chemical potential relation
\beq
\mu_{l_e} + \mu_{l_\tau} - \mu_\Phi  =  0.
\label{eq:eqZ2V}
\eeq
We can solve this pair of equations for $\delta n_{l_e}+\delta n_{l_\tau}$, taking $\delta n_{l_\mu} = \delta n^{(4)}_{l_\mu} = -\delta n^{(4)}_{l_e}$, because $l_\mu$ is not involved in the Z2V coupling.  This leads to a flavor-summed asymmetry
\be\label{eq:asymZ2V}
\sum_\alpha\delta n_{l_\alpha} \simeq 
-\delta n^{(4)}_{l_e}
\times
\begin{cases}
2/3\quad \quad \quad \quad \quad \quad\quad \quad T \gg M_{\Phi}\\
12 \left(\frac{M_\Phi}{2\pi T} \right)^{3/2}e^{-M_{\Phi}/T} \;\;\; T \ll M_{\Phi},
\end{cases}
\quad
\ee
where we use the relativistic or non-relativistic relation between $\mu_\Phi$ and $\delta n_\Phi$ depending on the temperature.  We see that the flavor-summed asymmetry effectively arises at $\mathcal{O}(F^4)$ when $\lambda$ comes into equilibrium. It makes sense that the $\mathcal{O}(F^4)$ asymmetry is Boltzmann suppressed at temperatures far below the $\Phi$ mass.  The $B-L$ charge in the SM sector is equal in magnitude to the sum of  the $B-L$ charge stored in $\chi/{\overline \chi}$, which is zero at $\mathcal{O}(F^4)$, and  the $B-L$ charge stored in $\Phi/\Phi^{*}$, which has essentially decayed away for  $T \ll \Phi$.   For sufficiently large $M_\Phi$,  the dominant contribution to the asymmetry arises at $\mathcal{O}(F^6)$.  As explained in Appendix~\ref{sec:YB6Z2V}, this contribution is distinct from the $\mathcal{O}(F^6)$ ARS one.  It arises from the generation of a $\Phi$ asymmetry at $\mathcal{O}(F^4)$, followed by generation of an $\mathcal{O}(F^6)$ $\chi$ asymmetry due to $\Phi^{(*)}$ decays to DM.

 In our simple example with only $e^c$ and $\mu^c$ coupled to DM, the asymmetry would be driven to zero if the Z2V coupling that came into equilibrium were  $\lambda_{12}$  instead of $\lambda_{13}$.  In that case, no asymmetry develops in the third-generation leptons, 
and Eq.~(\ref{eq:eqZ2V}) is replaced with $\mu_{l_e} + \mu_{l_\mu} - \mu_\Phi = 0$, which when combined with  Eq.~(\ref{eq:BmLZ2V}) forces all chemical potentials to zero.  More generally, for a generic $F$ matrix involving all active flavors, we should expect an $\mathcal{O}(F^4)$ asymmetry if one or two of the three Z2V couplings come into equilibrium.  If all three Z2V couplings come into equilibrium, then the chemical potentials of all three leptons are forced to be flavor-independent, preventing any asymmetry from being generated at all\footnote{This is true for the case of two DM mass eigenstates.  For three or more DM mass eigenstates, an asymmetry can arise at $\mathcal{O}(F^6)$ even if all active flavors have the same chemical potential \cite{Abada:2018oly};  this is the case in a model considered in Ref.~\cite{Shuve:2020evk} involving a single QCD-charged scalar.}.

 In our full calculation of the asymmetry, described in the following section,  we take into account that the $\lambda$ interactions continually process asymmetries as they are generated, and that SM spectator processes affect how the various asymmetries are related \cite{Shuve:2014zua}.

\begin{figure}
          \includegraphics[width=3.5in]{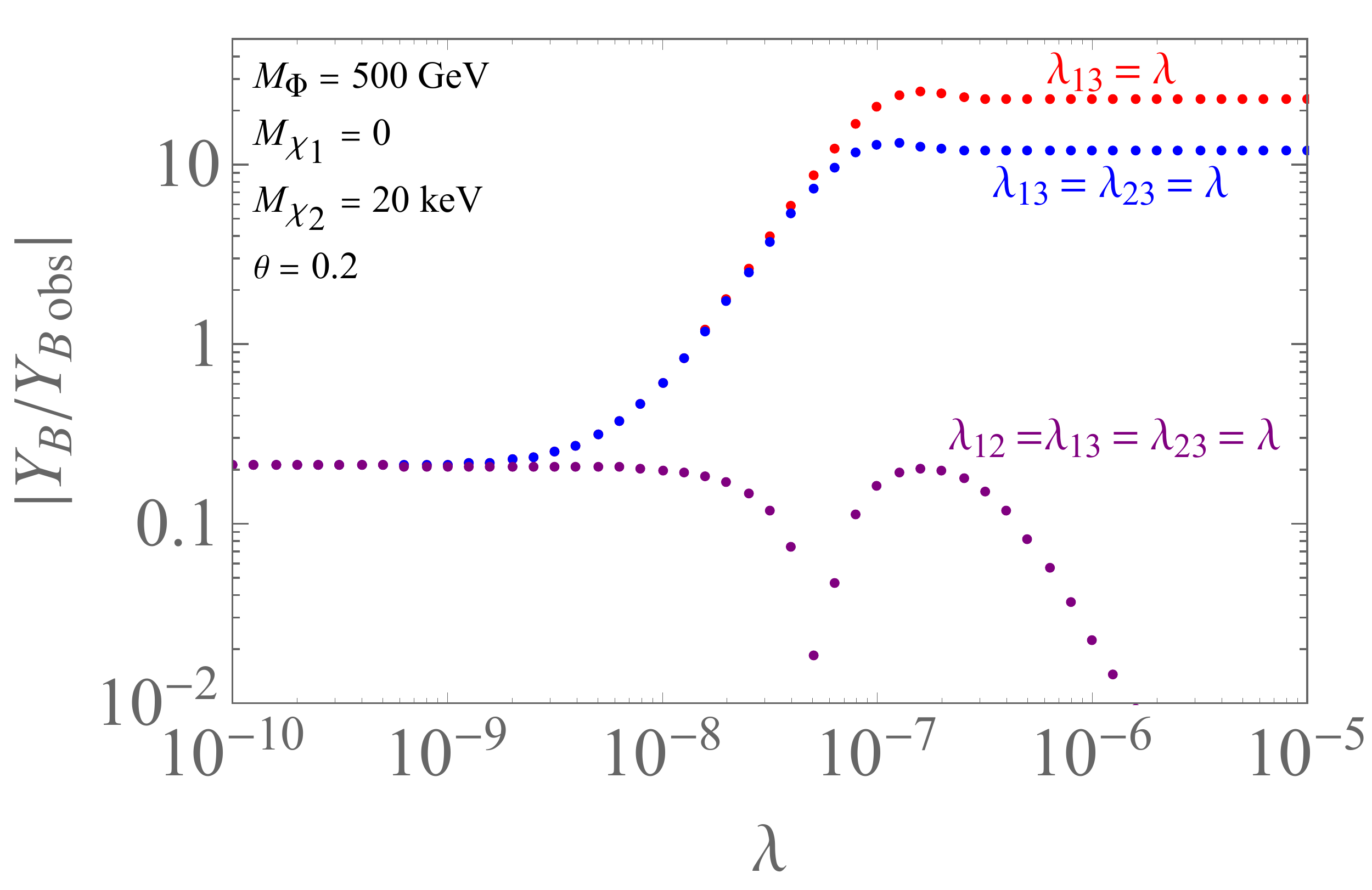}
   \caption{For a particular set of BSM particle masses and DM couplings, the baryon asymmetry as a function of Z2V coupling strength, with either one, two or all three independent Z2V couplings turned on.  The DM couplings are set to the Minimal Model benchmark,  Eq.~(\ref{eq:FMMbench}), with only the electron and muon coupling to DM, and with ${\rm Tr} F^\dagger F$ determined by the DM abundance constraint.
      }
   \label{fig:Z2Vequilibrate}
\end{figure}
\begin{figure*}
    \includegraphics[width=3in]{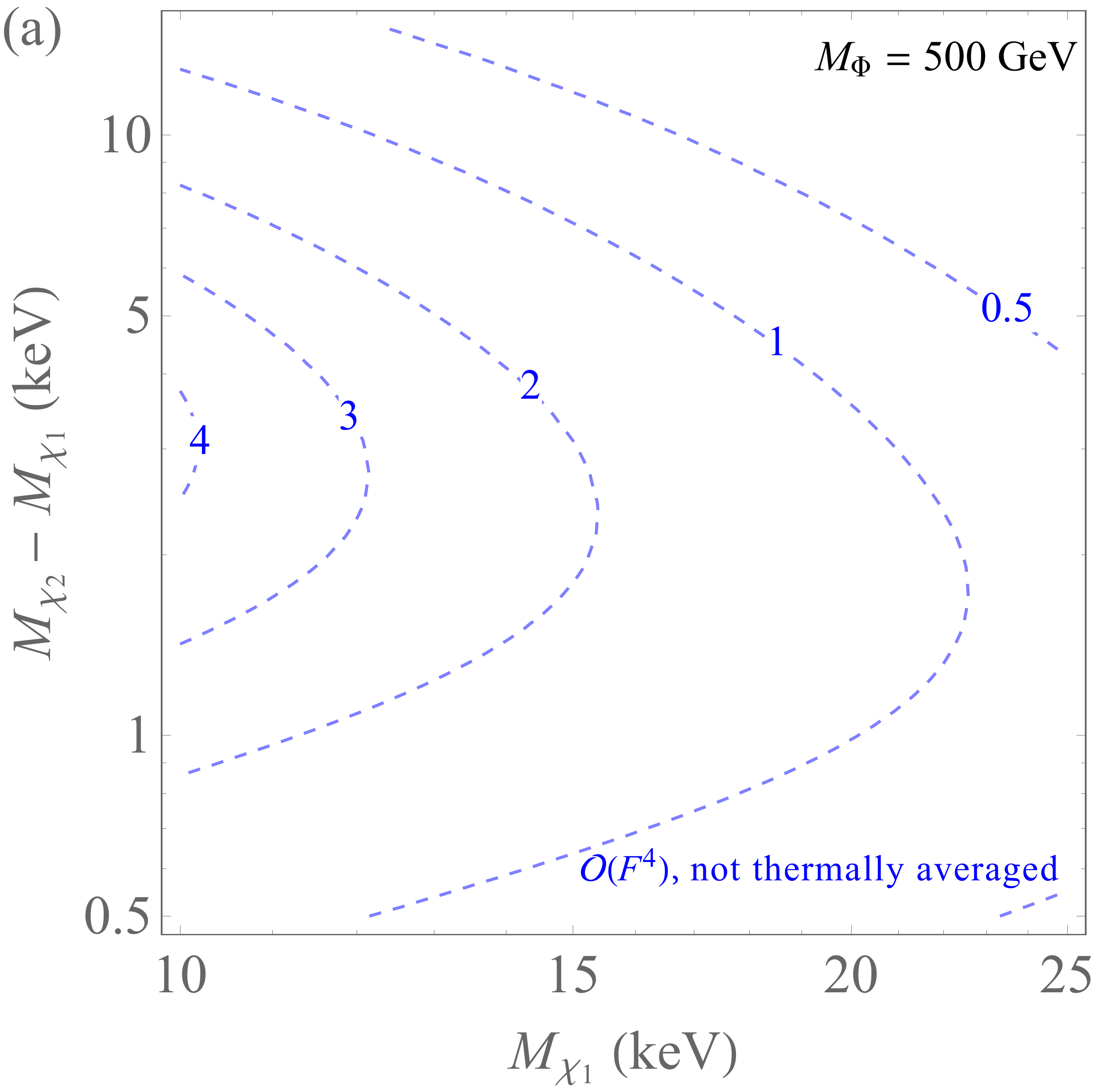}
           \quad 
          \includegraphics[width=3in]{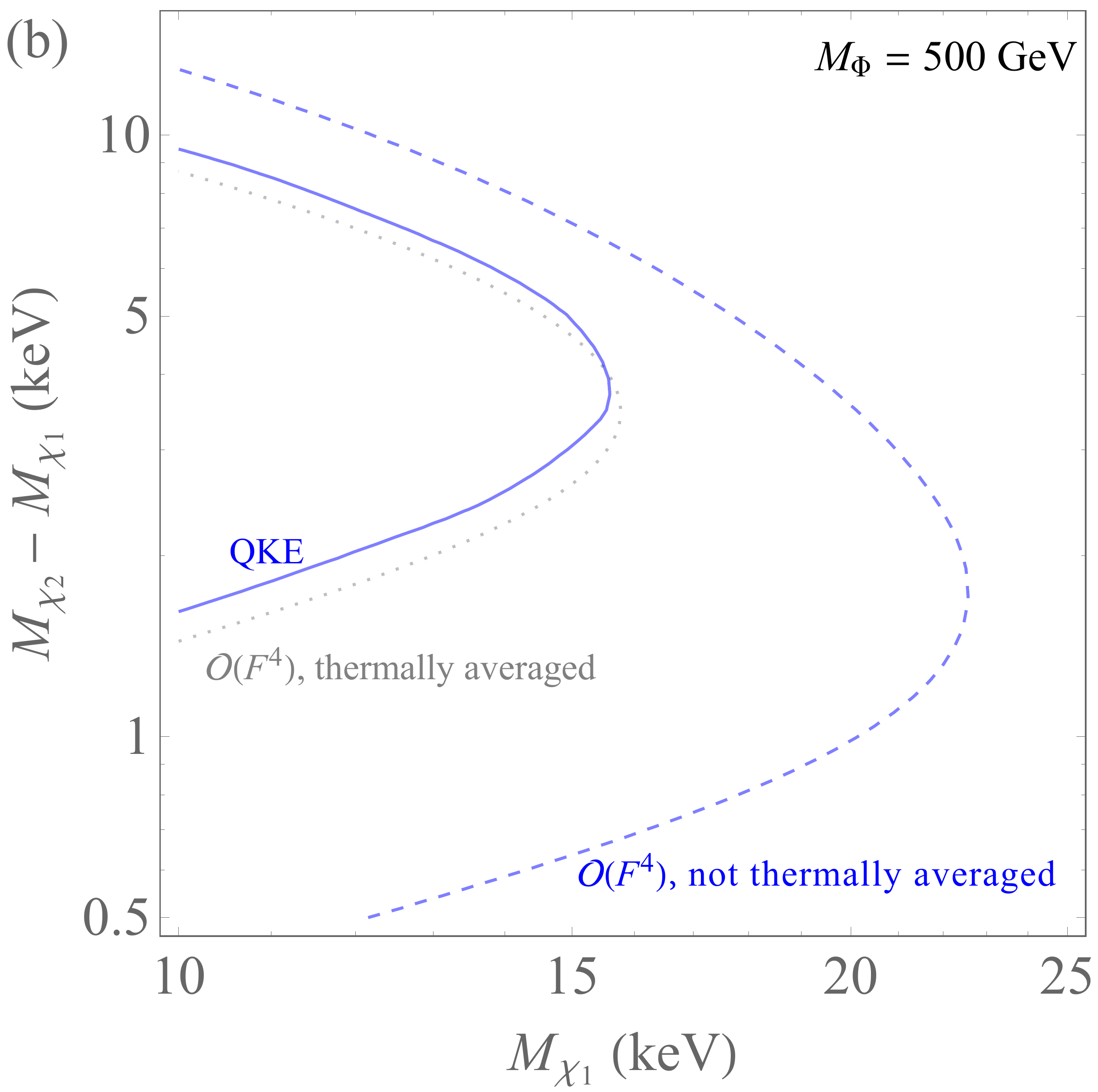}
   \caption{Baryon asymmetries in the larger-$M_1$ regime of the Z2V Model, with the $F$ matrix set to the Z2V benchmark form of Eq.~(\ref{eq:FZ2Vbench}).  We take $M_\Phi = 500$ GeV,  which roughly maximizes the range of viable $M_1$ values.    At each point in the plane, ${\rm Tr} \left[ F^\dagger F \right]$ and $\theta$ are chosen  to maximize $Y_B$ subject to the DM abundance constraint.  In (a), we show contours of $Y_B/Y_B^\text{obs}$, using the $\mathcal{O}(F^4)$ perturbative calculation of Eq.~(\ref{eq:YB4Z2V_mainbody1}), without thermal averaging.  In (b) we compare $Y_B=Y_B^\text{obs}$ contours based on the non-thermally-averaged perturbative calculation, the $\mathcal{O}(F^4)$ pertrubative calculation with thermal averaging (see the discussion leading to Eq.~(\ref{eq:I4ta})), and  numerical solution of the QKEs (see Appendix~\ref{sec:MIQKEs}).   }
   \label{fig:Z2V1}
\end{figure*}

\subsection{Calculating $Y_B$ in the Z2V Model}\label{sec:Z2Vquant}

In the QKEs of Appendix~\ref{sec:MIQKEs}, the Z2V interactions do not enter into the evolution equations for the DM density matrices $Y_\chi$ and $Y_{\overline \chi}$.  However, each Z2V interaction brings about a change in the $X_\alpha$ charges, so the evolution equations for the associated charge densities $Y_\alpha$ do get modified.  For example, if we have a single Z2V coupling $\lambda_{12}$, then each Z2V-induced $\Phi$ decay, $\Phi \rightarrow l_1 l_2$, produces the changes $\Delta X_1 = \Delta X_2 = -1$.  As shown in Eq.~(\ref{eq:YalphaQKE}),  both $dY_1/dt$ and $d Y_2/dt$  get contributions proportional to $|\lambda_{12}|^2 (\mu_{l_1} + \mu_{l_2}-\mu_\Phi)$ at leading order.    The chemical potentials $\mu_\Phi$ and $\mu_{l_\alpha}$ can be expressed in terms of $Y_\chi$, $Y_{\overline \chi}$, and $Y_\alpha$ using the results of Appendix~\ref{sec:appendixCPs} 
\footnote{The proportionality to $(\mu_{l_1} + \mu_{l_2}-\mu_\Phi)$ holds if $|\mu_\Phi/T| \ll 1$ is satisfied.   In our numerical work we allow for the possibility of highly asymmetric $\Phi/\Phi^{(*)}$ abundances using the approach described in Appendix~\ref{sec:appendixQKE}; see Eq.~(\ref{eq:don't_linearize_mu_phi}). This refinement is quantitatively unimportant in almost all  scenarios we study in this paper, however.}.

Fig.~\ref{fig:Z2Vequilibrate} shows the effect on the baryon asymmetry when one or more Z2V couplings are large enough to come into equilibrium.   For the chosen parameter point, $e^c$ and $\mu^c$ couple to DM, and a $Y_B$ value well below the observed one arises in the absence of Z2V couplings.  If a single Z2V coupling, $\lambda_{13}$, is large enough to come into equilibrium, $\lambda_{13} \gsim 10^{-7}$, the final baryon asymmetry is enhanced by 
about two orders of magnitude relative to the $Z_2$-preserving case.   If both $\lambda_{13}$ and $\lambda_{23}$ come into equilibrium, the final baryon asymmetry $Y_B$ is again enhanced by a large factor, although only about half as large as for the case with a single Z2V coupling.  Finally, if all three Z2V come into equilibrium, the asymmetry is strongly suppressed, as expected.  For the parameters chosen, the asymmetry happens to change sign as $\lambda$ is increased in this three-$\lambda$ case, because the $\mathcal{O}(F^6)$ and $\mathcal{O}(F^4 \lambda^2)$ contributions have opposite sign.

If one or two of the Z2V couplings come into equilibrium, the final baryon asymmetry is insensitive to the precise values of those couplings, as is evident in the $Y_B$ plateaus at larger $\lambda$ in Fig.~\ref{fig:Z2Vequilibrate}.  In this case, $Y_B$ effectively arises at $\mathcal{O}(F^4)$.  To write down an expression for $Y_B^{(4)}$, it is convenient to define $\beta$ to be the ``special'' flavor singled out by the Z2V couplings:~in the case of a single Z2V coupling in equilibrium, $\beta$ is the  flavor whose lepton doublet is not involved in that $Z_2$-violating coupling, while in the case of two Z2V couplings in equilibrium, it is the flavor whose lepton doublet is involved in both of those couplings.   In Appendix~\ref{sec:Z2Vpert} we show that $Y_B^{(4)}$ is determined by $Y_\beta^{(4)}$, the $\mathcal{O}(F^4)$ asymmetry  in the special flavor $\beta$, calculated in the {\em absence} of Z2V couplings, {\em i.e.} using the Minimal Model result of Eq.~(\ref{eq:Ya4}).  For example, if only $\lambda_{12}$ comes into equilibrium, we have $\beta = 3$, and we get a baryon asymmetry at $\mathcal{O}(F^4)$ if and only if one would calculate an $\mathcal{O}(F^4)$ asymmetry in $X_3$ in the absence of the Z2V coupling.

More precisely, we find 
\be\label{eq:YB4Z2V_mainbody1}
Y_B^{(4)}
&
\simeq
&
Y_\beta^{(4)}\times
 \frac{300 \;c_\Phi}{237+766 \;c_\Phi}  \;\;\;\quad\text{  one }\lambda \text{ in eq.}\\
 \label{eq:YB4Z2V_mainbody2}
 Y_B^{(4)}
&
\simeq
&
-Y_\beta^{(4)}\times
  \frac{150 \;c_\Phi}{237+529 \;c_\Phi}  \quad\text{  two }\lambda \text{'s in eq.},
\ee
where the $c_\Phi$ function is defined in Eqn.~(\ref{eq:cphi}) and plotted in Fig.~\ref{fig:cphi}.  The exponential suppression of $c_\Phi(x)$  at large $x$ leads to a Boltzmann-suppressed $Y_B^{(4)}$  for $T\ll M_\Phi$, as we anticipated in the discussion surrounding Eq.~(\ref{eq:asymZ2V}).  We evaluate $c_\Phi(x)$ at $x_\text{ew} \equiv M_\Phi/T_\text{ew}$ in Eqs.~(\ref{eq:YB4Z2V_mainbody1}, \ref{eq:YB4Z2V_mainbody2}) to get the final asymmetry.    For our numerical studies in the rest of this section, we adopt the Z2V benchmark $F$ matrix of Eq.~(\ref{eq:FZ2Vbench}) and take one $\lambda$ coupling in equilibrium, such that the $\mathcal{O}(F^4)$ asymmetry is maximized.

For Eq.~(\ref{eq:YB4Z2V_mainbody1}) or (\ref{eq:YB4Z2V_mainbody2}) to be a good approximation for the baryon asymmetry, three conditions must be satisfied.  First, one or two $\lambda$ couplings must be large enough to come into equilibrium, $\gsim 10^{-7}$ while the remaining $\lambda$ coupling(s) are small enough to remain well out of equilibrium, $\lesssim 10^{-8}$. Second, we need the $F$ couplings to be small enough to justify a perturbative treatment of the $F$ couplings, which means $Y_\chi \ll Y_{\chi}^\text{eq}$.  Third, the $\mathcal{O}(F^6)$ contributions to the asymmetry must be subdominant, which is not the case if the  $\mathcal{O}(F^4)$ asymmetry  is strongly Boltzmann suppressed.

In the large-$M_\Phi$ regime,  $Y_B$ is dominated by the $\mathcal{O}(F^6)$ contribution discussed earlier,  given in Eq.~(\ref{eq:YB6Z2Vbench}) for our Z2V benchmark $F$ matrix.   The  $\mathcal{O}(F^4)$ $\Phi$ asymmetry  leads to an $\mathcal{O}(F^6)$  $\chi$ asymmetry that persists after the $\Phi$ particles have disappeared, ensuring a surviving baryon asymmetry. The interplay between the $\mathcal{O}(F^4)$ and $\mathcal{O}(F^6)$ asymmetries will be evident in the results we present  in the following section.

\begin{figure*}
          \includegraphics[width=2.3in]{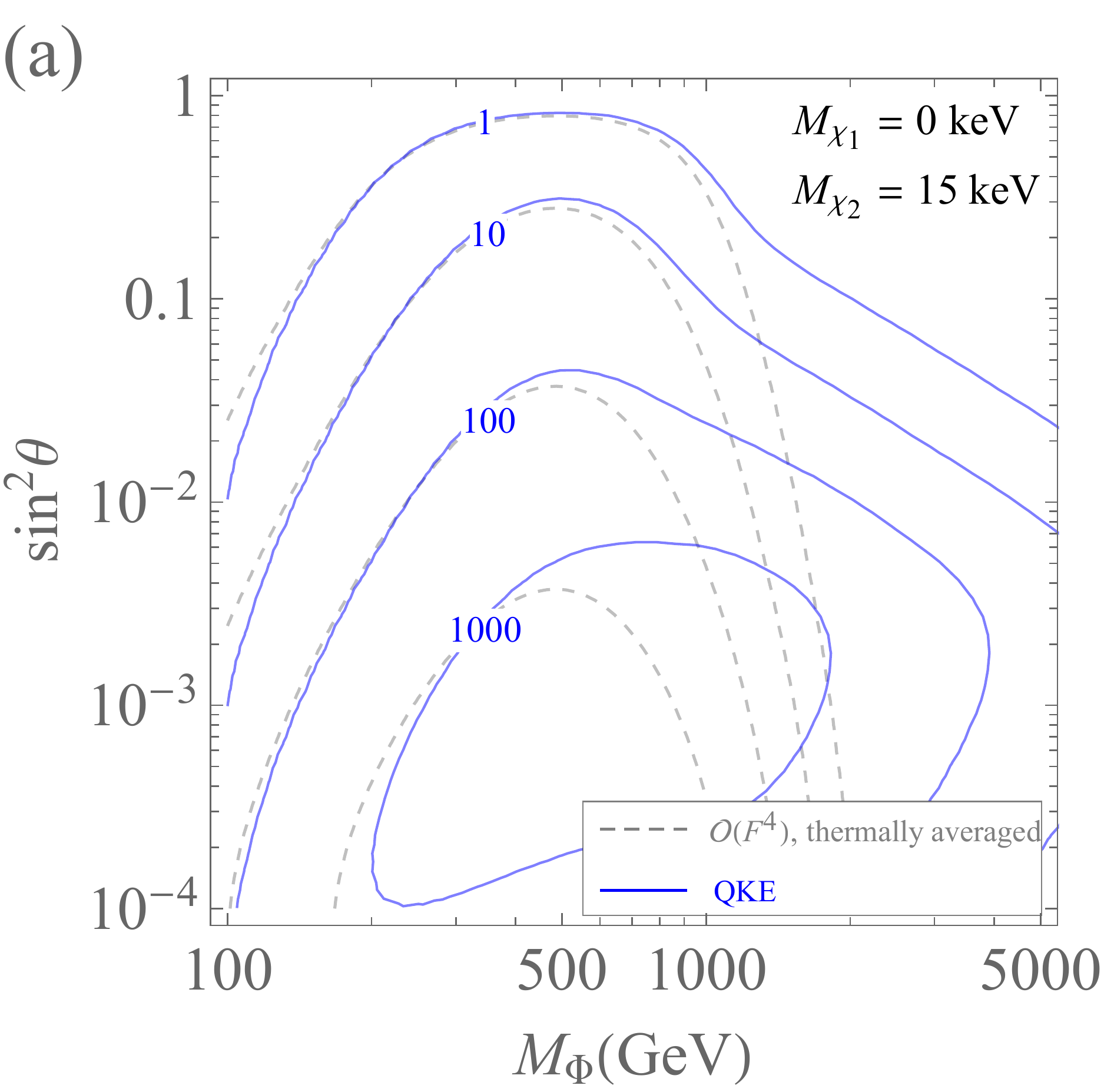}
             \includegraphics[width=2.3in]{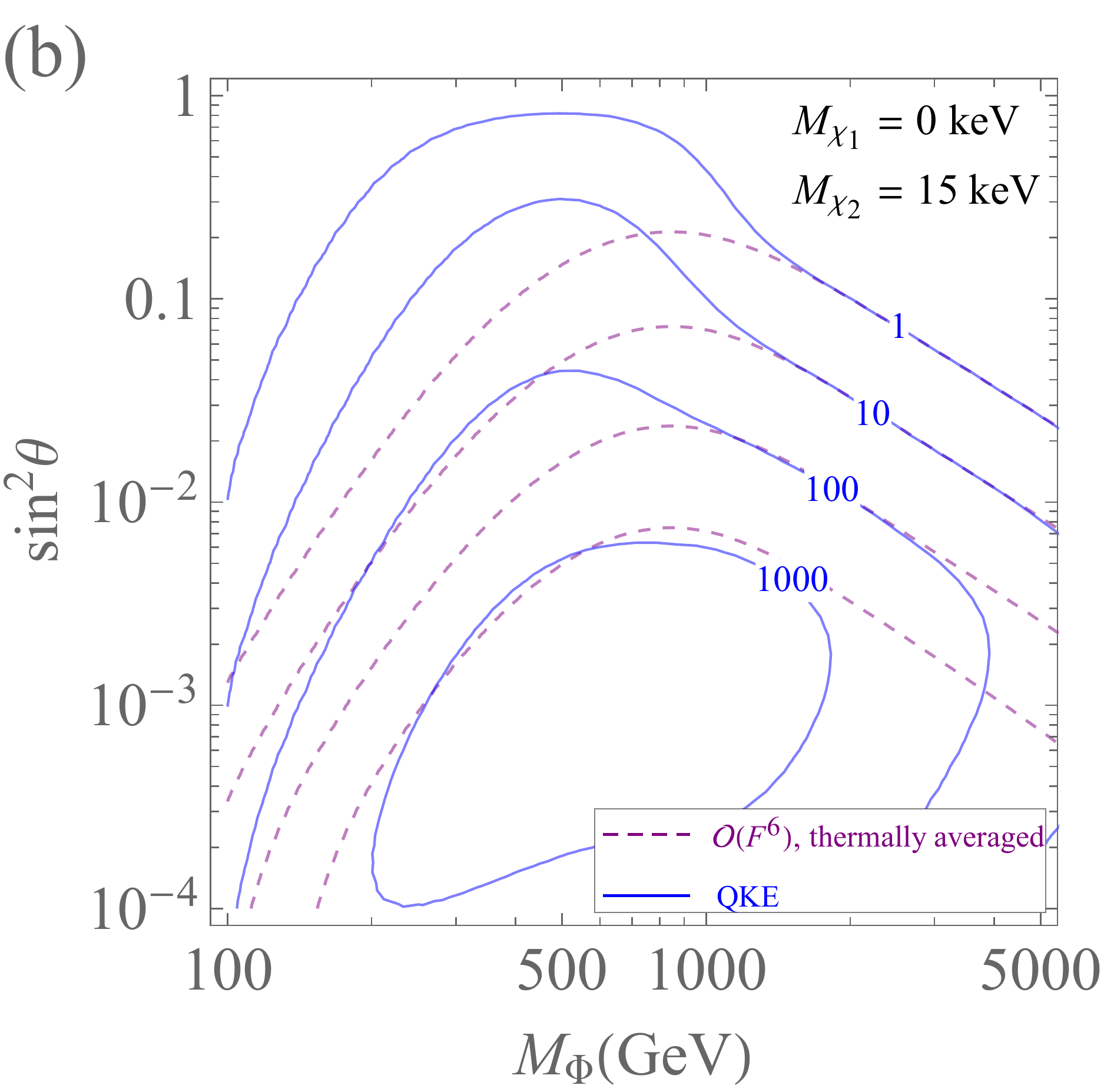}
   \includegraphics[width=2.3in]{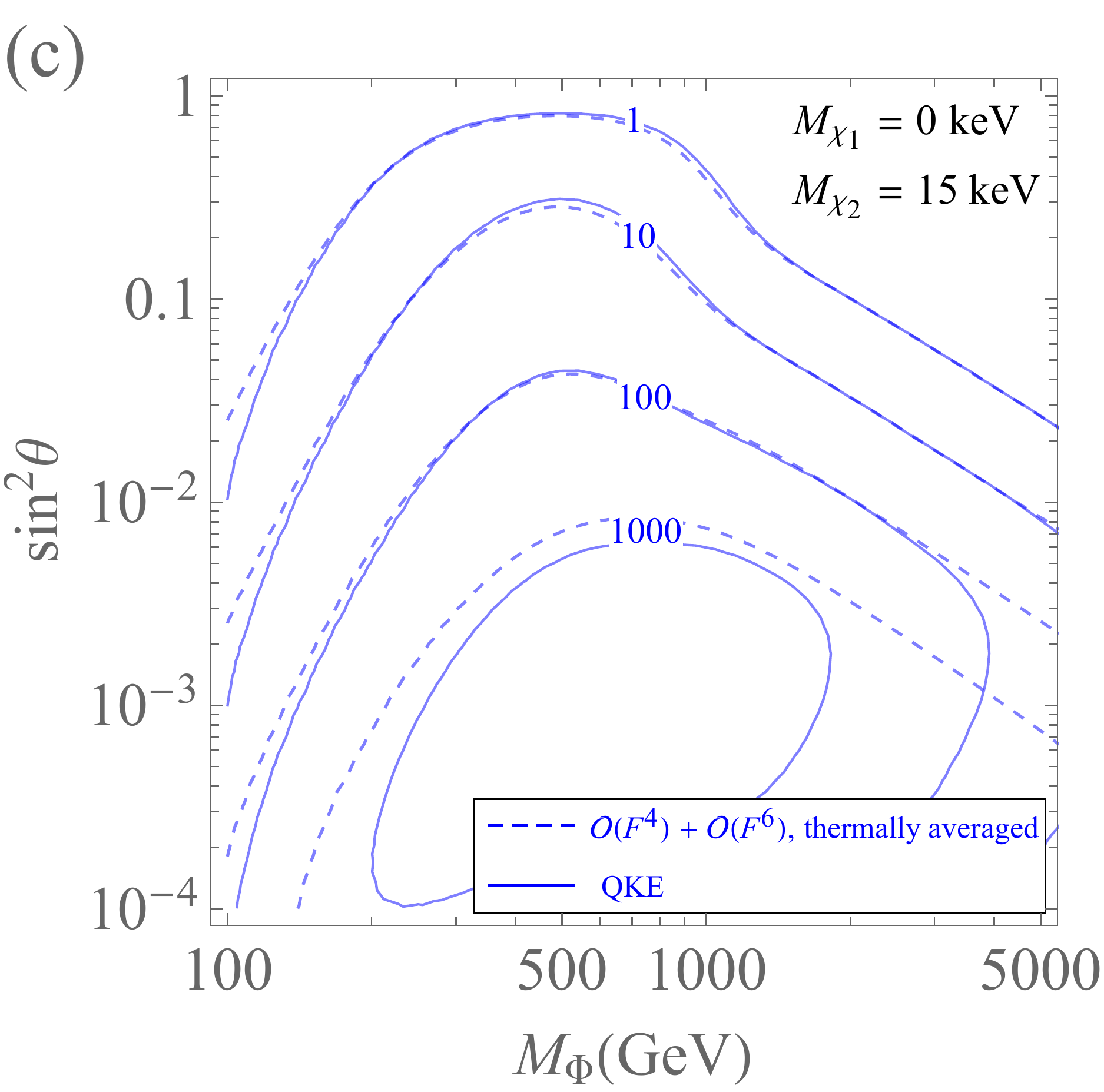}
   \\
          \includegraphics[width=2.3in]{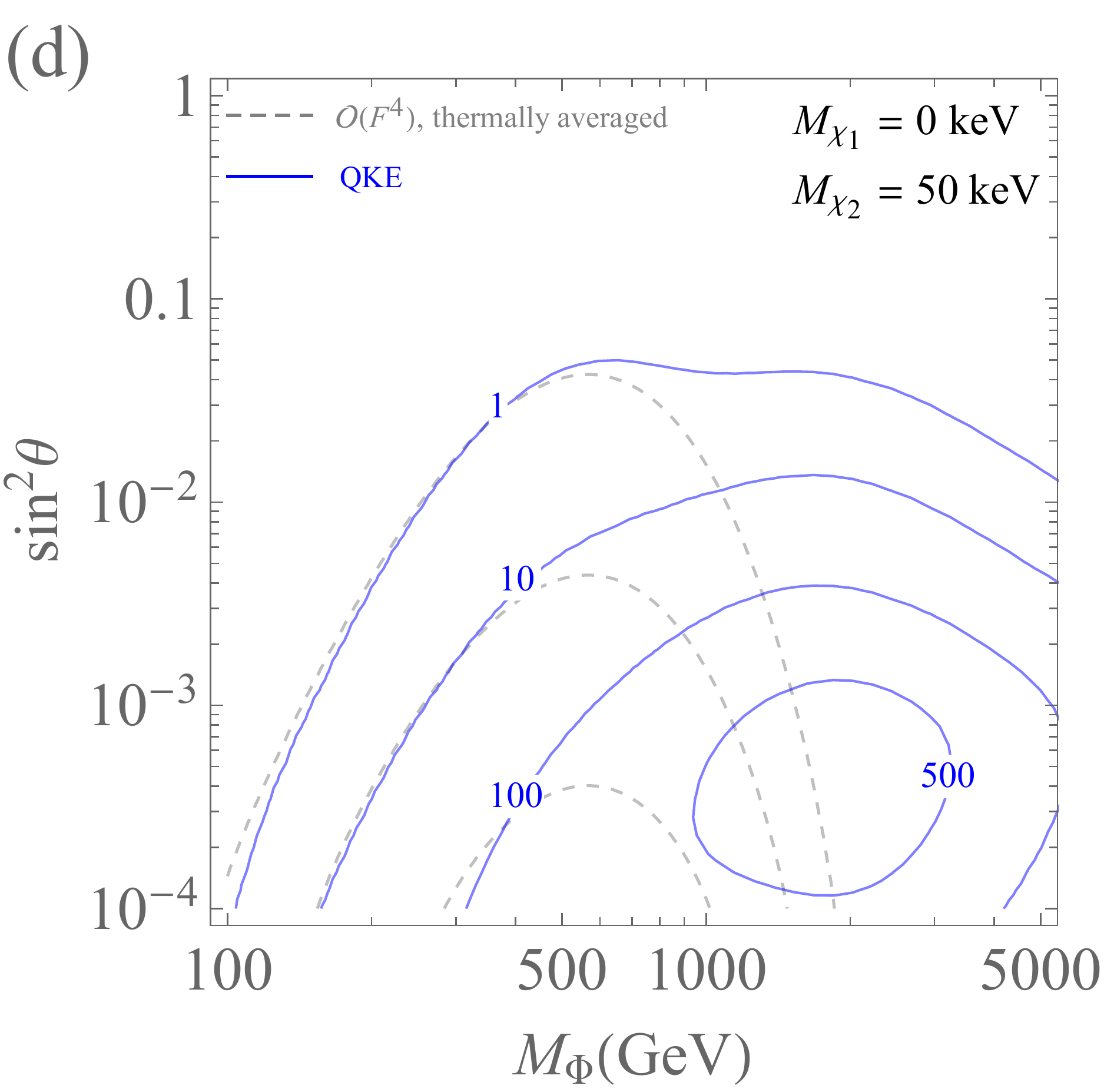}
             \includegraphics[width=2.3in]{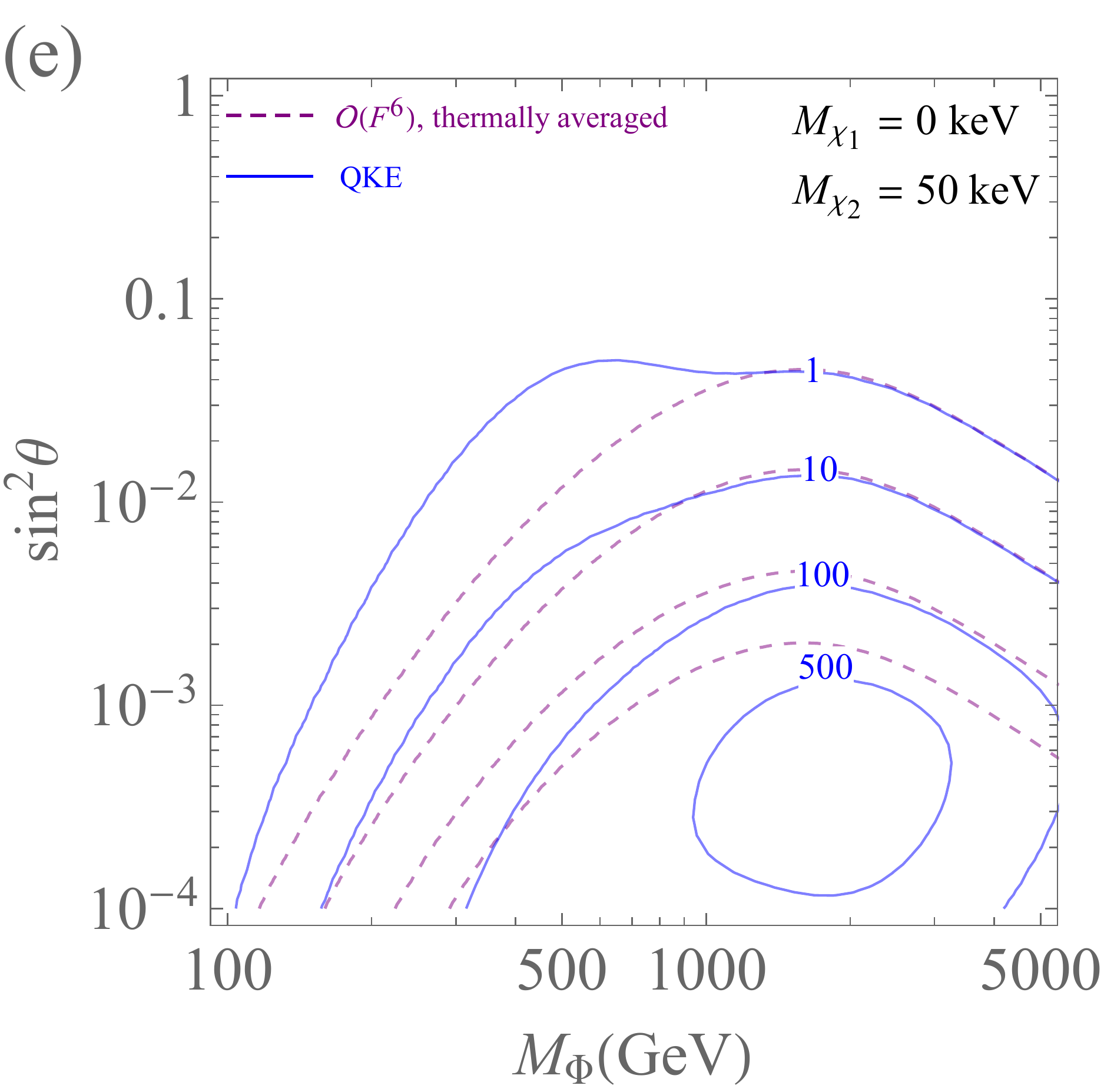}
   \includegraphics[width=2.3in]{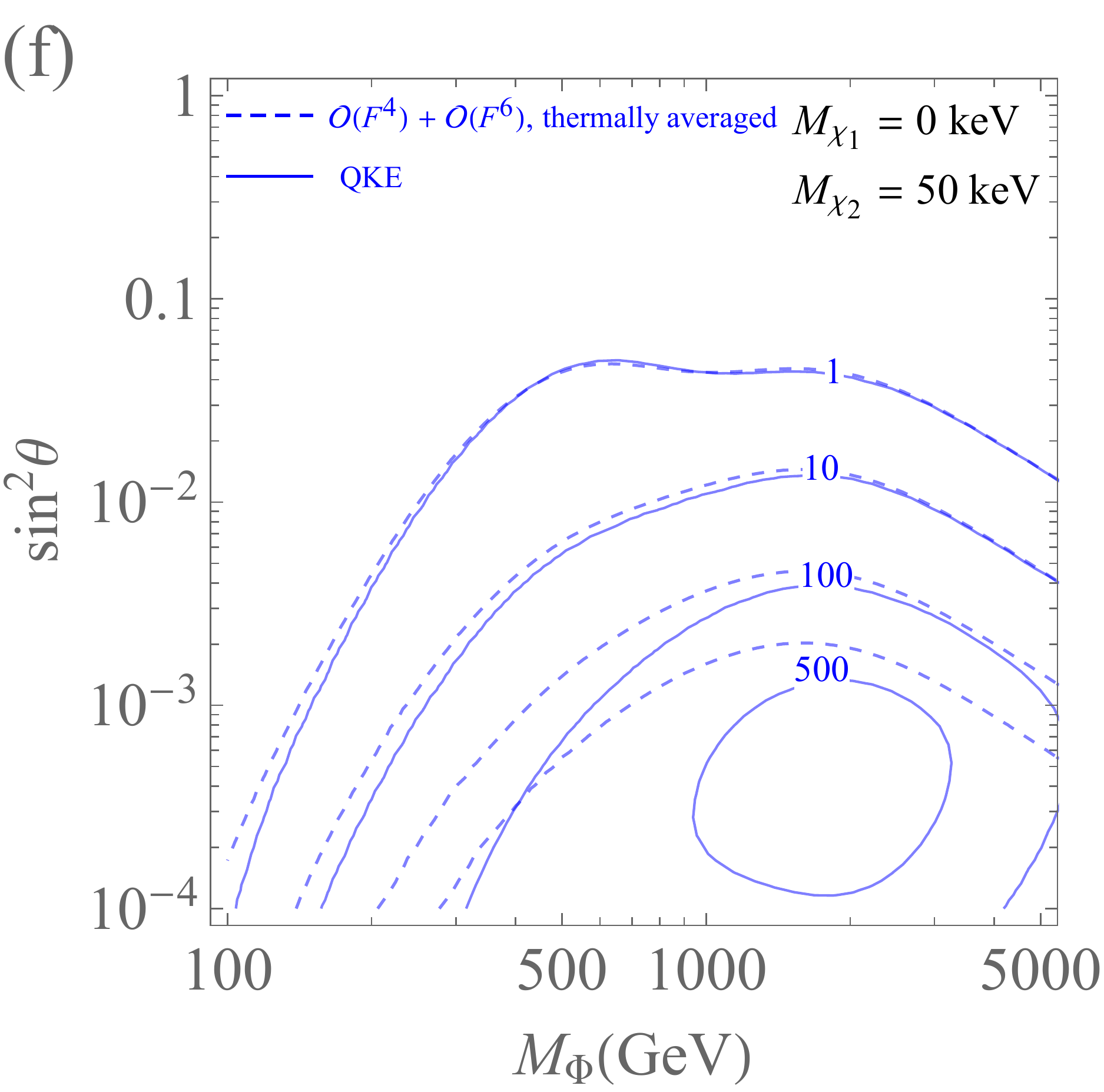}
   \caption{In the Z2V Model, contours of $Y_B/Y_B^\text{obs}$ in the $(M_\Phi,\; \sin^2\theta)$ plane for $M_2 = 15$ keV (a-c) and $M_2 = 50$ keV (d-f),  in the massless-$\chi_1$ limit, with the $F$ matrix set to the Z2V benchmark form of Eq.~(\ref{eq:FZ2Vbench}).  We compare the results from numerical integration of the QKEs with the $\mathcal{O}(F^4)$ contribution of Eq.~(\ref{eq:YB4Z2V_mainbody1}) (a, d), the $\mathcal{O}(F^6)$ contribution  of Eq.~(\ref{eq:YB6Z2Vbench}) (b, e), and the sum of both (c, f).  For the perturbative contributions, we adopt the same thermal ansatz for the DM momentum distribution as used to derive the QKEs, which means using the $\mathcal{I}^{(4)}_\text{t.a.}$ function defined in Eq.~(\ref{eq:I4ta}) in place of $\mathcal{I}^{(4)}$ in Eqs.~(\ref{eq:Ya4}) and (\ref{eq:I6_Z2V}), and we impose the DM constraint using the $\mathcal{O}(F^2)$ result for $\rho_\text{dm}$ based on Eqs.~(\ref{eq:YchiF2}-\ref{eq:aveDMmassapp}).    
    }
   \label{fig:Z2V2}
\end{figure*}
\begin{figure*}
             \includegraphics[width=2.3in]{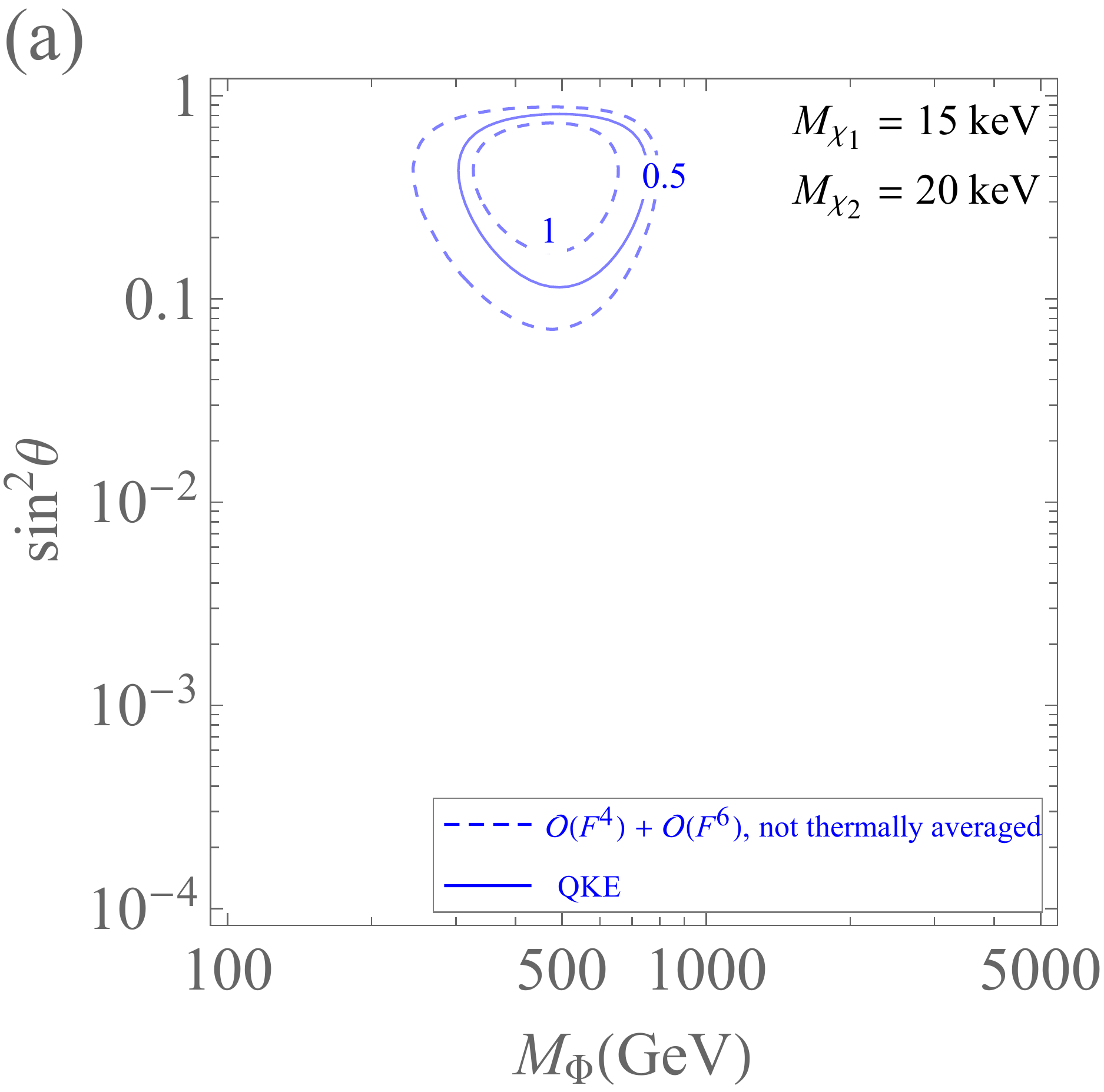}
             \includegraphics[width=2.3in]{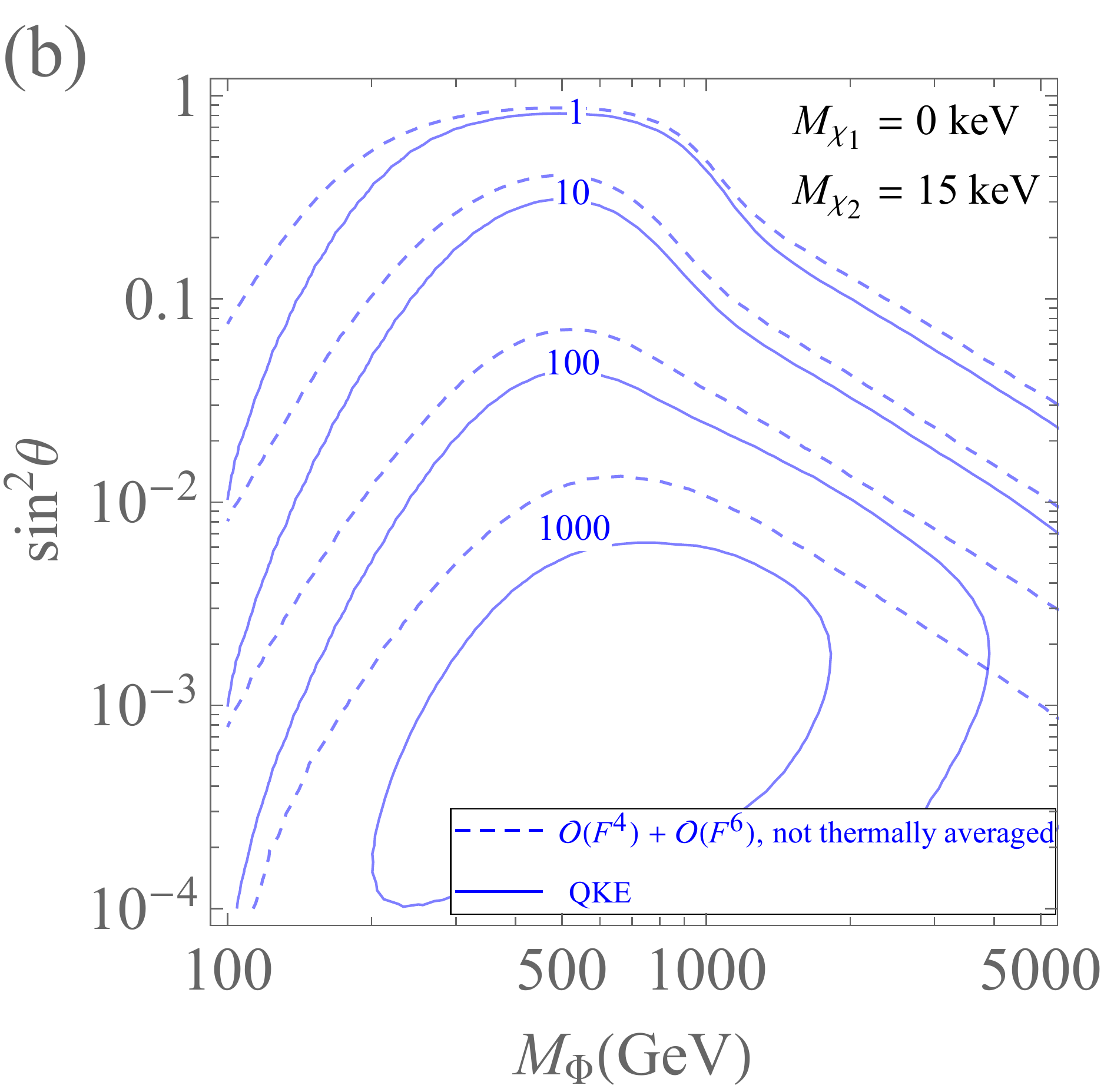}
             \includegraphics[width=2.3in]{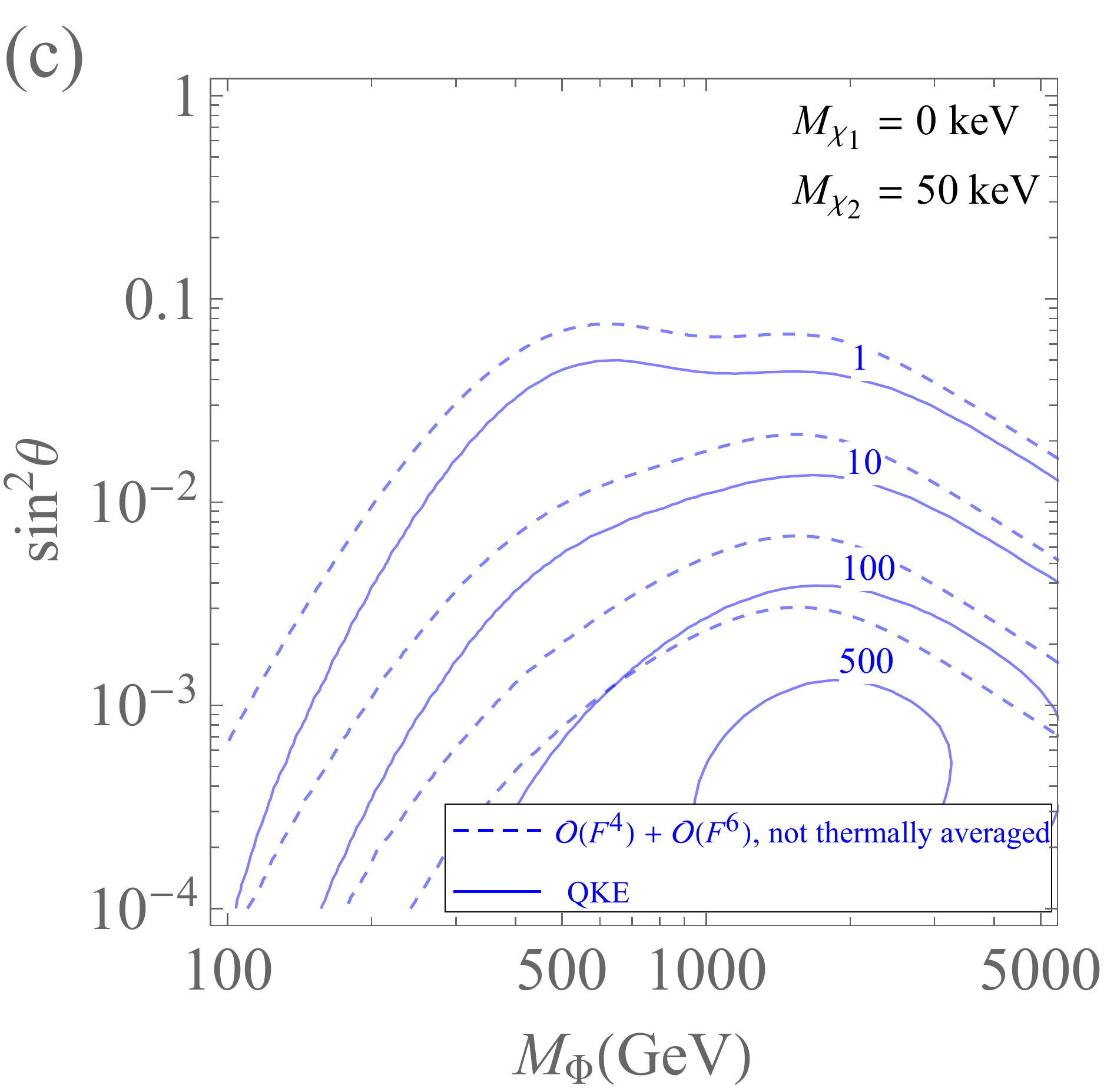}
   \caption{Similar to Fig.~\ref{fig:Z2V2}, except without thermal averaging in the perturbative calculation.  In (a), we take $M_1 = 15$ keV and $M_2 = 20$ keV to provide a  view of the larger-$M_1$ parameter space that complements Fig.~\ref{fig:Z2V1}.
   Comparison of (b) and (c) with Figs.~\ref{fig:Z2V2}(c, f) shows the effect of thermal averaging.
   }
   \label{fig:Z2V3}
\end{figure*}

\subsection{Results for the Z2V Model}\label{sec:Z2Vresults}
In Fig.~\ref{fig:Z2V1} we see that  the large enhancement of the asymmetry relative to the Minimal Model opens up a narrow region of viable parameter space with $M_1 > 15$ keV.  In this plot, $\theta$ and ${\rm Tr} F^\dagger  F$ are chosen to maximize the asymmetry subject to the DM constraint.  One finds that $\theta$ is relatively large, $\theta \sim 1$, for the parameter space shown.  We therefore expect that the perturbative result of Eq.~(\ref{eq:YB4Z2V_mainbody1}), which is used in Fig.~\ref{fig:Z2V1}(a), should be more accurate than 
QKE results based on a thermal ansatz for the DM momentum distribution.  The QKE calculation gives a smaller viable parameter space, but we see in  Fig.~\ref{fig:Z2V1}(b) that the difference is mostly due to the thermal averaging.  We show this by modifying the perturbative calculation to incorporate the same thermal ansatz, which amounts to using the $\mathcal{I}^{(4)}_\text{t.a.}$ function defined in Eq.~(\ref{eq:I4ta}) in place of $\mathcal{I}^{(4)}$ in Eq.~(\ref{eq:Ya4}).  The remaining discrepancy between the QKE and perturbative results is mostly due to the fact that our perturbative treatment approximates sphaleron decoupling  to occur instantaneously at $T=T_\text{ew}$, whereas we take into account that sphaleron decoupling is a gradual process when we numerically solve the QKEs, following the approach of Ref.~\cite{Eijima:2017cxr} (see Appendix~\ref{sec:MIQKEs} for details).

For the relatively large values of $M_1$ shown in Fig.~\ref{fig:Z2V1}, the $\mathcal{O}(F^4)$ contribution to the asymmetry dominates throughout the viable parameter space. We consider the massless-$\chi_1$ limit in Fig.~\ref{fig:Z2V2},  which opens up the parameter space to smaller $\theta$ and larger $M_\Phi$.  For Figs~\ref{fig:Z2V2}(a-c) we take $M_2=15$ keV and compare the QKE calculation with (a) the $\mathcal{O}(F^4)$ contribution, (b) the $\mathcal{O}(F^6)$ contribution, and (c) the sum of those two perturbative contributions.   
We see  the Boltzmann suppression of the $\mathcal{O}(F^4)$ contribution for $M_\Phi \gsim 
1$ TeV, with the $\mathcal{O}(F^6)$ contribution dominating at larger $M_\Phi$.  The same effect is evident in Figs.~\ref{fig:Z2V2}(d-f), for  which we take $M_2 =  50$ keV.  Here we see that increasing the DM mass shifts the parameter space for DM and leptogenesis to larger $M_\Phi$ and smaller $\theta$.  In Figs.~\ref{fig:Z2V2}(c,f)  the sum of the two perturbative contributions gives good agreement with the QKE result for sufficiently large $\theta$.

 For smaller $\theta$, higher-order effects in $F$ become important and the perturbative treatment is not reliable.  The relevant quantity for determining whether higher-order effects are important is $Y_\chi/Y_\chi^\text{eq}$, which is why the perturbative range for $\theta$ depends on $M_\Phi$. We get close agreement in the perturbative regime because we implement the same thermal ansatz for the DM momentum distribution in the perturbative calculation as in the QKE calculation.  The poorer agreement at lower $M_\Phi$ is mostly due to thermal mass effects, which are only significant in this region, and which are only included in the QKE calculation.

In Fig.~\ref{fig:Z2V2}, we include thermal averaging in our perturbative calculations as a consistency check with our QKE calculations.  However, at least in the perturbative regime, {\em not} performing thermal averaging should yield a  more accurate result.   In Fig.~\ref{fig:Z2V3}, we include the full momentum dependence in the perturbative calculation.  For $M_1 = 15$ keV, the $M_\Phi$ and $\theta$ parameter space that works for DM and leptogenesis is limited to $\theta \gsim 0.4$ and 300 GeV $\lesssim M_\Phi \lesssim$ 700 GeV, as we see in  Fig.~\ref{fig:Z2V3}(a).   Meanwhile, Figs.~\ref{fig:Z2V3}(b,c) should be compared with Figs.~\ref{fig:Z2V2}(c,f), which take the same DM masses but with thermal averaging.  We see that the agreement with  QKE results is not as close when we take thermal averaging out of the perturbative calculation.  The discrepancy is again largest for $M_\Phi \sim 100$ GeV, where thermal mass effects are also significant, leading to QKE asymmetries more than five times smaller than the non-thermally-averaged perturbative ones.    The QKE calculation nevertheless gives a reasonably accurate picture of the viable parameter space.    Because the perturbative calculation becomes unreliable for sufficiently small $\theta$, in the remainder of this section we use the QKE calculation to explore the full parameter space.

\begin{figure*}
   
          \includegraphics[width=3in]{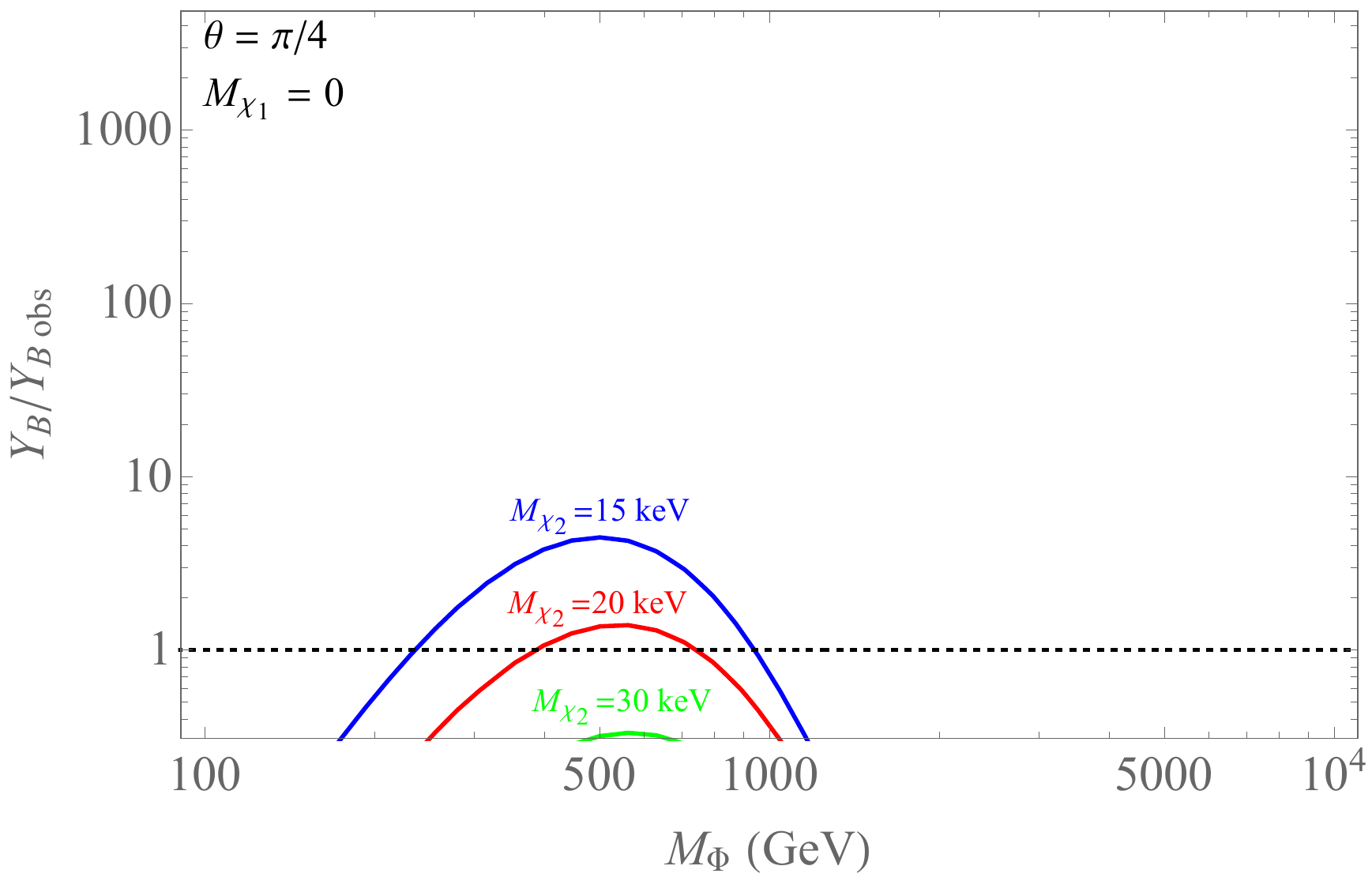}
           \quad \quad \quad
          \includegraphics[width=3in]{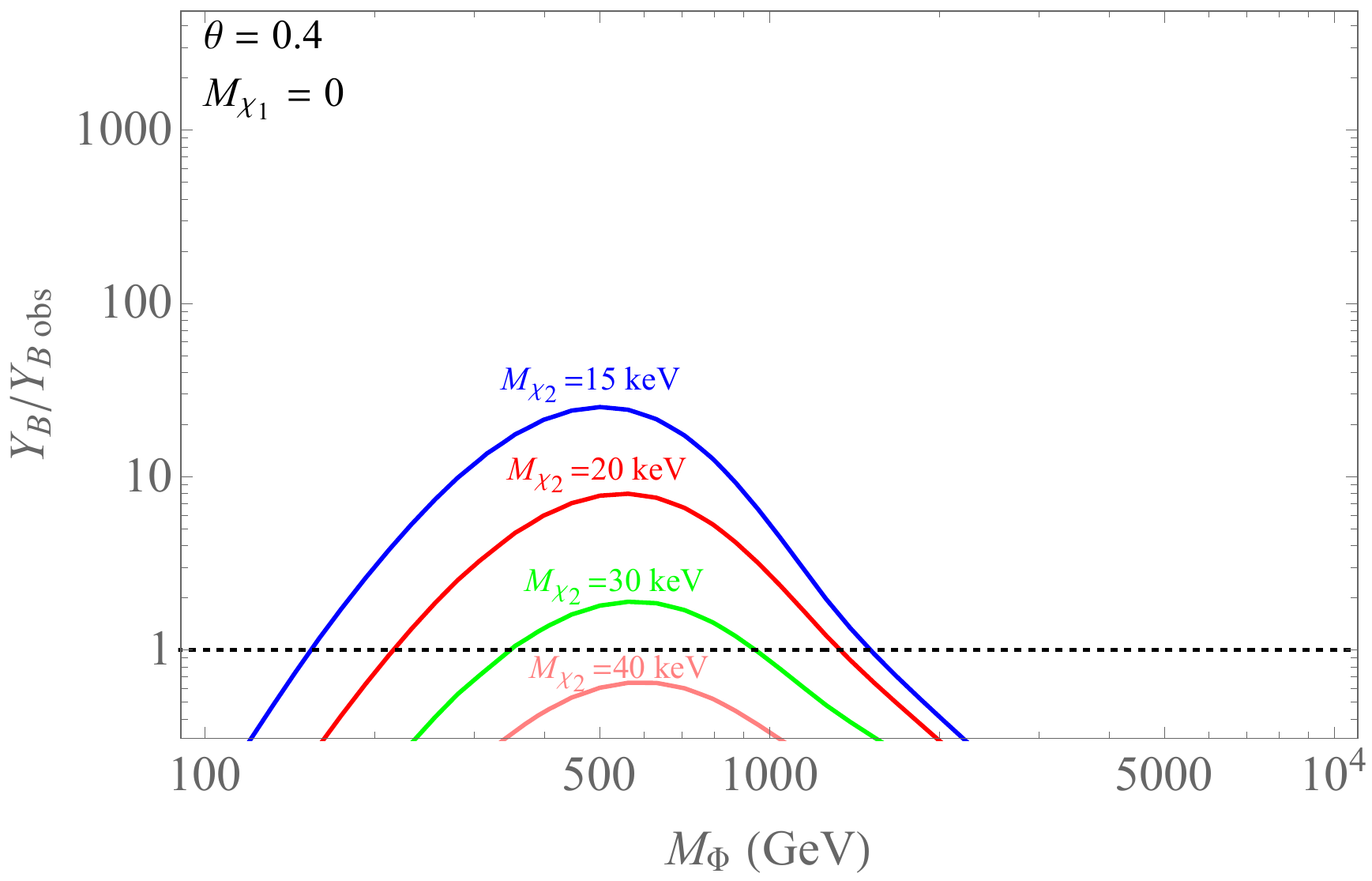}\\
          \includegraphics[width=3in]{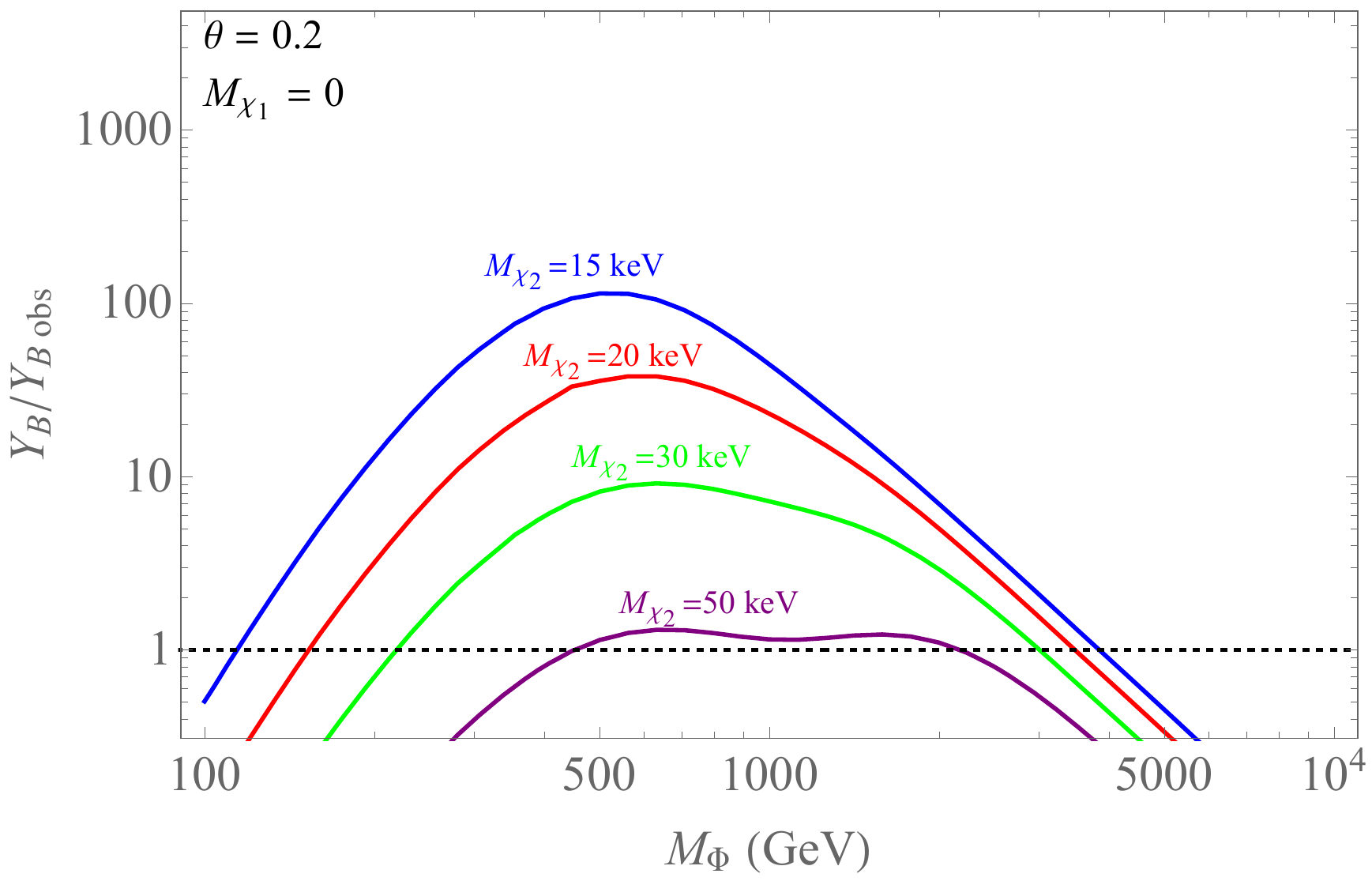}
           \quad \quad \quad
          \includegraphics[width=3in]{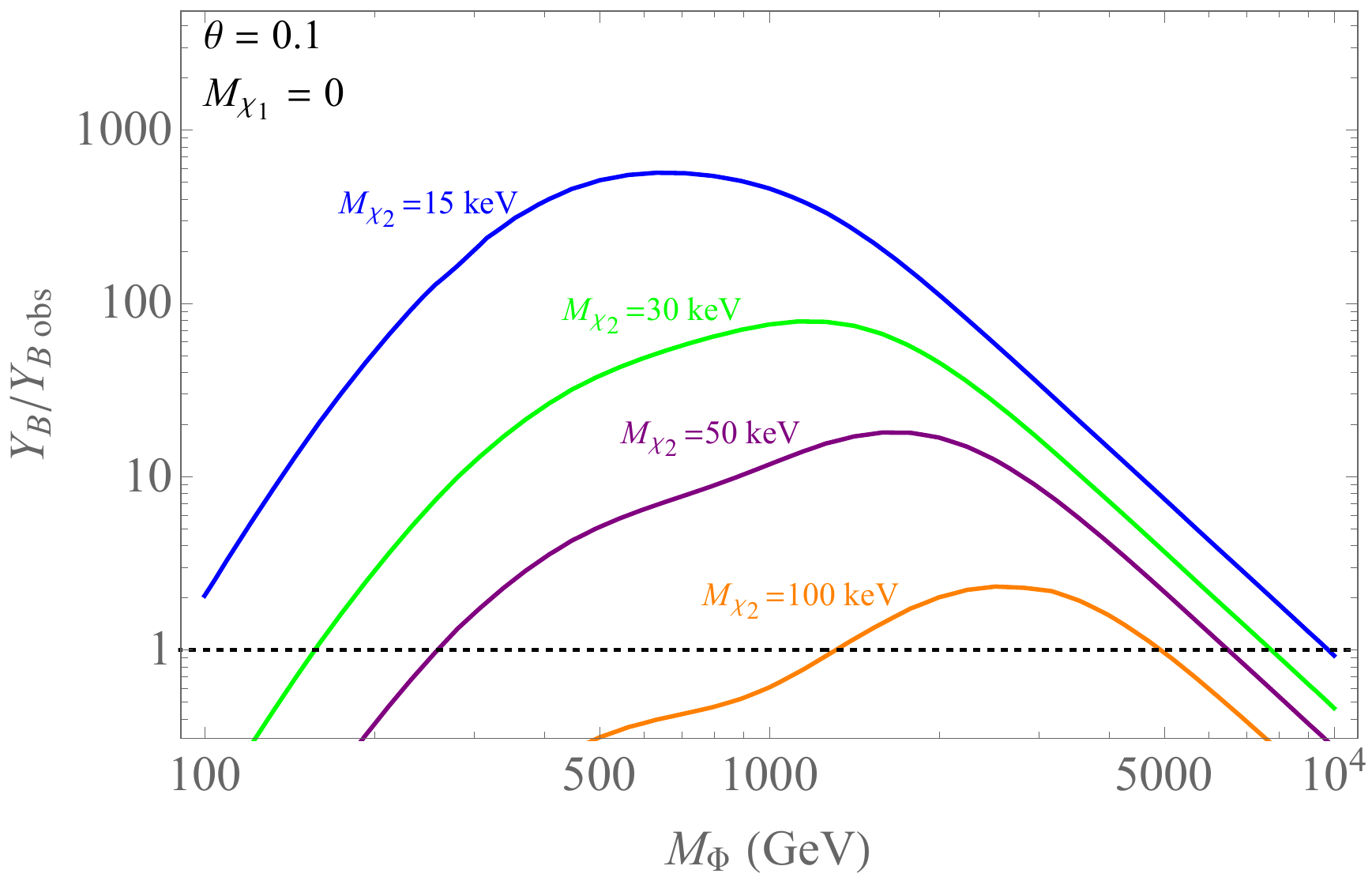}\\
          \includegraphics[width=3in]{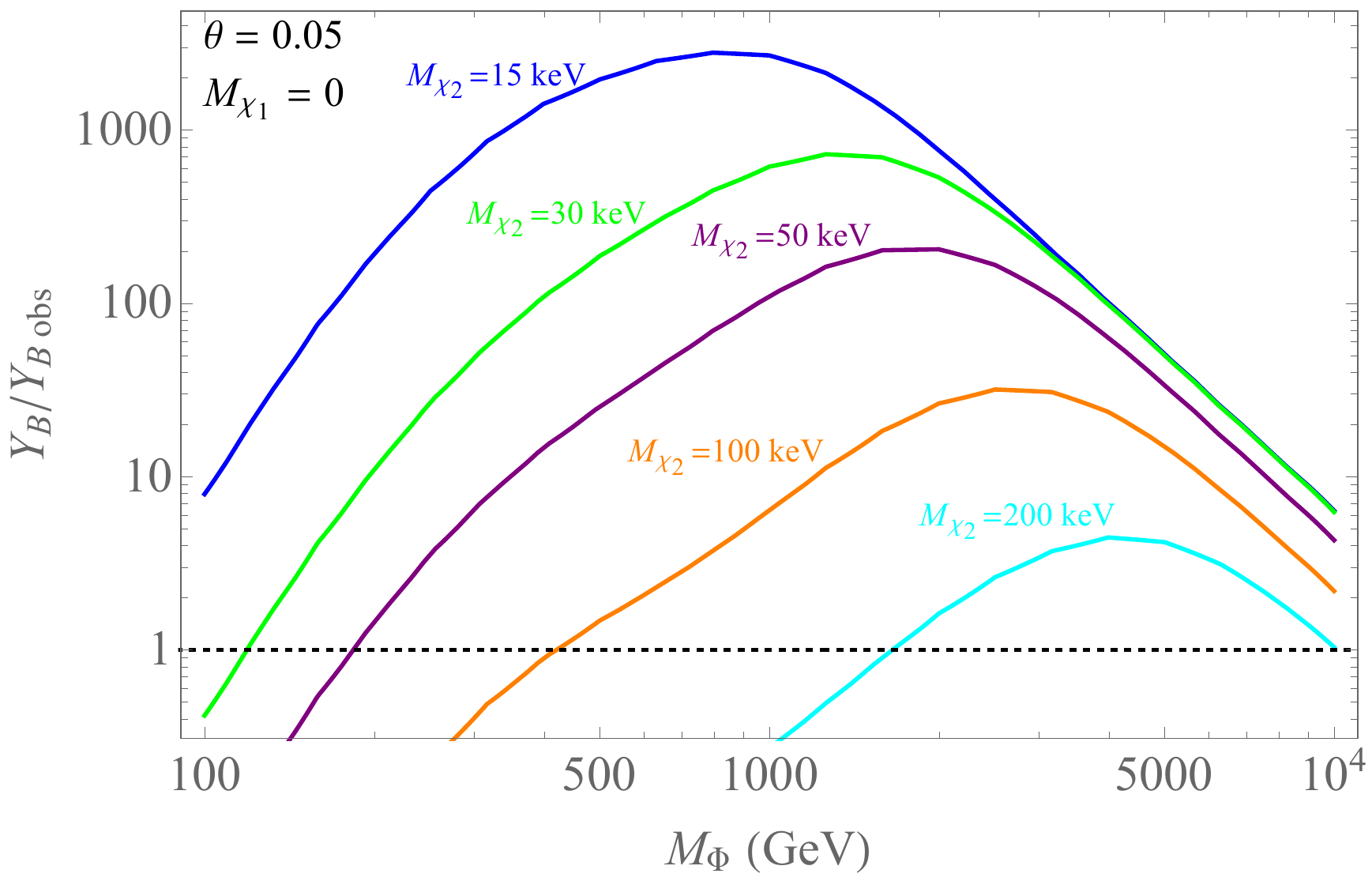}
           \quad \quad \quad
          \includegraphics[width=3in]{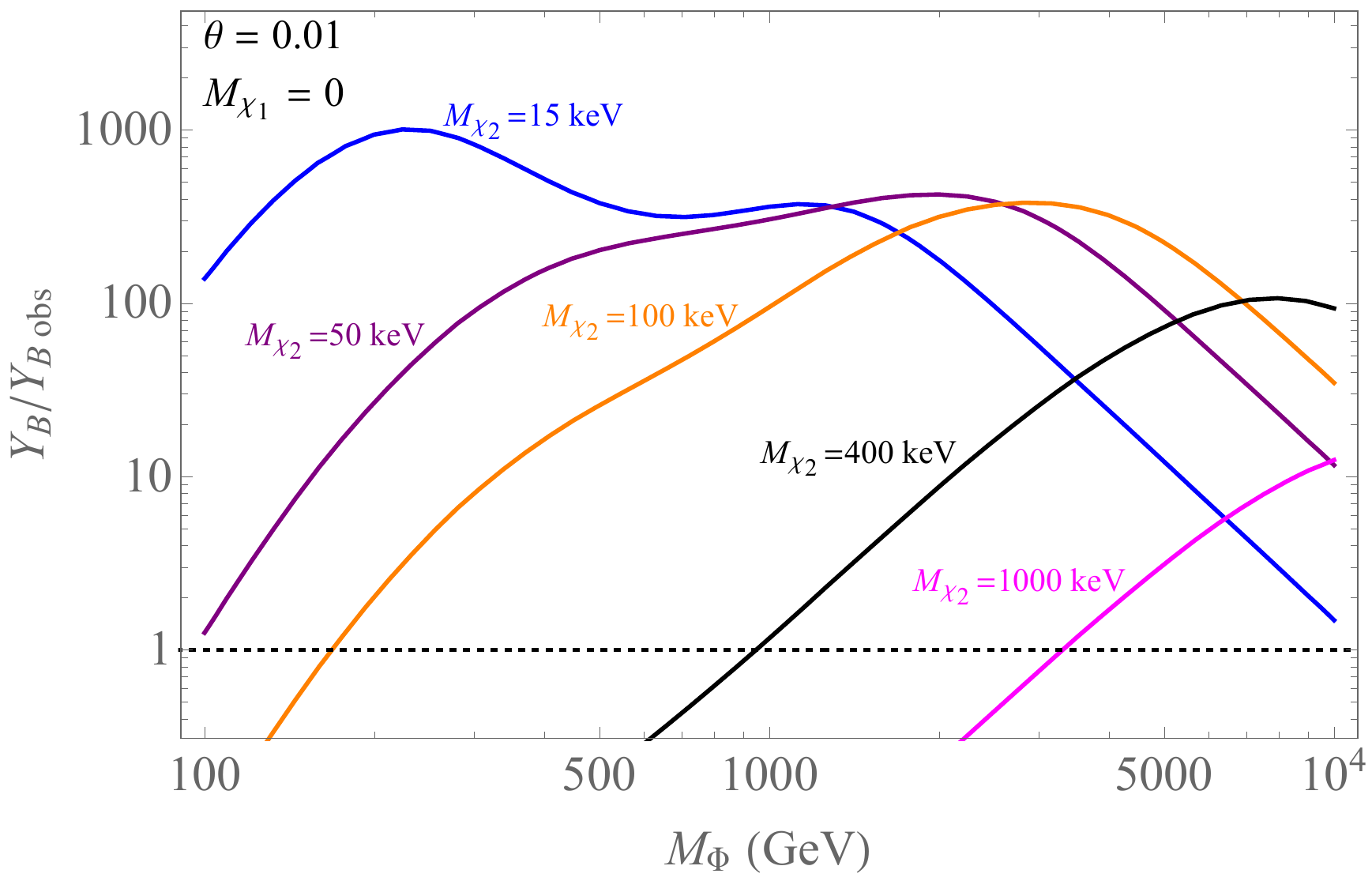}
\caption{
Similar to Fig.~\ref{fig:MM2}, except for the Z2V Model instead of the Minimal Model: in the massless-$\chi_1$ limit, $Y_B$ versus $M_\Phi$ for various $\theta$ and $M_2$, with the $F$ matrix set to the Z2V benchmark form of Eq.~(\ref{eq:FZ2Vbench}).   For each combination of parameters, ${\rm Tr} \left[ F^\dagger F \right]$ is chosen to satisfy the DM constraint.  
}
   \label{fig:Z2V4}
\end{figure*}
\begin{figure}
   \includegraphics[width=3in]{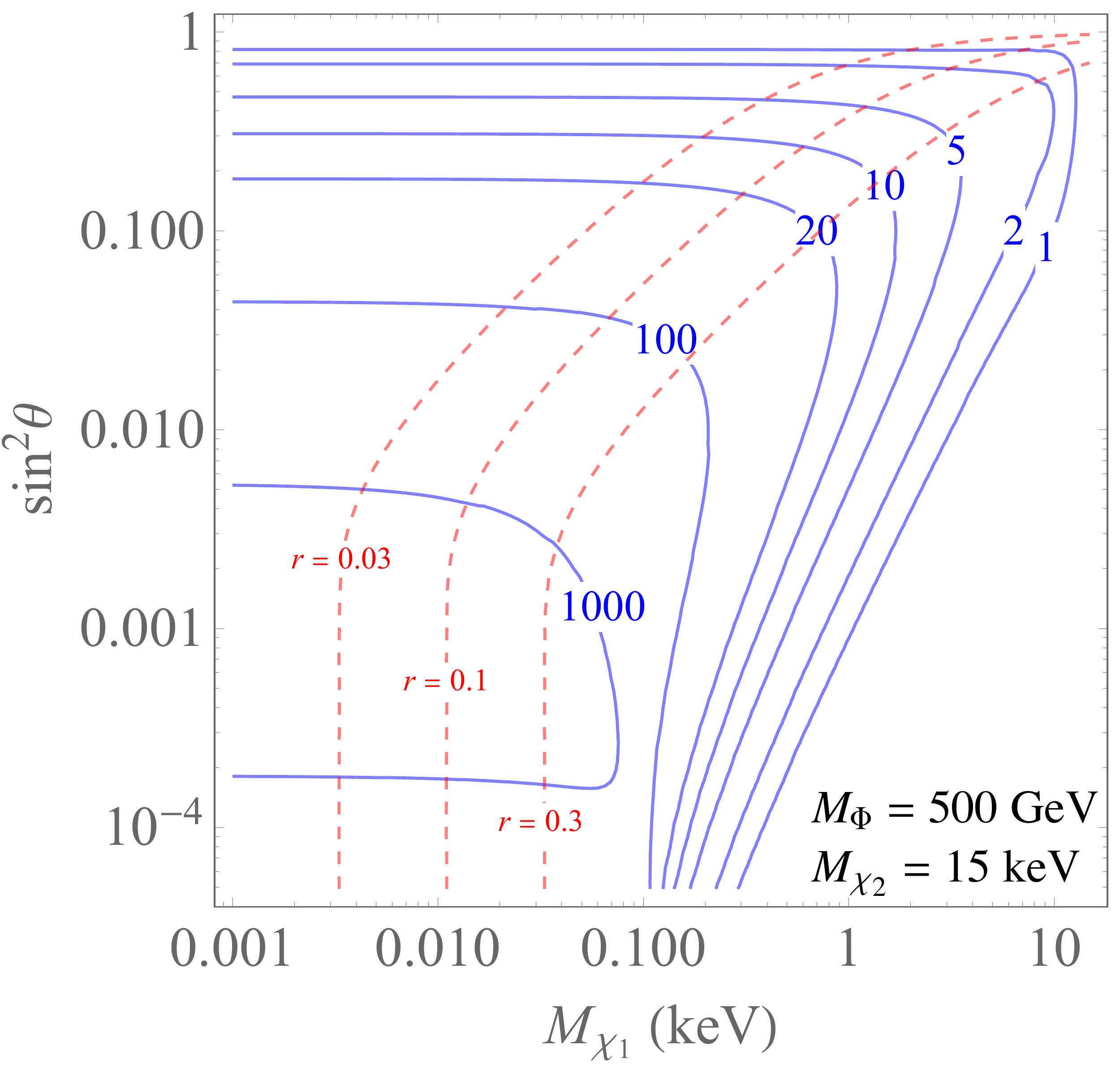}
\caption{
Similar to Fig.~\ref{fig:MM3}, except for the Z2V Model instead of the Minimal Model:  contours of $Y_B/Y_B^\text{obs}$ (blue, solid) and $r$ (red, dashed) in the $(M_1, \sin^2\theta)$ plane, for $M_\Phi = 500$ GeV and $M_2 = 15$ keV, with the $F$ matrix set to the Z2V benchmark form of Eq.~(\ref{eq:FZ2Vbench}). 
At each point in the plane, ${\rm Tr} \left[ F^\dagger F \right]$ is chosen to satisfy the DM constraint.  
}
   \label{fig:Z2V5}
\end{figure}
\begin{figure*}
   
          \includegraphics[width=2.3in]{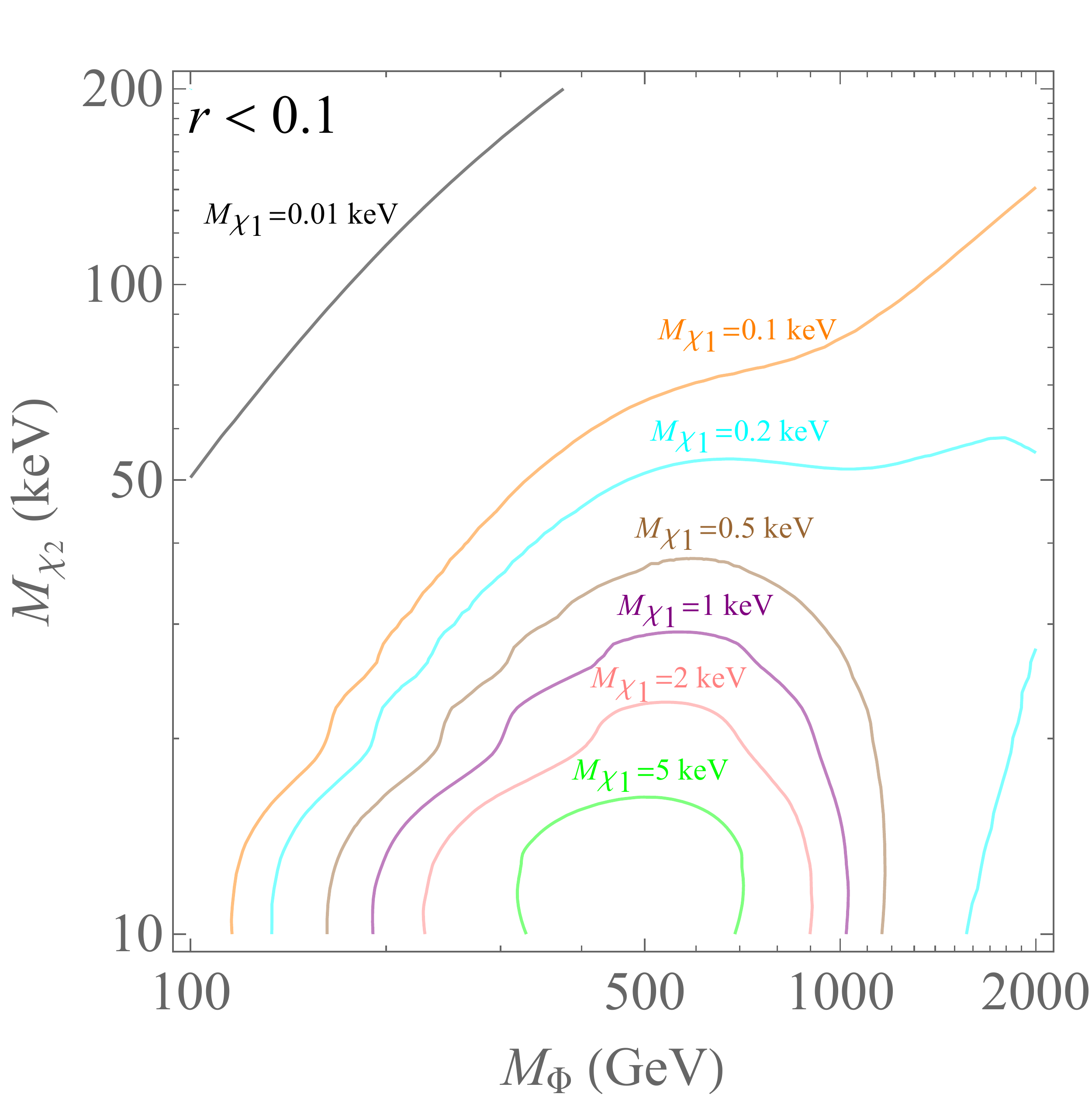}
          \includegraphics[width=2.3in]{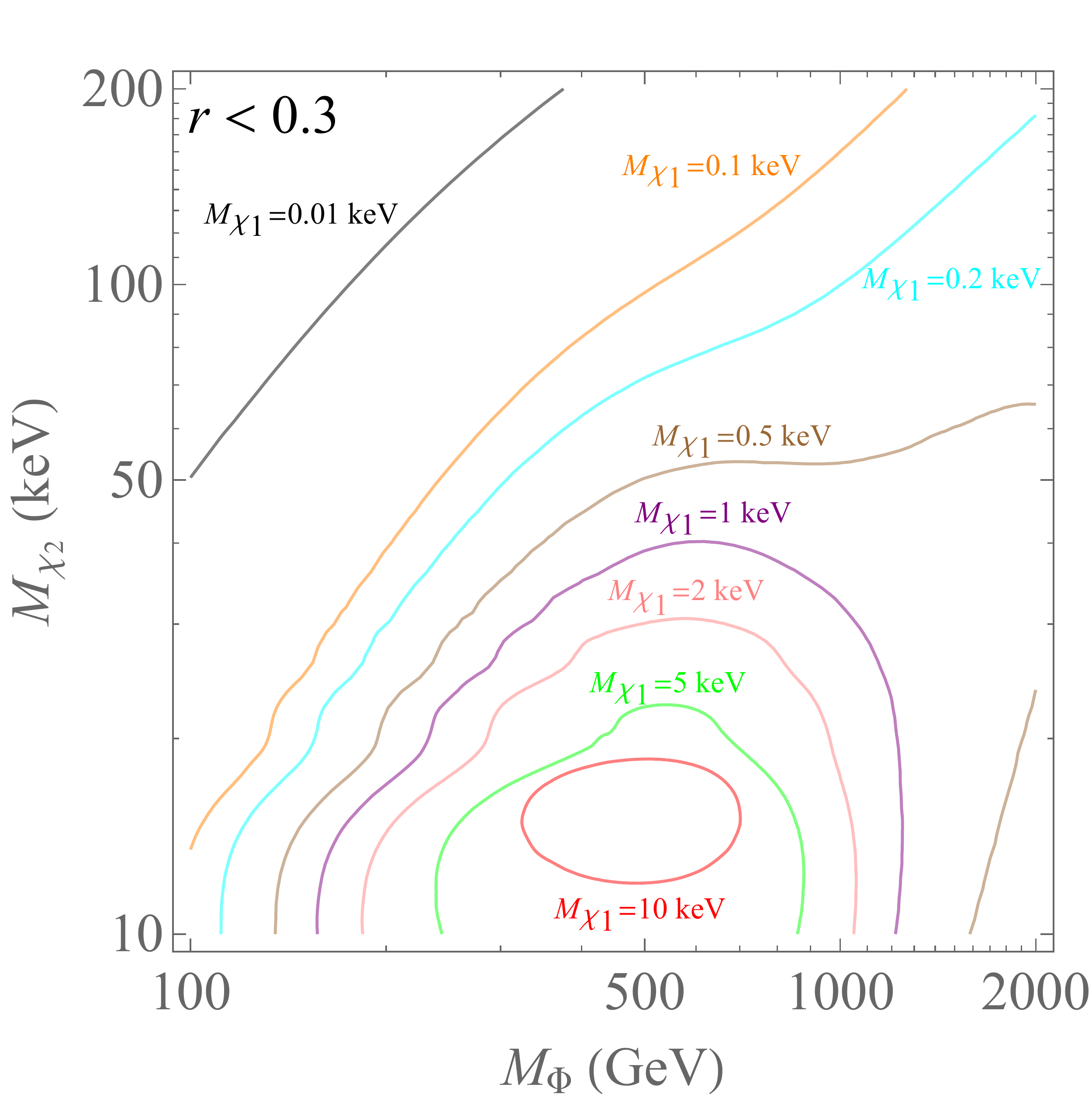}
          \includegraphics[width=2.3in]{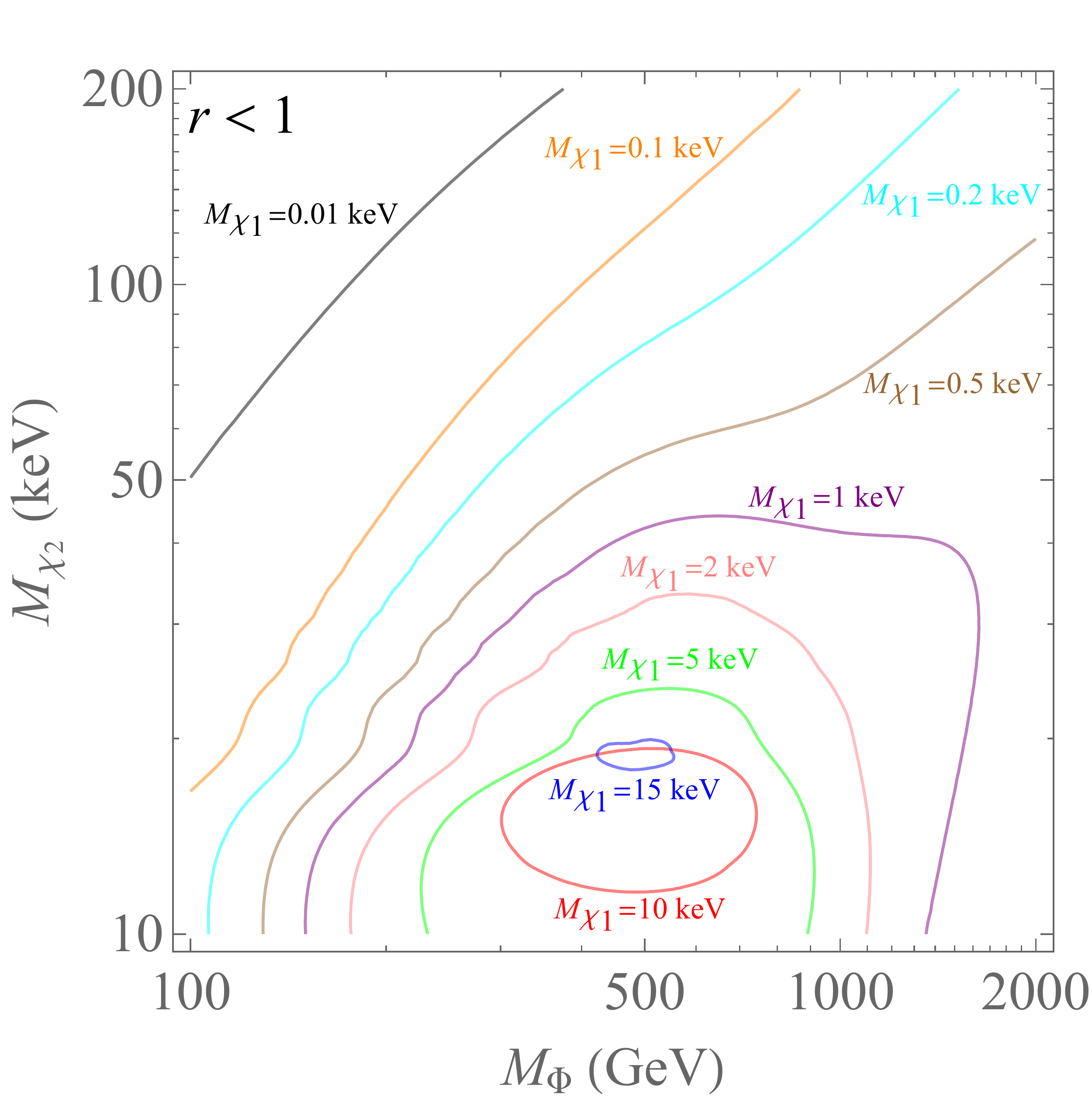}
\caption{
Similar to Fig.~\ref{fig:MM4}, except for the Z2V Model instead of the Minimal Model.  For various $M_1$, the contours enclose the $(M_\Phi,M_2)$ space that is viable for DM and leptogenesis in the Z2V Model, with the $F$ matrix set to the Z2V benchmark form of Eq.~(\ref{eq:FZ2Vbench}).   At each point in the plane, ${\rm Tr} \left[ F^\dagger F \right]$ and $\theta$ are chosen  to maximize $Y_B$ subject to both the DM abundance constraint and the upper bound on $r$ indicated; for the contours shown, that maximum $Y_B$ is equal to $Y_B^\text{obs}$. 
}   
   \label{fig:Z2V6}
\end{figure*}

Fig.~\ref{fig:Z2V4} shows the viable ranges of $\chi_2$ and $\Phi$ masses in the Z2V Model for various $\theta$, in the massless-$\chi_1$ limit.  We see larger $Y_B$ values compared with those in Fig.~\ref{fig:MM2} for Minimal Model. Maximal mixing is viable for $M_2\lesssim 20$ keV and $M_\Phi \lesssim$ 1 TeV, while for small $\theta$, DM and leptogenesis can work for $M_\Phi$ well beyond 10 TeV and $M_2$ well beyond 1 MeV.  

For  fixed $\theta$, the peaks of the contours in Fig.~\ref{fig:Z2V4} tend to shift to larger $M_\Phi$ as the DM mass is increased.  Larger DM mass means a shorter time scale for oscillations and asymmetry generation, which in turn means a larger optimal $M_\Phi$  for generating an asymmetry subject to the DM constraint.  This shift is obvious, for example, in the $\theta = 0.1$ and $\theta=0.05$ plots, in which the DM couplings are big enough that $\mathcal{O}(F^6)$ contributions tend to give the largest asymmetries, but not so big to completely invalidate the perturbative approximation.   The shift is less noticeable for the plots with  large mixing, $\theta = 0.4$ and $\theta=\pi/4$.  Here, the $\mathcal{O}(F^6)$ contribution is unimportant, and the Boltzmann suppression of the $\mathcal{O}(F^4)$ contribution prevents much movement to larger $M_\Phi$.

Because we take the massless-$\chi_1$ limit in Fig.~\ref{fig:Z2V4}, the DM couplings become larger for smaller $\theta$ and fixed $M_2$.  The $\mathcal{O}(F^6)$ contribution therefore increases in  importance relative to the $\mathcal{O}(F^4)$ one as $\theta$ is reduced.  For example, the features in the $\theta = 0.2$, $M_2 = 50$ keV contour reflect comparably sized, well separated peaks from the $\mathcal{O}(F^4)$ and $\mathcal{O}(F^6)$ contributions, whereas for $\theta = 0.1$ and $\theta = 0.05$, the peak in the $M_2 = 50$ keV contour is heavily dominated by the $\mathcal{O}(F^6)$ contribution.  For similar reasons, the $M_2 = 30$ keV  peak shifts to larger $M_\Phi$ for smaller $\theta$, but the movement for  $M_2 = 15$ keV is less pronounced because the $\mathcal{O}(F^6)$ contribution is itself peaked at lower $M_\Phi$, due to the longer oscillation timescale.

A power-law scaling of $Y_B$ with $M_\Phi$ is evident at large $M_\Phi$ in  Fig.~\ref{fig:Z2V4}.  From Eqs.~(\ref{eq:YB6Z2Vbench}) and (\ref{eq:asymp_I6_Z2V}) one can show that for fixed $\theta$, the $\mathcal{O}(F^6)$ asymmetry is proportional to $M_2^{-1} M_\Phi^{-3}$ in the regime in which only a small fraction of an oscillation has time to develop, $\beta_\text{osc} \ll 1$.  
In Appendix~\ref{sec:YB6Z2V} we show that the maximum $\mathcal{O}(F^6)$ asymmetry, optimized with respect to $M_2$ and $\theta$,  falls off as $M_\Phi^{-3/2}$ for large $M_\Phi$, and that this delicate tuning of parameters can allow DM and leptogenesis to work out to $M_\Phi \sim 100$ TeV and $M_2 \sim 10$ MeV while staying at least roughly within the perturbative regime.  

Finally, although many of the features of Fig.~\ref{fig:Z2V4} can be understood at the perturbative level, the $\theta = 0.01$ plot in particular is heavily impacted by effects higher-order in $F$.   For $\theta = 0.01$ and $M_2 = 15$ keV, for example,  we are in the strong-washout regime, and the interesting shape of the associated contour does not emerge at the perturbative level.  Moreover, the contours in the $\theta = 0.01$ plot clearly do not satisfy the perturbative relation $Y_B \propto 1/M_2$ for fixed $M_\Phi$ and  $\beta_\text{osc} \ll 1$.

The impact on the parameter space due to structure formation constraints, which come into play for $M_1 \gtrsim 0.01$ keV, can be seen in Figs.~\ref{fig:Z2V5} and ~\ref{fig:Z2V6}.  
For Fig.~\ref{fig:Z2V5} we take the same $M_\Phi$ and $M_2$ as for Fig.~\ref{fig:MM3} for the Minimal Model, where we found we needed $M_1 \lesssim 0.05$ keV for DM and leptogenesis to work while satisfying $r<0.1$.  Applying the same $r<0.1$ constraint to the Z2V Model, we see that $Y_B$ values two orders of magnitude larger the observed one are possible for $M_1\sim 0.05$ keV,  and that $M_1$ values up to several keV are viable.   
The appropriate bound on $r$ is dependent on the DM masses, the details of which we leave for future work.  Fig.~\ref{fig:Z2V6}, which can be compared with Fig.~\ref{fig:MM4} for the Minimal Model,  shows how parameter space with larger $M_1$ opens up as the upper bound on $r$ is relaxed.  

In this Section, we have seen that $Z_2$-violating interactions can qualitatively impact the parameter space for DM and leptogenesis.  If one or two of the three independent $\lambda$ couplings are large enough to come into equilibrium, the asymmetry is typically dramatically enhanced relative to the Minimal Model.  This leads, for example,  to viable parameter space with larger mixing angles $\theta \sim \pi/4$ and  $\chi_1$ masses, $M_1 \gtrsim 15$ keV.  Finally, for a Z2V coupling to come into equilibrium, we need
\be\label{eq:ctau_equilibration}
c \tau_\Phi  \lesssim 1/H_{T=M_\Phi} \sim 6 \text{ mm} \times  \left(\frac{500 \text{ GeV}}{M_\Phi} \right)^2.
\ee
For much of the Z2V-Model parameter space that works for DM and leptogenesis, $\lambda$ far exceeds the minimum value for equilibration, and we have $c \tau_\Phi \ll$ mm, in which case $\Phi$ qualifies as promptly decaying as far as collider searches are concerned.

\section{Phenomenology}\label{sec:pheno}
\subsection{Collider constraints}\label{sec:collider}
At the LHC, $\Phi$ pair production would be followed by $\Phi\rightarrow l\;+\;$invisible, where the invisible particle is either a neutrino (in the Z2V Model) or DM (in the Minimal and UVDM models). In much of the viable parameter space for the Minimal and Z2V models, $c\tau_\Phi$ is short enough for CMS and ATLAS searches for promptly decaying sleptons to be effective \cite{Aad:2019vnb,Sirunyan:2020eab,CMS:2018eqb,ATLAS:2014zve}.  In the UVDM Model, on the other hand, $c\tau_\Phi$ is typically  long enough to bring searches for long-lived particles (LLPs) into play, including the ATLAS displaced lepton search \cite{Aad:2020bay} and the CMS search for heavy stable charged particles (HSCPs) \cite{CMS-PAS-EXO-16-036}. Similar signatures have been proposed for other models of freeze-in  DM \cite{Co:2015pka} (for a recent study of LHC-friendly freeze-in models, see Ref.~\cite{Belanger:2018sti}).

\begin{figure}
          \includegraphics[width=3.3in]{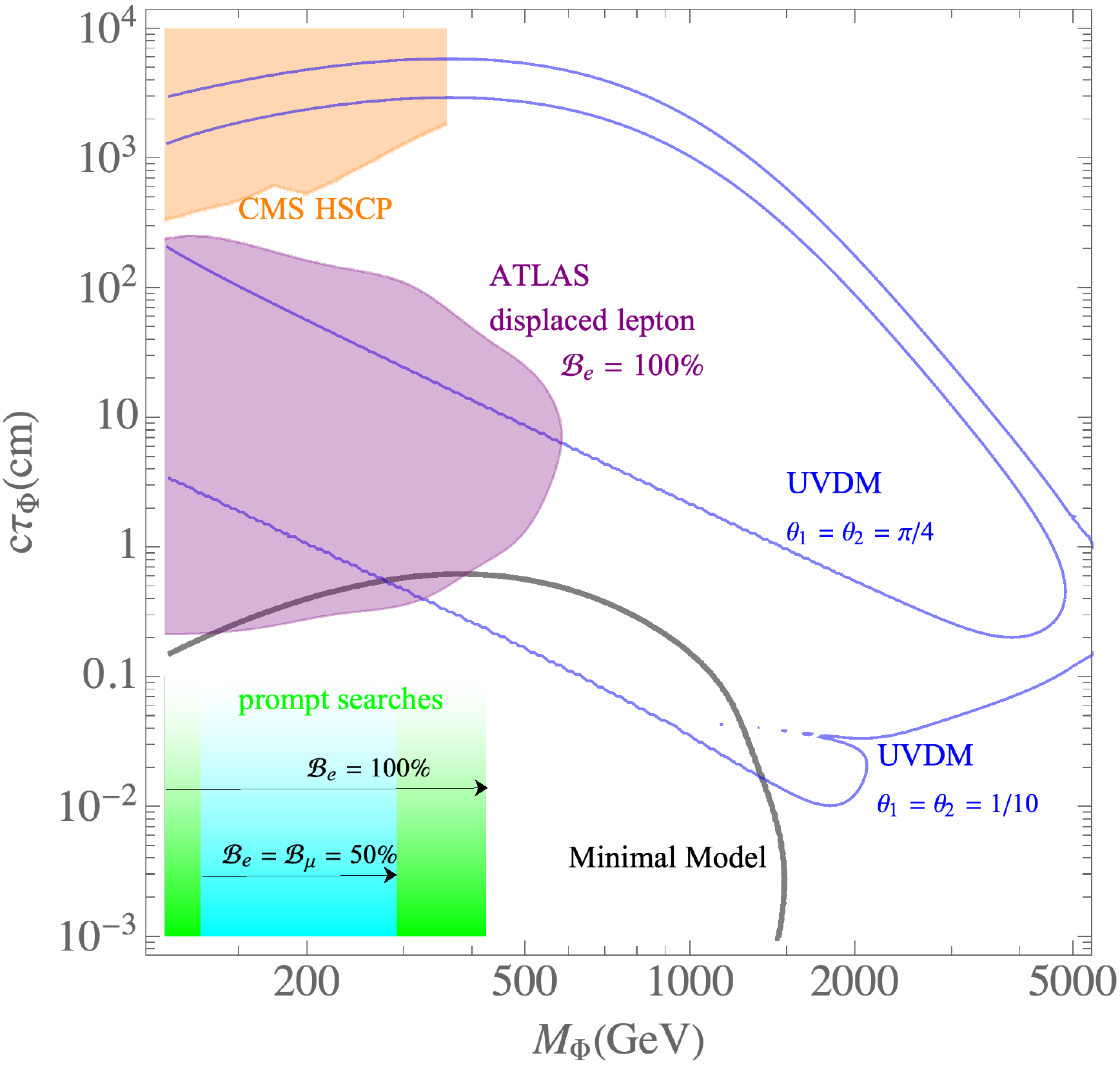}   \caption{LHC constraints from searches for promptly decaying sleptons \cite{Aad:2019vnb,Sirunyan:2020eab,CMS:2018eqb,ATLAS:2014zve}, displaced leptons \cite{Aad:2020bay}, and HSCPs \cite{CMS-PAS-EXO-16-036}, along with parameter space for the Minimal Model (from Fig.~\ref{fig:MM1}(a)) and for the UVDM Model, under two different DM coupling assumptions (from  Figs.~\ref{fig:UVDM1}(a) and (c)).  As discussed in the text, the green prompt and purple displaced excluded regions are for ${\mathcal B}_{e} = 100\%$; the corresponding limits for ${\mathcal B}_{\mu} = 100\%$  are roughly similar.  For the prompt case, the cyan region is a tentative estimate of the limits for ${\mathcal B}_e = {\mathcal B}_\mu=50\%$ (see the text).  The lifetime cutoffs for the prompt and  HSCP exclusion regions are highly approximate.  We take the CMS HSCP exclusion contour from Ref.~\cite{Junius_2019} and extend it from $M_\Phi = 270$ GeV to $M_\Phi =360$ GeV by assuming a similar behavior.
 }
   \label{fig:collider}
\end{figure}

Fig.~\ref{fig:collider} shows viable parameter space for the Minimal Model and, for selected parameters, the UVDM Model.  For our Z2V Model benchmarks, with one or two Z2V couplings coming into equilibrium, much (but not all) of the parameter space has $c\tau_\Phi \ll$ 1 mm, and the full $\Phi$ mass range shown in Fig.~\ref{fig:collider} is  viable; see Eq.~(\ref{eq:ctau_equilibration}) and  Fig.~\ref{fig:Z2V4}.  Fig.~\ref{fig:collider} also shows exclusion regions from LHC searches, with flavor-coupling assumptions and caveats specified below.  The essential takeaway is that while LHC analyses have already probed some of the interesting parameter space for the each of the models,  much of it remains open for exploration, particularly so if $\Phi$ has an appreciable  branching ratio to $\tau\;+\;$invisible.

While we have chosen models with an $SU(2)_w$-singlet $\Phi$ for detailed study, it is worth emphasizing that scenarios with an $SU(2)_w$-doublet $\Phi$ coupled to DM through an $F_{\alpha i}\Phi^*l_\alpha \chi_i$ interaction term can work equally well for DM and leptogenesis.  The collider phenomenology of a doublet $\Phi$ can differ significantly from the singlet, especially in the presence of Z2V couplings, which can have  arbitrary flavor structure provided that they are small enough to satisfy FCNC constraints.  For example, a $\Phi l e^c$ coupling can lead to  decays of the neutral component of $\Phi$  to opposite-sign, dileptons of any flavor combination.  Alternatively, $\Phi$ might decay hadronically via $\Phi q d^c$ and/or $\Phi q u^c$ couplings;  if these hadronic decays are displaced and/or produce top quarks, they might result in a detectable signal.  We set aside investigation of potential collider probes of the doublet case for future work, and now provide details on the various collider constraints on an $SU(2)_w$-singlet $\Phi$.

\subsubsection{Prompt searches}

The collider constraints on a promptly decaying $\Phi$ depend on its branching ratios ${\mathcal B}_{e,\mu, \tau}$, where for example ${\mathcal B}_\mu \equiv {\mathcal B}(\Phi \rightarrow \mu+\text{invisible})$.  In the models we consider, it is appropriate to regard the DM as massless in the collider context.  LHC searches \cite{Aad:2019vnb,Sirunyan:2020eab,CMS:2018eqb,ATLAS:2014zve} then rule out $110 \text{ GeV}< M_\Phi<425 \text{ GeV}$ and $100 \text{ GeV}< M_\Phi<450 \text{ GeV}$ for ${\mathcal B}_e=100\%$ and ${\mathcal B}_\mu=100\%$, respectively, while there is no LHC constraint for ${\mathcal B}_ \tau=100\%$ \cite{Aad:2019byo}.  
At the top of the green shaded region in Fig.~\ref{fig:collider}, which represents the exclusion for ${\mathcal B}_e=100\%$,
the gradual fade-out around $c\tau_\Phi \sim 1$ mm is meant to reflect our uncertainty regarding the degradation of prompt searches with increasing lifetime.

The leptogenesis mechanisms in the Minimal and Z2V models both require $\Phi$ to couple to two or more lepton flavors.  In fact, if Z2V couplings dominate $\Phi$ decay,  the {\em largest} branching ratio that any one flavor can have is $50\%$; specializing to the simple scenario in which a single Z2V coupling $\lambda_{\alpha \beta}$ dominates, we have ${\mathcal B}_\alpha = {\mathcal B}_\beta = 50\%$. Going beyond the single-flavor assumption in interpreting existing slepton search results is not completely straightforward.  Consider, for example, a scenario with ${\mathcal B}_e  ={\mathcal B}_\mu = 50\%$.  Because of the same-flavor requirement, the signal efficiency for $\Phi$ pair-production would be $\simeq$1/2 that for a mass-degenerate ${\tilde e}_R, {\tilde \mu}_R$ pair in SUSY. 
Ref.~\cite{Sirunyan:2020eab} gives cross-section limits for combined ${\tilde e}_{L,R}$ and ${\tilde \mu}_{L,R}$ production.  Bounding the NLO-NLL $\Phi$ pair-production cross sections~\cite{susyxsecwg,Bozzi:2007qr,Fuks:2013vua,Fuks:2013lya,Fiaschi:2018xdm,Beenakker:1999xh} to be below twice those limits,  we obtain the cyan exclusion region in Fig.~\ref{fig:collider}:  $130 \text{ GeV}< M_\Phi<290 \text{ GeV}$ \footnote{Performing the same steps with the cross-section limits reported in Ref.~\cite{Aad:2019vnb} does not appear to lead to any excluded region.}.  
This exercise ignores that $\Phi$ production would also lead, in the chosen scenario,  to opposite-sign, different-flavor dilepton events, which are used to define control regions for background estimation in the analysis of Ref.~\cite{Sirunyan:2020eab}.  For that reason, our ${\mathcal B}_e  ={\mathcal B}_\mu = 50\%$ exclusion region is tentative at best. We have also ignored that  kinematic differences between ${\tilde l}_L$ and ${\tilde l}_R$ pair-production might lead to different efficiencies.  However, the separate cross-section limits for LH and RH sleptons given in an earlier 13 TeV CMS analysis \cite{CMS:2018eqb} are similar enough to suggest that this issue is unimportant.

The most recent ATLAS and CMS slepton search results do not include separate cross-section upper limits for selectrons and smuons.  This complicates extracting limits when the $e$ or $\mu$ signals are suppressed by a significant branching ratio for $\Phi \rightarrow \tau\;+\;$ invisible, even if we simply neglect the potential contribution to the signal from events with taus\footnote{Light-flavor leptons produced from tau decays will differ kinematically from leptons directly produced by $\Phi$ decay, and such decay modes have a total branching fraction of only 35\%.  We leave an estimate of the potential impact of  $\Phi \rightarrow \tau\;+\;$invisible decays on searches in the $e$, $\mu$ channels for future work.}.  For example, consider the two cases ${\mathcal B}_e={\mathcal B}_\tau=50\%$ and ${\mathcal B}_\mu={\mathcal B}_\tau=50\%$.  Ignoring $\Phi$ events with $\tau$s, the signal efficiency is now $\simeq1/4$ that for slepton pair-production.  Taking into account the reported relative efficiencies for muons and electrons, the cross-section limits for combined ${\tilde e}_{L,R}$ and ${\tilde \mu}_{L,R}$ production given in Ref.~\cite{Sirunyan:2020eab} can be used to exclude narrow regions within the range $155 \text{ GeV}< M_\Phi<185 \text{ GeV}$ for ${\mathcal B}_\mu={\mathcal B}_\tau=50\%$, while no region is excluded  for  ${\mathcal B}_e={\mathcal B}_\tau=50\%$.  However, this exercise presumably underestimates the true sensitivity, because it includes background with both flavors of dilepton pairs even though the signal has a single flavor, electrons or muons.   The 35.9 fb$^{-1}$ CMS analysis \cite{CMS:2018eqb} does give separate cross-section upper limits for smuons and selectrons, but they are not strong enough to constrain the ${\mathcal B}_e={\mathcal B}_\tau=50\%$ and ${\mathcal B}_\mu={\mathcal B}_\tau=50\%$ scenarios, assuming we can neglect $\Phi$ decays to $\tau$s.

These various observations motivate dedicated LHC analyses that target intermediate-mass ``slepton-like'' particles  ($M \sim 100-300$ GeV)  with lower values of $\sigma\times {\mathcal B}(l^+l^- +\text{invisible})$ than those associated with the multiple flavors of LH and RH sleptons of SUSY, including the possibility of mixed flavor in the final-state dileptons.

\subsubsection{Displaced searches}

For the longer lifetimes typical of the UVDM Model, the ATLAS displaced lepton search \cite{Aad:2020bay} is relevant.  The purple excluded region in Fig.~\ref{fig:collider} applies to the case with ${\mathcal B}_e = 100\%$.  The limits for ${\mathcal B}_\mu = 100\%$ are comparable; for example, the maximum mass excluded (for the optimal $c\tau_\Phi \approx 6$ cm) lowers from $\simeq 580$ GeV to $\simeq 550$ GeV. For ${\mathcal B}_\tau = 100\%$, there appears to be no excluded region for the optimal $c\tau_\Phi \approx4.5$ cm (although the mass range from 100--200 GeV is on the borderline).

The asymmetry generation mechanism in the UVDM model does not require $\Phi_1$ to couple to multiple lepton flavors, but it is certainly allowed.   We use the cross-section upper limits provided by Ref.~\cite{Aad:2020bay} to obtain rough estimates of the mass reach in such scenarios. For ${\mathcal B}_e = 50\%$, and ignoring decays that do not involve electrons, the maximum excluded mass falls from $\simeq 580$ GeV to $\simeq 440$ GeV.  For ${\mathcal B}_\mu = 50\%$, and ignoring decays that do not involve muons, the maximum excluded mass falls from $\simeq 550$ GeV to $\simeq 410$ GeV.

\subsubsection{Searches for heavy stable charged particles}
Taking into account only direct stau pair production, a CMS search for HSCPs,  Ref.~\cite{CMS-PAS-EXO-16-036}, obtains a bound of approximately 360 GeV for a long-lived stau, for a SUSY parameter point in which the stau is mostly ${\tilde \tau}_R$.  This bound should apply to the model we consider for $c\tau_\Phi \gtrsim$  several  meters.  In Fig.~\ref{fig:collider} we roughly estimate the exclusion in the mass-lifetime plane by  using the results of Ref.~\cite{Junius_2019} up to $M_\Phi = 270$ GeV and extrapolating those results up until $M_\Phi = 360$ GeV.
The authors of Ref.~\cite{Junius_2019} find that when they apply their analysis to 8 TeV data, their results match reasonably well with  the earlier study of Ref.~\cite{ Evans:2016zau}.

\subsection{Astrophysical constraints}
\subsubsection{X-ray line constraints on the Z2V Model}\label{sec:xray}

In the presence of the $Z_2$-violating couplings of Eq.~(\ref{eq:LZ2V}), the DM is unstable, and the partial width for $\chi \rightarrow \nu \gamma,{\overline \nu} \gamma$ decays is constrained by X-ray observations.   The neutrino portal couplings induce $\chi_i$-neutrino mixing, with  mixing angle 
\begin{multline}\label{eq:DMnumix}
\theta_{\chi_i-\nu} \simeq \frac{\sqrt{\sum_\alpha |h_{\alpha i}|^2}v}{\sqrt{2}M_i}\\
\simeq
1.2 \times 10^{-6}
\left(\frac{\sqrt{\sum_\alpha |h_{\alpha i}|^2}}{10^{-13}} \right)
\left(\frac{15 \text{ keV}}{M_i} \right).
\end{multline}
The partial width for $\chi_i \rightarrow \nu \gamma,{\overline \nu} \gamma$ is \cite{Drewes:2016upu, Shrock:1974nd, Petcov:1976ff, Lee:1977tib, Marciano:1977wx, Pal:1981rm, Shrock:1982sc, Barger:1995ty} is
\be\label{eq:nuportalwidth}
\Gamma_i = \frac{9 \alpha G_F^2}{1028 \pi^4} \sin^2 2\theta_{\chi_i-\nu}  \;M_i^5.
\ee
Results based on NuSTAR \cite{Perez:2016tcq,Neronov:2016wdd,Ng:2019gch,Roach:2019ctw} and
INTEGRAL \cite{Laha:2020ivk} data then constrain $\theta_{\chi_i-\nu}  < 10^{-6}-10^{-8}$ for DM masses ranging from 10 keV to 100 keV, corresponding to upper bounds on the overall neutrino-portal coupling strength $\sqrt{\sum_\alpha |h_{\alpha i}|^2}$ of $\sim (7\times 10^{-14})- (5\times10^{-15})$.   These stringent limits justify  neglecting the neutrino portal couplings in our asymmetry calculations.

\begin{figure}
\begin{tikzpicture} \begin{feynman}
\vertex (a1) {\(\chi_i\)}; 
\vertex[right=1.5cm of a1] (a2);
\vertex[right=3cm of a2] (a3);
\vertex[right=1.5cm of a3] (a4) {\(\nu_\alpha\)};\vertex at ($(a2)!0.5!(a3)!1.5cm!90:(a3)$) (d); 
\vertex at ($(d)+ (1cm, 1cm)$) (c2);
\diagram* {
(a1) -- [fermion] (a2) -- [anti majorana,edge label=\(e^c_\beta  \quad\quad\quad e_\beta\)] (a3) -- [anti fermion] (a4),
(a2) -- [insertion=0.5] (a3),
(a2) -- [anti charged scalar, quarter left,edge label=\(\Phi\)] (d) -- [scalar, quarter left] (a3),
(d) -- [boson,  edge label=\(\gamma\)] (c2),
};
\end{feynman} \end{tikzpicture}
   \caption{Feynman diagram for DM decay induced by Z2V couplings. }
   \label{fig:DMdecay}
\end{figure}
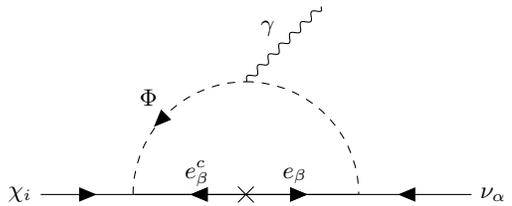

We now focus on the Z2V couplings $\lambda_{\alpha \beta}$, which induce DM decays  via diagrams such as that of Fig.~\ref{fig:DMdecay}, leading to a  $\chi_i \rightarrow \nu \gamma,{\overline \nu} \gamma$ partial width of
\be\label{eq:DMdecay}
\Gamma_{i} 
= \frac{\alpha}{512 \pi^4} \frac{M_i^3}{M_\Phi^4}
\sum_\alpha
\left|
\sum_\beta
\lambda_{\alpha \beta} F_{\beta i} M_\beta
\left(
\log
\frac{M_\Phi^2}{M_\beta^2}
-1
\right)
\right|^2,
\quad\quad
\ee
where $M_i$ is the $\chi_i$ mass and $M_\beta$ are the masses of the SM leptons.  

For purposes of illustration, we  specialize to benchmarks with a single Z2V coupling in equilibrium and an $F$ matrix of the form described in Sec.~\ref{sec:Z2VFbench}.  These scenarios  still allow a wide range of possible partial widths, depending on the SM lepton flavor running in the loop.  For example, if the single Z2V coupling involves $l_2$ and $l_3$ ($\lambda\equiv\lambda_{23}$) while the DM couples to $e^c_1$ and $e^c_3$, we get
\be\label{eq:DMdecaytau}
\Gamma_{2} 
= \frac{\alpha |\lambda|^2{\rm Tr}\left[F^\dagger F\right] \sin^2\theta}{1024 \pi^4}  \frac{M_2^3 M_\tau^2}{M_\Phi^4}
\left(\log
\frac{M_\Phi^2}{M_\tau^2}
-1
\right)^2,
\quad\quad
\ee
for the partial width of the heavier DM particle.  Permuting the flavors involved in the $\lambda$ and $F$ couplings leads to the same result except with $M_\tau$ replaced by $M_\mu$ or $M_e$.  
For these benchmark scenarios, the partial width of $\chi_1$ is given by replacing $\sin\theta \rightarrow \cos\theta$ in $\Gamma_2$, but because $\chi_1$ might be too light to contribute to a detectable X-ray signal, we focus on $\Gamma_2$.

\begin{figure*}
             \includegraphics[width=2.3in]{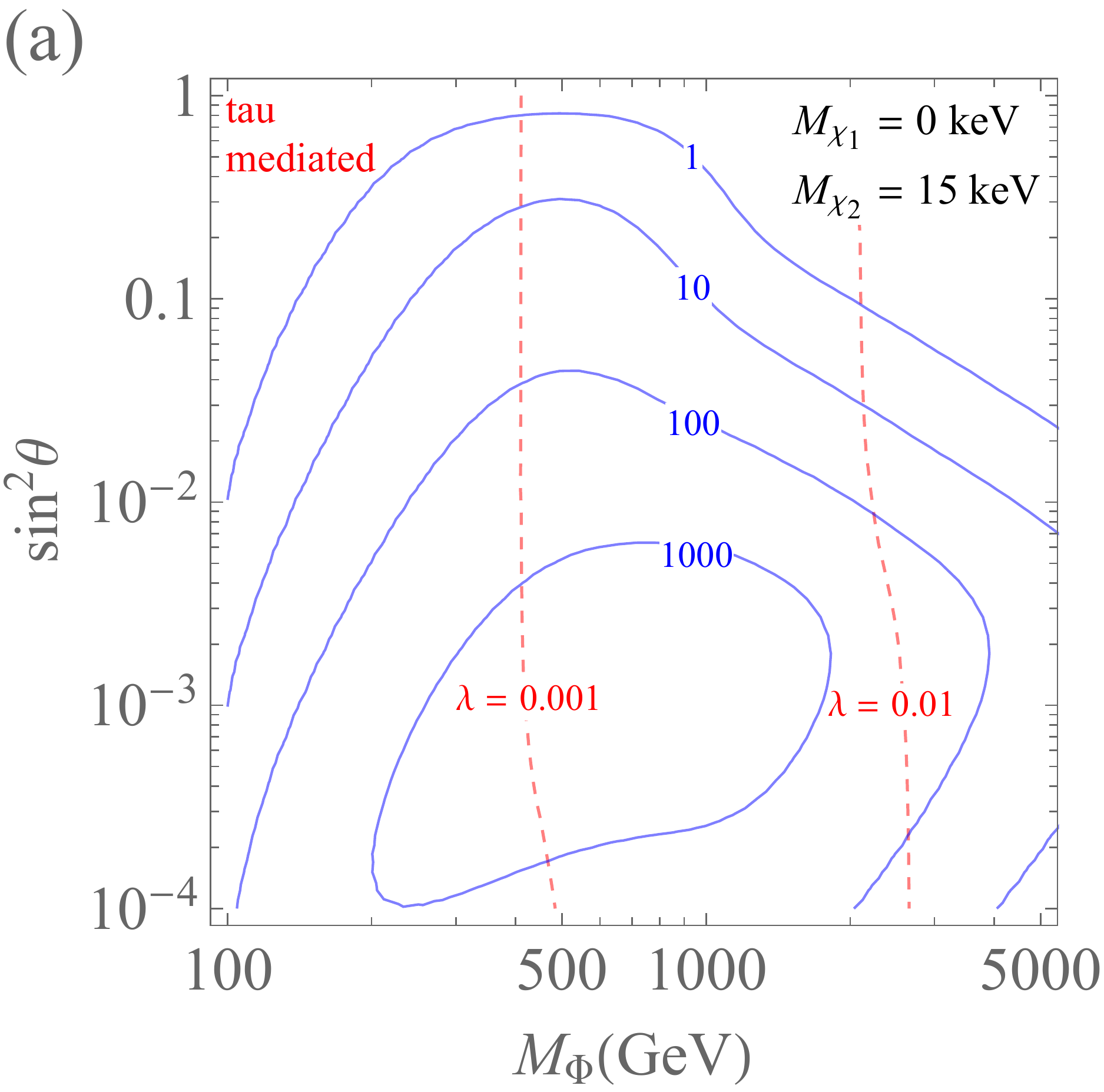}
             \includegraphics[width=2.3in]{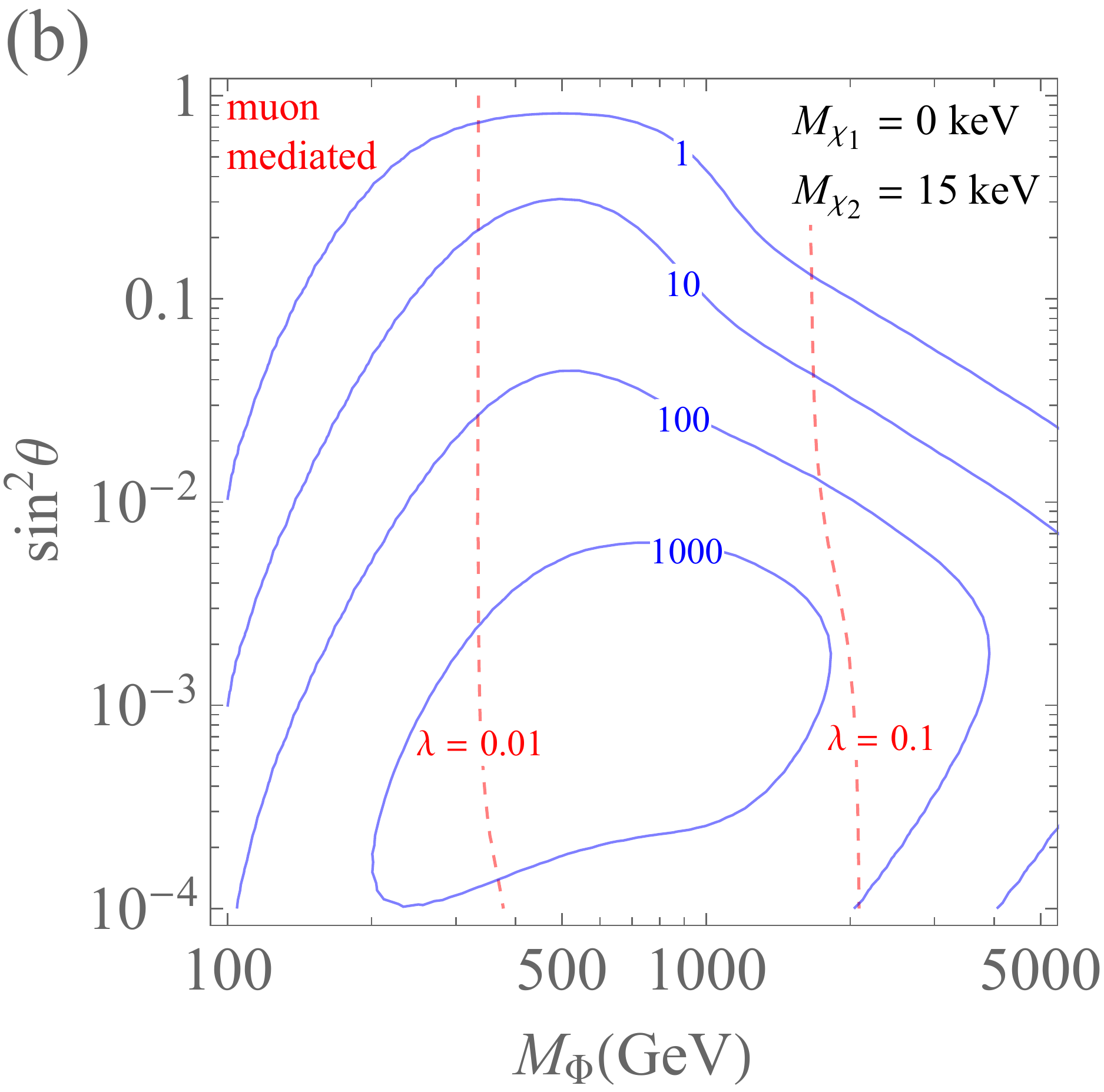}
             \includegraphics[width=2.3in]{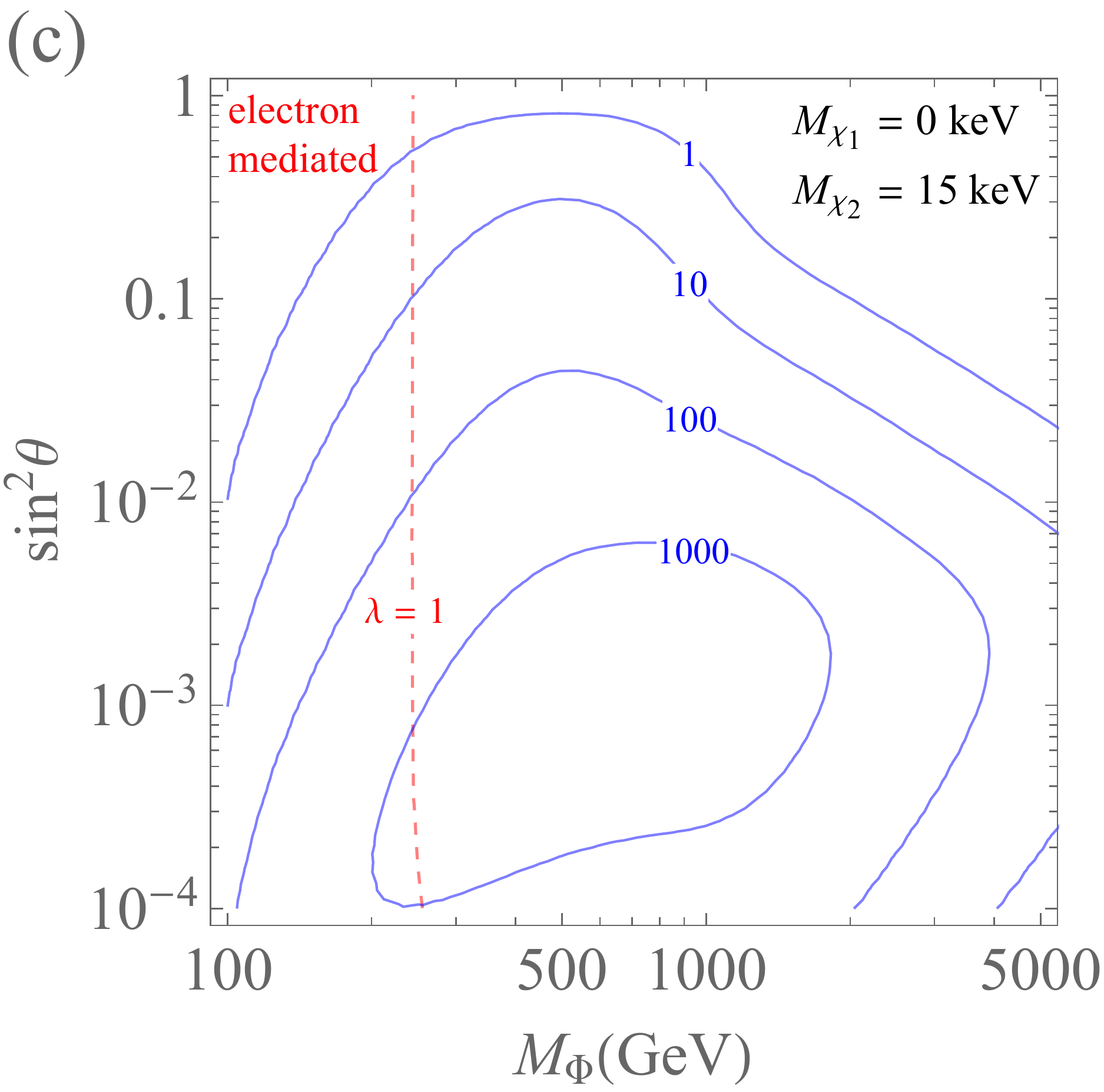}
\\
             \includegraphics[width=2.3in]{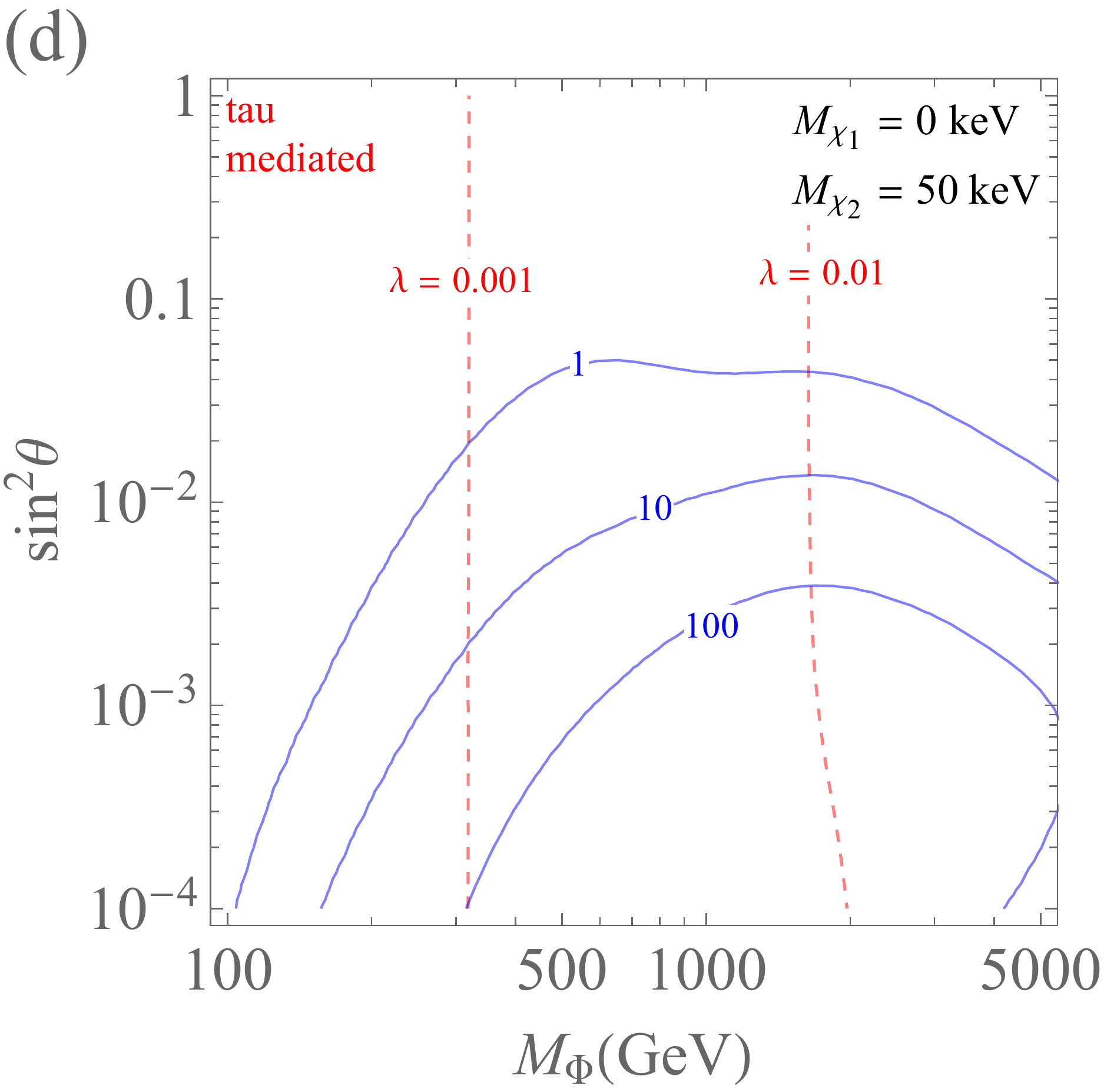}
             \includegraphics[width=2.3in]{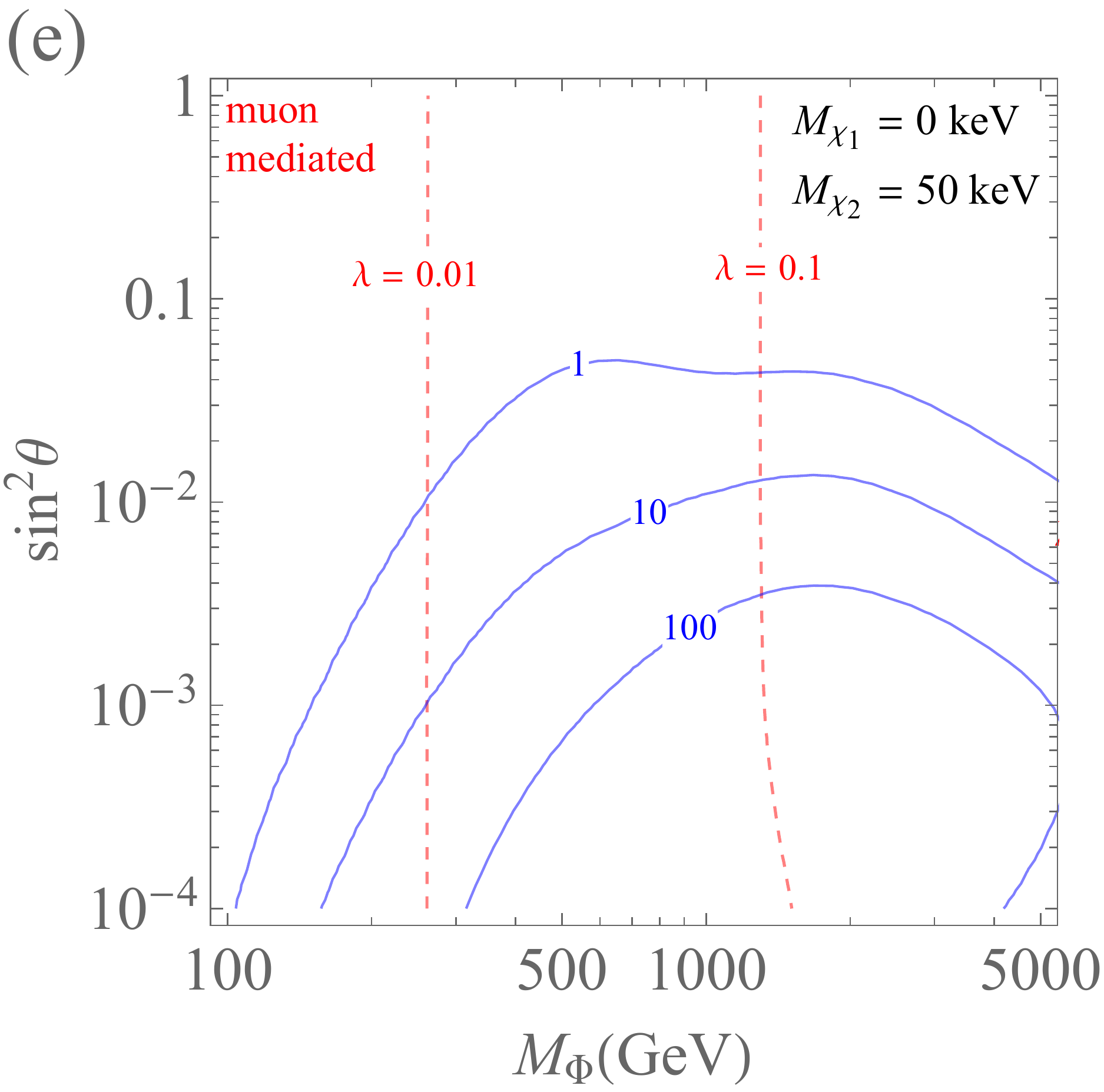}
             \includegraphics[width=2.3in]{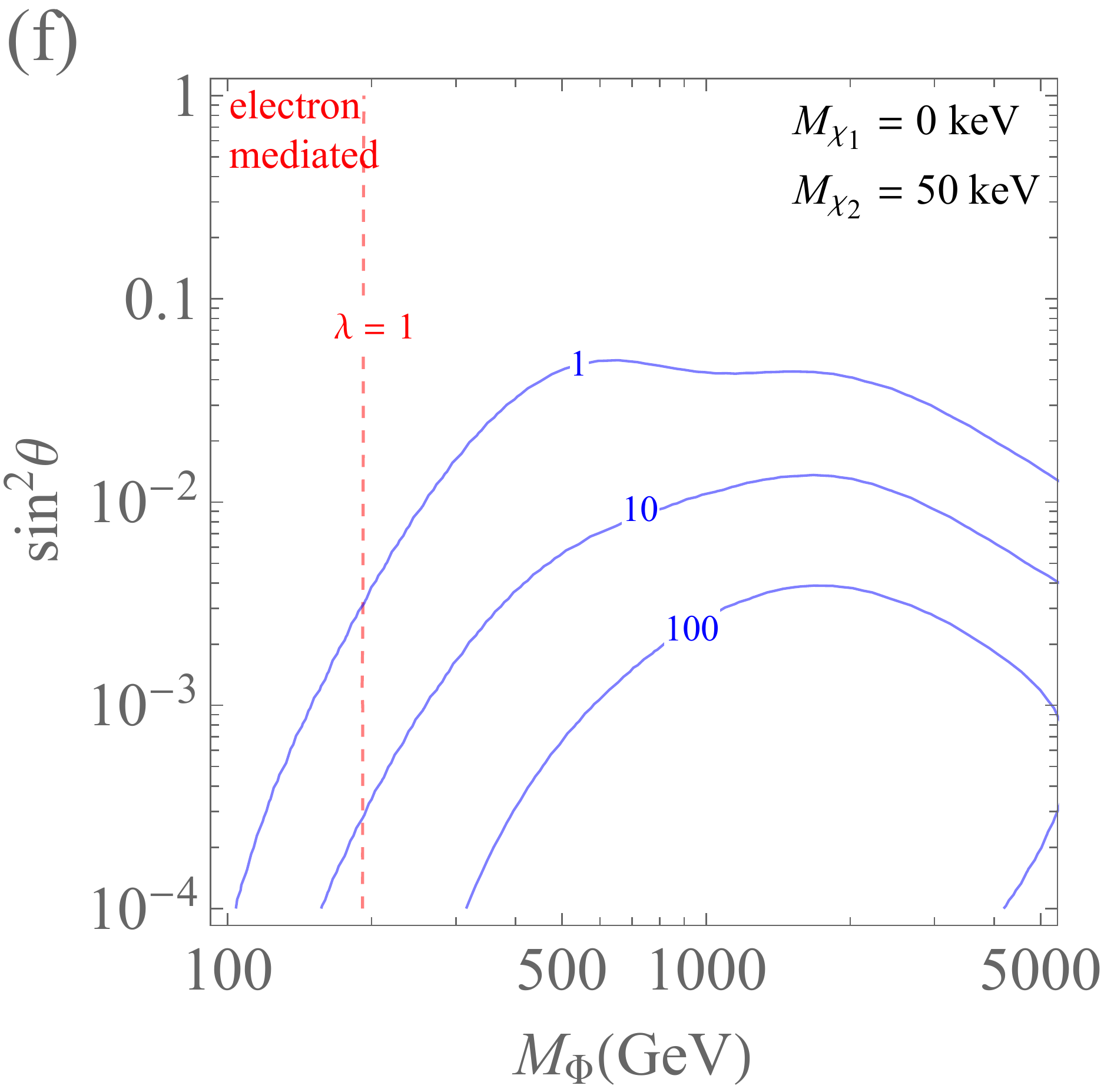}
\\
             \includegraphics[width=2.3in]{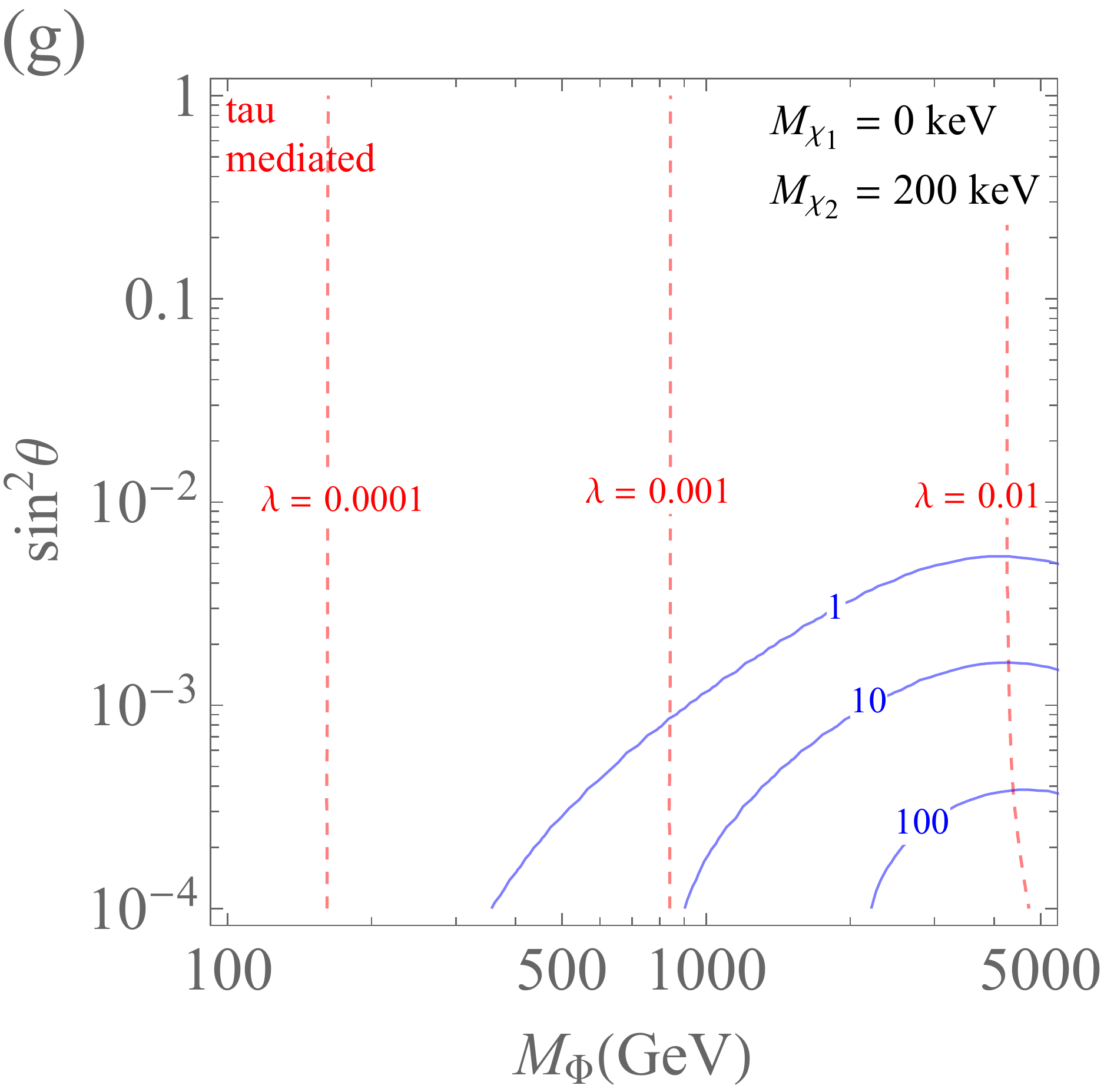}
             \includegraphics[width=2.3in]{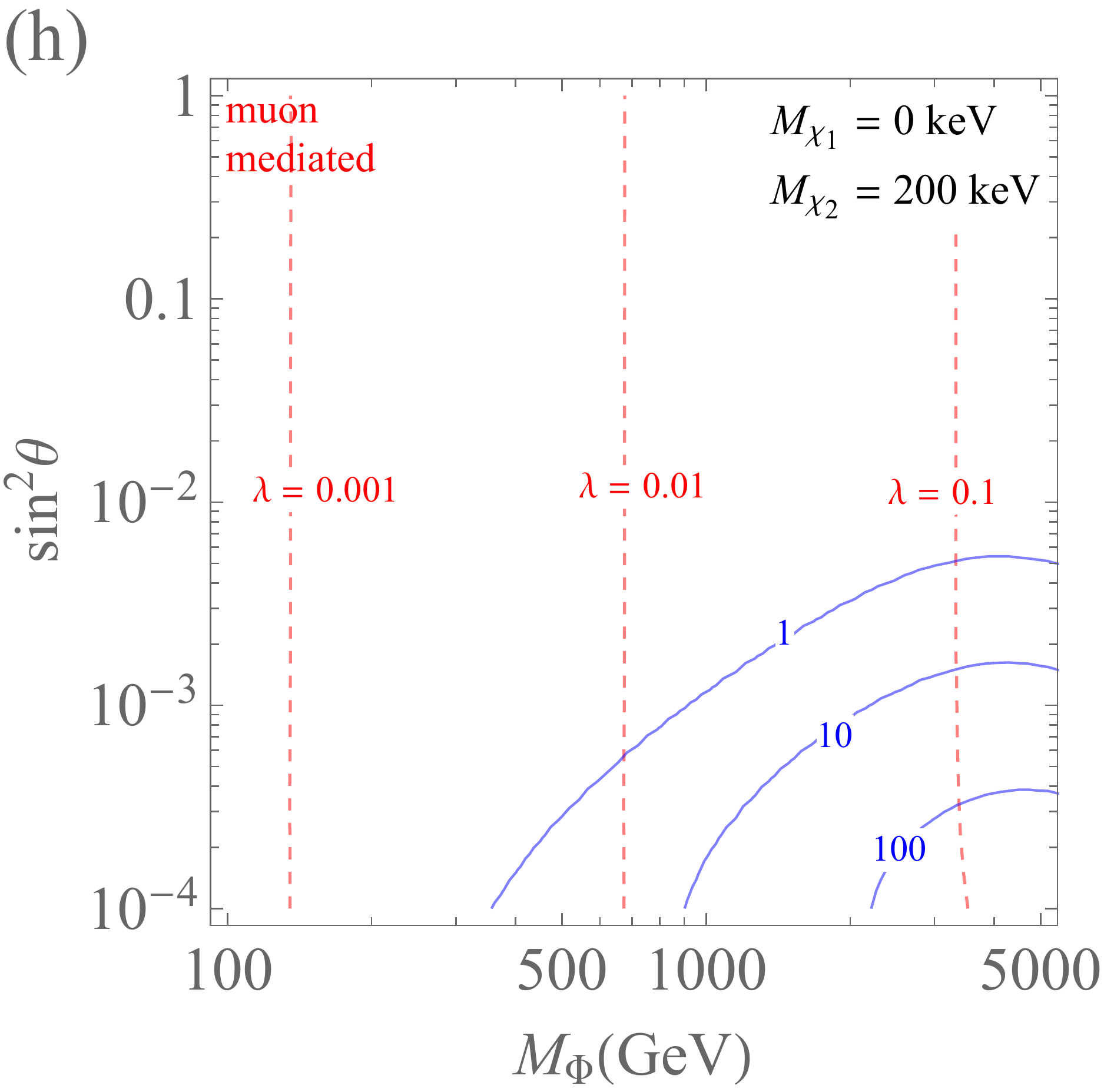}
             \includegraphics[width=2.3in]{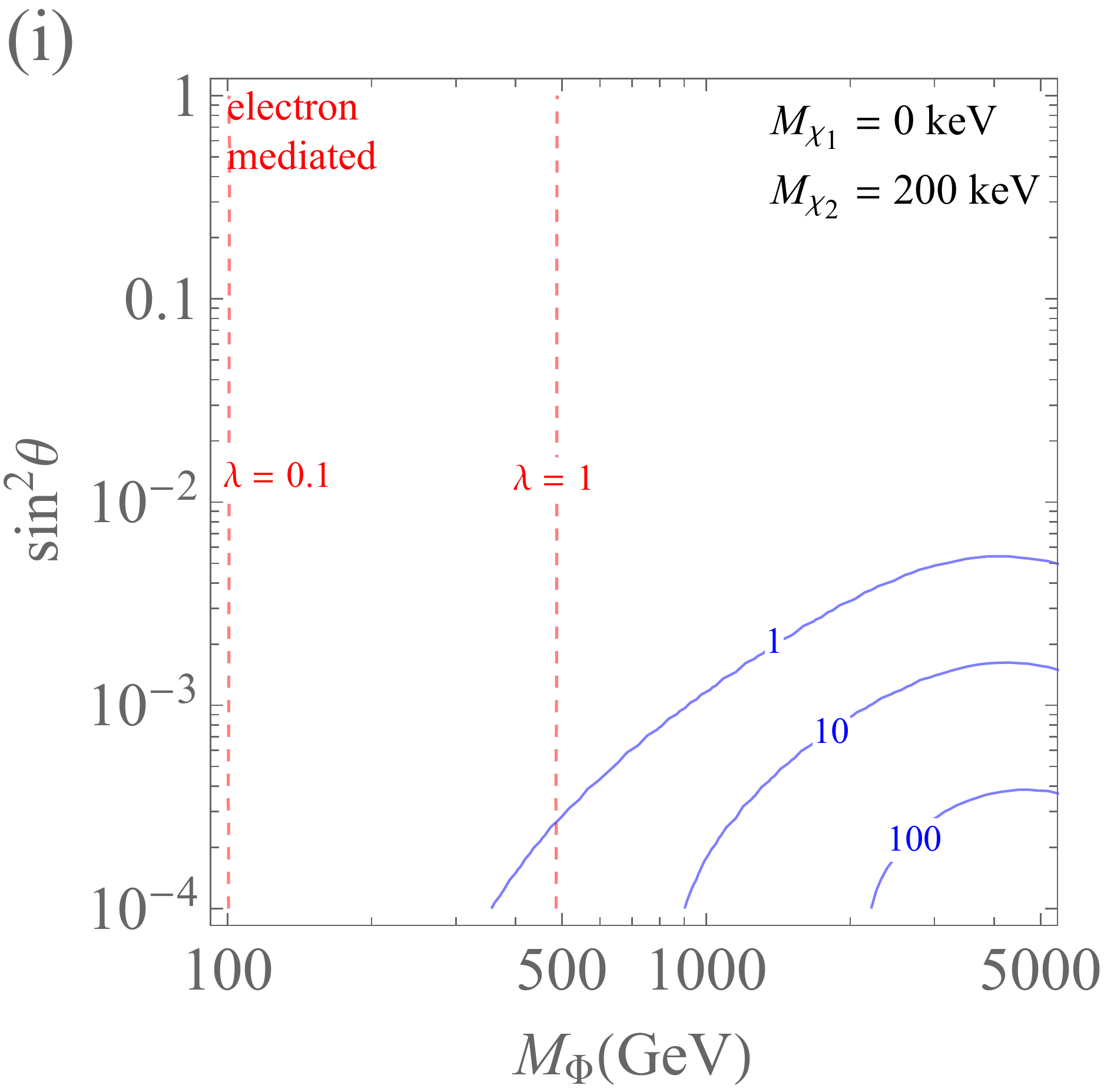}
\caption{
Contours of $Y_B/Y_B^\text{obs}$ (blue, solid) and X-ray bounds on $\lambda$ (red, dashed)  in the $(M_\Phi,\; \sin^2\theta)$ plane for $M_2 = 15$ keV (a-c), $M_2 = 50$ keV (d-f),
and $M_2 = 200$ keV (g-i). We take the massless-$\chi_1$ limit and adopt benchmark scenarios with a single Z2V coupling and the $F$ matrix chosen as described in Sec.~\ref{sec:Z2VFbench}.  In the left column  (plots a, d, and g), the flavor structure of the DM and Z2V couplings is such that the charged fermion in the diagram of Fig.~\ref{fig:DMdecay} is a $\tau$ lepton; the $\lambda$ bounds for the cases of $\mu$-mediated and $e$-mediated decays are shown in the middle column (plots b, e, and h) and right column (plots c, f, and i), respectively.
    }
   \label{fig:Z2Vxray}
\end{figure*}

The ``$\tau$-mediated'' partial width of Eq.~(\ref{eq:DMdecaytau}) leads to the X-ray bounds on $\lambda$ shown in  the left column of Fig.~\ref{fig:Z2Vxray} (plots a, d, and g), for various $M_2$ in the massless-$\chi_1$ limit.  At each point in the plane, $\text{Tr}[F^\dagger F]$ is fixed by the DM density constraint.  In  regions that work for DM and leptogenesis,  the upper bound on $\lambda$ is typically $\sim 10^{-2}-10^{-3}$ and never below $10^{-4}$.   The middle and right columns of Fig.~\ref{fig:Z2Vxray} show the weaker bounds on $\lambda$ that result when DM decays are instead mediated by muons or electrons, respectively, due to the flavor structure of the DM and Z2V couplings.

The  $\lambda$ couplings generate neutrino-portal couplings via the diagram of  Fig.~\ref{fig:DMdecay} with the photon line removed.  To assess the potential impact of this radiative effect on DM decay, we consider the case in which the $h_{\alpha i}$  are negligible at the cutoff scale $\Lambda$, leading to
\be\label{eq:radiativeh}
h_{\alpha i } \simeq - 
\sum_\beta
\frac{\lambda_{\alpha \beta} F_{\beta i} y_\beta}{8\pi^2}
\log
\frac{\Lambda}{M_\Phi}
\ee
at low energies, where $y_\beta$ is the SM Yukawa coupling for lepton flavor $\beta$.  We  continue to focus on benchmarks with a single Z2V coupling, so that only one term contributes to the sum in Eq.~(\ref{eq:radiativeh}), and only for one value of $\alpha$.  
The DM decay amplitude has a $1/M_\Phi^2$-suppressed ``IR'' contribution, which persists for $h_{\alpha i} =0$, and a ``UV'' contribution proportional to $\log(\Lambda/M_\Phi)$, arising from radiative generation of the neutrino-portal couplings. 

\begin{figure}
\includegraphics[width=3in]{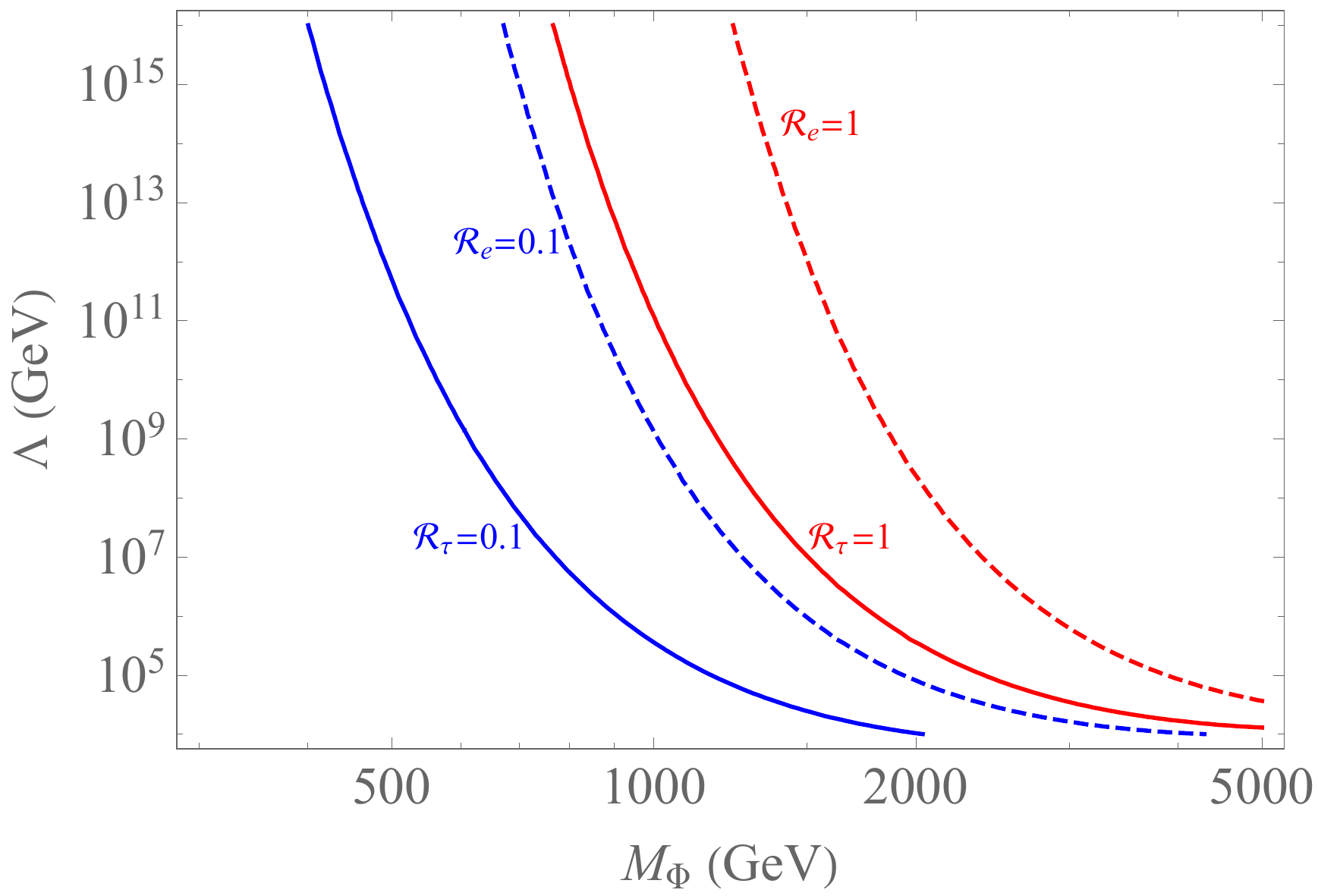}
\caption{Contours of ${\mathcal R} = 0.1$ (blue) and ${\mathcal R} = 1$ (red), where ${\mathcal R}$ is the ratio of decay widths defined in Eq.~(\ref{eq:widthratio}).  The solid and dashed contours are for the tau-mediated and electron-mediated scenarios, respectively.}
\label{fig:xrayratio}
\end{figure}

Although the IR and UV contributions should be combined at the amplitude level, for a simple comparison we evaluate the ratio
\be\label{eq:widthratio}
{\mathcal R}_\beta\equiv
\frac{
\Gamma_2^\text{UV} }{\Gamma_2^\text{IR}  }
=
\frac{9 G_F^2 M_\Phi^4}{32 \pi^4}
\left(\frac{\log
\frac{\Lambda}{M_\Phi}}{\log
\frac{M_\Phi^2}{M_\beta^2}
-1
}\right)^2,
\ee
where $\Gamma_2^\text{IR}$ is the $\chi_2$ partial width to $\nu\gamma,{\overline \nu}\gamma$ in the $h_{\alpha i}\rightarrow 0$ limit, while $\Gamma_2^\text{UV}$ is calculated using the neutrino portal couplings of Eq.~(\ref{eq:radiativeh}) and otherwise neglecting the $\lambda$ couplings (that is, ignoring 1/$M_\Phi^2$-suppressed contributions to the amplitude).   The ratio depends on the flavor $\beta$ of SM lepton running in the loop.  We show   contours of ${\mathcal R}_\tau$ and ${\mathcal R}_e$ in Fig.~\ref{fig:xrayratio}, corresponding to $\tau$-mediated and $e$-mediated scenarios. 
For $M_\Phi \lesssim 500$ GeV, the IR contribution dominates in either scenario, even for a Planck-scale cutoff.  For $M_\Phi \sim$ 1 TeV, the UV contribution dominates for $\Lambda \gsim 10^{11}$ GeV in the $\tau$-mediated scenario but is still subdominant in the $e$-mediated scenario up to $\Lambda \sim 10^{16}$ GeV.  Unlike the $\lambda$ upper bounds shown in Fig.~\ref{fig:Z2Vxray}, which are based on the IR contribution, the X-ray constraints on $\lambda$ one derives taking into account only the UV contribution get {\em stronger} with increasing $M_\Phi$, due to the larger DM couplings required by the DM abundance constraint.   However, the bound is never stronger than $\lambda < 10^{-4}$ for the parameters shown in Fig.~\ref{fig:Z2Vxray}, even for $\Lambda =  10^{16}$ GeV with $\tau$ in the loop.

\subsubsection{Supernova constraints}

Mixing of SM and sterile neutrinos is constrained by SN1987A.  If the mixing angle $\theta_{\nu N}$ is too large, an excessive fraction of the energy of the supernova explosion is carried away by the  sterile neutrinos.  One might expect  the couplings of the $\chi$ particles in our models to be similarly constrained, but this is not  the case in the Minimal and UVDM models.  Even if the DM couples to electrons, the cross section for the process $e^+ e^- \rightarrow \chi \chi$ inside the core of the supernova  is proportional to $F^4/M_\Phi^4$, versus the $G_F^2 \sin^2(2\theta_{\nu N})$ dependence for a weak interaction that produces a sterile neutrino.  The constraints in the sterile neutrino case are never stronger than $\sin^2(2\theta_{\nu N}) \lesssim 10^{-13}$ for any mass; see for example the ``no-feedback'' results of Ref.~\cite{Suliga:2020vpz}.  Meanwhile, we typically need $F\sim 10^{-7}$ for DM and leptogenesis, so that  $F^4$ is less than this upper bound by about fifteen orders of magnitude.  In the Z2V model, $\chi$ can be produced singly with cross sections that are proportional to $\lambda^2 F^2$ instead of $F^4$.  Here it is less obvious that the supernova cooling constraint can be ignored, although even a mild suppression from the $\lambda^2$ factor seems likely to be enough to evade it.  Moreover, these constraints become irrelevant if either the DM or Z2V couplings do not involve electrons, a scenario that works perfectly well for leptogenesis.

\subsection{Other Z2V phenomenology}\label{sec:otherZ2V}
\subsubsection{Constraints from low-energy probes}

Various low-energy measurements constrain the $\lambda - M_\Phi$ parameter space of the Z2V Model.  Some of the bounds can be inferred from earlier studies of  $LLE^c$ superpotential terms in the context of $R$-parity-violating supersymmetry, summarized in Ref.~\cite{Barbier:2004ez}. To interpret these  in the context of the Z2V Model, we replace the three flavors of right-handed sleptons  by a single $\Phi$ field and decouple the left-handed sleptons entirely.

Via tree-level $\Phi$ exchange, a non-zero $\lambda_{12}$ coupling affects the muon decay rate and therefore $G_F$, which in turn affects the extraction of $|V_{ud}|$, $|V_{us}|$, and $|V_{ub}|$ from nuclear $\beta$ decays, kaon decays, and charmless $B$ meson decays.  Refs. \cite{Barbier:2004ez, RPV1998} obtain the $2\sigma$ constraint 
\be
|\lambda_{12}| \lesssim 5\times10^{-2} \left(\frac{M_\Phi}{100\text{ GeV}} \right).
\ee
The constraint does not change appreciably when we update the analysis using current PDG values for the CKM matrix elements \cite{Zyla:2020zbs}.  Measurements of atomic parity violation in cesium lead to a similar constraint on $|\lambda_{12}|$, while neutrino-electron scattering gives a weaker bound.

Exchange of $\Phi$ can also lead to tree-level violation of lepton flavor universality.  Focusing for simplicity on scenarios in which a single $\lambda$ coupling dominates (neglecting possible cancelations), the current PDG average for the ratio $\Gamma(\tau\rightarrow \mu {\overline \nu}_\mu \nu_\tau)/\Gamma(\tau\rightarrow e {\overline \nu}_e \nu_\tau)$ leads to the constraints
\be
|\lambda_{13}| &\lesssim&  2\times10^{-2} \left(\frac{M_\Phi}{100\text{ GeV}} \right)
\ee
\be
|\lambda_{23}|& \lesssim &7\times10^{-2} \left(\frac{M_\Phi}{100\text{ GeV}} \right).
\ee
The upper bound on $|\lambda_{23}|$ is the same as reported in Refs.~\cite{Barbier:2004ez, RPV1998}, while the upper bound on $|\lambda_{13}|$ has strengthened  following the BaBar measurement of Ref.~\cite{BaBarLFU2010}.  Similarly, updating the analysis based on the ratio of partial widths $\Gamma(\tau\rightarrow \mu {\overline \nu}_\mu \nu_\tau)/\Gamma(\mu \rightarrow e {\overline \nu}_e \nu_\tau)$, we find
\be
|\lambda_{12}| &\lesssim  &1\times10^{-2} \left(\frac{M_\Phi}{100\text{ GeV}} \right)\\
|\lambda_{23}| &\lesssim &6\times10^{-2} \left(\frac{M_\Phi}{100\text{ GeV}} \right).
\ee
The current $2\sigma$ bound on $|\lambda_{12}|$ is significantly stronger than that obtained at the time of the original analysis ~\cite{Barbier:2004ez, RPV1998}, while the bound   on $|\lambda_{23}|$ has again not changed much.

If two or more $\lambda$ couplings are sufficiently large, Z2V interactions can lead to lepton flavor violating signals.
 Updating the bound given in Ref.~\cite{Barbier:2004ez} to take into the current experimental limit $B(\mu \rightarrow e \gamma)< 4.2 \times 10^{-13}$ (90\% CL)\cite{themegcollaboration2016search}, we obtain 
\be
|\lambda_{13}\lambda_{23}^*| \lesssim 4\times 10^{-5} \left(\frac{M_\Phi}{100\text{ GeV}} \right)^2
\ee
as the $\mu \rightarrow e\gamma$ constraint on the Z2V Model.

To obtain the bound from $\mu\rightarrow e$ conversion we follow Ref.~\cite{deGouvea:2000cf}.
The ratio of the conversion rate relative to the overall muon capture rate is 
\begin{multline}\label{eq:mue1}
R(\mu\rightarrow e)  =  \frac{4\alpha^5 Z_{\rm eff}^4|F(q)|^2 m_\mu^5}{Z\,\Gamma(\mu\,\,\mathrm{cap.})}\\
\times \left[\left|A_1^{\rm R}-A_2^{\rm L}\right|^2+Z^2\left|A_1^{\rm L}-A_2^{\rm R}\right|^2\right],
\end{multline}
where we neglect any direct coupling of $\Phi$ to quarks. The effective vertices are  \cite{deGouvea:2000cf}
\be
\label{eq:muevert1}
A_1^{\rm L} &=& \frac{\lambda_{13}\lambda^*_{23}}{288\pi^2M_\Phi^2}
\\
\label{eq:muevert2}
A_2^{\rm R}&=&\frac{\lambda_{13}\lambda^*_{23}}{192\pi^2 M_\Phi^2},
\ee
with the others zero. We therefore have
\beq
\label{eq:mue3}
R(\mu\rightarrow e) = \frac{\alpha^5 Z\,Z_{\rm eff}^4|F(q)|^2 m_\mu^5}{82944\pi^4M_\Phi^4\,\Gamma(\mu\,\,\mathrm{cap.})}|\lambda_{13}\lambda_{23}|^2.
\eeq
The strongest current limits are from the SINDRUM-II experiment, where gold ($Z=79$) was used as the target nucleus. The muon capture rate for gold is $\Gamma_{\rm Au}=13.1\times10^6\,\,\mathrm{s}=8.6\times10^{-18}\,\,\mathrm{GeV}$ \cite{Suzuki:1987jf}. The effective atomic number is $Z_{\rm eff}=33.64$ \cite{Suzuki:1987jf,Chiang:1993xz}. Finally, we need the nuclear form factor $|F(q)|$. Although Ref.~\cite{Chiang:1993xz} does not provide these form factors for gold, it does provide them for lead ($Z=82$):
\be
\label{eq:mueF1}
\overline{F}_p &=& 0.25,
\\
\label{eq:mueF2}
\overline{F}_n &=& 0.22.
\ee  
Ref.~\cite{deGouvea:2000cf} appears to take a weighted average of the form factors according to the number of protons and neutrons. This gives us $|F|=0.232$. Using the SINDRUM-II limit $R(\mu\rightarrow e) < 7\times10^{-13}$ \cite{Bertl:2006up}, we find the limit
\beq\label{eq:mue5}
\left|\lambda_{13}\lambda_{23}^*\right| \lesssim 1.8\times10^{-3}
\left(\frac{M_\Phi}{100\,\,\mathrm{GeV}}\right)^2.
\eeq
This constraint is much stronger if there exist both LH and RH sleptons due to log enhancements of the Feynman parameter integrals, but since we only have a RH-type scalar the bound is much weaker.

For an analysis of $\mu\rightarrow 3e$ decays, we again follow  Ref.~\cite{deGouvea:2000cf}. In terms of the same loop-induced FCNC photon vertices of Eqs.~(\ref{eq:muevert1},\ref{eq:muevert2}), the $\mu\rightarrow3e$ branching fraction is
\begin{multline}\label{eq:mu3e2}
\mathrm{BF}(\mu\rightarrow 3e) = \frac{6\pi^2\alpha^2}{G_{\rm F}^2}\\
\times \left[\left|A_1^{\rm L}\right|^2 + \frac{8}{3}\left(\log\frac{m_\mu^2}{m_e^2}-\frac{11}{4}\right)\left|A_2^{\rm R}\right|^2-4A_1^{\rm L}A_2^{\rm R}\right].
\end{multline}
The current PDG limit, $B(\mu \rightarrow e \gamma)< 4.2 \times 10^{-13}$, then gives
\beq\label{eq:mu3e4}
\left|\lambda_{13}\lambda_{23}^*\right| \lesssim 9.1\times10^{-4} \left(\frac{M_\Phi}{100\,\,\mathrm{GeV}}\right)^2,
\eeq
a limit that is slightly stronger than than for  $\mu \rightarrow e$ conversion but less stringent than $\mu \rightarrow e \gamma$.

\subsubsection{The $g_\mu-2$ anomaly}
Measurements of the muon anomalous magnetic moment, $a_\mu=(g_\mu-2)/2$, by the BNL E821 and Fermilab Muon $g-2$ experiments may indicate the need for new physics beyond the SM \cite{Muong-2:2006,Muong-2:2021ojo}.  Ref.~\cite{Muong-2:2021ojo} puts the discrepancy between experiment and SM theory at
\be
a_\mu(\text{Exp})-a_\mu(\text{SM}) = (251\pm 59)\times 10^{-11},
\ee
using the combined BNL and Fermilab measurements and the SM value determined by the Muon $g-2$ Theory Initiative \cite{Aoyama:2020ynm}.

In the Z2V Model of Sec.~\ref{sec:Z2V}, $a_\mu$ receives a {\em negative}  contribution from a one-loop diagram with $\Phi$ and SM leptons running in the loop.  We can extract the contribution from the results of Ref.~\cite{susymuongm2:2001}, which considered $a_\mu$ in the context of $R$-parity-violating supersymmetry.  
One finds \cite{susymuongm2:2001}
\be
(a_\mu)_{\Phi,\text{ one loop}} = -\frac{|\sum_\alpha\lambda_{\alpha 2}|^2}{48 \pi^2} \frac{M_\mu^2}{M_\Phi^2}.
\ee
In the model we have focused on in this paper, with an electroweak-singlet  $\Phi$, it is therefore not possible to explain the $g_\mu-2$ discrepancy.    

On the other hand, a $g_\mu-2$ explanation may be possible
in the model variation with an electroweak-doublet $\Phi$ for special arrangements of parameters.  Consider for concreteness a scenario in which DM couples only to the SM lepton doublets of the first two generations, while $\Phi$ has a single Z2V coupling,  to $l_2 e_3^c$:  
\be
\label{eq:Lfordoubletgm2}
\mathcal{L} \supset -F_{1 i}\Phi^*l_1\chi_i -F_{2 i}\Phi^*l_2\chi_i - \lambda \Phi l_2 e_3^c + \text{h.c.} 
\ee
Taking $\lambda$ to be large enough to come into equilibrium in the early universe, a baryon asymmetry is generated at $\mathcal{O}(F^4)$ via the mechanism discussed in Sec.~\ref{sec:Z2V}. The one-loop contribution to $a_\mu$ is~\cite{susymuongm2:2001}
\be
(a_\mu)_{\Phi,\text{ one loop}} & =  &+\frac{|\lambda|^2}{48 \pi^2} \frac{M_\mu^2}{M_{\Phi^0}^2}\\
& \simeq & (240\times 10^{-11})
\times |\lambda|^2 \left(\frac{100\text{ GeV}}{M_{\Phi^0}} \right)^2,
\quad\quad
\ee
where $M_{\Phi^0}$ is the mass of the electrically neutral $\Phi$ particle.  The $g-2$ discrepancy can thus be resolved for $\lambda \sim 1$ and $M_{\Phi^0} \sim 100$ GeV.

The special flavor structure assumed in Eq.~(\ref{eq:Lfordoubletgm2}) evades certain experimental constraints.  In the absence of additional sources of lepton flavor violation ({\em e.g.} neutrino masses), the operators $ l \chi H$, $l \sigma^{\mu \nu} \chi H B_{\mu \nu}$, and  $l \sigma^{\mu \nu} \chi  \sigma_a H W^a_{\mu \nu}$  are not generated radiatively.  This can be seen from a symmetry under equal phase rotations of  the fields $H$, $l_3$, $e^c_1$, and $e^c_2$, which is respected by the interactions of Eq.~(\ref{eq:Lfordoubletgm2}) and the SM charged-lepton Yukawa couplings.  The interactions responsible for neutrino masses presumably violate this symmetry and cause these operators to be induced at some level,  but these effects are model-dependent and generally come with additional loop and coupling suppressions\footnote{Radiative contributions to an  $H\Phi$ mass-squared term, which induces Higgs-$\Phi$ mixing,  come with similar suppressions.  As usual for a mass-squared term in the scalar potential, these radiative contributions are proportional to $\Lambda^2$, where $\Lambda$ is the cutoff of the theory.}.  It is thus possible to satisfy the X-ray constraints considered in Sec.~\ref{sec:xray}, even for the small $\Phi^0$ masses and large Z2V couplings required by the $g-2$ anomaly.  

At the LHC, $\Phi^0{\Phi^0}^*$ pair production would lead to events with $\mu\tau$ pairs.   A detailed study would be required to determine the status of this scenenario with respect to existing LHC searches.    The CMS search for LFV Higgs decays \cite{CMS_H_LFV_2021}, which constrains the branching ratio for $H\rightarrow \tau \mu$ to be below 0.15\% at 95\% CL, might be relevant, but there are various aspects of that analysis  that would seem to seriously reduce the signal efficiency for $\Phi$ pair production.  These include a veto on extra leptons and boosted decision tree variables chosen based on the kinematics of   $H\rightarrow\mu \tau$ events, which differ significantly from those of events with pair-production of particles that decay to $\mu \tau$.

\section{Conclusions}\label{sec:conclusions}

The freeze-in production and oscillation of DM provides a simple and well-motivated mechanism for baryogenesis. When DM couples to SM leptons, there must exist at least one new electroweak-charged scalar that can be as light as 100 GeV, and its couplings to the SM affect both the magnitude of the resulting asymmetry as well as the phenomenology. We have identified three benchmark models of interest that highlight the novel cosmology and signatures:~a minimal scenario in which there exists a single new scalar and all non-SM fields are charged under a $Z_2$ symmetry that stabilizes DM; a scenario where primordial production of DM through an unspecified mechanism can provide an enhancement to the asymmetry; and, a scenario in which the $Z_2$ symmetry is broken, and SM lepton flavor effects likewise enhance the asymmetry relative to the minimal model.

We have found the minimal model to be quite constrained, with the mass of the scalar, $\Phi$, required to lie below approximately 1.5 TeV such that high-energy collider searches could be sensitive to much of the parameter space favored by baryogenesis. In the other models, $\Phi$ can be heavier, although in that case the couplings are typically aligned to couple preferentially to the lighter DM state, which has a mass well below 1 keV to give a subdominant contribution to the dark matter energy density. Current collider constraints arise from searches for prompt or displaced leptons and missing transverse momentum motivated largely by supersymmetry, but  the different flavor structure in our models leads to a weakening of several of these constraints and motivates dedicated searches for scalars that can decay to multiple flavors of leptons.

Constraints from structure formation and dark radiation significantly impact the parameter spaces of all of the models, and so there are good prospects for observing a signal if the sensitivities can be improved.  
Furthermore, the model that violates the $Z_2$ symmetry can give an observable X-ray line from DM decay and/or signals at low-energy terrestrial experiments, including a possible explanation of the $(g_\mu-2)$ anomaly for the case of an electroweak-doublet $\Phi$.

\acknowledgments

We are grateful to Alex Kusenko, Jane Schlesinger, and Benny Weng for helpful conversations. The work of BS is supported by the U.S. National Science Foundation under Grant PHY-1820770 and by the Research Corporation for Science Advance through a Cottrell Scholar Award.

\appendix

\section{Benchmark $F$ matrices}\label{sec:appendixFbench}

\subsection{Minimal Model}\label{sec:MMFbench}
By appropriate phase transformations on the charged leptons, the $F_{\alpha i}$ matrix can be brought into the form
\be
\footnotesize{
\label{eq:Ffullparam}
    F =\sqrt{ \text{Tr}[F^\dag F]}\begin{pmatrix}
    \cos{\theta}\cos{\beta_1}\cos{\gamma_1} & \sin{\theta}\cos{\beta_2}\cos{\gamma_2}e^{i\phi_1} \\
    \cos{\theta}\sin{\beta_1}\cos{\gamma_1} & \sin{\theta}\sin{\beta_2}\cos{\gamma_2}e^{i\phi_2} \\
    \cos{\theta}\sin{\gamma_1} & \sin{\theta}\sin{\gamma_2}e^{i\phi_3}
    \end{pmatrix}\!,
    }
\quad\quad\;\;
\ee
with $0 \le \theta,\; \beta_1,\;\beta_2,\;\gamma_1,\;\gamma_2 \le \pi/2$ and $0 \le \phi_1,\;\phi_2,\;\phi_3 < 2\pi$.  
The partial widths of $\Phi$ to final states involving $\chi_1$ and $\chi_2$ are proportional to $(F^\dagger F)_{11} = \cos^2\theta\; \text{Tr}[F^\dag F]$ and $(F^\dagger F)_{22} = \sin^2\theta\; \text{Tr}[F^\dag F]$, respectively.  The DM energy density therefore only depends on $M_1$, $M_2$, $\text{Tr}[F^\dag F]$, and $\theta$, and not on any of the other angles or phases appearing in $F$. 

For our analysis of the Minimal Model, we adopt a benchmark in which only two lepton flavors couple.   We take $\gamma_1=\gamma_2=0$, $\phi_1 = \pi/2$, and $\phi_2=0$, with $\beta_1$ and $\beta_2$ set to a common value, $\beta_1=\beta_2=\beta$.  With these choices,  the $\mathcal{O}(F^6)$ asymmetry, which is given in Eq.~(\ref{eq:MMYB6}), is proportional to $\cos 2\beta \sin^2 2\beta$.  This factor is maximized for $\cos 2\beta =1/\sqrt{3}$, and so we adopt
\be
\label{eq:FMMbench}
\small{
    \frac{F}{\sqrt{ \text{Tr}[F^\dag F]}} =\begin{pmatrix}
    \cos{\theta}\cos{\beta} &i \sin{\theta}\cos{\beta} \\
    \cos{\theta}\sin{\beta}& \sin{\theta}\sin{\beta} \\
0 & 0
    \end{pmatrix};
   \;\;
    \cos 2\beta = \frac{1}{\sqrt{3}}
    }\quad\quad\;\;
\ee
as our benchmark $F$ matrix for the Minimal Model.  Having equal-strength couplings for the two active flavors,  $\beta=\pi/4$, gives zero asymmetry, because in that case the washout rate is the same for both flavors, and the ARS mechanism is spoiled.  

For this benchmark, one finds that the combination of couplings appearing in the $\mathcal{O}(F^6)$ baryon asymmetry evaluates to
\be\label{eq:MMbenchnum}
\frac{\sum_\alpha
(F F^\dagger)_{\alpha \alpha}
\mathrm{Im}\left[
F_{\alpha 1}^*  
F_{\alpha 2}
\left(F^\dagger F \right)_{21}
\right]}{(\text{Tr}[F^\dag F])^3}
\simeq  0.024 \sin^2 2\theta.
\quad\quad
\ee
Performing a full numerical optimization for three active flavors, we find that this two-active-flavor benchmark gives an $\mathcal{O}(F^6)$ asymmetry that is $\simeq 0.82$ and  $\simeq 0.69$ times the fully optimized symmetry for $\theta = \pi/4$ and  for $\theta \ll 1$, respectively.  So although our simple benchmark does not maximize the asymmetry (even at the perturbative level), it  gives a reasonable estimate of the full viable parameter space for DM and leptogenesis.  

\subsection{Benchmark $F$ matrices for the UVDM Model}\label{sec:UVDMFbench}
In the UVDM Model, the matrices $F^1$ and $F^2$ represent the couplings of DM to $\Phi_1$ and $\Phi_2$, respectively.   The leading-order baryon asymmetry is proportional to  
\begin{multline}\label{eq:UVDMcoup}
\mathrm{Im}\left[
\left({F^1}^\dagger 
F^1 \right)_{21}
\left({F^2}^\dagger F^2 \right)_{12}
\right] \\
=
\frac{1}{4}
\mathcal{J}\,\mathrm{Tr}\left[{F^1}^\dagger{F^1}\right]\mathrm{Tr}\left[{F^2}^\dagger {F^2}\right],
\quad\quad
\end{multline}
where $\mathcal{J}$ can be parametrized in terms of six mixing angles/phases \cite{Shuve:2020evk},
\be\label{eq:jarlskog_invariant}
\mathcal{J} =\sin2\theta_1\sin2\theta_2\cos\rho_1\cos\rho_2\sin(\phi_2-\phi_1),
\ee
with
\be
\label{eq:theta2s}
\cos\theta_i &=& \sqrt{\frac{({F^i}^\dagger {F^i})_{11}}{\mathrm{Tr}({F^i}^\dagger F^i)}},
\\
\label{eq:rho2s}
\cos\rho_i &=& \frac{|({F^i}^\dagger F^i)_{12}|}{\sqrt{({F^i}^\dagger F^i)_{11}({F^i}^\dagger F^i)_{22}}},
\\
\label{eq:phi2s}
\phi_i &=& \arg({F^i}^\dagger F^i)_{12}, \label{eq:phases}
\ee
and $0\le(\theta_i, \rho_i) \le \pi/2$.

The $\mathcal{O}(F^4)$ baryon asymmetry is maximized for $\rho_1 = \rho_2 = 0$ and $\phi_2 - \phi_1 = \pi/2$, choices which define our UVDM benchmark. These parameters are realized, for example, for the coupling matrices
\be\label{eq:UVDMbench}
    F^1 &= &\sqrt{ \text{Tr}[{F^1}^\dag F^1]}\begin{pmatrix}
    \cos{\theta_1}&  \sin{\theta_1} \\
    0& 0\\
   0 & 0
    \end{pmatrix}\\
        F^2 &= &\sqrt{ \text{Tr}[{F^2}^\dag F^2]}\begin{pmatrix}
    \cos{\theta_2}&  i\sin{\theta_2} \\
    0& 0\\
   0 & 0
    \end{pmatrix},
\ee
in which only a single flavor of charged lepton couples.

\subsection{Benchmark $F$ matrices for the Z2V Model}\label{sec:Z2VFbench}

For the Z2V Model of Sec.~\ref{sec:Z2V}, our benchmark $F$ matrix is motivated by scenarios in which either one or two of the Z2V couplings come into equilibrium.  In both cases there is one ``special'' SM flavor:~if a single Z2V coupling is in equilibrium, it is the flavor of SM lepton doublet that does {\em not} couple to $\Phi$; if two Z2V couplings are in equilibrium, it is the SM flavor involved in both of those couplings.  
For this discussion, we  label the special flavor with the index $\beta$. 

As we discuss in Appendix~\ref{sec:Z2Vpert},the $\mathcal{O}(F^4)$ baryon asymmetry turns out to be proportional to 
\be\label{eq:Z2Vfac}
Y^{(4)}_{\beta} \propto \mathrm{Im}\left[
F_{\beta 1}  
F_{\beta 2}^*
\left(F^\dagger F \right)_{12}
\right].  
\ee
We now refer back to the parametrization of the $F$ matrix in Eq.~(\ref{eq:Ffullparam}).  For fixed values of $\theta$ and $\text{Tr} [F^\dagger F ]$, the parameters relevant for the DM energy density, the maximum possible value of the quantity in Eq.~(\ref{eq:Z2Vfac}) is
\be\label{eq:Z2Vopt}
\frac{
\mathrm{Im}\left[
F_{\beta 1}  
F_{\beta 2}^*
\left(F^\dagger F \right)_{12}
\right]}{\left(\text{Tr}[F^\dagger F] \right)^2}
=
\frac{\sin^2 2\theta}{16}.
\ee
For example, for $\beta = 1$ the maximum value is realized for $\gamma_1=\gamma_2=0$, $\beta_1=\beta_2=\pi/4$, and $\phi_2-\phi_1=\pi/2$.
  That is, in scenarios with one or two Z2V couplings in equilibrium, having only two active flavors couple to DM turns out to be optimal for maximizing the asymmetry,  in the perturbative regime.  
  
 We take our benchmark $F$ matrices to be ones in which $e_\beta^c$ and one additional flavor of RH lepton couple, and in which Eq.~(\ref{eq:Z2Vopt}) is satisfied.  
For $\beta = 1$, one such $F$ matrix is
\be
\label{eq:FZ2Vbench}
    F =\sqrt{ \frac{\text{Tr}[F^\dag F]}{2}}\begin{pmatrix}
    \cos{\theta}&  \sin{\theta} \\
    \cos{\theta} &i \sin{\theta}\\
   0 & 0
    \end{pmatrix};
\ee
another would have zeros in the second row instead. 
For $\beta=2$, the row of zeros can be the first or the third, and for $\beta=3$ it can be the first or second.  These various flavor structures are equivalent as far as leptogenesis is concerned, although they have different collider implications.

\section{Chemical potential  relations}
\label{sec:appendixCPs}

\subsection{General relations}\label{sec:generalCPs}

Neglecting neutrino masses, SM interactions conserve the three charges $X_\alpha \equiv B/3-L_\alpha$, where $B$ is baryon number and the $L_\alpha$ are charges associated with the three lepton flavors.  We define the  $X_\alpha$ charges of the  BSM particles $\Phi$ and $\chi$ to be zero.
The DM and Z2V interactions are both $X_\alpha$-violating, which allows  non-zero $X_\alpha$ densities to evolve starting from what we assume is a neutral state after inflation. In Appendix~\ref{sec:appendixQKE}, we write down a set of quantum kinetic equations (QKEs) that model the evolution of the $X_\alpha$ and DM abundances, while in Appendix~\ref{sec:appendixpert} we provide a perturbative calculation of  these $X_\alpha$ densities that are valid in the weak-washout limit. 

To calculate a final baryon asymmetry, we need to take into account rapidly occurring SM spectator processes using the appropriate relations among the asymmetries of the various particle/antiparticle species.   In this appendix we derive those relations.  We include possible Z2V couplings from the start, which need not be in equilibrium, but do not include a neutrino-portal coupling for the DM.  The $X_\alpha$-violating interactions are then
\beq\label{eq:fullLint}
\mathcal{L} \supset -F_{\alpha i} e^c_\alpha \chi_i \Phi 
- \frac{\lambda_{\alpha \beta}}{2} l_\alpha l_\beta \Phi^* + \text{h.c.}
\eeq
In the UVDM model, there exists an additional scalar field $\Phi_2$ whose interactions can affect the chemical potential relations. However, in our work we restrict our discussion of the UVDM Model to the decoupled-$\Phi_2$ regime such that $\Phi_2$ is absent from the thermal bath at all cosmological times relevant for leptogenesis, and so its  couplings are not relevant here.  

Because the DM is light enough to be ultrarelativistic at sphaleron decoupling, we can neglect DM masses in this discussion.  This allows us to  treat the negative-helicity and positive-helicity DM states as a particle/antiparticle pair, $\chi$ and ${\overline \chi}$, carrying opposite values of conserved charges.  We define the $B$ and $L$ charges of $\chi$ and $\Phi$ to be $B_\chi = B_\Phi = 0$, $L_\chi = -1$, and $L_\Phi = 2$.  With these definitions,  the interactions of Eq.~(\ref{eq:fullLint}) conserve both $B$ and $L$.  We take $B-L$  to be conserved in general, with $B$ and $L$  separately conserved below the sphaleron decoupling temperature. The violation of $B-L$ induced by $\chi$ Majorana masses is inconsequential for the masses and timescales we consider. 

For particle species $i$, we define $Y_i \equiv n_i/s$ and $\delta Y_i \equiv Y_i - Y_{\overline i}$, where $n_i$ is the number density and $s$ is the entropy density.  The  $X_\alpha$ charge densities are  $Y_\alpha \equiv \sum_i (X_\alpha)_i \delta Y_i$, where $(X_\alpha)_i$ are the charges of particle species $i$.  In this appendix we will see that we can express all asymmetries, including the baryon asymmetry $Y_B \equiv \sum_i B_i \delta Y_i$, entirely in terms of the DM asymmetry $\delta Y_\chi$ and the three $Y_\alpha$.    

\begin{figure}
\includegraphics[width=3in]{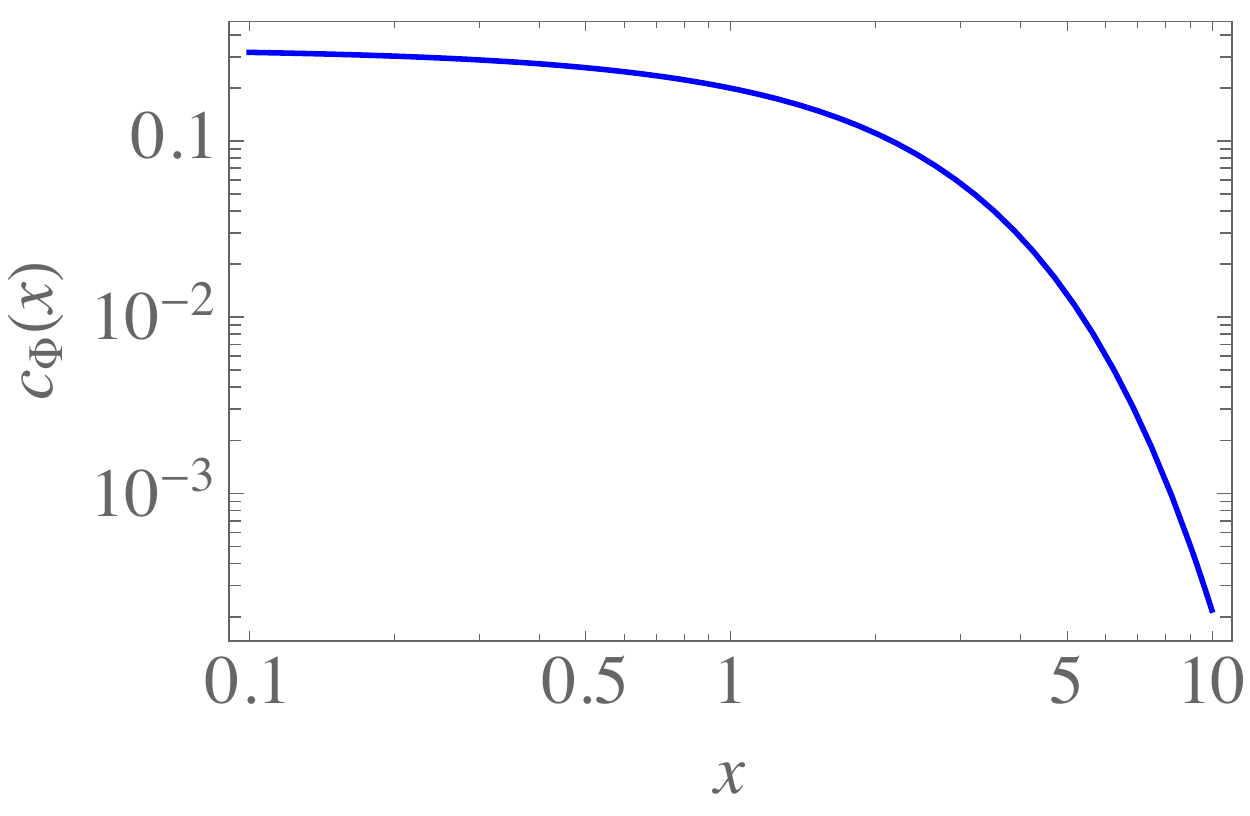}
\caption{The temperature-dependent function $c_\Phi$, defined in Eq.~(\ref{eq:cphi}). } 
\label{fig:cphi}
\end{figure}

In terms of the $\chi$ and ${\overline \chi}$ density matrices introduced in Appendix~\ref{sec:appendixQKE}, the DM asymmetry is $\delta Y_\chi = \text{Tr}\left[Y_\chi-Y_{\overline \chi} \right]$. There are eleven  additional particle/antiparticle asymmetries to consider, encoded in the chemical potentials $\mu_q$, $\mu_{u^c}$, $\mu_{d^c}$, $\mu_{l_\alpha}$, $\mu_{e^c_\alpha}$, $\mu_H$, and $\mu_\Phi$.  We take the quark chemical potentials to be flavor-independent due to flavor non-conservation in the quark sector.  Working in the electroweak symmetric phase and taking sphalerons to be fully in equilibrium, we have the following system of equations:
\be
\label{eq:upYuk}
\mu_q + \mu_{u^c} + \mu_H & = &0
\\
\label{eq:downYuk}
\mu_q + \mu_{d^c} - \mu_H & = &0 
\\
\label{eq:eYuk}
\mu_{l_\alpha} + \mu_{e^c_\alpha} - \mu_H & = &0 
\\
\label{eq:sph}
9 \mu_q+\sum_\alpha \mu_{l_\alpha}  & = &0 
\\
\label{eq:hyp} 
Y_y \equiv \sum_i y_i \delta Y_i  & = &0
\\
\label{eq:BmL}
Y_{B-L} \equiv \sum_\alpha Y_\alpha -2 \delta Y_\Phi + \delta Y_\chi  & = &0 
\\
\label{eq:Xalpha}
Y_\alpha - \sum_{i \in \text{ SM}} (B/3-L_\alpha)_i \delta Y_i & = &0. 
\ee
Eqs.~(\ref{eq:upYuk}-\ref{eq:sph}) are enforced by SM Yukawa interactions and sphalerons being in equilibrium.
Eq.~(\ref{eq:hyp}) follows from hypercharge neutrality, with $y_i$ being the hypercharge of particle species $i$,
Eq.~(\ref{eq:BmL}) similarly expresses neutrality under the conserved charge $B-L$, and Eq.~(\ref{eq:Xalpha}) simply reflects the definition of the $X_\alpha$ charges, with the sum over $i$ restricted to SM particles.   Eqs.~(\ref{eq:upYuk}-\ref{eq:Xalpha}) can be solved to express all asymmetries in terms of $Y_\alpha$ and $\delta Y_\chi$, using the appropriate relations between number-density asymmetries and chemical potentials\footnote{Here we linearize in all chemical potentials including $\mu_\Phi$.  In our numerical work we make the replacement $\mu_\Phi \rightarrow T \sinh(\mu_\Phi/T)$ to allow for the possibility of a highly asymmetric $\Phi$ population at temperatures well below $M_\Phi$;  see also the discussion leading to Eq.~(\ref{eq:don't_linearize_mu_phi}).}:
\be\label{eq:Yfrommu}
\delta Y_i = g_i c_i \frac{ \mu_iT^2}{s}.
\ee
Here, $s$ is the entropy density and $g_i$ counts gauge and flavor degrees of freedom ({\em e.g.} $g_q = 3\times 3\times 2 = 18$, while $g_{l_\alpha} = 2$ for each flavor $\alpha$). For simplicity, in Eq.~\ref{eq:Yfrommu} we neglect all masses (including thermal contributions) except for that of the $\Phi$ particle, giving $c_i = 1/6$ for SM fermions and $c_i =1/3$ for the SM Higgs doublet.   We take into account the potentially large mass of the  $\Phi$ particle, which leads to a temperature-dependent expression for $c_\Phi$.  Defining 
\be\label{eq:xdef}
x \equiv M_\Phi/T,
\ee
we find
\be
\label{eq:cphi}
c_\Phi & = & \frac{1}{\pi^2}\int_x^\infty\!dy\;y \sqrt{y^2-x^2} \frac{e^y}{(e^y-1)^{2}}
\\
\label{eq:cphiapprox}
& \simeq &
\begin{cases}
1/3  
\quad\quad\quad\quad\quad\quad
x \ll 1
\\
\frac{x^2}{\pi^2}
\mathcal{K}_2(x)
\quad\quad\quad\quad
x \gg 1.
\end{cases}
\ee
Fig.~\ref{fig:cphi} shows a plot of $c_\Phi$.

It is convenient to define $Y_\text{sm}$ to be the net $B-L$ charge stored in the SM sector:
\be\label{eq:Ysmdef}
Y_\text{sm} \equiv \sum_\alpha Y_\alpha.
\ee
Solving our system of equations then leads to
\be
\label{eq:ybspec}
Y_B & = & \frac{25}{79}\; Y_\text{sm}  - \frac{3}{79} \; \delta Y_\chi
\\
\label{eq:muespec}
\frac{\mu_{e^c_\alpha} T^2}{s} & = &
-\frac{59}{237}\;Y_\text{sm} 
  +2 \;Y_\alpha
  +   \frac{45}{237}\;\delta Y_\chi
\\
\label{eq:mulspec}
\frac{\mu_{l_\alpha}T^2}{s} & = &
 \frac{91}{474}\;Y_\text{sm} 
  -2 \;Y_\alpha
  +   \frac{27}{474}\;\delta Y_\chi
\\
\label{eq:muphispec}
\frac{\mu_\Phi T^2}{s}& = & \frac{1}{2 c_\Phi} \left( Y_\text{sm} + \delta Y_\chi \right).
\ee
In the QKEs, the washout terms depend on $\mu_{e^c_\alpha}$, $\mu_{l_\alpha}$, and $\mu_\Phi$.  We use Eqs.~(\ref{eq:muespec}-\ref{eq:muphispec}) to substitute in for these quantities, leading to a closed system of equations for tracking $\delta Y_\chi$ and the $Y_\alpha$.

Eqs.~(\ref{eq:ybspec}-\ref{eq:muphispec}) apply for all scenarios we consider, in the electroweak symmetric phase and while sphalerons remain in equilibrium.   For the perturbative calculations presented in Appendix~\ref{sec:appendixpert}, we take these relations to hold until sphaleron decoupling, which we approximate as occurring instantaneously at $T_\text{ew} = 131.7$ GeV \cite{DOnofrio:2014rug};  to calculate the final baryon asymmetry we evolve $Y_\alpha$ and $\delta Y_\chi$ from high temperatures to $T=T_\text{ew}$ and then evaluate Eq.~(\ref{eq:ybspec}) at $T=T_\text{ew}$.  When we numerically solve the QKEs of Appendix~\ref{sec:MIQKEs}, we take into account gradual sphaleron decoupling following the methods of Ref.~\cite{Eijima:2017cxr}, while continuing to use unbroken-phase relations.  Electroweak symmetry breaking could be taken into account following Ref.~\cite{Khlebnikov:1996vj}, but we expect these effects to be small.  For example, for $Z_2$-preserving scenarios we find $Y_B = (22/79) Y_\text{sm}$ in the unbroken phase (see the discussion leading to Eq.~(\ref{eq:ybspecZ2}) below), versus $Y_B = (22/74) Y_\text{sm}$ deep in the broken phase, a $\sim$7\% difference.

\subsection{Chemical potential relations in $Z_2$-preserving scenarios}\label{sec:Z2PCPs}
 In the Minimal and UVDM Models, we set the $Z_2$-violating couplings $\lambda_{\alpha \beta}$ in Eq.~(\ref{eq:fullLint}) to zero.  In this case we have an additional conserved $U(1)$ with only $\Phi$ and $\chi$ charged, oppositely.  The $\Phi$ and $\chi$ asymmetries are thus equal, and  Eq.~(\ref{eq:muphispec}) then gives  $Y_\text{sm} = \delta Y_\Phi = \delta Y_\chi $.  These relations can also be understood as direct consequences of neutrality under $B-L$, for two alternative 
BSM-particle lepton-number assignments, $\left\{L_\Phi = 1, \;L_\chi=0\right\}$ and $\left\{L_\Phi = 0, \;L_\chi=1\right\}$, both of which are consistent with $B-L$ conservation in the $Z_2$-preserving case (neglecting $\chi$ masses, for the second assignment). 
With this simplification, Eqs.~(\ref{eq:ybspec}-\ref{eq:muphispec}) become
\be
\label{eq:ybspecZ2} 
Y_B & = & \frac{22}{79}\; Y_\text{sm} 
\\
\label{eq:muespecZ2}
\frac{\mu_{e^c_\alpha} T^2}{s} & = &
-\frac{14}{237}\;Y_\text{sm} 
  +2 \;Y_\alpha
\\
\label{eq:mulspecZ2}
\frac{\mu_{l_\alpha}T^2}{s} & = &
 \frac{59}{237}\;Y_\text{sm} 
  -2 \;Y_\alpha
\\
\label{eq:muphispecZ2}
\frac{\mu_\Phi T^2}{s}& = & \frac{1}{ c_\Phi} Y_\text{sm} 
\ee
In these $Z_2$-preserving scenarios, the present-day baryon asymmetry is proportional to the $\Phi$ asymmetry at sphaleron decoupling.  For heavy $\Phi$ particles, $M_\Phi \gsim$  several TeV, the final baryon asymmetry thus suffers an exponential suppression if the $\Phi$ lifetime is much shorter than the age of the universe at sphaleron decoupling,  $c \tau_\Phi \ll t_{ew} \sim \text{cm}$.   In the Z2V Model, on the other hand, the $\Phi$ and $\chi$ asymmetries need not be equal, and this exponential suppression is not guaranteed.

\subsection{Chemical potential relations with one Z2V coupling in equilibrium}\label{sec:Z2VCPs1}
For the Z2V Model, we pay special attention to benchmarks in which one or two of the three independent $\lambda$ couplings are in equilibrium, with the remaining coupling(s) small enough to neglect.  

Consider the case in which exactly one Z2V coupling comes into equilibrium, and take the other Z2V couplings to be zero for simplicity.  There is then one particular lepton flavor $l_\beta$ that is {\em not} involved in Z2V interactions.  The chemical potentials of the two lepton flavors that do participate in Z2V interactions, which we  label as $l_\gamma$ and $l_\delta$, satisfy the equilibrium relation
\be\label{eq:Z2Vchempot1}
\mu_{l_\gamma}+\mu_{l_\delta} - \mu_\Phi = 0.
\ee
Using this relation along with Eqs.~(\ref{eq:ybspec}-\ref{eq:muphispec}), which apply generally, we can express $Y_B$ entirely in terms of $Y_\beta$ and $\delta Y_\chi$.  We find
\be\label{eq:Z2Vchempot2}
Y_B = \frac{300 \;c_\Phi \;Y_\beta - 12 \left[7+ c_\Phi\right] \delta Y_\chi}{237+766 \;c_\Phi}. 
\ee
Although Eq.~(\ref{eq:Z2Vchempot2}) applies whenever  the  $\lambda_{\gamma \delta}$ coupling is in equilibrium, (where $\beta\neq\gamma$ and $\beta  \neq \delta$), we will only use it when treating the DM couplings perturbatively, and under the  assumption that we can neglect Z2V couplings involving $l_\beta$.    In that context, $Y_\beta$ and $\delta Y_\chi$ can be calculated without taking into account Z2V couplings and then plugged into Eq.~(\ref{eq:Z2Vchempot2}) to get the baryon asymmetry.  Provided a perturbative treatment of the DM couplings is appropriate, this is a good approximation even for non-zero $\lambda_{\beta \gamma}$ and $\lambda_{\beta \delta}$, as long as those couplings are small enough not to come into equilibrium.    As discussed in Appendix~\ref{sec:Z2Vpert}, $Y_B$ effectively arises at either $\mathcal{O}(F^4)$ or  $\mathcal{O}(F^6)$ when one Z2V coupling comes into equilibrium, while the contributions to $Y_B$ from out-of-equilibrium Z2V couplings come with an additional suppression.

\subsection{Chemical potential relations with two Z2V couplings in equilibrium}\label{sec:Z2VCPs2}

 When two Z2V couplings come into equilibrium, we instead define $l_\beta$ as the lepton involved in {\em both} of the independent Z2V couplings, so that both $\lambda_{\beta \gamma}$ and $\lambda_{\beta \delta}$ are in equilibrium, but $\lambda_{\gamma\delta}$ is not.  In this case it is convenient to use the equilibrium relations
\be\label{eq:Z2Vchempot3}
\mu_{l_\beta}+\mu_{l_\gamma} - \mu_\Phi =\mu_{l_\beta}+\mu_{l_\delta} - \mu_\Phi =0
\ee
and  Eqs.~(\ref{eq:ybspec}-\ref{eq:muphispec}) to express the baryon asymmetry as
\be\label{eq:Z2Vchempot4}
Y_B = -\frac{75 \;c_\Phi \;(Y_\beta -Y_\gamma-Y_\delta)+3 \left[28+ c_\Phi\right] \delta Y_\chi}{237+529 \;c_\Phi}.
\quad\quad
\ee
The combination $Y_\beta -Y_\gamma-Y_\delta$ is unaffected by the in-equilibrium Z2V interactions.  In the perturbative context, we can  calculate that quantity and $\delta Y_\chi$  without taking into account the Z2V couplings and then plug those values into Eq.~(\ref{eq:Z2Vchempot4}) to get the baryon asymmetry, provided the remaining coupling $\lambda_{\gamma \delta}$ is small enough to remain out of equilibrium.

If all three Z2V couplings come into equilibrium, the lepton chemical potentials are flavor-universal, and one can  use Eqs.~(\ref{eq:upYuk}-\ref{eq:Xalpha}) to show that all asymmetries are proportional to $\delta Y_\chi$.  In this paper we restrict our attention to models with two $\chi$ mass eigenstates, in which case it turns out  all asymmetries vanish \cite{Abada:2018oly}.

\section{Reaction densities}\label{sec:appendixRDs}

In this appendix we provide expressions for reaction densities and related quantities that appear in the QKEs of Appendix~\ref{sec:appendixQKE} and in the perturbative results of Appendix~\ref{sec:appendixpert}.  Focusing first on interactions involving the DM, the momentum-integrated QKEs involve  generalized reaction density matrices in $\chi_i$ space, $\left[\gamma^X_\alpha\right]_{ij}$, where $\alpha$ labels the flavor of charge lepton involved, and  $X$ indicates whether the associated effects survive in the absence of  asymmetries ($X=0$), are driven by  a $\Phi - \Phi^*$ asymmetry ($X=\Phi 1,\Phi 2$)\footnote{Note that $\Phi1$ and $\Phi2$ label different reaction densities for a single scalar, rather than reaction densities for different scalars.}, or are driven by an $e^c_\alpha - \overline{e}^c_\alpha$ asymmetry ($X=e 1,e 2$).  
These reaction densities  have the form
\be\label{eq:RD0}
\left[\gamma^X_\alpha\right]_{ij}
=F_{\alpha i}^*F_{\alpha j}
\left( {\overline M}_{\Phi}^2  \!-\!  {\overline M}_{e}^2  \right)
\int\!\!
d\Pi
\;
\mathcal{F}^X,
\ee
where the phase space factor is
\begin{multline}\label{eq:ps}
d\Pi  = 
\frac{d^3 \bk}{(2\pi)^3}
\frac{1}{2E_\chi(\bk)}
 \frac{d^3 \bp}{(2\pi)^3}
 \frac{1}{2 E_e(\bp)} \\
 \times
\frac{d^3 \bq}{(2\pi)^3}
\frac{1}{2 E_{\Phi}(\bq)}
(2\pi)^4 \delta^4(q-p-k),
\end{multline}
with $E_\chi (\bk) = |\bk|$, $E_e (\bp) = ({\overline M}_{e}^2+|\bp|^2)^{1/2}$, and $E_{\Phi} (\bq) = ({\overline M}_{\Phi}^2+|\bq|^2)^{1/2}$. 
Here, the $\Phi$ and $e^c_\alpha$ masses-squared, including the leading thermal contributions from the $U(1)_y$ gauge coupling $g_y$, are
\be
\label{eq:thermalphimass}
{\overline M}_{\Phi}^2&  = & M_{\Phi}^2 + \frac{1}{4} g_y^2 T^2, 
\\
 \label{eq:thermalemass}
{\overline M}_{e}^2&  = &  \frac{1}{4} g_y^2 T^2,
\ee
where we neglect contributions from the charged-lepton Yukawa couplings and from possible $|\Phi|^2 |H|^2$ and $|\Phi|^4$  interaction terms.  

The integrands appearing in the reaction densities are
\be
\label{eq:RDint0}
\mathcal{F}^{0}
&=&
 f_-(y_\Phi) \left[1-f_+ (y_e)  \right] 
\\
\label{eq:RDinte1} 
 \mathcal{F}^{e1}
&=&
 f_-(y_\Phi) f_+ (y_e) \left[1-f_+ (y_e)  \right]
 \\
\label{eq:RDinte2} 
 \mathcal{F}^{e2}& = &f_+(y_\chi)  f_+ (y_e) \left[1-f_+ (y_e)  \right]  
\\
\label{eq:RDintphi1} 
  \mathcal{F}^{\Phi 1}& = &  f_-(y_\Phi) \left[1+f_- (y_\Phi)  \right] \left[1-f_+ (y_e)  \right] 
\\
\label{eq:RDintphi2} 
   \mathcal{F}^{\Phi 2}& = &f_+(y_\chi)  f_-(y_\Phi) \left[1+f_- (y_\Phi)  \right]  
\ee
with
\be\label{eq:ydef}
y_{\chi} = \frac{E_{\chi}({\bf k})}{T},
\quad
y_{e} = \frac{E_{e}({\bf p})}{T},
\;
\text{ and }
\;
y_{\Phi} = \frac{E_{\Phi}({\bf q})}{T},
\quad\quad
\ee
and where
\be\label{eq:QS}
f_\pm(y) \equiv (e^y\pm 1)^{-1}
\ee
are the standard Bose-Einstein ($f_-$) and Fermi-Dirac ($f_+$) distribution functions for vanishing chemical potential.  Because we label  $f_\pm$ by the $\pm$ sign appearing in the associated expression,  $f_+$ and $f_-$ apply to particles that are odd and even under exchange, respectively.  

Carrying out all integrations besides those over $E_{\Phi}$ and $E_\chi$, the reaction densities can be expressed as
\be\label{eq:RD1}
\!\!\!\!\!\!\!\!\!\!
\left[\gamma^X_{\alpha}\right]_{ij} 
 =  \frac{
  F_{\alpha i}^* F_{\alpha j}}
 {32 \pi^3}
\left( {\overline M}_{\Phi}^2 \! - \! {\overline M}_{e}^2  \right)
\!\!
\int_{\overline{M}_{\Phi_i}}^\infty\!\!\!
dE_{\Phi} \!
\int_{E_-}^{E_+}\!\!\!
dE_{\chi}\;
\mathcal{F}^{X}, 
\quad\quad
\ee
with the implied replacement $E_e \rightarrow E_{\Phi}-E_\chi$ in $\mathcal{F}^{X}$, and with
\be\label{eq:Epm}
E_\pm = \frac{\overline{M}_{\Phi}^2 - \overline{M}_e^2}{2 \overline{M}_{\Phi} }
\left(
\frac{E_{\Phi}}{\overline{M}_{\Phi}}
\pm
\sqrt{
\left( \frac{E_{\Phi}}{\overline{M}_{\Phi}} \right)^2 -1
}
\right).
\quad
\ee

In the Z2V Model, we also need to consider  $\Phi \leftrightarrow l_\alpha l_\beta$ processes and their CP conjugates.  The relevant reaction density (no longer a matrix in $\chi$ space), is
\be\label{eq:RDZ2V}
\gamma^\text{Z2V}_{\alpha \beta} = 2 |\lambda_{\alpha \beta}|^2
\left( {\overline M}_{\Phi}^2  \!-\! 2 {\overline M}_{l}^2  \right)
\int\!\!
d\Pi^\text{Z2V}
\;
\mathcal{F}^\text{Z2V},
\quad
\ee
where $\alpha$ and $\beta$ label the flavors of leptons involved, and where the overall factor of two arises from summation over $SU(2)_\text{w}$ gauge degrees of freedom.  Here we use the notation
\begin{multline}\label{eq:psZ2V}
d\Pi^\text{Z2V}  = 
\frac{d^3 \bk}{(2\pi)^3}
\frac{1}{2E_l(\bk)}
 \frac{d^3 \bp}{(2\pi)^3}
 \frac{1}{2 E_l(\bp)} \\
 \times
\frac{d^3 \bq}{(2\pi)^3}
\frac{1}{2 E_{\Phi}(\bq)}
(2\pi)^4 \delta^4(q-p-k)
\end{multline}
and 
\be\label{eq:RDintZ2V}
\mathcal{F}^\text{Z2V } = f_-(y_\Phi) 
\left[1 - f_+(y_l) \right]
\left[1 - f_+(y_\Phi - y_l) \right],
\quad\quad
\ee
where in addition to the $\Phi$-related quantities defined before,  we now have  $E_l (\bp) = ({\overline M}_{l}^2+|\bp|^2)^{1/2}$, $y_l = E_l({\bf p})/T$, and
\be\label{eq:thermallmass}
{\overline M}_{l}^2  =   \left(\frac{1}{16} g_y^2 +\frac{3}{16} g_w^2 \right)T^2,
\ee
where $g_w$ is the $SU(2)_w$ gauge coupling. (In an attempt to avoid notational confusion below, we have used energy conservation to express $\mathcal{F}^\text{Z2V }$ in terms of $E_\Phi$ and $E_l({\bf p})$, the energy of of one of the two leptons.)

Carrying out the integration over ${\bf k}$ and then the remaining angular integrations,  the Z2V reaction density can be expressed as
\be\label{eq:RDZ2VEE}
\!\!\!\!\!\!\!\!\!\!
\gamma^\text{Z2V}_{\alpha\beta}
 =  \frac{
  |\lambda_{\alpha \beta}|^2}
 {16 \pi^3}
\left( {\overline M}_{\Phi}^2 \! - 2\! {\overline M}_{l}^2  \right)
\!\!
\int_{\overline{M}_{\Phi_i}}^\infty\!\!\!
dE_{\Phi} \!
\int_{E_-}^{E_+}\!\!\!
dE_{l}\;
\mathcal{F}^\text{Z2V}, 
\quad\quad
\ee
where the limits of integration are now
\be\label{eq:EpmZ2V}
E_\pm =\frac{E_\Phi}{2}  \pm
\frac{\sqrt{E_\Phi^2-{\overline M}_\Phi^2} }{2}
\sqrt{1-4 \! {\overline M}_{l}^2/{\overline M}_\Phi^2 }.
\ee

The QKEs are  challenging to solve numerically and  we therefore use reaction densities averaged over $\chi$ momentum in these equations. In our perturbative calculation, however, we are able to determine the contributions of each $\chi$ momentum mode separately.   For simplicity, we neglect thermal masses in all of our perturbative calculations.  Reversing the order of integration in Eq.~(\ref{eq:RD1}) and carrying out the $E_\Phi$ integral gives
\be\label{eq:RD2}
\left[\gamma^0_\alpha\right]_{ij}
=
\frac{M_\Phi^4 x^{-2}}{32 \pi^3}
F_{\alpha i}^* F_{\alpha j}
\int_0^\infty
\! dy  \;g_0(x,y),
\ee
where 
\be\label{eq:g0}
g_0(x,y) =  
f_+(y)
\log
 \left(
 \frac
 {1+e^{- x^2/4y}}
 {1-e^{-(y+ x^2/4y)}}
 \right),
\ee
and where we continue to use the notation $x\equiv M_\Phi/T$.
The $g_0$ function will feature in the perturbative calculations of 
Appendix~\ref{sec:QKEpert} and Appendix~\ref{sec:appendixpert}, in which we calculate asymmetries and the DM energy density at leading order in $F$, while taking into account the full momentum dependence of the DM distribution function.  

The perturbative calculations of Appendix~\ref{sec:appendixpert} also involve the $\gamma^{e1}$ and $\gamma^{\Phi 1}$ reaction densities.  Because we neglect thermal masses  in our perturbative work, we can replace the $E_\chi$ integration in Eq.~(\ref{eq:RD1}) with an $E_e$ integration over the same range, and we find that these two reactions densities can be expressed as 
\be
\label{eq:pertRDe1}
\left[\gamma^{e1}_\alpha\right]_{ij}
& = &
\frac{M_\Phi^4 x^{-2}}{32 \pi^3}
F_{\alpha i}^* F_{\alpha j}
g_{e1} (x) 
\\
\label{eq:pertRDphi1}
\left[\gamma^{\Phi 1}_\alpha\right]_{ij}
& = &
\frac{M_\Phi^4 x^{-2}}{32 \pi^3}
F_{\alpha i}^* F_{\alpha j}
g_{\Phi1} (x),
\ee
with
\be
\label{eq:ge1}
g_{e1}(x)
& = & 
\int_{x}^{\infty}
\!\!\!\! dy_\Phi
\int_{y_-}^{y_+}
\!\!\!\! dy_e
\;
\frac{e^{y_e }}{(e^{y_e}+1)^{2}(e^{y_\Phi}-1)}
\\
\label{eq:gphi1}
g_{\Phi 1}(x) 
&=&  
\int_{x}^{\infty}
\!\!\!\!dy_\Phi
\int_{y_-}^{y_+}
\!\!\!\! dy_e
\;
\frac{e^{y_e }e^{y_\Phi}}{(e^{y_e}+1)(e^{y_\Phi}-1)^{2}}
\ee
and 
\be\label{eq:ypm}
y_\pm = \frac{1}{2} \left[y_\Phi \pm\sqrt{y_\Phi^2-x^2} \right].
\ee
%
%

\section{Quantum kinetic equations}\label{sec:appendixQKE}
%
%
Consider a single comoving DM mode, characterized by a particular value of $y \equiv |{\bf k}|/T$.  For this mode, information about the $\chi_1$ and $\chi_2$ occupation numbers and quantum coherence in $\{\chi_1, \chi_2\}$ space can be encoded in a  $2\times 2$   matrix, $f_\chi$.  Following Refs.~\cite{Hambye:2017elz,Abada:2018oly}, we take the quantum kinetic equation (QKE) for $f_\chi$ to have the form
\begin{multline}\label{eq:QKEsinglemode}
\frac{df_\chi}{dt}
=
-i \left[E_\chi, f_\chi \right]
\\
-\frac{1}{2} \sum_\alpha
\left(
\left\{
\frac{\Gamma_\alpha^>}{2E_\chi}
,f_\chi 
\right\}
-
\left\{
\frac{\Gamma_\alpha^<}{2E_\chi}
,1-f_\chi 
\right\}
\right).
\end{multline}
In the absorption and emission terms, we neglect DM masses by taking $E_\chi = |{\bf k}| = yT(t)$, whereas in the commutator term we define $E_\chi$ to be diagonal $2\times 2$ matrix whose non-zero entries are $E_i = (M_i^2 +y^2 T(t)^2)^{1/2}$.
As discussed in Sec.~\ref{sec:MM}, we leave for future work the inclusion of thermal corrections to the DM masses, whose effects are most important for large DM couplings and small values of $M_\Phi$ and $\Delta M^2$.  
We have chosen to work in terms of flavor-specific rates for $\chi$ absorption and emission, with the lepton flavor involved in the interaction labeled by $\alpha$.  In evaluating these rates we include only decay and inverse decay processes, $\Phi^* \leftrightarrow e^c_\alpha \chi_i$, which leads to
\begin{multline}\label{eq:emit}
\left[\Gamma_\alpha^< \right]_{ij}
= 
F_{\alpha i}^* F_{\alpha j}
\left( {\overline M}_{\Phi}^2 \! - \! {\overline M}_{e}^2  \right)
\\
\times
\int 
d\Pi^{(2)}
f_{\Phi^*}(\bq) \left[1-f_{e^c_\alpha}(\bp) \right]
\end{multline}
\begin{multline}
\label{eq:absorb}
\left[\Gamma_\alpha^> \right]_{ij}
= 
F_{\alpha i}^* F_{\alpha j}
\left( {\overline M}_{\Phi}^2 \! - \! {\overline M}_{e}^2  \right)
\\
\times
\int 
d\Pi^{(2)}
f_{e^c_\alpha}(\bp)\left[1+f_{\Phi^*}(\bq) \right],
\end{multline}
where in general the $\Phi$ and $e^c_\alpha$ chemical potentials enter into the distribution functions $f_{\Phi^*}$ and $f_{e^c_\alpha}$, and where we define the phase-space factor
\begin{multline}\label{eq:PS2}
d\Pi^{(2)} \equiv
 \frac{d^3 \bp}{(2\pi)^3}
 \frac{1}{2 E_e(\bp)} 
\frac{d^3 \bq}{(2\pi)^3}
\frac{1}{2 E_{\Phi}(\bq)}\\
\times
(2\pi)^4 \delta^4(q-p-k).
\end{multline}
As we did in Appendix~\ref{sec:appendixRDs}, we define $E_{\Phi} (\bq) = ({\overline M}_{\Phi}^2+|\bq|^2)^{1/2}$ and $E_e (\bp) = ({\overline M}_{e}^2+|\bp|^2)^{1/2}$, with ${\overline M}_{\Phi}$ and ${\overline M}_{e}$ defined in
Eqs.~(\ref{eq:thermalphimass}-\ref{eq:thermalemass}).  In Eq.~(\ref{eq:PS2}), we can  take $k$ to be any four-momentum satisfying $k^2=0$ and $|\bk| = yT$.

\subsection{Perturbative treatment of the QKEs}\label{sec:QKEpert}

We first analyze the QKEs  with the goal of calculating the $X_\alpha$ densities in the Minimal Model, at leading order. For this purpose, we can ignore chemical potentials in the distribution functions appearing in Eqs.~(\ref{eq:emit}-\ref{eq:absorb}).  We will also neglect thermal masses.

We start by taking  three steps to rewrite Eq.~(\ref{eq:QKEsinglemode}) in a more convenient form.
First, we use the relation
\be\label{eq:DB}
f_+(y)\;\Gamma^>_\alpha = \left[1-f_+(y)\right]\Gamma^<_\alpha, 
\ee
which applies in the absence of chemical potentials. Here $f_+$ is the Fermi-Dirac function defined in Eq.~(\ref{eq:QS}), and we continue to use $y = |\bk|/T$ for the momentum of our comoving DM mode. 

Second, we switch our independent variable from $t$  to 
\be\label{eq:xdef}
x \equiv M_\Phi/T.
\ee
The entropy density and Hubble parameter can then be written as  $s(x) =2\pi^2 g_* M_\Phi^3 x^{-3}/45$ and $H(x) = M_\Phi^2 x^{-2}/M_0$, where $g_*\simeq 106.75$ is the effective number of relativistic degrees of freedom and $M_0\simeq M_{\rm Pl}/(1.66\sqrt{g_*})\approx7.12\times10^{17}$ GeV.   We then have $t=(2H)^{-1} = x^2 M_0/(2 M_\Phi^2)$.

 Third, we define interaction-picture quantities
\be\label{eq:IP}
{\tilde f}_\chi  \equiv  U^\dagger f_\chi U
\quad\quad
{\tilde \Gamma^<}_\alpha  \equiv  U^\dagger \Gamma^< U,
\ee
where we take the time-evolution matrix to have the form
\be\label{EM}
U =  \text{diag}\left\{1, e^{-i\phi}\right\}.
\ee
The relative phase acquired between the two $\chi$ mass eigenstates is
\be\label{eq:relphase}
\phi  = \int^t\!dt\; (E_2-E_1) 
\simeq \int^t\! dt\; \frac{\Delta M^2}{2E_\chi} 
 =  \beta_\text{osc} x^3/y,
 \quad\quad
\ee
with $\beta_\text{osc}\equiv  \frac{M_0\Delta M^2}{6M_\Phi^3}$.

Following these three steps, the QKE for our comoving mode can be written as
\be\label{eq:QKEsinglemodeintpict}
H x \frac{d{\tilde f}_\chi}{dx}
=
\frac{1}{2} \sum_\alpha
\left\{
\frac{{\tilde \Gamma}_\alpha^<}{2E_\chi}
,
\;1-\frac{{\tilde f}_\chi}{f_+(y)} 
\right\}
.
\ee
To get the equation for ${\overline \chi}$, we replace  ${\tilde f}_{\chi} \rightarrow {\tilde f}_{\overline \chi}$  and $F \rightarrow F^*$.

 The number density of $\chi$ particles, $n_\chi$, is obtained by integrating $\tilde{f}_\chi$ over momentum and taking the trace.  We define the $2\times 2$ matrix ${\tilde Y}_\chi$ so that its trace is $n_\chi/s$:
\be\label{eq:Ychidef} 
{\tilde Y}_\chi = \frac{1}{s} \int\!\frac{d^3\bk}{(2\pi)^3}\; \tilde{f}_\chi
= \frac{45}{4\pi^4 g_*} \int_0^\infty\!\!dy \; y^2 {\tilde f}_\chi.
\ee
In a DM interaction involving lepton flavor $\alpha$, the changes in the $X_\alpha$ charge and in the $\chi$ and ${\overline \chi}$ populations are related by $\Delta X_\alpha = \Delta N_\chi - \Delta N_{\overline \chi}$.  We then have
\be\label{eq:YalphafromYchi}
\frac{dY_\alpha}{dx} = \text{Tr} \left[ \left. \frac{d{\tilde Y}_\chi}{dx} \right|_\alpha -  \left. \frac{d{\tilde Y}_{
\overline \chi}}{dx} \right|_\alpha\right],
\ee
where the $\alpha$ subscripts on the right-hand side specify that only contributions associated with lepton flavor $\alpha$ are to be included.  Integrating both sides of Eq.~(\ref{eq:QKEsinglemodeintpict}) over momentum, we obtain
\be\label{eq:QKEsm2}
  \left. \frac{d{\tilde Y}_\chi}{dx} \right|_\alpha
=
 \frac{45}{8\pi^4 g_*} \frac{1}{xH}
 \int_0^\infty\!\!\!dy \; y^2
\left\{
\frac{{\tilde \Gamma}_\alpha^<}{2E_\chi}
,
\;1-\frac{{\tilde f}_\chi}{f_+(y)} 
\right\}
.\quad\quad\quad
\ee
To evaluate $Y_\alpha$ at leading order, we plug Eq.~(\ref{eq:QKEsm2}) and the corresponding equation for ${\overline \chi}$ into Eq.~(\ref{eq:YalphafromYchi}), using the $\mathcal{O}(F^2)$ expressions  for ${\tilde f}_\chi$ and ${\tilde f}_{\overline \chi}$; integrating Eq.~(\ref{eq:QKEsinglemodeintpict}) gives
\be\label{eq:fchi2}
{\tilde f}_\chi^{(2)}
(x)
=
\sum_\beta
\int_0^x
\!\!
\frac{dx'}{x'H(x')}
\;
\frac{{\tilde \Gamma}_\beta^<}{2E_\chi}(x'),
\ee
while for ${\tilde f}_{\overline \chi}^{(2)}$ we replace $F\rightarrow F^*$.  After these steps, we get the following $\mathcal{O}(F^4)$ expression for the $X_\alpha$ asymmetry:
\begin{multline}\label{eq:Ya41}
\!\!\!\!Y_\alpha^{(4)}(x)=
- \frac{45}{8\pi^4 g_*}
\sum_\beta
\int_0^\infty 
\!\!\frac{dy \;y^2}{f_+(y)}
\int_0^x
\!\!\!\!
\frac{dx_1}{x_1H(x_1)} 
\int_0^{x_1}
\!\!
\frac{dx_2}{x_2H(x_2)}\\
\!\!\!\!\times
\text{Tr}
\left[
\left\{
\frac{{\tilde \Gamma}_\alpha^<}{2E_\chi}(x_1)
\;,
\frac{{\tilde \Gamma}_\beta^<}{2E_\chi}(x_2)
\right\}
-(F\rightarrow F^*)
\right].
\end{multline}

Because we ignore chemical potentials in the emission rate, we can express
$\Gamma_\alpha^<$  in terms of the $g_0$ function introduced in Eq.~(\ref{eq:g0}):
\be\label{eq:Gg2}
 \frac{\left[ \Gamma_\alpha^< \right]_{ij}}{2E_\chi}= \frac{M_\Phi }{16 \pi}F_{\alpha i}^* F_{\alpha j} \frac{x}{y^2} g_0(x,y).
\ee
We can verify this relation by carrying out the integrations in Eq.~(\ref{eq:emit}).  Equivalently, we can start with the fact that, in the absence of chemical potentials, $\Gamma_\alpha^<$ is related in a simple way to the $\gamma^0_\alpha$ reaction density defined in Eq.~(\ref{eq:RD0}), 
\be\label{eq:Gg1}
\!\!
\gamma^0_\alpha(x)=\int \!\! \frac{d^3\bk}{(2\pi)^3}  \frac{ \Gamma_\alpha^<}{2E_\chi}(x)
=\frac{M_\Phi^3 x^{-3}}{2\pi^2}
\!\!
\int_0^\infty \!\!\! dy \;y^2 \frac{ \Gamma_\alpha^<}{2E_\chi}(x).
\quad\quad\quad
\ee
Using the expression for $\gamma^0_\alpha$ given in Eq.~(\ref{eq:RD2}) again leads to Eq.~(\ref{eq:Gg2}).

In Eqn.~(\ref{eq:Ya41}), time evolution matrices appear through the interaction-picture matrices ${\tilde \Gamma}_\alpha^<$ and ${\tilde \Gamma}_\beta^<$. Defining the matrices
\be\label{eq:FFM}
\left[M_\alpha \right]_{ij}\equiv F_{\alpha i}^* F_{\alpha j}
\quad
\text{and}
\quad
{\tilde M}_\alpha(x) \equiv 
U^\dagger(x)
M_\alpha
U(x),
\quad\quad\quad
\ee
the trace involves
\begin{multline}\label{eq:FFMtr}
\sum_\beta 
\text{Tr}
\left[\left\{
{\tilde M}_\alpha(x_1),
{\tilde M}_\beta(x_2)
\right\}-(F\rightarrow F^*)
\right]\\
= 8\; \text{Im}
\left[
F_{\alpha 1}^*
F_{\alpha 2}
(F^\dagger F)_{21}
\right]
\sin
\left[
\frac{\beta_\text{osc}}{y}(x_1^3-x_2^3)
\right].
\end{multline}
These steps lead to our final expression for the leading-order $X_\alpha$ asymmetry,
\begin{multline}\label{eq:Ya42}
Y_\alpha^{(4)}(x)=\frac{45}{256 g_*\pi^6} \left(\frac{ M_{\rm 0}}{M_{\Phi}}\right)^2
\mathrm{Im}\left[
F_{\alpha 1}  
F_{\alpha 2}^*
\left(F^\dagger F \right)_{12}
\right] \\
\times
\int_0^\infty 
\!\!\frac{dy}{y^2 f_+(y)}
\int_0^{x}\!\!\!dx_1\,
 x_1^2\;
 g_0(x_1,y)  \\
 \times
\int_0^{x_1}\!\!\!dx_2\,
x_2^2\;
g_0(x_2,y)
\;
\sin\left[\frac{\beta_\text{osc}}{y}(x_1^3-x_2^3)\right],
\end{multline}
which is used extensively in our perturbative analyses. 

\subsection{Momentum-integrated QKEs}
\label{sec:MIQKEs}

To go beyond a perturbative treatment of the QKEs, we start by integrating Eq.~(\ref{eq:QKEsinglemode}) over momentum, assuming a thermal DM momentum distribution:
\beq\label{eq:TA}
\left[f_\chi \right]_{ij}(y) \rightarrow f_+(y) \frac{\left[Y_\chi\right]_{ij}}{Y_\chi^\text{eq}}
\eeq
where $Y^\text{eq}_\chi =135 \zeta(3)/(8 \pi^4 g_*)\simeq 1.95\times 10^{-3}$.  The commutator term can then be expressed in terms of the diagonal matrix
\beq\label{eq:aveEchi}
\mathcal{E}_\chi = \text{diag}\left(0,\;  \frac{\Delta M^2}{2T} \left\langle \frac{T}{E_\chi}\right \rangle \right),
\eeq
where  $\langle T/{E_\chi} \rangle =\frac{\pi^2}{18 \zeta (3)} \simeq 0.46 $.  Linearizing in chemical potentials, we find
\be\label{eq:chiQKE}
s H x \frac{dY_\chi}{dx}
=
-is \left[\mathcal{E}_\chi, Y_\chi \right]
+
\sum_\alpha
\left( 
\gamma^1_\alpha 
-\frac{1}{2} 
\left\{
\gamma^2_\alpha,
\frac{Y_\chi}{Y_\chi^\text{eq}}
\right\}
\right),
\quad\quad\quad
\ee
where $\gamma_\alpha^1$ and $\gamma_\alpha^2$ can be expressed in terms of the reaction densities introduced in Appendix~\ref{sec:appendixRDs}:
\be
\label{eq:gamma1QKE}
\gamma^1_\alpha & = & \gamma^0_\alpha - \frac{\mu_{e^c_\alpha}}{T} \gamma^{e1}_\alpha -  \frac{\mu_\Phi}{T} \gamma^{\Phi1}_\alpha 
\\
\label{eq:gamma2QKE}
\gamma^2_\alpha & = & \gamma^0_\alpha + \frac{\mu_{e^c_\alpha}}{T} \gamma^{e2}_\alpha -  \frac{\mu_\Phi}{T} \gamma^{\Phi2}_\alpha
\ee
The kinetic equations for $\overline{\chi}$ are obtained via the substitutions $Y_\chi \rightarrow Y_{\overline \chi}$, $F_{\alpha i} \rightarrow F^*_{\alpha i}$,  $\mu_{\Phi} \rightarrow -\mu_{\Phi}$,  and $\mu_{e^c_\alpha} \rightarrow -\mu_{e^c_\alpha}$:  
\be\label{eq:chibarQKE}
s H x \frac{dY_{\overline \chi}}{dx}
&=&
-i s\left[\mathcal{E}_\chi, Y_{\overline \chi} \right]
+
\sum_\alpha
\left( 
{\overline \gamma}^1_\alpha 
-\frac{1}{2} 
\left\{
{\overline \gamma}^2_\alpha,
\frac{Y_{\overline \chi}}{Y_\chi^\text{eq}}
\right\}
\right),
\quad\quad\quad
\ee
with
\be
\label{eq:gammabar1QKE}
{\overline \gamma}^1_\alpha & = & {\gamma^0_\alpha}^* + \frac{\mu_{e^c_\alpha}}{T} {\gamma^{e1}_\alpha}^* +  \frac{\mu_\Phi}{T} {\gamma^{\Phi1}_\alpha}^* 
\\
\label{eq:gammabar2QKE}
{\overline \gamma}^2_\alpha & = & {\gamma^0_\alpha }^*- \frac{\mu_{e^c_\alpha}}{T} {\gamma^{e2}_\alpha}^* + \frac{\mu_\Phi}{T} {\gamma^{\Phi2}_\alpha}^*
\ee

The $X_\alpha$ densities change only due to DM interactions and Z2V interactions.  Each DM interaction involving SM flavor $\alpha$ produces equal changes to the $X_\alpha$ charge density and the $\chi/{\overline \chi}$ number-density asymmetry, $\Delta Y_\alpha =  \Delta {\rm Tr} \left[ Y_{\chi} -   Y_{\overline {\chi}}\right]$. Each 
Z2V interaction that changes the number of $l_\alpha$ (${\overline l}_\alpha$) particles produces an opposite (equal) change in $X_\alpha$.  Working from these facts, we find
\begin{multline}\label{eq:YalphaQKE}
sH x \frac{d Y_\alpha}{dx}
= 
- {\rm Tr} 
\left[
 {\gamma^0_\alpha}
\frac{Y_\chi}{Y_\chi^\text{eq}}
-
 {\gamma^0_\alpha}^*
\frac{Y_{\overline \chi}}{Y_\chi^\text{eq}}
\right]\\
-\frac{2\mu_{e^c_\alpha}}{T} {\rm Tr} \left[\gamma^{e1}_\alpha \right]
-\frac{2\mu_{\Phi}}{T}{\rm Tr} \left[\gamma^{\Phi1}_\alpha \right]\\
-
\frac{\mu_{e^c_\alpha}}{T}
{\rm Tr}\left[
 {\gamma^{e2}_\alpha}
\frac{Y_\chi}{Y_\chi^\text{eq}}
+
 {\gamma^{e2}_\alpha}^*
\frac{Y_{\overline \chi}}{Y_\chi^\text{eq}}
\right]\\
+
\frac{\mu_{\Phi}}{T}
{\rm Tr}\left[
 {\gamma^{\Phi2}_\alpha}
\frac{Y_\chi}{Y_\chi^\text{eq}}
+
 {\gamma^{\Phi 2}_\alpha}^*
\frac{Y_{\overline \chi}}{Y_\chi^\text{eq}}
\right]\\
-2\sum_\beta
\left(\frac{\mu_\Phi}{T}
-\frac{\mu_{l_\alpha}}{T}
-\frac{\mu_{l_\beta}}{T} \right) \gamma^\text{Z2V}_{\alpha\beta},
\end{multline}
where $\gamma^{Z2V}$ is defined in Eq.~(\ref{eq:RDZ2V}).

As written, the QKEs presented in this appendix are valid assuming it is appropriate to linearize the $\Phi^{(*)}$ distribution functions in chemical potential:
\beq\label{eq:linearize_mu_phi}
f_{\Phi/\Phi^{*}} = f_- \pm \frac{\mu_\Phi}{T}  f_-(1+f_-).
\eeq
Although  $|\mu_\Phi/T| \ll 1$ is almost always satisfied in the scenarios of interest in this paper, we replace Eq.~(\ref{eq:linearize_mu_phi}) with
\beq\label{eq:don't_linearize_mu_phi}
f_{\Phi/\Phi^{*}} =  \cosh(\mu_\Phi/T) f_- \pm   \sinh(\mu_\Phi/T) f_-(1+f_-)
\eeq
in our numerical work, a recipe that effectively takes into account the possibility of a highly asymmetric $\Phi/\Phi^{(*)}$ background left behind at $T \ll M_\Phi$, while remaining valid in the  $|\mu_\Phi/T| \ll 1$ regime.  

We can express the chemical potentials appearing in the QKEs  in terms of $Y_\alpha$ and  $\text{Tr}[Y_\chi-Y_{\overline \chi}]$, using the results of Appendix~\ref{sec:appendixCPs}.  Then we numerically solve 
the system of QKEs to obtain $Y_\text{sm}(x) \equiv \sum_\alpha Y_\alpha(x)$ for temperatures down to the $T_\text{ew}$.  Having done that, we determine the final baryon asymmetry by following Ref.~\cite{Eijima:2017cxr}, which takes into account gradual sphaleron decoupling.  That is, we take
\be
sHx \frac{d Y_B}{d x}=-\Gamma_B(x)\left[Y_B(x)-Y_B^\text{eq}(x)\right],
\ee
where $Y_B^\text{eq}(x)$ is calculated from $Y_\text{sm}(x)$ using the relations of Appendix~\ref{sec:appendixCPs}, which assume sphalerons to be in equilibrium.  We use the SM result for $\Gamma_B$, because no new chiral states couple to sphalerons in our model.

\section{Perturbative results}\label{sec:appendixpert}
\subsection{Minimal Model}\label{sec:MMpert}
\subsubsection{$\mathcal{O}(F^2)$ DM density in the Minimal Model}

To calculate the DM abundance at leading order, we ignore inverse $\Phi^{(*)}$ decays, neglect Pauli-blocking due to the $\chi/\overline{\chi}$ abundance, and neglect thermal mass effects.
Using the leading-order calculation of Appendix~\ref{sec:QKEpert},
the $\mathcal{O}(F^2)$ $\chi$ number density divided by entropy density (which is equal to that for  ${\overline \chi}$) is
\be\label{eq:DMfromQKE1}
Y_\chi^{(2)}(x)
& = &
\frac{1}{s(x)}\int\!\frac{d^3\bk}{(2\pi)^3}\;  \text{Tr} \left[{\tilde f}_\chi^{(2)} (x) \right]\\
& = &
\label{eq:DMfromQKE2}
\frac{45}{4\pi^4 g_*} \int_0^\infty\!\!dy \; y^2\;
\text{Tr} \left[{\tilde f}_\chi^{(2)} (x)\right],
\ee
where ${\tilde f}_\chi^{(2)}$ is given in Eq.~(\ref{eq:fchi2}).  Here and for all of our perturbative results, we switch to a notation in which $Y_\chi$ is number density divided by entropy density, a number rather than a matrix.  We restrict the notation in which $Y_\chi$ represents a matrix to Appendix~\ref{sec:appendixQKE}.  
We continue to use  $x \equiv M_\Phi/T$ as our independent variable; see the paragraph containing Eq.~(\ref{eq:xdef}).

Using Eqs.~\ref{eq:DMfromQKE2},~\ref{eq:fchi2}, and~\ref{eq:Gg2}, and taking $x\rightarrow \infty$ to get the final abundance, we find
\be\label{eq:YchiF2gamma}
Y_\chi^{(2)} =  
\label{eq:YchiF2}
\frac{45}{64 \pi^5 g_*} {\rm Tr} F^\dagger F \frac{M_0}{M_\Phi}\int_0^\infty \!\!\! dx \; x^2  \;
\tilde{g}_0(x),
\ee
where we have defined the momentum-integrated quantity
\be\label{eq:gtilde}
\tilde{g}_0(x) \equiv \int_0^\infty \!dy\; g_0(x,y),
\ee
with $g_0$ itself defined in Eq.~(\ref{eq:g0}).

The dark matter energy density, $\rho_\text{dm}$, is determined by $Y_\chi$, the DM masses $M_1$ and $M_2$, and the 
$\theta$ parameter appearing in Eq.~(\ref{eq:Ffullparam}), which determines the relative abundances of $\chi_1$ and $\chi_2$.  We have
\be\label{eq:rhobasic}
\frac{\rho^{(2)}_\text{dm}}{s} =2 { \overline M} Y_\chi^{(2)},
\ee
where 
\be\label{eq:aveDMmassapp}
{\overline M} = \cos^2\theta M_1+\sin^2 \theta M_2
\ee
 is the average DM mass for the $\chi$ particles produced in $\Phi$ decays, and where the factor of two takes into account the two DM helicity states, $\chi$ and ${\overline \chi}$.  Taking the observed DM energy density to be $\rho_\text{dm}^\text{obs}/s = 4.3\times 10^{-4}$ keV \cite{Zyla:2020zbs}, we find
\be\label{eq:rhoGamma}
\frac{\rho_\text{dm}^{(2)}}{\rho_\text{dm}^\text{obs}} \simeq
22
\left(
\frac{\Gamma_\Phi}{H_\text{ew}}
\right)
\left(
\frac{{\overline M}}{15 \text{ keV}}
\right)
\left(
\frac{500 \text{ GeV}}{M_\Phi}
\right)^2,
\ee
where $H_\text{ew} = T_\text{ew}^2/M_0$ is the Hubble parameter at sphaleron decoupling and 
\be\label{eq:width_appendix}
\Gamma_\Phi = \frac{{\rm Tr} \left[F^\dagger F \right]}{16 \pi} M_\Phi
\ee
is the $\Phi$ decay width.

\subsubsection{$\mathcal{O}(F^4)$ $X_\alpha$ asymmetry in the Minimal Model}
In the Maxwell-Boltzmann approximation, the $\mathcal{O}(F^4)$ flavor asymmetry in $X_\alpha \equiv B/3 - L_\alpha$ can determined via a relatively simple and physically motivated calculation that separately considers $\Phi^{(*)}$ decays,  inverse decays, and DM oscillations in between.  This is the approach introduced in Ref.~\cite{Shuve:2020evk}.  In the present work, we fully incorporate quantum statistics and find it more convenient to obtain the desired result via a perturbative analysis of the QKEs, which we perform in Appendix~\ref{sec:QKEpert},  neglecting thermal mass effects. There we find that the $\mathcal{O}(F^4)$  $X_\alpha$ asymmetry, as a function of $x\equiv M_\Phi/T$,  is 
\begin{multline}\label{eq:Ya4}
Y_\alpha^{(4)} (x) = \frac{45}{256 g_*\pi^6} \left(\frac{ M_{\rm 0}}{M_{\Phi}}\right)^2
\\
\times
\mathrm{Im}\left[
F_{\alpha 1}  
F_{\alpha 2}^*
\left(F^\dagger F \right)_{12}
\right] 
\;
 \mathcal{I}^{(4)}(x,\beta_\text{osc}),
 \end{multline}
 with
 \begin{multline}\label{eq:final_I4_fn_1}
 \mathcal{I}^{(4)}(x,\beta) = \int_0^\infty\!
 dy
 \,
 \int_0^{x}\!dx_1\,
 x_1^2\,
 g_0(x_1,y) 
 \\
 \times
\int_0^{x_1}\!dx_2\,
x_2^2\,
\;
 \frac{g_0(x_2,y)}{y^2 f_+(y)}
\;
\sin\left[\frac{\beta}{y}(x_1^3-x_2^3)\right]
\quad\quad
\end{multline}
and
\be\label{eq:betaapp}
\beta_{\rm osc} & \equiv & \frac{M_0\Delta M^2}{6M_\Phi^3} \\
& \simeq &
0.214
\times
\left( \frac{500 \text{ GeV}}{M_\Phi}\right)^3
\frac{\Delta M^2}{\left( 15 \text{ keV}\right)^2}.
\ee
Starting from the parametrization of Eq.~(\ref{eq:Ffullparam}), one can derive the bound (for any $\alpha$)
\be\label{eq:F4coupbound}
\left|\mathrm{Im}\left[
F_{\alpha 1}^*  
F_{\alpha 2}
\left(F^\dagger F \right)_{21}
\right] 
\right|
&\le&
\sin^2 2\theta
\left( 
\frac{\text{Tr}\left[F^\dagger F\right]}{4}
\right)^2 \quad\quad\\
& =  &\sin^2 2\theta\left( 
\frac{4 \pi \Gamma_\Phi}{M_\Phi}
\right)^2.
\ee
Using this inequality and evaluating Eq.~(\ref{eq:Ya4}) at the sphaleron decoupling temperature, $T=T_\text{ew}$, 
we find
\begin{multline}\label{eq:MMYa4ineq}
\frac{Y_\alpha^{(4)} (x_\text{ew})}{Y_B^\text{obs}}
\lesssim
(1.5\times 10^4)
\;
\sin^2 2\theta 
\\
\times
\left(
\frac{\Gamma_\Phi}{H_\text{ew}}
\right)^2
\left(
\frac{500 \text{ GeV}}{M_\Phi}
\right)^4
\mathcal{I}^{(4)}(x_\text{ew},\beta_\text{osc}),
\end{multline}
where 
\be\label{eq:xewdef}
x_\text{ew} \equiv M_\Phi/T_\text{ew},
\ee
and where $Y_B^\text{obs}=8.7\times 10^{-11}$ is the observed baryon asymmetry \cite{Zyla:2020zbs}.
We use Eq.~(\ref{eq:MMYa4ineq}) in the perturbative analysis of Sec.~(\ref{sec:F4pert}).

\begin{figure*}
\includegraphics[height=0.27\textwidth]{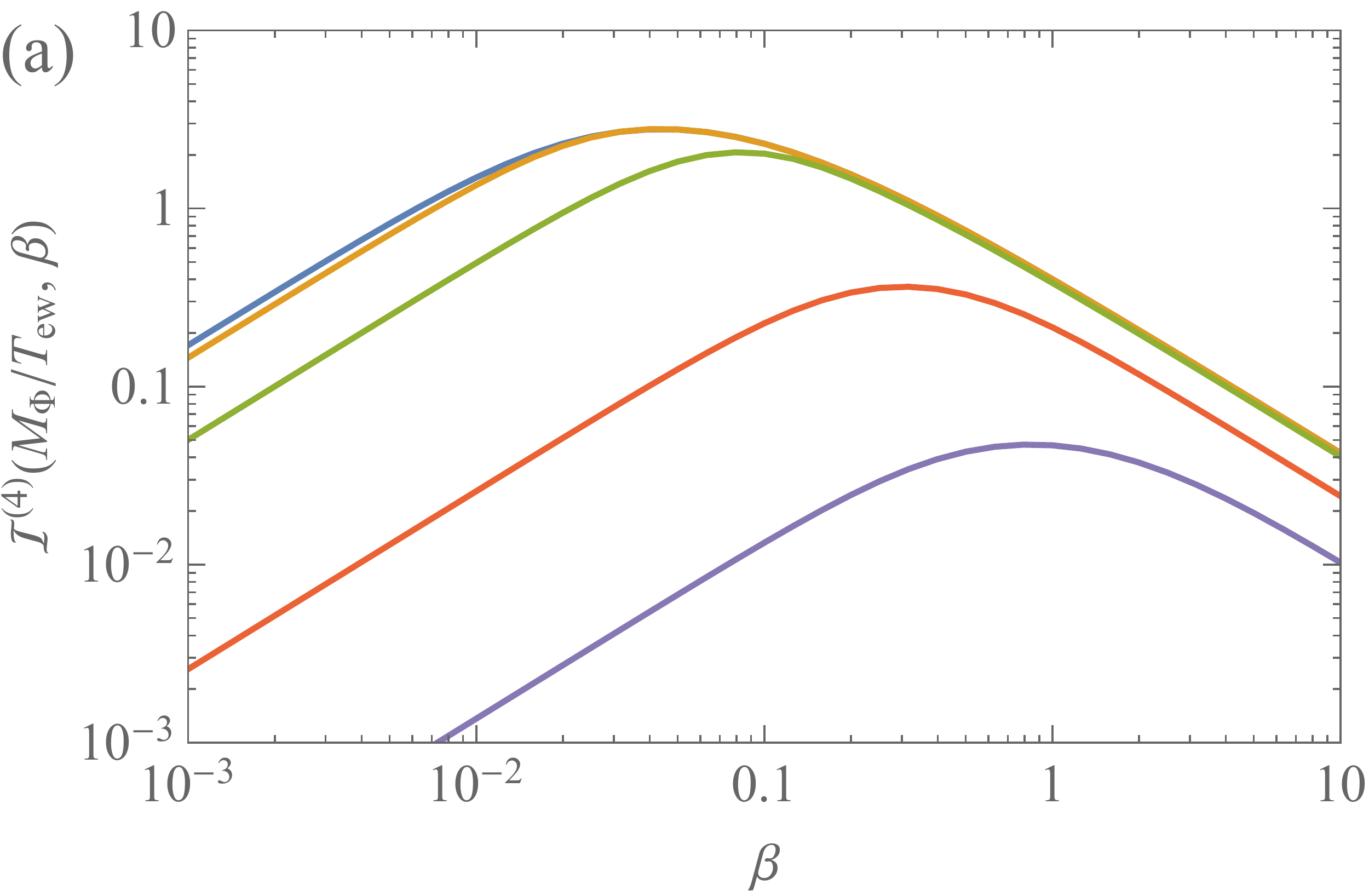}
\includegraphics[height=0.27\textwidth]{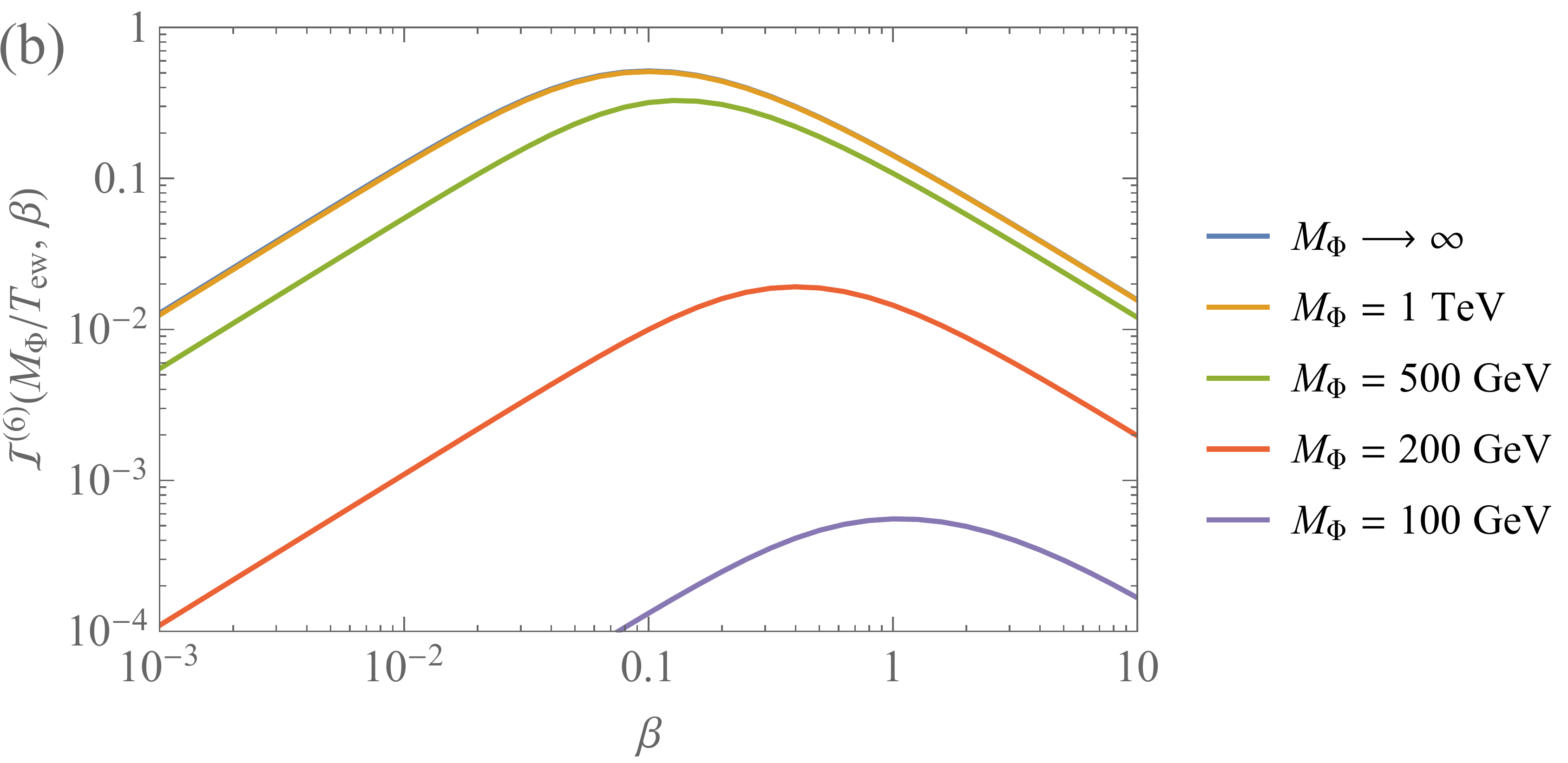}\\
\includegraphics[height=0.27\textwidth]{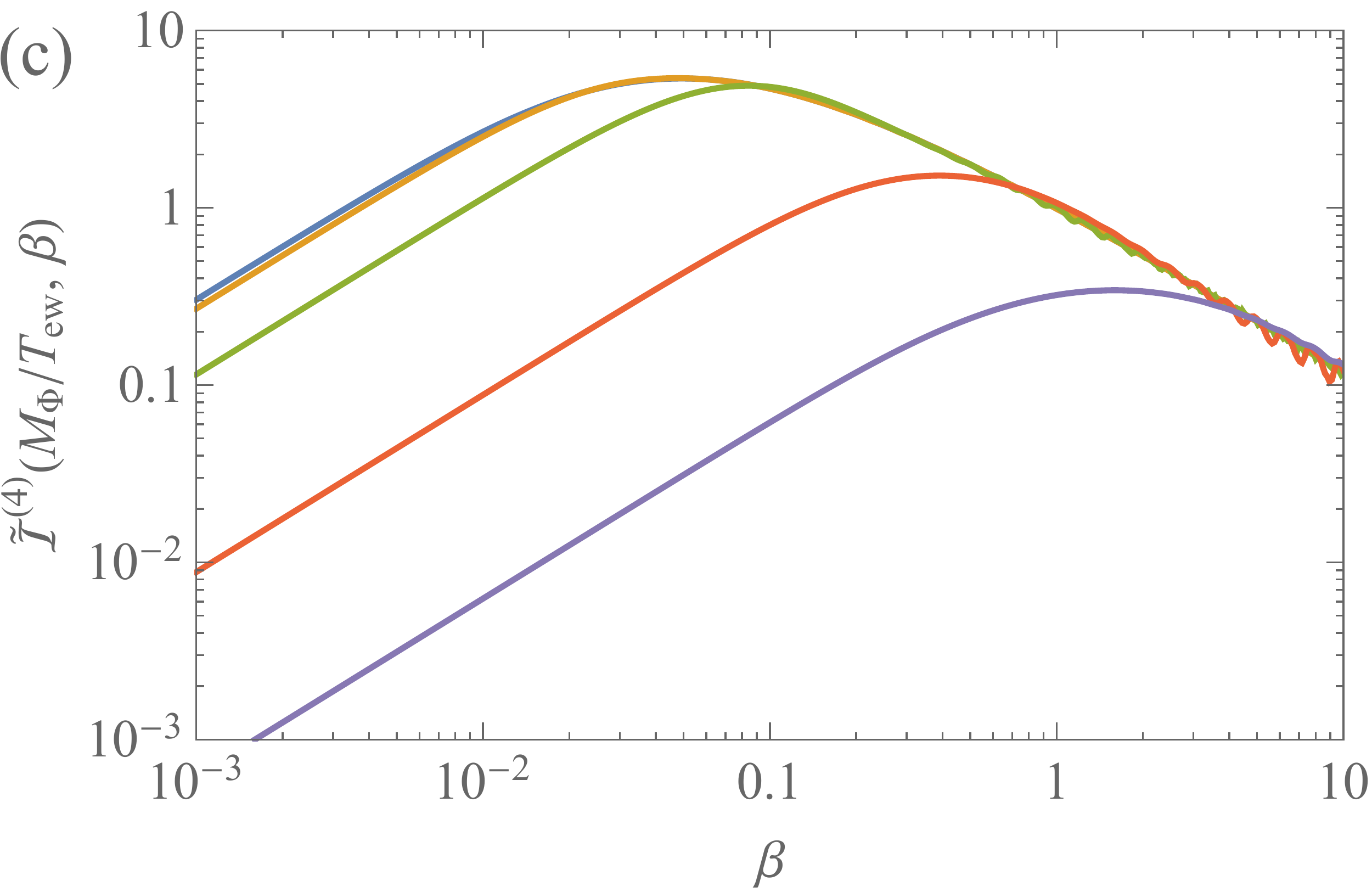}
\includegraphics[height=0.27\textwidth]{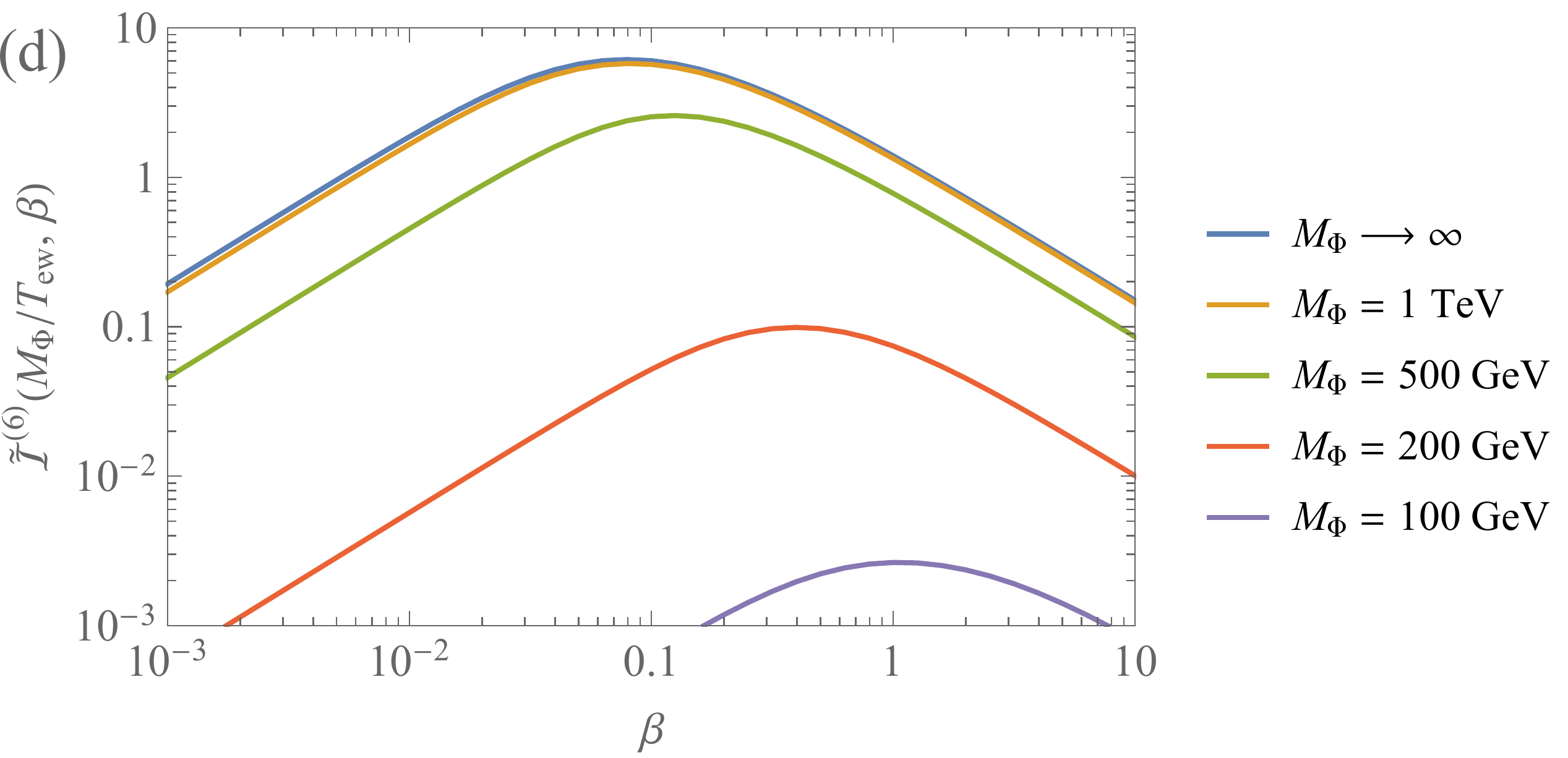}
\caption{For various $M_\Phi$,  $\mathcal{I}^{(4)} $,   $\mathcal{I}^{(6)}$,   $\tilde{\mathcal{I}}^{(4)} $, and   $\tilde{\mathcal{I}}^{(6)} $ plotted against $\beta$.}
  \label{fig:I4}
\end{figure*}

For $x \gg 1$, corresponding to $M_\Phi \gg T$, the asymptotic behavior of $\mathcal{I}^{(4)}$ for large and small $\beta$ is
\be\label{eq:asymp_I4}
\mathcal{I}^{(4)} (\infty,\beta) \simeq 
\begin{cases}
1.71
\times 10^2 \;\beta &  \beta \ll1 \\
0.424 / \beta& \beta \gg 1.
\end{cases}
\ee
 In Fig.~\ref{fig:I4}(a) we see that peak values for ${\mathcal I}^{(4)}(x,\beta)$  range from $ \sim 3$ for  $x \gg 1$ (at $\beta \sim 4 \times 10^{-2}$) to $\sim 0.05$ for  $x \sim 1$ (at $\beta \sim 1$). 
  \

The momentum-integrated QKEs of Appendix~\ref{sec:MIQKEs} incorporate a thermal ansatz for the DM momentum distribution; see Eq~(\ref{eq:TA}).  Adopting the same ansatz in the perturbative context amounts to modifying Eq.~(\ref{eq:final_I4_fn_1}) by the replacements
\be\label{eq:TArep1}
\sin\left[\frac{\beta}{y}(x_1^3-x_2^3)\right]  \rightarrow   \sin\left[\beta \left\langle \frac{1}{y}\right\rangle(x_1^3-x_2^3)\right]
\ee
and
\be\label{eq:TArep2}
 \frac{g_0(x_2,y)}{y^2 f_+(y)}
  \rightarrow 
  \frac{\int_0^\infty \!dy \;g_0(x_2,y)}{\int_0^\infty \!dy \;y^2 f_+(y)}
  =\frac{2}{3\zeta(3)} \tilde{g}_0(x_2),
\ee
where in the first replacement, $\left\langle 1/y \right\rangle = \frac{\pi^2}{18 \zeta (3)} \simeq 0.46$ is the average $T/E_\chi$ for a thermal distribution, while the second replacement involves the momentum-integrated function $\tilde{g}_0$ introduced in Eq.~(\ref{eq:gtilde}).
After making these substitutions and carrying out the $y$ integration, Eq.~(\ref{eq:final_I4_fn_1}) becomes
\begin{multline}\label{eq:I4ta}
\mathcal{I}^{(4)}_\text{t.a.}(x,\beta) =
 \frac{2}{3 \zeta(3)}
 \int_0^{x}\!dx_1\,
 x_1^2\,
 \tilde{g}_0(x_1)
 \\
 \times
\int_0^{x_1}\!dx_2\,
x_2^2\,
\;
 \tilde{g}_0(x_2)
\;
\sin\left[\frac{\pi^2 \beta}{18 \zeta(3)}(x_1^3-x_2^3)\right],
\quad\quad
\end{multline}
where ``t.a.'' specifies that a thermal ansatz has been adopted for the DM momentum distribution.

\subsubsection{$\mathcal{O}(F^6)$ baryon asymmetry in the Minimal Model}
In the Minimal Model, one can use Eq.~(\ref{eq:Ya4}) to show that  the flavor-summed asymmetry vanishes at $\mathcal{O}(F^4)$:  $Y_\text{sm}^{(4)} \equiv \sum_\alpha Y^{(4)}_\alpha = 0$.  As discussed in Appendix~(\ref{sec:Z2PCPs}), we have $\delta Y_\chi = Y_\Phi = Y_\text{sm}$ in the Minimal Model, so the  $\Phi$ and $\chi$ asymmetries vanish at $\mathcal{O}(F^4)$ as well.  

A flavor-summed asymmetry arises at $\mathcal{O}(F^6)$, via the standard ARS mechanism \cite{Akhmedov:1998qx,Asaka:2005pn}.   In the presence of an $\mathcal{O}(F^4)$ chemical potential $\mu^{(4)}_{e^c_\alpha}$ for a particular flavor lepton (but  with $\mu^{(4)}_\Phi = 0$, and continuing to neglect the DM abundance),  the difference between the rates per volume for $\Phi^*\rightarrow \chi+e^c_\alpha$ and  $\Phi\rightarrow \overline{\chi} +\overline {e^c_\alpha}$ decays is $-2(\mu_{e^c_\alpha}/T) {\rm Tr} \left[ \gamma^{e1}_\alpha \right]$, where the $\gamma^{e1}$ reaction density is given by Eq.~(\ref{eq:pertRDe1}) when thermal mass effects are neglected.   Each decay produces a change $\Delta X_\alpha = \pm1$, so at $\mathcal{O}(F^6)$, the flavor-summed asymmetry at sphaleron decoupling is 
\be\label{eq:Ysm6}
Y^{(6)}_\text{sm}  =  
-2
\int_0^{x_\text{ew}} 
\!\frac{dx}{s H x}
\sum_\alpha
\frac{\mu^{(4)}_{e^c_\alpha}}{T}
 {\rm Tr} \left[ \gamma^{e1}_\alpha (x) \right].
\ee
Eq.~(\ref{eq:muespecZ2}), with $Y_\text{sm}^{(4)} = 0$, allows us to rewrite this as
\be\label{eq:ARSF6}
Y^{(6)}_\text{sm} = -\frac{4 M_0}{M_\Phi^5}\sum_\alpha  \int_0^{x_\text{ew}} \!\!\!\!   dx \; x^4 \;Y_\alpha^{(4)} (x) \; {\rm Tr} \left[ \gamma^{e1}_\alpha (x) \right].
\quad\quad\quad
\ee
Finally, from Eq.~(\ref{eq:ybspecZ2}) we have $Y_B={\mathcal K}_B Y_\text{sm}$, with ${\mathcal K}_B = 22/79$.   Using Eqs.~(\ref{eq:Ya4}) and (\ref{eq:final_I4_fn_1}), we can write the final $\mathcal{O}(F^6)$ baryon asymmetry as
\begin{multline} \label{eq:MMYB6}
Y_B^{(6)}
=
\sum_\alpha
(F F^\dagger)_{\alpha \alpha}
\mathrm{Im}\left[
F_{\alpha 1}^*  
F_{\alpha 2}
\left(F^\dagger F \right)_{21}
\right] 
\\
\times
\frac{45{\mathcal K}_B}{2048 \pi^9 g_*}
\left(\frac{M_0}{M_\Phi}\right)^3
 \mathcal{I}^{(6)}(x_\text{ew},\beta_\text{osc}),
 \end{multline}
 with
 \be\label{eq:I6def}
 \mathcal{I}^{(6)}(x,\beta) = \int_0^x \! dx' \; x'^2\; g_{e1} (x') \;   \mathcal{I}^{(4)}(x',\beta).
 \ee
For the numerical studies of Sec.~\ref{sec:MM}, we adopt the benchmark $F$ matrix introduced in Sec~(\ref{sec:MMFbench}), so that Eq.~(\ref{eq:MMbenchnum}) applies.  For this benchmark we find
\begin{multline}\label{eq:Y6constraint1app}
\frac{Y_B^{(6)}}{Y_B^\text{obs}}
\simeq
23
\;
\sin^2 2\theta
\\
\times
\left(
\frac{\Gamma_\Phi}{H_\text{ew}}
\right)^3
\left(
\frac{500 \text{ GeV}}{M_\Phi}
\right)^6
\mathcal{I}^{(6)}(x_\text{ew},\beta_\text{osc}).
\end{multline}
For $x \gg 1$, the asymptotic behavior of $\mathcal{I}^{(6)}$ for large and small $\beta$ is
\be\label{eq:asymp_I6}
\mathcal{I}^{(6)} (\infty,\beta) \simeq 
\begin{cases}
12.8 \;\beta &  \beta \ll1 \\
0.157 / \beta& \beta \gg 1.
\end{cases}
\ee
Fig.~\ref{fig:I4}(b) shows $\mathcal{I}^{(6)}(x_\text{ew},\beta)$ versus $\beta$ for various $M_\Phi$.

\subsection{Perturbative results in the UVDM Model}\label{sec:UVDMpert}

In the UVDM Model, we include a ``primordial'' population of $\chi$ particles produced at high temperatures.   For  concreteness, we adopt a model with two scalars $\Phi_1$ and $\Phi_2$, which have hierarchical masses, $M_{\Phi_2} \gg M_{\Phi_1}$, and independent DM coupling matrices $F^1$ and $F^2$.  At $\mathcal{O}(F^2)$, the DM energy density can be expressed as
\begin{multline}\label{eq:DM2s}
\frac{\rho_\text{dm}^{(2)}}{\rho_\text{dm}^\text{obs}}= 
22
\left(
\frac{\Gamma_\Phi}{H_\text{ew}}
\right)
\left(
\frac{\overline M^{(1)}}{15 \text{ keV}}
\right)
\left(
\frac{500 \text{ GeV}}{M_{\Phi_1}}
\right)^2
\\
+
135
\left(
\frac{Y^\text{UV}_{\chi}}{Y^\text{eq}_\chi}
\right)
\left(
\frac{\overline M^{(2)}}{15 \text{ keV}}
\right).
\end{multline}
Here, ${\overline M^{(1)}}$ and ${\overline M^{(2)}}$ are the average masses for DM produced in $\Phi_1$ and $\Phi_2$ decays, respectively.  Generalizing Eq.~(\ref{eq:aveDMmassapp}), we have
\be\label{eq:aveDMmass2s}
{\overline M^{(i)}}= \cos^2\theta_i M_1 +\sin^2\theta_i M_2
\ee
for $i=1,2$; $\theta_1$ and $\theta_2$ are determined from the coupling matrices $F^1$ and $F^2$ analogously to Eq.~(\ref{eq:Ffullparam}) for the Minimal Model.  The first term in Eq.~(\ref{eq:DM2s}) gives the DM energy density due to $\Phi_1$ decays, reproducing the result from Eq.~(\ref{eq:rhoGamma}) for the Minimal Model.  The second term gives the DM energy density due to $\Phi_2$ decays, which we  express in terms of $Y^\text{UV}_{\chi}$, the number-density  divided by entropy density for DM particles from $\Phi_2$ decays (defined to include  both DM mass eigenstates but only one helicity state: $\chi$ or ${\overline \chi}$, not both), and the reference value $Y^\text{eq}_\chi =135 \zeta(3)/8 \pi^4 g_*\simeq 1.95\times 10^{-3}$,  the equilibrium abundance for an individual  helicity and mass eigenstate of $\chi$ particle.  We express the $\Phi_2$ contribution in terms of $Y^\text{UV}_{\chi}$ because we take $\Phi_2$ to be decoupled as far as phenomenology is concerned.  Its only impact is through the coherent $\chi$ background its decays leave behind.   

In the UVDM Model, a baryon asymmetry arises at   $\mathcal{O}(F^4)$, due to $\chi$ production by $\Phi_2$ decay followed by inverse decays to $\Phi_1$.  We can obtain an expression for the leading-order baryon asymmetry by appropriate modification of the flavor-summed version of Eq.~(\ref{eq:Ya4}), which applies in the Minimal Model.  We get
\begin{multline}\label{eq:YB4_UVDM}
Y_B^{(4)} =\frac{45 \mathcal{K}_B}{256 g_*\pi^6}\,\frac{ M_{\rm 0}^2}{M_{\Phi_1} M_{\Phi_2}}\\
\times
\mathrm{Im}\left[
\left({F^1}^\dagger 
F^1 \right)_{21}
\left({F^2}^\dagger F^2 \right)_{12}
\right] 
\;
\tilde{\mathcal{I}}^{(4)}(x_\text{ew},\beta_\text{osc}),
\end{multline}
 with
 \begin{multline}
 \label{eq:final_I4_fn_2}
 \tilde{\mathcal{I}}^{(4)}(x,\beta) = \int_0^\infty\!
 dy
 \int_0^{\infty}\!dx_2\,
x_2^2\,
\;
 \frac{g_0(x_2,y)}{y^2 f_+(y)}
\\
\times
 \int_0^{x}\!dx_1\,
 x_1^2\,
 g_0(x_1,y)
 \;
 \sin\left[\beta x_1^3/y\right],
\quad\quad
\end{multline}
where we define $x_\text{ew} \equiv M_{\Phi_1}/T_\text{ew}$ in the context of the UVDM Model.  Using the relations between $F^2$ and  $Y^\text{UV}_{\chi}$ and $F^1$ and $\Gamma_{\Phi_1}$, the final baryon asymmetry can be reexpressed as
\begin{multline}\label{eq:YBnumUVDM}
\frac{Y_B^{(4)}}{Y_B^\text{obs}} \simeq
(1.03 \times10^{5})
\;
\mathcal{J}
\;
\left(
\frac{Y^\text{UV}_{\chi}}{Y^\text{eq}_\chi}
\right)\\
\times
\left(
\frac{\Gamma_{\Phi_1}}{H_\text{ew}}
\right)
\left(
\frac{500 \text{ GeV}}{M_{\Phi_1}}
\right)^2
\;
\tilde{\mathcal{I}}^{(4)}(x_\text{ew},\beta_\text{osc}),
\end{multline}
where  $\mathcal{J}$, which is defined in Appendix~\ref{sec:UVDMFbench}, satisfies $\mathcal{J} \le \sin2\theta_1 \sin 2\theta_2$.  We take the bound to be saturated for our UVDM benchmark scenario, as it is for the $F$ matrices of Eq.~(\ref{eq:UVDMbench}) for example.  

For $x \gg 1$, the asymptotic behavior of $\tilde{\mathcal{I}}^{(4)} $ for large and small $\beta$ is
\be\label{eq:asymp_I4UVDM}
\tilde{\mathcal{I}}^{(4)}  (\infty,\beta) \simeq 
\begin{cases}
303 \;\beta &  \beta \ll1 \\
1.18/ \beta& \beta \gg 1.
\end{cases}
\ee
Fig.~\ref{fig:I4}(c) shows $\tilde{\mathcal{I}}^{(4)}(x_\text{ew},\beta)$ versus $\beta$ for various $M_{\Phi_1}$.  
 
\subsection{Perturbative results in the Z2V Model}\label{sec:Z2Vpert}

We restrict our perturbative analysis of the Z2V Model to cases in which one or two of the three independent Z2V couplings come into equilibrium, and we neglect higher-order corrections induced by Z2V couplings that remain out of equilibrium.  If all three Z2V couplings come into equilibrium the baryon asymmetry essentially vanishes, as discussed in Appendix~\ref{sec:appendixCPs}.
In the Z2V Model, the DM abundance at $\mathcal{O}(F^2)$ is given by  Eq.~(\ref{eq:YchiF2}), as  in the Minimal Model.  However, because we assume that the Z2V couplings dominate the $\Phi$ decay width,  Eq.~(\ref{eq:rhoGamma}), which expresses the Minimal-Model connection between the dark matter abundance and the $\Phi$ lifetime, no longer applies.

\subsubsection{$Y_B$  at $\mathcal{O}(F^4)$ in the Z2V Model}\label{sec:YB4Z2V}
{\em One Z2V coupling in equilibrium:\;\;} Consider the case of a single Z2V coupling $\lambda_{\gamma \delta} = -\lambda_{\delta \gamma}$  in equilibrium.  Then, for the one lepton flavor  that does {\em not} participate in $Z_2$-violating interactions, which we call $\beta$, the $\mathcal{O} (F^4)$ result for $Y_\beta$ is unaffected by the Z2V coupling, allowing one to use Eq.~(\ref{eq:Ya4}) for that flavor. Furthermore, at $\mathcal{O} (F^4)$, there is no $\chi$ asymmetry, $\delta Y^{(4)}_\chi = 0$.  This is the case in the Minimal Model, as discussed in Appendix~\ref{sec:Z2PCPs}, and because the  Z2V couplings do not involve $\chi$, it remains true in the Z2V Model.  Eq.~(\ref{eq:Z2Vchempot2}) then allows us to write the  $\mathcal{O} (F^4)$ baryon asymmetry in this scenario as
\beq\label{eq:YBsingleZ2V}
Y_B^{(4)} = \frac{300 \;c_\Phi}{237+766 \;c_\Phi}Y_\beta^{(4)},
\eeq
with $Y^{(4)}_\beta$ calculated using Eq.~(\ref{eq:Ya4}).   The final baryon asymmetry is given by evaluating all quantities at $x = x_\text{ew}$.

In fact, using $\delta Y^{(4)}_\chi = 0$, the equilibrium relation Eq.~(\ref{eq:Z2Vchempot1}), and the general relations Eqs.~(\ref{eq:ybspec}-\ref{eq:muphispec}), one can express all $\mathcal{O} (F^4)$ asymmetries entirely in terms of $Y^{(4)}_\beta$ and $Y^{(4)}_\gamma - Y^{(4)}_\delta$, the two combinations of $X_\alpha$ asymmetries that are not affected by the Z2V couplings and so can be calculated using Eq.~(\ref{eq:Ya4}).  For example, we find
\be
\label{eq:mue1oneZ2V}
\frac{\mu_{e^c_\beta}^{(4)} T^2}{s} & = &
\frac{474 +1296 \;c_\Phi}{237+766 \;c_\Phi}Y_\beta^{(4)} 
\\
\label{eq:mue2oneZ2V}
\frac{\mu_{e^c_\gamma}^{(4)} T^2}{s} & = &
Y_\gamma^{(4)}-Y_\delta^{(4)}
-
\frac{
237+54 \;c_\Phi
}
{237+766 \;c_\Phi}
Y_\beta^{(4)} 
\\
\label{eq:mue3oneZ2V}
\frac{\mu_{e^c_\delta}^{(4)} T^2}{s} & = &
-(Y_\gamma^{(4)}-Y_\delta^{(4)})
-
\frac{
237+54 \;c_\Phi
}
{237+766 \;c_\Phi}
Y_\beta^{(4)} 
\quad
\\
\label{eq:oneZ2VF4CP}
\frac{\mu_\Phi^{(4)} T^2}{s}& = & 
\frac{
474
}
{237+766 \;c_\Phi}
Y_\beta^{(4)},
\ee
which we use below.

{\em Two Z2V couplings in equilibrium:\;\;} 
Similarly, for the case with two Z2V couplings $\lambda_{\beta \gamma}$ and $\lambda_{\beta \delta}$ in equilibrium, and neglecting the remaining coupling $\lambda_{\gamma \delta}$, Eq.~(\ref{eq:Z2Vchempot4}) leads to
\beq\label{eq:YBdoubleZ2Vprep}
Y_B^{(4)} = -\frac{75 \;c_\Phi}{237+529 \;c_\Phi} (Y_\beta^{(4)}-Y^{(4)}_\gamma-Y^{(4)}_\delta),
\eeq
where we again use the result from the Minimal Model Eq.~(\ref{eq:Ya4}), to evaluate each of the three terms in the combination $Y_\beta^{(4)}-Y^{(4)}_\gamma-Y^{(4)}_\delta$, which is not affected by Z2V couplings to the extent that we can neglect $\lambda_{\gamma \delta}$.  We can express this combination entirely in terms of the Minimal-Model result for $Y_\beta^{(4)}$, using the fact that $Y^{(4)}_\text{sm} = \sum_\alpha Y_\alpha^{(4)}=0$ in the Minimal Model, giving 
\beq\label{eq:YBdoubleZ2V}
Y_B^{(4)} = -\frac{150 \;c_\Phi}{237+529 \;c_\Phi} Y_\beta^{(4)}.
\eeq
In this equation, $Y_\beta^{(4)}$ is to be calculated using the Minimal-Model result, Eq.~(\ref{eq:Ya4}), even though that expression does {\em not} in fact give correct result for $Y_\beta^{(4)}$ in the particular Z2V scenario under consideration. 

Using $\delta Y^{(4)}_\chi = 0$, the equilibrium relation Eq.~(\ref{eq:Z2Vchempot3}), and  Eqs.~(\ref{eq:ybspec}-\ref{eq:muphispec}), one can also express the other $\mathcal{O} (F^4)$ asymmetries in terms of $Y_\beta^{(4)}-Y^{(4)}_\gamma-Y^{(4)}_\delta$, which we can again replace with twice the Minimal-Model result for $Y_\beta^{(4)}$.  We find
\be
\label{eq:mue1twoZ2V}
\frac{\mu_{e^c_\beta}^{(4)} T^2}{s} & = &
\frac{474 +702 \;c_\Phi}
{237+529 \;c_\Phi}
Y_\beta^{(4)} 
\\
\label{eq:mue2twoZ2V}
\frac{\mu_{e^c_\gamma}^{(4)} T^2}{s} =\frac{\mu_{e^c_\delta}^{(4)} T^2}{s} & = &
-\frac{
237+648 \;c_\Phi
}
{237+529 \;c_\Phi}
Y_\beta^{(4)}
\\
\label{eq:twoZ2VF4CP}
\frac{\mu_\Phi^{(4)} T^2}{s}& = & 
-\frac{
237
}
{237+529 \;c_\Phi}
Y_\beta^{(4)}.
\ee

We now take a moment to review the notation used in Eqs.~(\ref{eq:YBsingleZ2V}--\ref{eq:twoZ2VF4CP}).
In Eqs.~(\ref{eq:YBsingleZ2V}--\ref{eq:oneZ2VF4CP}), which apply in the case of a single Z2V coupling in equilibrium, $\beta$ is the special flavor whose lepton doublet is not involved in that Z2V coupling.   In Eqs.~(\ref{eq:YBdoubleZ2V}--\ref{eq:twoZ2VF4CP}), which apply in the case of two Z2V couplings in equilibrium, $\beta$ is the special flavor whose lepton doublet is involved both of those Z2V couplings. Both sets of equations apply when both DM mass eigenstates and the remaining Z2V couplings remain well out of equilibrium: $\lambda_{\beta \gamma}$ and $\lambda_{\beta \delta}$ for Eqs.~(\ref{eq:YBsingleZ2V}--\ref{eq:oneZ2VF4CP}) and $\lambda_{\gamma \delta}$ for Eqs.~(\ref{eq:YBdoubleZ2V}--\ref{eq:twoZ2VF4CP}).  In both sets of equations, the Minimal-Model result, Eq.~(\ref{eq:Ya4}), is to be used to calculate the asymmetries appearing on the right-hand sides.

\subsubsection{$Y_B$  at $\mathcal{O}(F^6)$ in the Z2V Model}\label{sec:YB6Z2V}
For large $\Phi$ massses, $M_\Phi \gg T_\text{ew}$, the overall factors of $c_\Phi$ in Eqs.~(\ref{eq:YBsingleZ2V}) and ~(\ref{eq:YBdoubleZ2V}) exponentially suppress the $\mathcal{O}(F^4)$ baryon asymmetry (see Fig.~\ref{fig:cphi}).   This is a consequence of the fact there is no $\chi$ asymmetry at this order, $\delta Y^{(4)}_\chi = 0$.  The final baryon asymmetry is proportional to the $B-L$ charge in the SM sector at sphaleron decoupling. 
For $x_\text{ew} \gg 1$, the $\Phi$ abundance becomes Boltzmann suppressed before sphaleron decoupling, along with any $\Phi$ asymmetry\footnote{ In our perturbative analysis of the Z2V Model, we assume that at least one $\lambda$ coupling comes into equilibrium, which guarantees that $\mu_\Phi/T \ll 1$ always holds.  In $Z_2$-preserving scenarios, it is possible for a $\Phi$ asymmetry to survive at temperatures $T\ll M_\Phi$. In the Z2V Model,  $\Phi^{(*)}$ decays via Z2V interactions prevent this from happening.
 }.  
When the surviving $\delta Y_\Phi$ at sphaleron decoupling  is negligible, 
conservation of $B-L$, expressed by Eq.~(\ref{eq:BmL}), tells us that the surviving $B-L$ charge in the SM sector is determined by the $\chi$ asymmetry at $T=T_\text{ew}$.    

A non-zero $\delta Y_\chi$ is generated by $\Phi^{(*)}$ decays to DM in the presence of $\mathcal{O}(F^4)$ chemical potentials for $e_c^\alpha$ and/or $\Phi$.  For sufficiently large $M_\Phi$, then, the final baryon asymmetry effectively arises at $\mathcal{O}(F^6)$. 

 The $\mu_{{e^c_\alpha}}$-induced contribution to $Y_\chi^{(6)}$ is essentially the standard ARS one, equal to the right-hand side of Eq.~(\ref{eq:ARSF6}).  However, when considering this contribution in the Z2V Model there are two differences that arise relative to the Minimal Model.  
 
 First, the Z2V couplings complicate the dependence of the final baryon asymmetry on the DM couplings.  To calculate the quantity in Eq.~(\ref{eq:ARSF6}), one needs to express $\mu_{{e^c_\alpha}}$ in terms of $X_\alpha$, using the temperature-dependent relations in
 Eqs.~(\ref{eq:mue1oneZ2V}--\ref{eq:mue3oneZ2V})
 or Eqs.~(\ref{eq:mue1twoZ2V}--\ref{eq:mue2twoZ2V}).  The final expression is a mess;  in particular the coupling dependence cannot be summarized in terms of the usual $\mathcal{O}(F^6)$ ARS factor $\sum_\alpha
(F F^\dagger)_{\alpha \alpha}
\mathrm{Im}\left[
F_{\alpha 1}^*  
F_{\alpha 2}
\left(F^\dagger F \right)_{21}
\right]$.
 For our Z2V benchmark $F$ matrix, given in Eq.~(\ref{eq:FZ2Vbench}), the $\mu_{{e^c_\alpha}}$-induced contribution to $\delta Y^{(6)}_\chi$ is not generally zero, even though the standard $\mathcal{O}(F^6)$ ARS coupling factor vanishes.  
 
Second, and more importantly, in the Z2V case the baryon asymmetry does not necessarily disappear when the $\Phi$ particles do, and consequently it  possible to get a sufficiently large asymmetry for  larger $\Phi$ masses than in the Minimal Model.  In the Minimal Model, the final baryon asymmetry is proportional to the $\Phi$ asymmetry at sphaleron decoupling, and Eq.~(\ref{eq:MMYB6}) is a good approximation for the final baryon asymmetry only to the extent that we can ignore washout of $\delta Y_\Phi$ by $\Phi$ decay, $\Gamma_\Phi \lesssim H_\text{ew}$.  In the Z2V Model, the baryon asymmetry survives even in the absence of a $\Phi$ asymmetry at sphaleron decoupling, provided a $\chi$ asymmetry has been generated.  For the Z2V perturbative result to be valid we only need the DM abundances to remain well below their equilibrium values, and in particular it is not necessary for 
$\Gamma(\Phi \rightarrow {\overline \chi} \;\overline{e_c}) \lesssim H_\text{ew}$ to be satisfied.  For large $\Phi$ masses, $M_\Phi \gg T_\text{ew}$, the perturbativity criterion based on the DM abundance is the less restrictive one.

The $\mu_{\Phi}$-induced contribution  is special to the Z2V Model, because in the Minimal Model we have $\mu_\Phi^{(4)} = 0$.  In fact, the $\mu_{\Phi}$-induced contribution tends to dominate the $\mu_{{e^c_\alpha}}$-induced one in the Z2V case, due to the different combinations of distribution functions appearing in the $\gamma^{\Phi 1}$ and $\gamma^{e1}$ reaction densities.  The new contribution to $\delta Y_\chi$ is given by the analogue to Eqn~(\ref{eq:Ysm6}),
\be\label{eq:YB6Z2Vmuphi}
\delta Y^{(6, \text{ from }\mu_\Phi)}_\chi = 
-2
\int_0^{x_\text{ew}} 
\!\!\!\frac{dx}{s H x}
\frac{\mu^{(4)}_{\Phi}}{T}
 \sum_\alpha {\rm Tr} \left[ \gamma^{\Phi1}_\alpha  \right].
 \quad\quad 
\ee
Using either Eq.~(\ref{eq:oneZ2VF4CP}) or Eq.~(\ref{eq:twoZ2VF4CP}) to express $\mu_\Phi$ in terms of $Y_\beta^{(4)}$, which we evaluate using Eq.~(\ref{eq:Ya4}), we obtain
\begin{multline}\label{eq:Z2VYB6}
\delta Y^{(6, \text{ from }\mu_\Phi)}_\chi 
=
\mp\frac{45}{2048 \pi^9 g_*}
\;
\frac{M_0^3}{M_\Phi^3}
\;
{\rm Tr}\left[F F^\dagger \right]\\
\times
\mathrm{Im}\left[
F_{\beta 1}  
F_{\beta 2}^*
\left(F^\dagger F \right)_{12}
\right] 
\;
 \mathcal{{\tilde I}}^{(6)}(x_\text{ew},\beta_\text{osc}),
 \end{multline}
 with
 \be\label{eq:I6_Z2V}
 \mathcal{{\tilde I}}^{(6)}(x,\beta) = \int_0^x \! dx' \; x'^2\; A_\Phi(x')\; g_{\Phi 1} (x') \;   \mathcal{I}^{(4)}(x',\beta).
 \quad\quad\quad
 \ee
Here, the upper and lower signs apply to the cases of one and two Z2V couplings in equilibrium, respectively,  $g_{\Phi 1}$ is the function defined in Eq.~(\ref{eq:gphi1}), and $A_\Phi$ takes into account the temperature-dependent conversion from $\mu_\Phi^{(4)}$ to $Y_\beta^{(4)}$:
  \beq\label{eq:Aphi}
 A_\Phi = \begin{cases}
 (1+3.23 \;c_\Phi)^{-1}
\quad \;
\text{one Z2V couplings}
\\
 \frac{1}{2}
 (1+2.23 \;c_\Phi)^{-1}
 \;\;
\text{two Z2V couplings.}
 \end{cases}
 \eeq
As before,  the flavor $\beta$ appearing in Eq.~(\ref{eq:Z2VYB6}) is the one that is either not involved in the Z2V coupling or the one that is involved in both Z2V couplings, for the cases of one or two  Z2V couplings in equilibrium, respectively.  
 
 Taking a single Z2V coupling to be in equilibrium and $x \gg 1$, the asymptotic behavior of $\mathcal{I}^{(6)}$ for large and small $\beta$ is
\beq\label{eq:asymp_I6_Z2V}
\mathcal{{\tilde I}}^{(6)} (\infty,\beta) \simeq 
\begin{cases}
194 \;\beta &  \beta \ll1 \\
1.50 / \beta& \beta \gg 1.
\end{cases}
\eeq
Fig.~\ref{fig:I4}(d) shows $\mathcal{{\tilde I}}^{(6)}(x_\text{ew},\beta)$ versus $\beta$ for various $M_\Phi$.

For $M_\Phi \gg T_\text{ew}$,which is the case in which the  $\mathcal{O}(F^6)$ contributions to the baryon asymmetry dominate, we can set $c_\Phi = 0$ when evaluating Eqs.~(\ref{eq:Z2Vchempot2}) or  Eqs.~(\ref{eq:Z2Vchempot4}) at sphaleron decoupling to get the final baryon asymmetry, giving $Y_B=-(84/237)\; \delta Y_\chi$. 
Using this relation and specializing to the Z2V benchmark $F$ matrix defined in Eq.~(\ref{eq:FZ2Vbench}), we obtain
\begin{multline}\label{eq:YB6Z2Vbench}
\left|\frac{Y_B^{(6, \text{ from }\mu_\Phi)}}{Y_B^\text{obs}}\right|
\simeq(1.76\times 10^4)\; \left(\frac{Y^{(2)}_\chi}{Y_\chi^\text{eq}} \right)^3 \\
\times
\sin^2 2\theta\;
 \mathcal{{\tilde I}}^{(6)}(x_\text{ew},\beta_\text{osc}).
\end{multline}
  We have used Eq.~(\ref{eq:YchiF2}) to express the asymmetry in terms of the number density of dark matter particles rather than the DM coupling strength.   Eq.~(\ref{eq:YB6Z2Vbench}) applies for large $\Phi$ masses, $M_\Phi \gg T_\text{ew}$, and relies on a perturbative expansion that is valid for $Y^{(2)}_\chi \ll Y_\chi^\text{eq}$.  
  
To get a sense of how large $M_\Phi$ can be consistent with the DM and baryon asymmetry constraints, take $M_1 = 0$ and make the small angle approximation for $\theta$.  Imposing the DM constraint, one finds 
\begin{multline}\label{eq:YB6Z2Vmasslesschi1smalltheta}
\left|\frac{Y_B^{(6, \text{ from }\mu_\Phi)}}{Y_B^\text{obs}}\right|
\simeq
240 \; \left(\frac{Y^{(2)}_\chi}{Y_\chi^\text{eq}} \right)^2 \\
\times
\left(
\frac{500 \text{ GeV}}{M_{\Phi}}
\right)^{3/2}\;
\beta_\text{osc}^{-1/2} \;
\mathcal{{\tilde I}}^{(6)}(x_\text{ew},\beta_\text{osc}).
\end{multline}  
For large $\Phi$ masses, $x_\text{ew}\gg 1$,  the combination $\beta_\text{osc}^{-1/2} \mathcal{{\tilde I}}^{(6)}$ has a maximum of $\simeq 26$ at $\beta_\text{osc} \simeq 3.7 \times 10^{-2} $.  Aggressively taking   $Y^{(2)}_\chi = Y_\chi^\text{eq}$, one finds that the observed baryon asymmetry can be achieved masses for $\Phi$ masses up to $M_\Phi \sim 170$ TeV, requiring a $\chi_2$ mass of $M_2 \sim 40$ MeV and  $\theta \sim 2 \times 10^{-3}$.    For the more perturbative case  $Y^{(2)}_\chi = Y_\chi^\text{eq}/3$,  the observed baryon asymmetry can be achieved with masses up to $M_\Phi \sim 40$ TeV for  $M_2 \sim 4$ MeV and  $\theta \sim 9 \times10^{-3}$.

\bibliography{biblio}

\begin{thebibliography}{77}%
\makeatletter
\providecommand \@ifxundefined [1]{%
 \@ifx{#1\undefined}
}%
\providecommand \@ifnum [1]{%
 \ifnum #1\expandafter \@firstoftwo
 \else \expandafter \@secondoftwo
 \fi
}%
\providecommand \@ifx [1]{%
 \ifx #1\expandafter \@firstoftwo
 \else \expandafter \@secondoftwo
 \fi
}%
\providecommand \natexlab [1]{#1}%
\providecommand \enquote  [1]{``#1''}%
\providecommand \bibnamefont  [1]{#1}%
\providecommand \bibfnamefont [1]{#1}%
\providecommand \citenamefont [1]{#1}%
\providecommand \href@noop [0]{\@secondoftwo}%
\providecommand \href [0]{\begingroup \@sanitize@url \@href}%
\providecommand \@href[1]{\@@startlink{#1}\@@href}%
\providecommand \@@href[1]{\endgroup#1\@@endlink}%
\providecommand \@sanitize@url [0]{\catcode `\\12\catcode `\$12\catcode
  `\&12\catcode `\#12\catcode `\^12\catcode `\_12\catcode `\%12\relax}%
\providecommand \@@startlink[1]{}%
\providecommand \@@endlink[0]{}%
\providecommand \url  [0]{\begingroup\@sanitize@url \@url }%
\providecommand \@url [1]{\endgroup\@href {#1}{\urlprefix }}%
\providecommand \urlprefix  [0]{URL }%
\providecommand \Eprint [0]{\href }%
\providecommand \doibase [0]{http://dx.doi.org/}%
\providecommand \selectlanguage [0]{\@gobble}%
\providecommand \bibinfo  [0]{\@secondoftwo}%
\providecommand \bibfield  [0]{\@secondoftwo}%
\providecommand \translation [1]{[#1]}%
\providecommand \BibitemOpen [0]{}%
\providecommand \bibitemStop [0]{}%
\providecommand \bibitemNoStop [0]{.\EOS\space}%
\providecommand \EOS [0]{\spacefactor3000\relax}%
\providecommand \BibitemShut  [1]{\csname bibitem#1\endcsname}%
\let\auto@bib@innerbib\@empty
\bibitem [{\citenamefont {McDonald}(2002)}]{McDonald:2001vt}%
  \BibitemOpen
  \bibfield  {author} {\bibinfo {author} {\bibfnamefont {J.}~\bibnamefont
  {McDonald}},\ }\href {\doibase 10.1103/PhysRevLett.88.091304} {\bibfield
  {journal} {\bibinfo  {journal} {Phys. Rev. Lett.}\ }\textbf {\bibinfo
  {volume} {88}},\ \bibinfo {pages} {091304} (\bibinfo {year} {2002})},\
  \Eprint {http://arxiv.org/abs/hep-ph/0106249} {arXiv:hep-ph/0106249 [hep-ph]}
  \BibitemShut {NoStop}%
\bibitem [{\citenamefont {Choi}\ and\ \citenamefont
  {Roszkowski}(2005)}]{Choi:2005vq}%
  \BibitemOpen
  \bibfield  {author} {\bibinfo {author} {\bibfnamefont {K.-Y.}\ \bibnamefont
  {Choi}}\ and\ \bibinfo {author} {\bibfnamefont {L.}~\bibnamefont
  {Roszkowski}},\ }\bibfield  {booktitle} {\emph {\bibinfo {booktitle}
  {{Proceedings on 11th International Symposium on Particles, Strings and
  Cosmology (PASCOS 2005): Gyeongju, Korea, 30 May - 4 June 2005}}},\ }\href
  {\doibase 10.1063/1.2149672} {\bibfield  {journal} {\bibinfo  {journal} {AIP
  Conf. Proc.}\ }\textbf {\bibinfo {volume} {805}},\ \bibinfo {pages} {30}
  (\bibinfo {year} {2005})},\ \Eprint {http://arxiv.org/abs/hep-ph/0511003}
  {arXiv:hep-ph/0511003 [hep-ph]} \BibitemShut {NoStop}%
\bibitem [{\citenamefont {Kusenko}(2006)}]{Kusenko:2006rh}%
  \BibitemOpen
  \bibfield  {author} {\bibinfo {author} {\bibfnamefont {A.}~\bibnamefont
  {Kusenko}},\ }\href {\doibase 10.1103/PhysRevLett.97.241301} {\bibfield
  {journal} {\bibinfo  {journal} {Phys. Rev. Lett.}\ }\textbf {\bibinfo
  {volume} {97}},\ \bibinfo {pages} {241301} (\bibinfo {year} {2006})},\
  \Eprint {http://arxiv.org/abs/hep-ph/0609081} {arXiv:hep-ph/0609081 [hep-ph]}
  \BibitemShut {NoStop}%
\bibitem [{\citenamefont {Petraki}\ and\ \citenamefont
  {Kusenko}(2008)}]{Petraki:2007gq}%
  \BibitemOpen
  \bibfield  {author} {\bibinfo {author} {\bibfnamefont {K.}~\bibnamefont
  {Petraki}}\ and\ \bibinfo {author} {\bibfnamefont {A.}~\bibnamefont
  {Kusenko}},\ }\href {\doibase 10.1103/PhysRevD.77.065014} {\bibfield
  {journal} {\bibinfo  {journal} {Phys. Rev.}\ }\textbf {\bibinfo {volume}
  {D77}},\ \bibinfo {pages} {065014} (\bibinfo {year} {2008})},\ \Eprint
  {http://arxiv.org/abs/0711.4646} {arXiv:0711.4646 [hep-ph]} \BibitemShut
  {NoStop}%
\bibitem [{\citenamefont {Hall}\ \emph {et~al.}(2010)\citenamefont {Hall},
  \citenamefont {Jedamzik}, \citenamefont {March-Russell},\ and\ \citenamefont
  {West}}]{Hall:2009bx}%
  \BibitemOpen
  \bibfield  {author} {\bibinfo {author} {\bibfnamefont {L.~J.}\ \bibnamefont
  {Hall}}, \bibinfo {author} {\bibfnamefont {K.}~\bibnamefont {Jedamzik}},
  \bibinfo {author} {\bibfnamefont {J.}~\bibnamefont {March-Russell}}, \ and\
  \bibinfo {author} {\bibfnamefont {S.~M.}\ \bibnamefont {West}},\ }\href
  {\doibase 10.1007/JHEP03(2010)080} {\bibfield  {journal} {\bibinfo  {journal}
  {JHEP}\ }\textbf {\bibinfo {volume} {03}},\ \bibinfo {pages} {080} (\bibinfo
  {year} {2010})},\ \Eprint {http://arxiv.org/abs/0911.1120} {arXiv:0911.1120
  [hep-ph]} \BibitemShut {NoStop}%
\bibitem [{\citenamefont {Bernal}\ \emph {et~al.}(2017)\citenamefont {Bernal},
  \citenamefont {Heikinheimo}, \citenamefont {Tenkanen}, \citenamefont
  {Tuominen},\ and\ \citenamefont {Vaskonen}}]{Bernal:2017kxu}%
  \BibitemOpen
  \bibfield  {author} {\bibinfo {author} {\bibfnamefont {N.}~\bibnamefont
  {Bernal}}, \bibinfo {author} {\bibfnamefont {M.}~\bibnamefont {Heikinheimo}},
  \bibinfo {author} {\bibfnamefont {T.}~\bibnamefont {Tenkanen}}, \bibinfo
  {author} {\bibfnamefont {K.}~\bibnamefont {Tuominen}}, \ and\ \bibinfo
  {author} {\bibfnamefont {V.}~\bibnamefont {Vaskonen}},\ }\href {\doibase
  10.1142/S0217751X1730023X} {\bibfield  {journal} {\bibinfo  {journal} {Int.
  J. Mod. Phys.}\ }\textbf {\bibinfo {volume} {A32}},\ \bibinfo {pages}
  {1730023} (\bibinfo {year} {2017})},\ \Eprint
  {http://arxiv.org/abs/1706.07442} {arXiv:1706.07442 [hep-ph]} \BibitemShut
  {NoStop}%
\bibitem [{\citenamefont {Shuve}\ and\ \citenamefont
  {Tucker-Smith}(2020)}]{Shuve:2020evk}%
  \BibitemOpen
  \bibfield  {author} {\bibinfo {author} {\bibfnamefont {B.}~\bibnamefont
  {Shuve}}\ and\ \bibinfo {author} {\bibfnamefont {D.}~\bibnamefont
  {Tucker-Smith}},\ }\href {\doibase 10.1103/PhysRevD.101.115023} {\bibfield
  {journal} {\bibinfo  {journal} {Phys. Rev. D}\ }\textbf {\bibinfo {volume}
  {101}},\ \bibinfo {pages} {115023} (\bibinfo {year} {2020})},\ \Eprint
  {http://arxiv.org/abs/2004.00636} {arXiv:2004.00636 [hep-ph]} \BibitemShut
  {NoStop}%
\bibitem [{\citenamefont {Akhmedov}\ \emph {et~al.}(1998)\citenamefont
  {Akhmedov}, \citenamefont {Rubakov},\ and\ \citenamefont
  {Smirnov}}]{Akhmedov:1998qx}%
  \BibitemOpen
  \bibfield  {author} {\bibinfo {author} {\bibfnamefont {E.~K.}\ \bibnamefont
  {Akhmedov}}, \bibinfo {author} {\bibfnamefont {V.~A.}\ \bibnamefont
  {Rubakov}}, \ and\ \bibinfo {author} {\bibfnamefont {A.~{\relax Yu}.}\
  \bibnamefont {Smirnov}},\ }\href {\doibase 10.1103/PhysRevLett.81.1359}
  {\bibfield  {journal} {\bibinfo  {journal} {Phys. Rev. Lett.}\ }\textbf
  {\bibinfo {volume} {81}},\ \bibinfo {pages} {1359} (\bibinfo {year}
  {1998})},\ \Eprint {http://arxiv.org/abs/hep-ph/9803255}
  {arXiv:hep-ph/9803255 [hep-ph]} \BibitemShut {NoStop}%
\bibitem [{\citenamefont {Asaka}\ and\ \citenamefont
  {Shaposhnikov}(2005)}]{Asaka:2005pn}%
  \BibitemOpen
  \bibfield  {author} {\bibinfo {author} {\bibfnamefont {T.}~\bibnamefont
  {Asaka}}\ and\ \bibinfo {author} {\bibfnamefont {M.}~\bibnamefont
  {Shaposhnikov}},\ }\href {\doibase 10.1016/j.physletb.2005.06.020} {\bibfield
   {journal} {\bibinfo  {journal} {Phys. Lett.}\ }\textbf {\bibinfo {volume}
  {B620}},\ \bibinfo {pages} {17} (\bibinfo {year} {2005})},\ \Eprint
  {http://arxiv.org/abs/hep-ph/0505013} {arXiv:hep-ph/0505013 [hep-ph]}
  \BibitemShut {NoStop}%
\bibitem [{\citenamefont {Drewes}\ \emph {et~al.}(2018)\citenamefont {Drewes},
  \citenamefont {Garbrecht}, \citenamefont {Hernandez}, \citenamefont {Kekic},
  \citenamefont {Lopez-Pavon}, \citenamefont {Racker}, \citenamefont {Rius},
  \citenamefont {Salvado},\ and\ \citenamefont {Teresi}}]{Drewes:2017zyw}%
  \BibitemOpen
  \bibfield  {author} {\bibinfo {author} {\bibfnamefont {M.}~\bibnamefont
  {Drewes}}, \bibinfo {author} {\bibfnamefont {B.}~\bibnamefont {Garbrecht}},
  \bibinfo {author} {\bibfnamefont {P.}~\bibnamefont {Hernandez}}, \bibinfo
  {author} {\bibfnamefont {M.}~\bibnamefont {Kekic}}, \bibinfo {author}
  {\bibfnamefont {J.}~\bibnamefont {Lopez-Pavon}}, \bibinfo {author}
  {\bibfnamefont {J.}~\bibnamefont {Racker}}, \bibinfo {author} {\bibfnamefont
  {N.}~\bibnamefont {Rius}}, \bibinfo {author} {\bibfnamefont {J.}~\bibnamefont
  {Salvado}}, \ and\ \bibinfo {author} {\bibfnamefont {D.}~\bibnamefont
  {Teresi}},\ }\href {\doibase 10.1142/S0217751X18420022} {\bibfield  {journal}
  {\bibinfo  {journal} {Int. J. Mod. Phys. A}\ }\textbf {\bibinfo {volume}
  {33}},\ \bibinfo {pages} {1842002} (\bibinfo {year} {2018})},\ \Eprint
  {http://arxiv.org/abs/1711.02862} {arXiv:1711.02862 [hep-ph]} \BibitemShut
  {NoStop}%
\bibitem [{\citenamefont {Goudelis}\ \emph {et~al.}()\citenamefont {Goudelis},
  \citenamefont {Papachristou},\ and\ \citenamefont
  {Spanos}}]{Goudelis:2021lra}%
  \BibitemOpen
  \bibfield  {author} {\bibinfo {author} {\bibfnamefont {A.}~\bibnamefont
  {Goudelis}}, \bibinfo {author} {\bibfnamefont {P.}~\bibnamefont
  {Papachristou}}, \ and\ \bibinfo {author} {\bibfnamefont {V.~C.}\
  \bibnamefont {Spanos}},\ }\href@noop {} {\ }\Eprint
  {http://arxiv.org/abs/2111.05740} {arXiv:2111.05740 [hep-ph]} \BibitemShut
  {NoStop}%
\bibitem [{\citenamefont {Hall}\ \emph {et~al.}()\citenamefont {Hall},
  \citenamefont {March-Russell},\ and\ \citenamefont {West}}]{Hall:2010jx}%
  \BibitemOpen
  \bibfield  {author} {\bibinfo {author} {\bibfnamefont {L.~J.}\ \bibnamefont
  {Hall}}, \bibinfo {author} {\bibfnamefont {J.}~\bibnamefont {March-Russell}},
  \ and\ \bibinfo {author} {\bibfnamefont {S.~M.}\ \bibnamefont {West}},\
  }\href@noop {} {\ }\Eprint {http://arxiv.org/abs/1010.0245} {arXiv:1010.0245
  [hep-ph]} \BibitemShut {NoStop}%
\bibitem [{\citenamefont {Hook}(2011)}]{Hook:2011tk}%
  \BibitemOpen
  \bibfield  {author} {\bibinfo {author} {\bibfnamefont {A.}~\bibnamefont
  {Hook}},\ }\href {\doibase 10.1103/PhysRevD.84.055003} {\bibfield  {journal}
  {\bibinfo  {journal} {Phys. Rev.}\ }\textbf {\bibinfo {volume} {D84}},\
  \bibinfo {pages} {055003} (\bibinfo {year} {2011})},\ \Eprint
  {http://arxiv.org/abs/1105.3728} {arXiv:1105.3728 [hep-ph]} \BibitemShut
  {NoStop}%
\bibitem [{\citenamefont {Unwin}(2014)}]{Unwin:2014poa}%
  \BibitemOpen
  \bibfield  {author} {\bibinfo {author} {\bibfnamefont {J.}~\bibnamefont
  {Unwin}},\ }\href {\doibase 10.1007/JHEP10(2014)190} {\bibfield  {journal}
  {\bibinfo  {journal} {JHEP}\ }\textbf {\bibinfo {volume} {10}},\ \bibinfo
  {pages} {190} (\bibinfo {year} {2014})},\ \Eprint
  {http://arxiv.org/abs/1406.3027} {arXiv:1406.3027 [hep-ph]} \BibitemShut
  {NoStop}%
\bibitem [{\citenamefont {D'Onofrio}\ \emph {et~al.}(2014)\citenamefont
  {D'Onofrio}, \citenamefont {Rummukainen},\ and\ \citenamefont
  {Tranberg}}]{DOnofrio:2014rug}%
  \BibitemOpen
  \bibfield  {author} {\bibinfo {author} {\bibfnamefont {M.}~\bibnamefont
  {D'Onofrio}}, \bibinfo {author} {\bibfnamefont {K.}~\bibnamefont
  {Rummukainen}}, \ and\ \bibinfo {author} {\bibfnamefont {A.}~\bibnamefont
  {Tranberg}},\ }\href {\doibase 10.1103/PhysRevLett.113.141602} {\bibfield
  {journal} {\bibinfo  {journal} {Phys. Rev. Lett.}\ }\textbf {\bibinfo
  {volume} {113}},\ \bibinfo {pages} {141602} (\bibinfo {year} {2014})},\
  \Eprint {http://arxiv.org/abs/1404.3565} {arXiv:1404.3565 [hep-ph]}
  \BibitemShut {NoStop}%
\bibitem [{\citenamefont {Asaka}\ \emph {et~al.}(2017)\citenamefont {Asaka},
  \citenamefont {Eijima}, \citenamefont {Ishida}, \citenamefont {Minogawa},\
  and\ \citenamefont {Yoshii}}]{Asaka:2017rdj}%
  \BibitemOpen
  \bibfield  {author} {\bibinfo {author} {\bibfnamefont {T.}~\bibnamefont
  {Asaka}}, \bibinfo {author} {\bibfnamefont {S.}~\bibnamefont {Eijima}},
  \bibinfo {author} {\bibfnamefont {H.}~\bibnamefont {Ishida}}, \bibinfo
  {author} {\bibfnamefont {K.}~\bibnamefont {Minogawa}}, \ and\ \bibinfo
  {author} {\bibfnamefont {T.}~\bibnamefont {Yoshii}},\ }\href {\doibase
  10.1103/PhysRevD.96.083010} {\bibfield  {journal} {\bibinfo  {journal} {Phys.
  Rev.}\ }\textbf {\bibinfo {volume} {D96}},\ \bibinfo {pages} {083010}
  (\bibinfo {year} {2017})},\ \Eprint {http://arxiv.org/abs/1704.02692}
  {arXiv:1704.02692 [hep-ph]} \BibitemShut {NoStop}%
\bibitem [{\citenamefont {Asaka}\ \emph {et~al.}(2005)\citenamefont {Asaka},
  \citenamefont {Blanchet},\ and\ \citenamefont {Shaposhnikov}}]{Asaka:2005an}%
  \BibitemOpen
  \bibfield  {author} {\bibinfo {author} {\bibfnamefont {T.}~\bibnamefont
  {Asaka}}, \bibinfo {author} {\bibfnamefont {S.}~\bibnamefont {Blanchet}}, \
  and\ \bibinfo {author} {\bibfnamefont {M.}~\bibnamefont {Shaposhnikov}},\
  }\href {\doibase 10.1016/j.physletb.2005.09.070} {\bibfield  {journal}
  {\bibinfo  {journal} {Phys. Lett. B}\ }\textbf {\bibinfo {volume} {631}},\
  \bibinfo {pages} {151} (\bibinfo {year} {2005})},\ \Eprint
  {http://arxiv.org/abs/hep-ph/0503065} {arXiv:hep-ph/0503065} \BibitemShut
  {NoStop}%
\bibitem [{\citenamefont {Garzilli}\ \emph {et~al.}(2019)\citenamefont
  {Garzilli}, \citenamefont {Ruchayskiy}, \citenamefont {Magalich},\ and\
  \citenamefont {Boyarsky}}]{garzilli2019warm}%
  \BibitemOpen
  \bibfield  {author} {\bibinfo {author} {\bibfnamefont {A.}~\bibnamefont
  {Garzilli}}, \bibinfo {author} {\bibfnamefont {O.}~\bibnamefont
  {Ruchayskiy}}, \bibinfo {author} {\bibfnamefont {A.}~\bibnamefont
  {Magalich}}, \ and\ \bibinfo {author} {\bibfnamefont {A.}~\bibnamefont
  {Boyarsky}},\ }\href@noop {} {\enquote {\bibinfo {title} {How warm is too
  warm? towards robust lyman-$\alpha$ forest bounds on warm dark matter},}\ }
  (\bibinfo {year} {2019}),\ \Eprint {http://arxiv.org/abs/1912.09397}
  {arXiv:1912.09397 [astro-ph.CO]} \BibitemShut {NoStop}%
\bibitem [{\citenamefont {Palanque-Delabrouille}\ \emph
  {et~al.}(2020)\citenamefont {Palanque-Delabrouille}, \citenamefont {Yèche},
  \citenamefont {Schöneberg}, \citenamefont {Lesgourgues}, \citenamefont
  {Walther}, \citenamefont {Chabanier},\ and\ \citenamefont
  {Armengaud}}]{Palanque_Delabrouille_2020}%
  \BibitemOpen
  \bibfield  {author} {\bibinfo {author} {\bibfnamefont {N.}~\bibnamefont
  {Palanque-Delabrouille}}, \bibinfo {author} {\bibfnamefont {C.}~\bibnamefont
  {Yèche}}, \bibinfo {author} {\bibfnamefont {N.}~\bibnamefont {Schöneberg}},
  \bibinfo {author} {\bibfnamefont {J.}~\bibnamefont {Lesgourgues}}, \bibinfo
  {author} {\bibfnamefont {M.}~\bibnamefont {Walther}}, \bibinfo {author}
  {\bibfnamefont {S.}~\bibnamefont {Chabanier}}, \ and\ \bibinfo {author}
  {\bibfnamefont {E.}~\bibnamefont {Armengaud}},\ }\href {\doibase
  10.1088/1475-7516/2020/04/038} {\bibfield  {journal} {\bibinfo  {journal}
  {Journal of Cosmology and Astroparticle Physics}\ }\textbf {\bibinfo {volume}
  {2020}},\ \bibinfo {pages} {038–038} (\bibinfo {year} {2020})}\BibitemShut
  {NoStop}%
\bibitem [{\citenamefont {Baur}\ \emph {et~al.}(2016)\citenamefont {Baur},
  \citenamefont {Palanque-Delabrouille}, \citenamefont {Yèche}, \citenamefont
  {Magneville},\ and\ \citenamefont {Viel}}]{Baur_2016}%
  \BibitemOpen
  \bibfield  {author} {\bibinfo {author} {\bibfnamefont {J.}~\bibnamefont
  {Baur}}, \bibinfo {author} {\bibfnamefont {N.}~\bibnamefont
  {Palanque-Delabrouille}}, \bibinfo {author} {\bibfnamefont {C.}~\bibnamefont
  {Yèche}}, \bibinfo {author} {\bibfnamefont {C.}~\bibnamefont {Magneville}},
  \ and\ \bibinfo {author} {\bibfnamefont {M.}~\bibnamefont {Viel}},\ }\href
  {\doibase 10.1088/1475-7516/2016/08/012} {\bibfield  {journal} {\bibinfo
  {journal} {Journal of Cosmology and Astroparticle Physics}\ }\textbf
  {\bibinfo {volume} {2016}},\ \bibinfo {pages} {012–012} (\bibinfo {year}
  {2016})}\BibitemShut {NoStop}%
\bibitem [{\citenamefont {Baur}\ \emph {et~al.}(2017)\citenamefont {Baur},
  \citenamefont {Palanque-Delabrouille}, \citenamefont {Yèche}, \citenamefont
  {Boyarsky}, \citenamefont {Ruchayskiy}, \citenamefont {Armengaud},\ and\
  \citenamefont {Lesgourgues}}]{Baur_2017}%
  \BibitemOpen
  \bibfield  {author} {\bibinfo {author} {\bibfnamefont {J.}~\bibnamefont
  {Baur}}, \bibinfo {author} {\bibfnamefont {N.}~\bibnamefont
  {Palanque-Delabrouille}}, \bibinfo {author} {\bibfnamefont {C.}~\bibnamefont
  {Yèche}}, \bibinfo {author} {\bibfnamefont {A.}~\bibnamefont {Boyarsky}},
  \bibinfo {author} {\bibfnamefont {O.}~\bibnamefont {Ruchayskiy}}, \bibinfo
  {author} {\bibfnamefont {E.}~\bibnamefont {Armengaud}}, \ and\ \bibinfo
  {author} {\bibfnamefont {J.}~\bibnamefont {Lesgourgues}},\ }\href {\doibase
  10.1088/1475-7516/2017/12/013} {\bibfield  {journal} {\bibinfo  {journal}
  {Journal of Cosmology and Astroparticle Physics}\ }\textbf {\bibinfo {volume}
  {2017}},\ \bibinfo {pages} {013–013} (\bibinfo {year} {2017})}\BibitemShut
  {NoStop}%
\bibitem [{\citenamefont {Iršič}\ \emph {et~al.}(2017)\citenamefont
  {Iršič}, \citenamefont {Viel}, \citenamefont {Haehnelt}, \citenamefont
  {Bolton}, \citenamefont {Cristiani}, \citenamefont {Becker}, \citenamefont
  {D’Odorico}, \citenamefont {Cupani}, \citenamefont {Kim}, \citenamefont
  {Berg},\ and\ \citenamefont {et~al.}}]{Ir_i__2017}%
  \BibitemOpen
  \bibfield  {author} {\bibinfo {author} {\bibfnamefont {V.}~\bibnamefont
  {Iršič}}, \bibinfo {author} {\bibfnamefont {M.}~\bibnamefont {Viel}},
  \bibinfo {author} {\bibfnamefont {M.~G.}\ \bibnamefont {Haehnelt}}, \bibinfo
  {author} {\bibfnamefont {J.~S.}\ \bibnamefont {Bolton}}, \bibinfo {author}
  {\bibfnamefont {S.}~\bibnamefont {Cristiani}}, \bibinfo {author}
  {\bibfnamefont {G.~D.}\ \bibnamefont {Becker}}, \bibinfo {author}
  {\bibfnamefont {V.}~\bibnamefont {D’Odorico}}, \bibinfo {author}
  {\bibfnamefont {G.}~\bibnamefont {Cupani}}, \bibinfo {author} {\bibfnamefont
  {T.-S.}\ \bibnamefont {Kim}}, \bibinfo {author} {\bibfnamefont {T.~A.}\
  \bibnamefont {Berg}}, \ and\ \bibinfo {author} {\bibnamefont {et~al.}},\
  }\href {\doibase 10.1103/physrevd.96.023522} {\bibfield  {journal} {\bibinfo
  {journal} {Physical Review D}\ }\textbf {\bibinfo {volume} {96}} (\bibinfo
  {year} {2017}),\ 10.1103/physrevd.96.023522}\BibitemShut {NoStop}%
\bibitem [{\citenamefont {Kamada}\ and\ \citenamefont
  {Yanagi}(2019)}]{Kamada_2019}%
  \BibitemOpen
  \bibfield  {author} {\bibinfo {author} {\bibfnamefont {A.}~\bibnamefont
  {Kamada}}\ and\ \bibinfo {author} {\bibfnamefont {K.}~\bibnamefont
  {Yanagi}},\ }\href {\doibase 10.1088/1475-7516/2019/11/029} {\bibfield
  {journal} {\bibinfo  {journal} {Journal of Cosmology and Astroparticle
  Physics}\ }\textbf {\bibinfo {volume} {2019}},\ \bibinfo {pages} {029–029}
  (\bibinfo {year} {2019})}\BibitemShut {NoStop}%
\bibitem [{\citenamefont {Ballesteros}\ \emph {et~al.}(2021)\citenamefont
  {Ballesteros}, \citenamefont {Garcia},\ and\ \citenamefont
  {Pierre}}]{Ballesteros_2021}%
  \BibitemOpen
  \bibfield  {author} {\bibinfo {author} {\bibfnamefont {G.}~\bibnamefont
  {Ballesteros}}, \bibinfo {author} {\bibfnamefont {M.~A.}\ \bibnamefont
  {Garcia}}, \ and\ \bibinfo {author} {\bibfnamefont {M.}~\bibnamefont
  {Pierre}},\ }\href {\doibase 10.1088/1475-7516/2021/03/101} {\bibfield
  {journal} {\bibinfo  {journal} {Journal of Cosmology and Astroparticle
  Physics}\ }\textbf {\bibinfo {volume} {2021}},\ \bibinfo {pages} {101}
  (\bibinfo {year} {2021})}\BibitemShut {NoStop}%
\bibitem [{\citenamefont {Boyarsky}\ \emph {et~al.}(2009)\citenamefont
  {Boyarsky}, \citenamefont {Lesgourgues}, \citenamefont {Ruchayskiy},\ and\
  \citenamefont {Viel}}]{Boyarsky_2009}%
  \BibitemOpen
  \bibfield  {author} {\bibinfo {author} {\bibfnamefont {A.}~\bibnamefont
  {Boyarsky}}, \bibinfo {author} {\bibfnamefont {J.}~\bibnamefont
  {Lesgourgues}}, \bibinfo {author} {\bibfnamefont {O.}~\bibnamefont
  {Ruchayskiy}}, \ and\ \bibinfo {author} {\bibfnamefont {M.}~\bibnamefont
  {Viel}},\ }\href {\doibase 10.1088/1475-7516/2009/05/012} {\bibfield
  {journal} {\bibinfo  {journal} {Journal of Cosmology and Astroparticle
  Physics}\ }\textbf {\bibinfo {volume} {2009}},\ \bibinfo {pages} {012–012}
  (\bibinfo {year} {2009})}\BibitemShut {NoStop}%
\bibitem [{\citenamefont {Kamada}\ \emph {et~al.}(2016)\citenamefont {Kamada},
  \citenamefont {Inoue},\ and\ \citenamefont {Takahashi}}]{Kamada_2016}%
  \BibitemOpen
  \bibfield  {author} {\bibinfo {author} {\bibfnamefont {A.}~\bibnamefont
  {Kamada}}, \bibinfo {author} {\bibfnamefont {K.~T.}\ \bibnamefont {Inoue}}, \
  and\ \bibinfo {author} {\bibfnamefont {T.}~\bibnamefont {Takahashi}},\ }\href
  {\doibase 10.1103/physrevd.94.023522} {\bibfield  {journal} {\bibinfo
  {journal} {Physical Review D}\ }\textbf {\bibinfo {volume} {94}} (\bibinfo
  {year} {2016}),\ 10.1103/physrevd.94.023522}\BibitemShut {NoStop}%
\bibitem [{\citenamefont {Brust}\ \emph {et~al.}(2013)\citenamefont {Brust},
  \citenamefont {Kaplan},\ and\ \citenamefont {Walters}}]{Brust:2013ova}%
  \BibitemOpen
  \bibfield  {author} {\bibinfo {author} {\bibfnamefont {C.}~\bibnamefont
  {Brust}}, \bibinfo {author} {\bibfnamefont {D.~E.}\ \bibnamefont {Kaplan}}, \
  and\ \bibinfo {author} {\bibfnamefont {M.~T.}\ \bibnamefont {Walters}},\
  }\href {\doibase 10.1007/JHEP12(2013)058} {\bibfield  {journal} {\bibinfo
  {journal} {JHEP}\ }\textbf {\bibinfo {volume} {12}},\ \bibinfo {pages} {058}
  (\bibinfo {year} {2013})},\ \Eprint {http://arxiv.org/abs/1303.5379}
  {arXiv:1303.5379 [hep-ph]} \BibitemShut {NoStop}%
\bibitem [{\citenamefont {Zyla}\ \emph {et~al.}(2020)\citenamefont {Zyla} \emph
  {et~al.}}]{Zyla:2020zbs}%
  \BibitemOpen
  \bibfield  {author} {\bibinfo {author} {\bibfnamefont {P.}~\bibnamefont
  {Zyla}} \emph {et~al.} (\bibinfo {collaboration} {Particle Data Group}),\
  }\href {\doibase 10.1093/ptep/ptaa104} {\bibfield  {journal} {\bibinfo
  {journal} {PTEP}\ }\textbf {\bibinfo {volume} {2020}},\ \bibinfo {pages}
  {083C01} (\bibinfo {year} {2020})}\BibitemShut {NoStop}%
\bibitem [{\citenamefont {Hambye}\ and\ \citenamefont
  {Teresi}(2017)}]{Hambye:2017elz}%
  \BibitemOpen
  \bibfield  {author} {\bibinfo {author} {\bibfnamefont {T.}~\bibnamefont
  {Hambye}}\ and\ \bibinfo {author} {\bibfnamefont {D.}~\bibnamefont
  {Teresi}},\ }\href {\doibase 10.1103/PhysRevD.96.015031} {\bibfield
  {journal} {\bibinfo  {journal} {Phys. Rev.}\ }\textbf {\bibinfo {volume}
  {D96}},\ \bibinfo {pages} {015031} (\bibinfo {year} {2017})},\ \Eprint
  {http://arxiv.org/abs/1705.00016} {arXiv:1705.00016 [hep-ph]} \BibitemShut
  {NoStop}%
\bibitem [{\citenamefont {Abada}\ \emph {et~al.}(2019)\citenamefont {Abada},
  \citenamefont {Arcadi}, \citenamefont {Domcke}, \citenamefont {Drewes},
  \citenamefont {Klaric},\ and\ \citenamefont {Lucente}}]{Abada:2018oly}%
  \BibitemOpen
  \bibfield  {author} {\bibinfo {author} {\bibfnamefont {A.}~\bibnamefont
  {Abada}}, \bibinfo {author} {\bibfnamefont {G.}~\bibnamefont {Arcadi}},
  \bibinfo {author} {\bibfnamefont {V.}~\bibnamefont {Domcke}}, \bibinfo
  {author} {\bibfnamefont {M.}~\bibnamefont {Drewes}}, \bibinfo {author}
  {\bibfnamefont {J.}~\bibnamefont {Klaric}}, \ and\ \bibinfo {author}
  {\bibfnamefont {M.}~\bibnamefont {Lucente}},\ }\href {\doibase
  10.1007/JHEP01(2019)164} {\bibfield  {journal} {\bibinfo  {journal} {JHEP}\
  }\textbf {\bibinfo {volume} {01}},\ \bibinfo {pages} {164} (\bibinfo {year}
  {2019})},\ \Eprint {http://arxiv.org/abs/1810.12463} {arXiv:1810.12463
  [hep-ph]} \BibitemShut {NoStop}%
\bibitem [{\citenamefont {Shuve}\ and\ \citenamefont
  {Yavin}(2014)}]{Shuve:2014zua}%
  \BibitemOpen
  \bibfield  {author} {\bibinfo {author} {\bibfnamefont {B.}~\bibnamefont
  {Shuve}}\ and\ \bibinfo {author} {\bibfnamefont {I.}~\bibnamefont {Yavin}},\
  }\href {\doibase 10.1103/PhysRevD.89.075014} {\bibfield  {journal} {\bibinfo
  {journal} {Phys. Rev. D}\ }\textbf {\bibinfo {volume} {89}},\ \bibinfo
  {pages} {075014} (\bibinfo {year} {2014})},\ \Eprint
  {http://arxiv.org/abs/1401.2459} {arXiv:1401.2459 [hep-ph]} \BibitemShut
  {NoStop}%
\bibitem [{\citenamefont {Eijima}\ \emph {et~al.}(2017)\citenamefont {Eijima},
  \citenamefont {Shaposhnikov},\ and\ \citenamefont
  {Timiryasov}}]{Eijima:2017cxr}%
  \BibitemOpen
  \bibfield  {author} {\bibinfo {author} {\bibfnamefont {S.}~\bibnamefont
  {Eijima}}, \bibinfo {author} {\bibfnamefont {M.}~\bibnamefont
  {Shaposhnikov}}, \ and\ \bibinfo {author} {\bibfnamefont {I.}~\bibnamefont
  {Timiryasov}},\ }\href {\doibase 10.1088/1475-7516/2017/11/030} {\bibfield
  {journal} {\bibinfo  {journal} {JCAP}\ }\textbf {\bibinfo {volume} {1711}},\
  \bibinfo {pages} {030} (\bibinfo {year} {2017})},\ \Eprint
  {http://arxiv.org/abs/1709.07834} {arXiv:1709.07834 [hep-ph]} \BibitemShut
  {NoStop}%
\bibitem [{\citenamefont {Aad}\ \emph {et~al.}(2020{\natexlab{a}})\citenamefont
  {Aad} \emph {et~al.}}]{Aad:2019vnb}%
  \BibitemOpen
  \bibfield  {author} {\bibinfo {author} {\bibfnamefont {G.}~\bibnamefont
  {Aad}} \emph {et~al.} (\bibinfo {collaboration} {ATLAS}),\ }\href {\doibase
  10.1140/epjc/s10052-019-7594-6} {\bibfield  {journal} {\bibinfo  {journal}
  {Eur. Phys. J. C}\ }\textbf {\bibinfo {volume} {80}},\ \bibinfo {pages} {123}
  (\bibinfo {year} {2020}{\natexlab{a}})},\ \Eprint
  {http://arxiv.org/abs/1908.08215} {arXiv:1908.08215 [hep-ex]} \BibitemShut
  {NoStop}%
\bibitem [{\citenamefont {Sirunyan}\ \emph {et~al.}()\citenamefont {Sirunyan}
  \emph {et~al.}}]{Sirunyan:2020eab}%
  \BibitemOpen
  \bibfield  {author} {\bibinfo {author} {\bibfnamefont {A.~M.}\ \bibnamefont
  {Sirunyan}} \emph {et~al.} (\bibinfo {collaboration} {CMS}),\ }\href@noop {}
  {\ }\Eprint {http://arxiv.org/abs/2012.08600} {arXiv:2012.08600 [hep-ex]}
  \BibitemShut {NoStop}%
\bibitem [{\citenamefont {Sirunyan}\ \emph {et~al.}(2019)\citenamefont
  {Sirunyan} \emph {et~al.}}]{CMS:2018eqb}%
  \BibitemOpen
  \bibfield  {author} {\bibinfo {author} {\bibfnamefont {A.~M.}\ \bibnamefont
  {Sirunyan}} \emph {et~al.} (\bibinfo {collaboration} {CMS}),\ }\href
  {\doibase 10.1016/j.physletb.2019.01.005} {\bibfield  {journal} {\bibinfo
  {journal} {Phys. Lett. B}\ }\textbf {\bibinfo {volume} {790}},\ \bibinfo
  {pages} {140} (\bibinfo {year} {2019})},\ \Eprint
  {http://arxiv.org/abs/1806.05264} {arXiv:1806.05264 [hep-ex]} \BibitemShut
  {NoStop}%
\bibitem [{\citenamefont {Aad}\ \emph {et~al.}(2014)\citenamefont {Aad} \emph
  {et~al.}}]{ATLAS:2014zve}%
  \BibitemOpen
  \bibfield  {author} {\bibinfo {author} {\bibfnamefont {G.}~\bibnamefont
  {Aad}} \emph {et~al.} (\bibinfo {collaboration} {ATLAS}),\ }\href {\doibase
  10.1007/JHEP05(2014)071} {\bibfield  {journal} {\bibinfo  {journal} {JHEP}\
  }\textbf {\bibinfo {volume} {05}},\ \bibinfo {pages} {071} (\bibinfo {year}
  {2014})},\ \Eprint {http://arxiv.org/abs/1403.5294} {arXiv:1403.5294
  [hep-ex]} \BibitemShut {NoStop}%
\bibitem [{\citenamefont {Aad}\ \emph {et~al.}()\citenamefont {Aad} \emph
  {et~al.}}]{Aad:2020bay}%
  \BibitemOpen
  \bibfield  {author} {\bibinfo {author} {\bibfnamefont {G.}~\bibnamefont
  {Aad}} \emph {et~al.} (\bibinfo {collaboration} {ATLAS}),\ }\href@noop {} {\
  }\Eprint {http://arxiv.org/abs/2011.07812} {arXiv:2011.07812 [hep-ex]}
  \BibitemShut {NoStop}%
\bibitem [{CMS(2016)}]{CMS-PAS-EXO-16-036}%
  \BibitemOpen
  \href {https://cds.cern.ch/record/2205281} {\emph {\bibinfo {title} {{Search
  for heavy stable charged particles with $12.9~\mathrm{fb}^{-1}$ of 2016
  data}}}},\ \bibinfo {type} {Tech. Rep.}\ (\bibinfo  {institution} {CERN},\
  \bibinfo {address} {Geneva},\ \bibinfo {year} {2016})\BibitemShut {NoStop}%
\bibitem [{\citenamefont {Co}\ \emph {et~al.}(2015)\citenamefont {Co},
  \citenamefont {D'Eramo}, \citenamefont {Hall},\ and\ \citenamefont
  {Pappadopulo}}]{Co:2015pka}%
  \BibitemOpen
  \bibfield  {author} {\bibinfo {author} {\bibfnamefont {R.~T.}\ \bibnamefont
  {Co}}, \bibinfo {author} {\bibfnamefont {F.}~\bibnamefont {D'Eramo}},
  \bibinfo {author} {\bibfnamefont {L.~J.}\ \bibnamefont {Hall}}, \ and\
  \bibinfo {author} {\bibfnamefont {D.}~\bibnamefont {Pappadopulo}},\ }\href
  {\doibase 10.1088/1475-7516/2015/12/024} {\bibfield  {journal} {\bibinfo
  {journal} {JCAP}\ }\textbf {\bibinfo {volume} {1512}},\ \bibinfo {pages}
  {024} (\bibinfo {year} {2015})},\ \Eprint {http://arxiv.org/abs/1506.07532}
  {arXiv:1506.07532 [hep-ph]} \BibitemShut {NoStop}%
\bibitem [{\citenamefont {Bélanger}\ \emph {et~al.}(2019)\citenamefont
  {Bélanger} \emph {et~al.}}]{Belanger:2018sti}%
  \BibitemOpen
  \bibfield  {author} {\bibinfo {author} {\bibfnamefont {G.}~\bibnamefont
  {Bélanger}} \emph {et~al.},\ }\href {\doibase 10.1007/JHEP02(2019)186}
  {\bibfield  {journal} {\bibinfo  {journal} {JHEP}\ }\textbf {\bibinfo
  {volume} {02}},\ \bibinfo {pages} {186} (\bibinfo {year} {2019})},\ \Eprint
  {http://arxiv.org/abs/1811.05478} {arXiv:1811.05478 [hep-ph]} \BibitemShut
  {NoStop}%
\bibitem [{\citenamefont {Junius}\ \emph {et~al.}(2019)\citenamefont {Junius},
  \citenamefont {Lopez-Honorez},\ and\ \citenamefont {Mariotti}}]{Junius_2019}%
  \BibitemOpen
  \bibfield  {author} {\bibinfo {author} {\bibfnamefont {S.}~\bibnamefont
  {Junius}}, \bibinfo {author} {\bibfnamefont {L.}~\bibnamefont
  {Lopez-Honorez}}, \ and\ \bibinfo {author} {\bibfnamefont {A.}~\bibnamefont
  {Mariotti}},\ }\href {\doibase 10.1007/jhep07(2019)136} {\bibfield  {journal}
  {\bibinfo  {journal} {Journal of High Energy Physics}\ }\textbf {\bibinfo
  {volume} {2019}} (\bibinfo {year} {2019}),\
  10.1007/jhep07(2019)136}\BibitemShut {NoStop}%
\bibitem [{\citenamefont {Aad}\ \emph {et~al.}(2020{\natexlab{b}})\citenamefont
  {Aad} \emph {et~al.}}]{Aad:2019byo}%
  \BibitemOpen
  \bibfield  {author} {\bibinfo {author} {\bibfnamefont {G.}~\bibnamefont
  {Aad}} \emph {et~al.} (\bibinfo {collaboration} {ATLAS}),\ }\href {\doibase
  10.1103/PhysRevD.101.032009} {\bibfield  {journal} {\bibinfo  {journal}
  {Phys. Rev. D}\ }\textbf {\bibinfo {volume} {101}},\ \bibinfo {pages}
  {032009} (\bibinfo {year} {2020}{\natexlab{b}})},\ \Eprint
  {http://arxiv.org/abs/1911.06660} {arXiv:1911.06660 [hep-ex]} \BibitemShut
  {NoStop}%
\bibitem [{sus()}]{susyxsecwg}%
  \BibitemOpen
  \href@noop {} {}\bibinfo {howpublished}
  {\url{https://twiki.cern.ch/twiki/bin/view/LHCPhysics/SUSYCrossSections13TeVslepslep}},\
  \bibinfo {note} {{L}HC SUSY Cross Section Working Group}\BibitemShut
  {NoStop}%
\bibitem [{\citenamefont {Bozzi}\ \emph {et~al.}(2007)\citenamefont {Bozzi},
  \citenamefont {Fuks},\ and\ \citenamefont {Klasen}}]{Bozzi:2007qr}%
  \BibitemOpen
  \bibfield  {author} {\bibinfo {author} {\bibfnamefont {G.}~\bibnamefont
  {Bozzi}}, \bibinfo {author} {\bibfnamefont {B.}~\bibnamefont {Fuks}}, \ and\
  \bibinfo {author} {\bibfnamefont {M.}~\bibnamefont {Klasen}},\ }\href
  {\doibase 10.1016/j.nuclphysb.2007.03.052} {\bibfield  {journal} {\bibinfo
  {journal} {Nucl. Phys. B}\ }\textbf {\bibinfo {volume} {777}},\ \bibinfo
  {pages} {157} (\bibinfo {year} {2007})},\ \Eprint
  {http://arxiv.org/abs/hep-ph/0701202} {arXiv:hep-ph/0701202} \BibitemShut
  {NoStop}%
\bibitem [{\citenamefont {Fuks}\ \emph {et~al.}(2013)\citenamefont {Fuks},
  \citenamefont {Klasen}, \citenamefont {Lamprea},\ and\ \citenamefont
  {Rothering}}]{Fuks:2013vua}%
  \BibitemOpen
  \bibfield  {author} {\bibinfo {author} {\bibfnamefont {B.}~\bibnamefont
  {Fuks}}, \bibinfo {author} {\bibfnamefont {M.}~\bibnamefont {Klasen}},
  \bibinfo {author} {\bibfnamefont {D.~R.}\ \bibnamefont {Lamprea}}, \ and\
  \bibinfo {author} {\bibfnamefont {M.}~\bibnamefont {Rothering}},\ }\href
  {\doibase 10.1140/epjc/s10052-013-2480-0} {\bibfield  {journal} {\bibinfo
  {journal} {Eur. Phys. J. C}\ }\textbf {\bibinfo {volume} {73}},\ \bibinfo
  {pages} {2480} (\bibinfo {year} {2013})},\ \Eprint
  {http://arxiv.org/abs/1304.0790} {arXiv:1304.0790 [hep-ph]} \BibitemShut
  {NoStop}%
\bibitem [{\citenamefont {Fuks}\ \emph {et~al.}(2014)\citenamefont {Fuks},
  \citenamefont {Klasen}, \citenamefont {Lamprea},\ and\ \citenamefont
  {Rothering}}]{Fuks:2013lya}%
  \BibitemOpen
  \bibfield  {author} {\bibinfo {author} {\bibfnamefont {B.}~\bibnamefont
  {Fuks}}, \bibinfo {author} {\bibfnamefont {M.}~\bibnamefont {Klasen}},
  \bibinfo {author} {\bibfnamefont {D.~R.}\ \bibnamefont {Lamprea}}, \ and\
  \bibinfo {author} {\bibfnamefont {M.}~\bibnamefont {Rothering}},\ }\href
  {\doibase 10.1007/JHEP01(2014)168} {\bibfield  {journal} {\bibinfo  {journal}
  {JHEP}\ }\textbf {\bibinfo {volume} {01}},\ \bibinfo {pages} {168} (\bibinfo
  {year} {2014})},\ \Eprint {http://arxiv.org/abs/1310.2621} {arXiv:1310.2621
  [hep-ph]} \BibitemShut {NoStop}%
\bibitem [{\citenamefont {Fiaschi}\ and\ \citenamefont
  {Klasen}(2018)}]{Fiaschi:2018xdm}%
  \BibitemOpen
  \bibfield  {author} {\bibinfo {author} {\bibfnamefont {J.}~\bibnamefont
  {Fiaschi}}\ and\ \bibinfo {author} {\bibfnamefont {M.}~\bibnamefont
  {Klasen}},\ }\href {\doibase 10.1007/JHEP03(2018)094} {\bibfield  {journal}
  {\bibinfo  {journal} {JHEP}\ }\textbf {\bibinfo {volume} {03}},\ \bibinfo
  {pages} {094} (\bibinfo {year} {2018})},\ \Eprint
  {http://arxiv.org/abs/1801.10357} {arXiv:1801.10357 [hep-ph]} \BibitemShut
  {NoStop}%
\bibitem [{\citenamefont {Beenakker}\ \emph {et~al.}(1999)\citenamefont
  {Beenakker}, \citenamefont {Klasen}, \citenamefont {Kramer}, \citenamefont
  {Plehn}, \citenamefont {Spira},\ and\ \citenamefont
  {Zerwas}}]{Beenakker:1999xh}%
  \BibitemOpen
  \bibfield  {author} {\bibinfo {author} {\bibfnamefont {W.}~\bibnamefont
  {Beenakker}}, \bibinfo {author} {\bibfnamefont {M.}~\bibnamefont {Klasen}},
  \bibinfo {author} {\bibfnamefont {M.}~\bibnamefont {Kramer}}, \bibinfo
  {author} {\bibfnamefont {T.}~\bibnamefont {Plehn}}, \bibinfo {author}
  {\bibfnamefont {M.}~\bibnamefont {Spira}}, \ and\ \bibinfo {author}
  {\bibfnamefont {P.~M.}\ \bibnamefont {Zerwas}},\ }\href {\doibase
  10.1103/PhysRevLett.100.029901} {\bibfield  {journal} {\bibinfo  {journal}
  {Phys. Rev. Lett.}\ }\textbf {\bibinfo {volume} {83}},\ \bibinfo {pages}
  {3780} (\bibinfo {year} {1999})},\ \bibinfo {note} {[Erratum: Phys.Rev.Lett.
  100, 029901 (2008)]},\ \Eprint {http://arxiv.org/abs/hep-ph/9906298}
  {arXiv:hep-ph/9906298} \BibitemShut {NoStop}%
\bibitem [{\citenamefont {Evans}\ and\ \citenamefont
  {Shelton}(2016)}]{Evans:2016zau}%
  \BibitemOpen
  \bibfield  {author} {\bibinfo {author} {\bibfnamefont {J.~A.}\ \bibnamefont
  {Evans}}\ and\ \bibinfo {author} {\bibfnamefont {J.}~\bibnamefont
  {Shelton}},\ }\href {\doibase 10.1007/JHEP04(2016)056} {\bibfield  {journal}
  {\bibinfo  {journal} {JHEP}\ }\textbf {\bibinfo {volume} {04}},\ \bibinfo
  {pages} {056} (\bibinfo {year} {2016})},\ \Eprint
  {http://arxiv.org/abs/1601.01326} {arXiv:1601.01326 [hep-ph]} \BibitemShut
  {NoStop}%
\bibitem [{\citenamefont {Drewes}\ \emph {et~al.}(2017)\citenamefont {Drewes}
  \emph {et~al.}}]{Drewes:2016upu}%
  \BibitemOpen
  \bibfield  {author} {\bibinfo {author} {\bibfnamefont {M.}~\bibnamefont
  {Drewes}} \emph {et~al.},\ }\href {\doibase 10.1088/1475-7516/2017/01/025}
  {\bibfield  {journal} {\bibinfo  {journal} {JCAP}\ }\textbf {\bibinfo
  {volume} {01}},\ \bibinfo {pages} {025} (\bibinfo {year} {2017})},\ \Eprint
  {http://arxiv.org/abs/1602.04816} {arXiv:1602.04816 [hep-ph]} \BibitemShut
  {NoStop}%
\bibitem [{\citenamefont {Shrock}(1974)}]{Shrock:1974nd}%
  \BibitemOpen
  \bibfield  {author} {\bibinfo {author} {\bibfnamefont {R.}~\bibnamefont
  {Shrock}},\ }\href {\doibase 10.1103/PhysRevD.9.743} {\bibfield  {journal}
  {\bibinfo  {journal} {Phys. Rev. D}\ }\textbf {\bibinfo {volume} {9}},\
  \bibinfo {pages} {743} (\bibinfo {year} {1974})}\BibitemShut {NoStop}%
\bibitem [{\citenamefont {Petcov}(1977)}]{Petcov:1976ff}%
  \BibitemOpen
  \bibfield  {author} {\bibinfo {author} {\bibfnamefont {S.~T.}\ \bibnamefont
  {Petcov}},\ }\href@noop {} {\bibfield  {journal} {\bibinfo  {journal} {Sov.
  J. Nucl. Phys.}\ }\textbf {\bibinfo {volume} {25}},\ \bibinfo {pages} {340}
  (\bibinfo {year} {1977})},\ \bibinfo {note} {[Erratum: Sov.J.Nucl.Phys. 25,
  698 (1977), Erratum: Yad.Fiz. 25, 1336 (1977)]}\BibitemShut {NoStop}%
\bibitem [{\citenamefont {Lee}\ and\ \citenamefont
  {Shrock}(1977)}]{Lee:1977tib}%
  \BibitemOpen
  \bibfield  {author} {\bibinfo {author} {\bibfnamefont {B.~W.}\ \bibnamefont
  {Lee}}\ and\ \bibinfo {author} {\bibfnamefont {R.~E.}\ \bibnamefont
  {Shrock}},\ }\href {\doibase 10.1103/PhysRevD.16.1444} {\bibfield  {journal}
  {\bibinfo  {journal} {Phys. Rev. D}\ }\textbf {\bibinfo {volume} {16}},\
  \bibinfo {pages} {1444} (\bibinfo {year} {1977})}\BibitemShut {NoStop}%
\bibitem [{\citenamefont {Marciano}\ and\ \citenamefont
  {Sanda}(1977)}]{Marciano:1977wx}%
  \BibitemOpen
  \bibfield  {author} {\bibinfo {author} {\bibfnamefont {W.~J.}\ \bibnamefont
  {Marciano}}\ and\ \bibinfo {author} {\bibfnamefont {A.~I.}\ \bibnamefont
  {Sanda}},\ }\href {\doibase 10.1016/0370-2693(77)90377-X} {\bibfield
  {journal} {\bibinfo  {journal} {Phys. Lett. B}\ }\textbf {\bibinfo {volume}
  {67}},\ \bibinfo {pages} {303} (\bibinfo {year} {1977})}\BibitemShut
  {NoStop}%
\bibitem [{\citenamefont {Pal}\ and\ \citenamefont
  {Wolfenstein}(1982)}]{Pal:1981rm}%
  \BibitemOpen
  \bibfield  {author} {\bibinfo {author} {\bibfnamefont {P.~B.}\ \bibnamefont
  {Pal}}\ and\ \bibinfo {author} {\bibfnamefont {L.}~\bibnamefont
  {Wolfenstein}},\ }\href {\doibase 10.1103/PhysRevD.25.766} {\bibfield
  {journal} {\bibinfo  {journal} {Phys. Rev. D}\ }\textbf {\bibinfo {volume}
  {25}},\ \bibinfo {pages} {766} (\bibinfo {year} {1982})}\BibitemShut
  {NoStop}%
\bibitem [{\citenamefont {Shrock}(1982)}]{Shrock:1982sc}%
  \BibitemOpen
  \bibfield  {author} {\bibinfo {author} {\bibfnamefont {R.~E.}\ \bibnamefont
  {Shrock}},\ }\href {\doibase 10.1016/0550-3213(82)90273-5} {\bibfield
  {journal} {\bibinfo  {journal} {Nucl. Phys. B}\ }\textbf {\bibinfo {volume}
  {206}},\ \bibinfo {pages} {359} (\bibinfo {year} {1982})}\BibitemShut
  {NoStop}%
\bibitem [{\citenamefont {Barger}\ \emph {et~al.}(1995)\citenamefont {Barger},
  \citenamefont {Phillips},\ and\ \citenamefont {Sarkar}}]{Barger:1995ty}%
  \BibitemOpen
  \bibfield  {author} {\bibinfo {author} {\bibfnamefont {V.~D.}\ \bibnamefont
  {Barger}}, \bibinfo {author} {\bibfnamefont {R.~J.~N.}\ \bibnamefont
  {Phillips}}, \ and\ \bibinfo {author} {\bibfnamefont {S.}~\bibnamefont
  {Sarkar}},\ }\href {\doibase 10.1016/0370-2693(95)00486-5} {\bibfield
  {journal} {\bibinfo  {journal} {Phys. Lett. B}\ }\textbf {\bibinfo {volume}
  {352}},\ \bibinfo {pages} {365} (\bibinfo {year} {1995})},\ \bibinfo {note}
  {[Erratum: Phys.Lett.B 356, 617--617 (1995)]},\ \Eprint
  {http://arxiv.org/abs/hep-ph/9503295} {arXiv:hep-ph/9503295} \BibitemShut
  {NoStop}%
\bibitem [{\citenamefont {Perez}\ \emph {et~al.}(2017)\citenamefont {Perez},
  \citenamefont {Ng}, \citenamefont {Beacom}, \citenamefont {Hersh},
  \citenamefont {Horiuchi},\ and\ \citenamefont {Krivonos}}]{Perez:2016tcq}%
  \BibitemOpen
  \bibfield  {author} {\bibinfo {author} {\bibfnamefont {K.}~\bibnamefont
  {Perez}}, \bibinfo {author} {\bibfnamefont {K.~C.~Y.}\ \bibnamefont {Ng}},
  \bibinfo {author} {\bibfnamefont {J.~F.}\ \bibnamefont {Beacom}}, \bibinfo
  {author} {\bibfnamefont {C.}~\bibnamefont {Hersh}}, \bibinfo {author}
  {\bibfnamefont {S.}~\bibnamefont {Horiuchi}}, \ and\ \bibinfo {author}
  {\bibfnamefont {R.}~\bibnamefont {Krivonos}},\ }\href {\doibase
  10.1103/PhysRevD.95.123002} {\bibfield  {journal} {\bibinfo  {journal} {Phys.
  Rev. D}\ }\textbf {\bibinfo {volume} {95}},\ \bibinfo {pages} {123002}
  (\bibinfo {year} {2017})},\ \Eprint {http://arxiv.org/abs/1609.00667}
  {arXiv:1609.00667 [astro-ph.HE]} \BibitemShut {NoStop}%
\bibitem [{\citenamefont {Neronov}\ \emph {et~al.}(2016)\citenamefont
  {Neronov}, \citenamefont {Malyshev},\ and\ \citenamefont
  {Eckert}}]{Neronov:2016wdd}%
  \BibitemOpen
  \bibfield  {author} {\bibinfo {author} {\bibfnamefont {A.}~\bibnamefont
  {Neronov}}, \bibinfo {author} {\bibfnamefont {D.}~\bibnamefont {Malyshev}}, \
  and\ \bibinfo {author} {\bibfnamefont {D.}~\bibnamefont {Eckert}},\ }\href
  {\doibase 10.1103/PhysRevD.94.123504} {\bibfield  {journal} {\bibinfo
  {journal} {Phys. Rev. D}\ }\textbf {\bibinfo {volume} {94}},\ \bibinfo
  {pages} {123504} (\bibinfo {year} {2016})},\ \Eprint
  {http://arxiv.org/abs/1607.07328} {arXiv:1607.07328 [astro-ph.HE]}
  \BibitemShut {NoStop}%
\bibitem [{\citenamefont {Ng}\ \emph {et~al.}(2019)\citenamefont {Ng},
  \citenamefont {Roach}, \citenamefont {Perez}, \citenamefont {Beacom},
  \citenamefont {Horiuchi}, \citenamefont {Krivonos},\ and\ \citenamefont
  {Wik}}]{Ng:2019gch}%
  \BibitemOpen
  \bibfield  {author} {\bibinfo {author} {\bibfnamefont {K.~C.~Y.}\
  \bibnamefont {Ng}}, \bibinfo {author} {\bibfnamefont {B.~M.}\ \bibnamefont
  {Roach}}, \bibinfo {author} {\bibfnamefont {K.}~\bibnamefont {Perez}},
  \bibinfo {author} {\bibfnamefont {J.~F.}\ \bibnamefont {Beacom}}, \bibinfo
  {author} {\bibfnamefont {S.}~\bibnamefont {Horiuchi}}, \bibinfo {author}
  {\bibfnamefont {R.}~\bibnamefont {Krivonos}}, \ and\ \bibinfo {author}
  {\bibfnamefont {D.~R.}\ \bibnamefont {Wik}},\ }\href {\doibase
  10.1103/PhysRevD.99.083005} {\bibfield  {journal} {\bibinfo  {journal} {Phys.
  Rev. D}\ }\textbf {\bibinfo {volume} {99}},\ \bibinfo {pages} {083005}
  (\bibinfo {year} {2019})},\ \Eprint {http://arxiv.org/abs/1901.01262}
  {arXiv:1901.01262 [astro-ph.HE]} \BibitemShut {NoStop}%
\bibitem [{\citenamefont {Roach}\ \emph {et~al.}(2020)\citenamefont {Roach},
  \citenamefont {Ng}, \citenamefont {Perez}, \citenamefont {Beacom},
  \citenamefont {Horiuchi}, \citenamefont {Krivonos},\ and\ \citenamefont
  {Wik}}]{Roach:2019ctw}%
  \BibitemOpen
  \bibfield  {author} {\bibinfo {author} {\bibfnamefont {B.~M.}\ \bibnamefont
  {Roach}}, \bibinfo {author} {\bibfnamefont {K.~C.~Y.}\ \bibnamefont {Ng}},
  \bibinfo {author} {\bibfnamefont {K.}~\bibnamefont {Perez}}, \bibinfo
  {author} {\bibfnamefont {J.~F.}\ \bibnamefont {Beacom}}, \bibinfo {author}
  {\bibfnamefont {S.}~\bibnamefont {Horiuchi}}, \bibinfo {author}
  {\bibfnamefont {R.}~\bibnamefont {Krivonos}}, \ and\ \bibinfo {author}
  {\bibfnamefont {D.~R.}\ \bibnamefont {Wik}},\ }\href {\doibase
  10.1103/PhysRevD.101.103011} {\bibfield  {journal} {\bibinfo  {journal}
  {Phys. Rev. D}\ }\textbf {\bibinfo {volume} {101}},\ \bibinfo {pages}
  {103011} (\bibinfo {year} {2020})},\ \Eprint
  {http://arxiv.org/abs/1908.09037} {arXiv:1908.09037 [astro-ph.HE]}
  \BibitemShut {NoStop}%
\bibitem [{\citenamefont {Laha}\ \emph {et~al.}(2020)\citenamefont {Laha},
  \citenamefont {Mu\~noz},\ and\ \citenamefont {Slatyer}}]{Laha:2020ivk}%
  \BibitemOpen
  \bibfield  {author} {\bibinfo {author} {\bibfnamefont {R.}~\bibnamefont
  {Laha}}, \bibinfo {author} {\bibfnamefont {J.~B.}\ \bibnamefont {Mu\~noz}}, \
  and\ \bibinfo {author} {\bibfnamefont {T.~R.}\ \bibnamefont {Slatyer}},\
  }\href {\doibase 10.1103/PhysRevD.101.123514} {\bibfield  {journal} {\bibinfo
   {journal} {Phys. Rev. D}\ }\textbf {\bibinfo {volume} {101}},\ \bibinfo
  {pages} {123514} (\bibinfo {year} {2020})},\ \Eprint
  {http://arxiv.org/abs/2004.00627} {arXiv:2004.00627 [astro-ph.CO]}
  \BibitemShut {NoStop}%
\bibitem [{\citenamefont {Suliga}\ \emph {et~al.}(2020)\citenamefont {Suliga},
  \citenamefont {Tamborra},\ and\ \citenamefont {Wu}}]{Suliga:2020vpz}%
  \BibitemOpen
  \bibfield  {author} {\bibinfo {author} {\bibfnamefont {A.~M.}\ \bibnamefont
  {Suliga}}, \bibinfo {author} {\bibfnamefont {I.}~\bibnamefont {Tamborra}}, \
  and\ \bibinfo {author} {\bibfnamefont {M.-R.}\ \bibnamefont {Wu}},\ }\href
  {\doibase 10.1088/1475-7516/2020/08/018} {\bibfield  {journal} {\bibinfo
  {journal} {JCAP}\ }\textbf {\bibinfo {volume} {08}},\ \bibinfo {pages} {018}
  (\bibinfo {year} {2020})},\ \Eprint {http://arxiv.org/abs/2004.11389}
  {arXiv:2004.11389 [astro-ph.HE]} \BibitemShut {NoStop}%
\bibitem [{\citenamefont {Barbier}\ \emph {et~al.}(2005)\citenamefont {Barbier}
  \emph {et~al.}}]{Barbier:2004ez}%
  \BibitemOpen
  \bibfield  {author} {\bibinfo {author} {\bibfnamefont {R.}~\bibnamefont
  {Barbier}} \emph {et~al.},\ }\href {\doibase 10.1016/j.physrep.2005.08.006}
  {\bibfield  {journal} {\bibinfo  {journal} {Phys. Rept.}\ }\textbf {\bibinfo
  {volume} {420}},\ \bibinfo {pages} {1} (\bibinfo {year} {2005})},\ \Eprint
  {http://arxiv.org/abs/hep-ph/0406039} {arXiv:hep-ph/0406039} \BibitemShut
  {NoStop}%
\bibitem [{\citenamefont {Ledroit}\ and\ \citenamefont
  {Sajot}(1998)}]{RPV1998}%
  \BibitemOpen
  \bibfield  {author} {\bibinfo {author} {\bibfnamefont {F.}~\bibnamefont
  {Ledroit}}\ and\ \bibinfo {author} {\bibfnamefont {G.}~\bibnamefont
  {Sajot}},\ }\href@noop {} {\enquote {\bibinfo {title} {{Indirect limits on
  SUSY Rp violating couplings lambda and lambda’}},}\ } (\bibinfo {year}
  {1998}),\ \bibinfo {note}
  {\url{http://hal.in2p3.fr/in2p3-00362621}}\BibitemShut {NoStop}%
\bibitem [{\citenamefont {Aubert}\ \emph {et~al.}(2010)\citenamefont {Aubert},
  \citenamefont {Karyotakis}, \citenamefont {Lees}, \citenamefont {Poireau},
  \citenamefont {Prencipe}, \citenamefont {Prudent}, \citenamefont {Tisserand},
  \citenamefont {Garra~Tico}, \citenamefont {Grauges}, \citenamefont
  {Martinelli},\ and\ \citenamefont {et~al.}}]{BaBarLFU2010}%
  \BibitemOpen
  \bibfield  {author} {\bibinfo {author} {\bibfnamefont {B.}~\bibnamefont
  {Aubert}}, \bibinfo {author} {\bibfnamefont {Y.}~\bibnamefont {Karyotakis}},
  \bibinfo {author} {\bibfnamefont {J.~P.}\ \bibnamefont {Lees}}, \bibinfo
  {author} {\bibfnamefont {V.}~\bibnamefont {Poireau}}, \bibinfo {author}
  {\bibfnamefont {E.}~\bibnamefont {Prencipe}}, \bibinfo {author}
  {\bibfnamefont {X.}~\bibnamefont {Prudent}}, \bibinfo {author} {\bibfnamefont
  {V.}~\bibnamefont {Tisserand}}, \bibinfo {author} {\bibfnamefont
  {J.}~\bibnamefont {Garra~Tico}}, \bibinfo {author} {\bibfnamefont
  {E.}~\bibnamefont {Grauges}}, \bibinfo {author} {\bibfnamefont
  {M.}~\bibnamefont {Martinelli}}, \ and\ \bibinfo {author} {\bibnamefont
  {et~al.}},\ }\href {\doibase 10.1103/physrevlett.105.051602} {\bibfield
  {journal} {\bibinfo  {journal} {Physical Review Letters}\ }\textbf {\bibinfo
  {volume} {105}} (\bibinfo {year} {2010}),\
  10.1103/physrevlett.105.051602}\BibitemShut {NoStop}%
\bibitem [{\citenamefont {Baldini}\ \emph {et~al.}(2016)\citenamefont {Baldini}
  \emph {et~al.}}]{themegcollaboration2016search}%
  \BibitemOpen
  \bibfield  {author} {\bibinfo {author} {\bibfnamefont {A.~M.}\ \bibnamefont
  {Baldini}} \emph {et~al.} (\bibinfo {collaboration} {MEG}),\ }\href {\doibase
  10.1140/epjc/s10052-016-4271-x} {\bibfield  {journal} {\bibinfo  {journal}
  {Eur. Phys. J. C}\ }\textbf {\bibinfo {volume} {76}},\ \bibinfo {pages} {434}
  (\bibinfo {year} {2016})},\ \Eprint {http://arxiv.org/abs/1605.05081}
  {arXiv:1605.05081 [hep-ex]} \BibitemShut {NoStop}%
\bibitem [{\citenamefont {de~Gouvea}\ \emph {et~al.}(2001)\citenamefont
  {de~Gouvea}, \citenamefont {Lola},\ and\ \citenamefont
  {Tobe}}]{deGouvea:2000cf}%
  \BibitemOpen
  \bibfield  {author} {\bibinfo {author} {\bibfnamefont {A.}~\bibnamefont
  {de~Gouvea}}, \bibinfo {author} {\bibfnamefont {S.}~\bibnamefont {Lola}}, \
  and\ \bibinfo {author} {\bibfnamefont {K.}~\bibnamefont {Tobe}},\ }\href
  {\doibase 10.1103/PhysRevD.63.035004} {\bibfield  {journal} {\bibinfo
  {journal} {Phys. Rev. D}\ }\textbf {\bibinfo {volume} {63}},\ \bibinfo
  {pages} {035004} (\bibinfo {year} {2001})},\ \Eprint
  {http://arxiv.org/abs/hep-ph/0008085} {arXiv:hep-ph/0008085} \BibitemShut
  {NoStop}%
\bibitem [{\citenamefont {Suzuki}\ \emph {et~al.}(1987)\citenamefont {Suzuki},
  \citenamefont {Measday},\ and\ \citenamefont {Roalsvig}}]{Suzuki:1987jf}%
  \BibitemOpen
  \bibfield  {author} {\bibinfo {author} {\bibfnamefont {T.}~\bibnamefont
  {Suzuki}}, \bibinfo {author} {\bibfnamefont {D.~F.}\ \bibnamefont {Measday}},
  \ and\ \bibinfo {author} {\bibfnamefont {J.~P.}\ \bibnamefont {Roalsvig}},\
  }\href {\doibase 10.1103/PhysRevC.35.2212} {\bibfield  {journal} {\bibinfo
  {journal} {Phys. Rev. C}\ }\textbf {\bibinfo {volume} {35}},\ \bibinfo
  {pages} {2212} (\bibinfo {year} {1987})}\BibitemShut {NoStop}%
\bibitem [{\citenamefont {Chiang}\ \emph {et~al.}(1993)\citenamefont {Chiang},
  \citenamefont {Oset}, \citenamefont {Kosmas}, \citenamefont {Faessler},\ and\
  \citenamefont {Vergados}}]{Chiang:1993xz}%
  \BibitemOpen
  \bibfield  {author} {\bibinfo {author} {\bibfnamefont {H.~C.}\ \bibnamefont
  {Chiang}}, \bibinfo {author} {\bibfnamefont {E.}~\bibnamefont {Oset}},
  \bibinfo {author} {\bibfnamefont {T.~S.}\ \bibnamefont {Kosmas}}, \bibinfo
  {author} {\bibfnamefont {A.}~\bibnamefont {Faessler}}, \ and\ \bibinfo
  {author} {\bibfnamefont {J.~D.}\ \bibnamefont {Vergados}},\ }\href {\doibase
  10.1016/0375-9474(93)90259-Z} {\bibfield  {journal} {\bibinfo  {journal}
  {Nucl. Phys. A}\ }\textbf {\bibinfo {volume} {559}},\ \bibinfo {pages} {526}
  (\bibinfo {year} {1993})}\BibitemShut {NoStop}%
\bibitem [{\citenamefont {Bertl}\ \emph {et~al.}(2006)\citenamefont {Bertl}
  \emph {et~al.}}]{Bertl:2006up}%
  \BibitemOpen
  \bibfield  {author} {\bibinfo {author} {\bibfnamefont {W.~H.}\ \bibnamefont
  {Bertl}} \emph {et~al.} (\bibinfo {collaboration} {SINDRUM II}),\ }\href
  {\doibase 10.1140/epjc/s2006-02582-x} {\bibfield  {journal} {\bibinfo
  {journal} {Eur. Phys. J. C}\ }\textbf {\bibinfo {volume} {47}},\ \bibinfo
  {pages} {337} (\bibinfo {year} {2006})}\BibitemShut {NoStop}%
\bibitem [{\citenamefont {Bennett}\ \emph {et~al.}(2006)\citenamefont
  {Bennett}, \citenamefont {Bousquet}, \citenamefont {Brown}, \citenamefont
  {Bunce}, \citenamefont {Carey}, \citenamefont {Cushman}, \citenamefont
  {Danby}, \citenamefont {Debevec}, \citenamefont {Deile}, \citenamefont
  {Deng},\ and\ \citenamefont {et~al.}}]{Muong-2:2006}%
  \BibitemOpen
  \bibfield  {author} {\bibinfo {author} {\bibfnamefont {G.~W.}\ \bibnamefont
  {Bennett}}, \bibinfo {author} {\bibfnamefont {B.}~\bibnamefont {Bousquet}},
  \bibinfo {author} {\bibfnamefont {H.~N.}\ \bibnamefont {Brown}}, \bibinfo
  {author} {\bibfnamefont {G.}~\bibnamefont {Bunce}}, \bibinfo {author}
  {\bibfnamefont {R.~M.}\ \bibnamefont {Carey}}, \bibinfo {author}
  {\bibfnamefont {P.}~\bibnamefont {Cushman}}, \bibinfo {author} {\bibfnamefont
  {G.~T.}\ \bibnamefont {Danby}}, \bibinfo {author} {\bibfnamefont {P.~T.}\
  \bibnamefont {Debevec}}, \bibinfo {author} {\bibfnamefont {M.}~\bibnamefont
  {Deile}}, \bibinfo {author} {\bibfnamefont {H.}~\bibnamefont {Deng}}, \ and\
  \bibinfo {author} {\bibnamefont {et~al.}},\ }\href {\doibase
  10.1103/physrevd.73.072003} {\bibfield  {journal} {\bibinfo  {journal}
  {Physical Review D}\ }\textbf {\bibinfo {volume} {73}} (\bibinfo {year}
  {2006}),\ 10.1103/physrevd.73.072003}\BibitemShut {NoStop}%
\bibitem [{\citenamefont {Abi}\ \emph {et~al.}(2021)\citenamefont {Abi} \emph
  {et~al.}}]{Muong-2:2021ojo}%
  \BibitemOpen
  \bibfield  {author} {\bibinfo {author} {\bibfnamefont {B.}~\bibnamefont
  {Abi}} \emph {et~al.} (\bibinfo {collaboration} {Muon g-2}),\ }\href
  {\doibase 10.1103/PhysRevLett.126.141801} {\bibfield  {journal} {\bibinfo
  {journal} {Phys. Rev. Lett.}\ }\textbf {\bibinfo {volume} {126}},\ \bibinfo
  {pages} {141801} (\bibinfo {year} {2021})},\ \Eprint
  {http://arxiv.org/abs/2104.03281} {arXiv:2104.03281 [hep-ex]} \BibitemShut
  {NoStop}%
\bibitem [{\citenamefont {Aoyama}\ \emph {et~al.}(2020)\citenamefont {Aoyama}
  \emph {et~al.}}]{Aoyama:2020ynm}%
  \BibitemOpen
  \bibfield  {author} {\bibinfo {author} {\bibfnamefont {T.}~\bibnamefont
  {Aoyama}} \emph {et~al.},\ }\href {\doibase 10.1016/j.physrep.2020.07.006}
  {\bibfield  {journal} {\bibinfo  {journal} {Phys. Rept.}\ }\textbf {\bibinfo
  {volume} {887}},\ \bibinfo {pages} {1} (\bibinfo {year} {2020})},\ \Eprint
  {http://arxiv.org/abs/2006.04822} {arXiv:2006.04822 [hep-ph]} \BibitemShut
  {NoStop}%
\bibitem [{\citenamefont {Kim}\ \emph {et~al.}(2001)\citenamefont {Kim},
  \citenamefont {Kyae},\ and\ \citenamefont {Lee}}]{susymuongm2:2001}%
  \BibitemOpen
  \bibfield  {author} {\bibinfo {author} {\bibfnamefont {J.~E.}\ \bibnamefont
  {Kim}}, \bibinfo {author} {\bibfnamefont {B.}~\bibnamefont {Kyae}}, \ and\
  \bibinfo {author} {\bibfnamefont {H.~M.}\ \bibnamefont {Lee}},\ }\href
  {\doibase 10.1016/s0370-2693(01)01134-0} {\bibfield  {journal} {\bibinfo
  {journal} {Physics Letters B}\ }\textbf {\bibinfo {volume} {520}},\ \bibinfo
  {pages} {298–306} (\bibinfo {year} {2001})}\BibitemShut {NoStop}%
\bibitem [{\citenamefont {Sirunyan}\ \emph {et~al.}(2021)\citenamefont
  {Sirunyan}, \citenamefont {Tumasyan}, \citenamefont {Adam}, \citenamefont
  {Andrejkovic}, \citenamefont {Bergauer}, \citenamefont {Chatterjee},
  \citenamefont {Dragicevic}, \citenamefont {Escalante Del~Valle},
  \citenamefont {Frühwirth}, \citenamefont {Jeitler},\ and\ \citenamefont
  {et~al.}}]{CMS_H_LFV_2021}%
  \BibitemOpen
  \bibfield  {author} {\bibinfo {author} {\bibfnamefont {A.}~\bibnamefont
  {Sirunyan}}, \bibinfo {author} {\bibfnamefont {A.}~\bibnamefont {Tumasyan}},
  \bibinfo {author} {\bibfnamefont {W.}~\bibnamefont {Adam}}, \bibinfo {author}
  {\bibfnamefont {J.}~\bibnamefont {Andrejkovic}}, \bibinfo {author}
  {\bibfnamefont {T.}~\bibnamefont {Bergauer}}, \bibinfo {author}
  {\bibfnamefont {S.}~\bibnamefont {Chatterjee}}, \bibinfo {author}
  {\bibfnamefont {M.}~\bibnamefont {Dragicevic}}, \bibinfo {author}
  {\bibfnamefont {A.}~\bibnamefont {Escalante Del~Valle}}, \bibinfo {author}
  {\bibfnamefont {R.}~\bibnamefont {Frühwirth}}, \bibinfo {author}
  {\bibfnamefont {M.}~\bibnamefont {Jeitler}}, \ and\ \bibinfo {author}
  {\bibnamefont {et~al.}},\ }\href {\doibase 10.1103/physrevd.104.032013}
  {\bibfield  {journal} {\bibinfo  {journal} {Physical Review D}\ }\textbf
  {\bibinfo {volume} {104}} (\bibinfo {year} {2021}),\
  10.1103/physrevd.104.032013}\BibitemShut {NoStop}%
\bibitem [{\citenamefont {Khlebnikov}\ and\ \citenamefont
  {Shaposhnikov}(1996)}]{Khlebnikov:1996vj}%
  \BibitemOpen
  \bibfield  {author} {\bibinfo {author} {\bibfnamefont {S.~Y.}\ \bibnamefont
  {Khlebnikov}}\ and\ \bibinfo {author} {\bibfnamefont {M.~E.}\ \bibnamefont
  {Shaposhnikov}},\ }\href {\doibase 10.1016/0370-2693(96)01116-1} {\bibfield
  {journal} {\bibinfo  {journal} {Phys. Lett. B}\ }\textbf {\bibinfo {volume}
  {387}},\ \bibinfo {pages} {817} (\bibinfo {year} {1996})},\ \Eprint
  {http://arxiv.org/abs/hep-ph/9607386} {arXiv:hep-ph/9607386} \BibitemShut
  {NoStop}%
\end{thebibliography}%

\end{document}